\documentclass[aps,preprint,prd,superscriptaddress,preprintnumbers,nofootinbib]{revtex4}[12pt]
\pdfoutput=1

\usepackage{amsmath,amssymb,graphicx, graphics, color,verbatim}
\usepackage{rotating}
\usepackage{multirow}
\usepackage{latexsym}
\usepackage{cancel}
\usepackage{enumerate}
\usepackage[normalem]{ulem} 
\usepackage {slashed}
\usepackage{epsfig}
\usepackage{url}
\usepackage{hyperref}
\usepackage{booktabs}
\usepackage{pstricks}
\usepackage{fancyvrb}
\usepackage{amsthm}
\usepackage{bm}
\usepackage{epstopdf}
\usepackage{wrapfig}
\usepackage{sidecap}
\usepackage{bbold}
\usepackage{fancyhdr}
\usepackage{comment}
\usepackage{array}
\usepackage{titlesec}

\graphicspath{{figs/}}

\newcommand{\nc}{\newcommand}
 
\nc{\beq}{\begin{equation}}
\nc{\eeq}{\end{equation}}
\nc{\beqa}{\begin{eqnarray}}
\nc{\eeqa}{\end{eqnarray}}
\nc{\bea}{\begin{eqnarray}}
\nc{\eea}{\end{eqnarray}}
\nc{\ra}{\rightarrow}
\nc{\Tr}{{\rm Tr}}
\nc{\slsh}{\slash\hspace*{-0.22cm}}

\def\be{\begin{equation}}
\def\ee{\end{equation}}
\def\bea{\begin{eqnarray}}
\def\eea{\end{eqnarray}}
\def\bit{\begin{itemize}}
\def\eit{\end{itemize}}
\def\cc{{\rm c.c.}}
\nc{\barray}{\begin{eqnarray}}
\nc{\earray}{\end{eqnarray}}
\nc{\barrayn}{\begin{eqnarray*}}
\nc{\earrayn}{\end{eqnarray*}}
\nc{\mc}{\mathcal}
\nc{\M}{\mathcal{M}}

\newcommand{\tref}[1]{Table~\ref{#1}}
\newcommand{\Eref}[1]{Eq.~(\ref{#1})}

\newcommand{\fb}{\,{\rm fb}^{-1}}

\newcommand{\gev}{{\ \rm GeV}}
\newcommand{\tev}{{\ \rm TeV}}
\newcommand{\mev}{{\ \rm MeV}}

\nc{\h}{$h$}
\nc{\infinity}{\infty}

\def\ben{\begin{enumerate}}
\def\een{\end{enumerate}}

\newcommand{\Eqref}[1]{Eq.~(\ref{eq:#1})}
\newcommand{\secref}[1]{\S\ref{sec:#1}}

\newcommand{\figref}[1]{Fig.~\ref{fig:#1}}

\newcommand{\Figref}[1]{Fig.~\ref{fig:#1}}
\newcommand{\tabref}[1]{Table~\ref{tab:#1}}

\newcommand{\met}{E\!\!\!\!/_T}
\newcommand{\MET}{$E\!\!\!\!/_T$}
\def\ifb{{\ \rm fb}^{-1}}
\def\ipb{{\ \rm pb}^{-1}}

\def\lt{\left}
\def\rt{\right}
\def\ord{\mathcal{O}}

\def\to{\rightarrow}

\def\pb{{\ \rm pb}}

\newcommand{\Br}{{\mathrm{Br}}}
\newcommand{\BR}{{\mathrm{Br}}}

\newcommand{\hsm}{h}

\def\sing{S}
\def\acplg{k}
\def\half{\frac{1}{2}}

\def\secsize{\normalsize}
\def\secbold#1{\mathbf{#1}}

\def\lowerGreekBold{\pmb}

\renewcommand\thesection{\arabic{section}}
\renewcommand\thesubsection{\arabic{section}.\arabic{subsection}}
\renewcommand\thesubsubsection{\arabic{section}.\arabic{subsection}.\arabic{subsubsection}}
\makeatletter
\def\p@subsection{}
\def\p@subsubsection{}
\makeatother

\begin{document}

\begin{flushright}
YITP-13-47, PITT PACC 1314
\end{flushright}

\title{Exotic Decays of the 125~GeV Higgs Boson}

\author{David~Curtin}
\affiliation{C.N.~Yang Institute for Theoretical Physics, Stony Brook University, Stony Brook, NY 11794, USA}

\author{Rouven~Essig}
\affiliation{C.N.~Yang Institute for Theoretical Physics, Stony Brook University, Stony Brook, NY 11794, USA}

\author{Stefania~Gori}
\affiliation{Enrico Fermi Institute and Department of Physics, University of Chicago, Chicago, IL,
60637, USA}
\affiliation{HEP Division, Argonne National Laboratory, 9700 Cass Ave., Argonne, IL 60439, USA}
\affiliation{Perimeter Institute for Theoretical Physics, Waterloo, Ontario, Canada}

\author{Prerit~Jaiswal}
\affiliation{Department of Physics, Florida State University, Tallahassee, FL 32306}

\author{Andrey~Katz}
\affiliation{Center for the Fundamental Laws of Nature, Harvard University, Cambridge, MA 02138, USA}

\author{Tao~Liu}
\affiliation{Department of Physics, The Hong Kong University of
  Science and Technology, Clear Water Bay, Kowloon, Hong Kong}

\author{ Zhen~Liu}
\affiliation{PITT PACC,
Department of Physics and Astronomy, \\ University of Pittsburgh,
3941 O'Hara St., Pittsburgh, PA 15260, USA}

\author{David~McKeen}
\affiliation{Department of Physics and Astronomy, University of Victoria,
Victoria, BC V8P 5C2, Canada}
\affiliation{Department of Physics, University of Washington, Seattle, WA 98195, USA}

\author{Jessie~Shelton}
\affiliation{Center for the Fundamental Laws of Nature, Harvard University, Cambridge, MA 02138, USA}

\author{Matthew~Strassler}
\affiliation{Center for the Fundamental Laws of Nature, Harvard University, Cambridge, MA 02138, USA}

\author{Ze'ev~Surujon}
\affiliation{C.N.~Yang Institute for Theoretical Physics, Stony Brook University, Stony Brook, NY 11794, USA}

\author{Brock~Tweedie}
\affiliation{PITT PACC,
Department of Physics and Astronomy, \\ University of Pittsburgh,
3941 O'Hara St., Pittsburgh, PA 15260, USA}
\affiliation{Physics Department, Boston University, Boston, MA 02215, USA}

\author{Yi-Ming~Zhong}
 \thanks{david.curtin@stonybrook.edu, rouven.essig@stonybrook.edu, sgori@perimeterinstitute.ca, \\ prerit.jaiswal@hep.fsu.edu, andrey@physics.harvard.edu, taoliu@ust.hk, zhl61@pitt.edu, \\ dmckeen@uw.edu, jshelton@physics.harvard.edu, strassler@physics.harvard.edu, \\ zeev.surujon@stonybrook.edu, bat42@pitt.edu, yiming.zhong@stonybrook.edu}
\affiliation{C.N.~Yang Institute for Theoretical Physics, Stony Brook University, Stony Brook, NY 11794, USA}

\begin{abstract}
We perform an extensive survey
of non-standard Higgs decays
  that are consistent with the 125~GeV Higgs-like resonance.  Our aim
  is to motivate a large set of new experimental analyses on the
  existing and forthcoming data from the Large Hadron Collider (LHC).
  The explicit search for exotic Higgs decays presents a largely
  untapped discovery opportunity for the LHC collaborations, as such
  decays may be easily missed by other searches.
   We emphasize that the Higgs is uniquely sensitive to the potential existence of new weakly coupled particles and provide a unified discussion of a large class of both simplified and complete models that give rise to characteristic patterns of exotic Higgs decays. We assess the status of exotic Higgs decays after LHC Run I. In many cases we are able to set new nontrivial constraints by reinterpreting existing experimental analyses. We point out that improvements are possible with dedicated analyses and perform some preliminary collider studies.
  We prioritize the analyses according to their theoretical motivation
  and their experimental feasibility.  This document is accompanied by
  a website that will be continuously updated with further
  information: \href{http://exotichiggs.physics.sunysb.edu}{\texttt{exotichiggs.physics.sunysb.edu}}.

\end{abstract}

\maketitle

\tableofcontents

\newpage

\section{Introduction and Overview}

The discovery at the Large Hadron Collider (LHC) of a Higgs-like
particle near 125~GeV~\cite{Aad:2012tfa,Chatrchyan:2012ufa} (referred
to as ``the Higgs'', $\hsm$, for simplicity in this paper) is a
triumph for
theoretical~\cite{Englert:1964et,Higgs:1964ia,Higgs:1964pj,Guralnik:1964eu,Higgs:1966ev,Kibble:1967sv,Glashow:1961tr,Weinberg:1967tq,Salam:1968}
and experimental particle physics, and marks the culmination of
several decades of experimental search.  However, the experimental
investigation of this new state has only just begun.  The Higgs plays
an essential role in the Standard Model (SM) of particle physics, and
impacts a wide range of new physics beyond the SM (BSM).  The
discovery of this new state presents us with a rich experimental
program that includes the precise measurement of its couplings to SM
particles, the search for additional Higgs-like states, and the focus
of this paper: the search for ``exotic'' decays, i.e.~decays that
involve new light states beyond the SM.

The aim of this document is to provide a summary and overview of the
theoretical motivation and basis for a large set of new analyses that
could be done by the LHC experimentalists.  In the course of doing so
we provide a thorough and unified description of a large class of
models that generate exotic Higgs decays, and perform numerous
original collider studies to assess the current status and discovery
potential of different modes.

Non-standard Higgs decays have always been a well-motivated
possibility as evidenced by an extensive existing, and growing,
literature.  They {\em remain} a well-motivated possibility even with
the discovery of a Higgs particle that is consistent with the simplest
SM expectations.  Indeed, they may provide our {\em only} window into
BSM physics at the LHC and must be searched for explicitly as they are
often unconstrained by other analyses.  The search for non-standard
Higgs decays should form an important component of the experimental
program of the LHC and future colliders.

Our focus here will be on the existing LHC data at 7 and 8~TeV
(``LHC7'' and ``LHC8'').  However, many signatures will remain
unconstrained by this dataset and should be searched for during future
runs of the LHC and at other colliders.  While this document may be
periodically updated, we note that it is accompanied by the website
\bea
\textrm{\href{http://exotichiggs.physics.sunysb.edu}{\texttt{exotichiggs.physics.sunysb.edu}}}\,,
\nonumber 
\eea 
which will serve as a centralized repository of information about new
collider studies and experimental analyses.

This document is structured as follows.  In
\S\ref{subsec:general-motivation}, we provide a general motivation for
non-standard Higgs decays.  In \S\ref{subsec:list}, we then detail the
decay modes considered in the subsequent sections.  We then summarize
several simplified and complete models in \S\ref{subsec:theory} that
illustrate the ease with which non-standard Higgs decays arise without
being in conflict with the current LHC data. (Two Appendices contain
some additional details).  The remaining sections,
\S\ref{sec:MET}--\S\ref{tatamet}, each treat one exotic Higgs decay in
detail and contain additional comments on theory motivation, existing
(theoretical) collider studies, limits from existing collider searches
(including our own reinterpretations of studies not aimed at Higgs
decays), and in some cases our own preliminary collider studies
outlining new search proposals at the LHC.  A summary in \S
\ref{sec:conclusions} considers the relative sensitivity of possible
analyses, and concludes with a suggested priority list for future
analyses of both Run I and Run II data, a brief discussion of Run II
triggering issues, and a short catalogue of research areas deserving
further investgation in the short term.

\subsection{General Motivation to Search for Exotic Higgs Decays}
\label{subsec:general-motivation}

\begin{table}
\begin{center}
\begin{tabular}{|l||c|c||c|c||c|c|}
\hline 
Production & $\sigma_{7~{\rm TeV}}$ (pb) & $N_{\rm ev}^{10\%}$, 5~fb$^{-1}$ & $\sigma_{8~{\rm TeV}}$ (pb) & $N_{\rm ev}^{10\%}$, 20~fb$^{-1}$ & $\sigma_{14~{\rm TeV}}$ (pb)& $N_{\rm ev}^{10\%}$, 300~fb$^{-1}$ \\
\hline 
\hline 
ggF  & 15.13 & 7,600 & 19.27 & 38,500  & 49.85  & $1.5\times 10^6$\\
\hline 
VBF  & 1.22 & 610 & 1.58 &  3,200 & 4.18 & 125,000 \\
\hline
$h W^\pm $  & 0.58 & 290  & $0.70 $ & 1,400  & $1.5$ & 45,000  \\  \cline{2-7}
$hW^\pm(\ell^\pm\nu)$  & $0.58 \cdot 0.21$ & 62 & $0.70 \cdot 0.21$ &  300  & $1.5 \cdot 0.21$ & 9,600 \\  \cline{2-7}
\hline
$h Z$  & $0.34 $ & 170  & $0.42 $ & 830 & $0.88 $ & 26,500 \\ \cline{2-7}
$h Z(\ell^+\ell^-)$  & $0.34 \cdot 0.067$ & 11 & $0.42 \cdot 0.067$ & 56  & $0.88 \cdot 0.067$ & 1,800 \\
\hline
 $t \bar{t} h$  & 0.086 & 43 & 0.13 & 260 & 0.61 & 18,300 \\
\hline
\end{tabular}
\caption{The number of exotic Higgs decays in existing LHC data, per
  experiment, at 7~TeV (5~fb$^{-1}$) and 8~TeV (20~fb$^{-1}$), and at
  a future 14~TeV run (300~fb$^{-1}$), assuming the Standard Model
  production cross section of a 125 GeV Higgs
  boson~\cite{Dittmaier:2011ti} and a branching ratio of Br$(h\to {\rm
    BSM}) = 10\%$ for various production channels: gluon-gluon fusion
  (ggF), vector-boson fusion (VBF), associated production ($hW^\pm$
  and $hZ $, with and without branching ratios $W^\pm\to \ell^\pm \nu$
  or $Z\to \ell^+\ell^-$, where $\ell = e, \mu$, included), and
  through radiation off the top-quark ($t\bar th$).}
\label{tab:expected-exotic-decays}
\end{center}
\end{table}

In this subsection, we review the reasons why searches for exotic
Higgs decays are a particularly rich and fruitful way to search for
new physics.

The data collected at LHC7 and LHC8 may easily contain
$\mathcal{O}(50,000)$ exotic Higgs decays per experiment, presenting
us with a large discovery potential for new physics, of a kind which
is mostly unconstrained by existing analyses.  Indeed, as we will
explain in more detail in the following, the current data allows the
branching ratio ($\mathrm{Br}$) of the 125 GeV Higgs boson into BSM
states to be as large as $\mathcal{O}(20\%-50\%)$, which includes
constraints from observing the Higgs boson in various SM channels.
Table~\ref{tab:expected-exotic-decays} lists the number of exotic
Higgs decays that could be contained in the LHC7 and LHC8 data,
assuming $\mathrm{Br}(h\to{\rm BSM}) = 10\%$; we list these numbers
separately for each Higgs production channel.  Of course these are
only the number of events {\it produced}; the trigger efficiency
depends strongly on the final states that appear in the exotic decay.
Nevertheless, the table makes it clear that, for exotic final states
where triggering is not disastrously inefficient, a dedicated search
has the potential for a spectacular discovery.

Several theoretical and experimental studies have constrained the
possible $\mathrm{Br}$ into an invisible or an (as yet) undetected
final state by fitting for the couplings of the Higgs to SM states.
These ``coupling fits'' constrain $\mathrm{Br}(h\to\mbox{BSM})\lesssim
20\%$ at $95\%$ CL if the Higgs is produced with SM strength; a larger
BSM branching fraction, $\mathrm{Br}(h\to\mbox{BSM})\lesssim 30\%$, is
possible if new physics is allowed to modify the loop-induced Higgs
couplings to both $gg$ and $\gamma \gamma$ (see for example
\cite{Belanger:2013kya, Giardino:2013bma, Ellis:2013lra,
  Cheung:2013kla} for some more recent fits).  Fits that take more
conservative approaches for the theoretical uncertainty on the SM
Higgs production cross-sections can leave room for larger ($\lesssim
60\%$) BSM branching fractions \cite{Djouadi:2013qya}. This result is
similar to the one obtained by the ATLAS and CMS collaborations
\cite{CMS:yva,ATLAS:2013sla}.  Bounds can be further relaxed for
models with Higgs couplings to gauge bosons larger than in the SM
\cite{Dobrescu:2012td}.  Future projections for the LHC suggest an
ultimate precision on this indirect measurement of
$\mathrm{Br}(h\to\mbox{BSM})$ of $ \mathcal{O}(5-10\%)$, see
e.g.~\cite{Peskin:2012we,CMS:2013xfa,ATLAS:2013hta}.  Branching
fractions of $\mathcal{O}(10\%)$ into exotic decay modes are therefore
not only still allowed by existing data but {\em will remain
  reasonable targets for the duration of the physics program of the
  LHC.}

In the right columns of Table~\ref{tab:expected-exotic-decays} we show
the possible number of exotic Higgs decays in the anticipated LHC14
dataset with 300~fb$^{-1}$, again assuming $\mathrm{Br}(h\to{\rm BSM})
= 10\%$. The large rates for producing these exotic states suggest
that branching fractions as small as $\mathcal{O}(10^{-6})$ could be
detected, if the decay signature is both visible and clean.

As for any newly discovered particle, a detailed experimental
characterization of the Higgs is imperative. Such an experimental
characterization must necessarily include an exhaustive study of its
decay modes.  These programs have been established for other
particles, such as the top quark, the $Z$-boson, $B$-hadrons etc., as
rare decay modes of SM particles are prime places for new physics to
appear.  However, it is worth emphasizing that the Higgs boson is a
special case. The tiny natural width of the SM Higgs boson, together
with the ease with which the Higgs can mediate interactions with new
physics, make exotic Higgs decays a natural and expected signature of
a very broad class of theories beyond the SM.

A SM-like Higgs boson with a mass of $m_\hsm = 125$~GeV has an
extremely narrow width, $\Gamma_\hsm \simeq 4.07$~MeV, so that
$\Gamma_\hsm/m_\hsm \simeq 3.3 \times 10^{-5}$.  The reason is that
tree-level decays to SM fermions are suppressed by the small Yukawa
couplings, e.g.~$y_{b,\tau}\lesssim \mathcal{O}(10^{-2})$, decays to
two photons ($\gamma\gamma$), two gluons ($gg$), and $Z\gamma$ are
suppressed by loop factors, and decays to $WW^*$ and $Z Z^*$ are
suppressed by multibody phase space.  Since the dominant decay, to two
$b$-quarks, is controlled by a coupling with a size of only $\sim
0.017$ (this assumes a running $b$-quark mass $m_b (125 \gev) =
2.91$~GeV evaluated in the $\overline{\rm MS}$ scheme), even a small
coupling to another light state can easily open up additional sizable
decay modes
\cite{Shrock:1982kd,Gunion:1984yn,Li:1985hy,Gunion:1986nh}.

\begin{figure}[t]
\begin{center}
\includegraphics[width=0.48\textwidth]{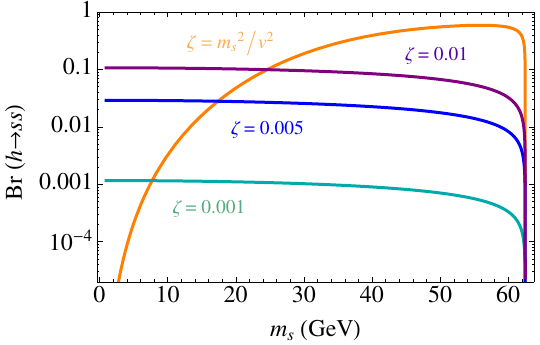}
~~~
\includegraphics[width=0.48\textwidth]{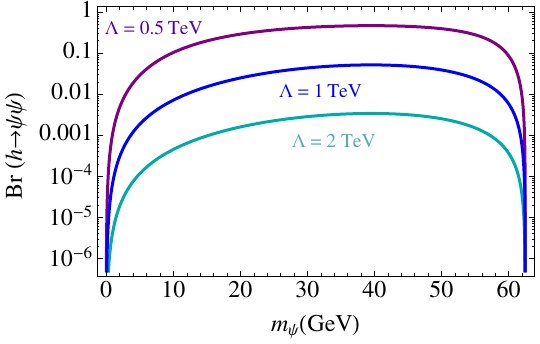}
\caption{ \small Sensitivity of a 125 GeV Higgs to light weakly
  coupled particles.  {\bf Left:} Exotic Higgs branching fraction to a
  singlet scalar $s$ versus the singlet's mass $m_s$, assuming the
  interaction Eq.~(\ref{eq:HtoS}) is solely responsible for the $h\to
  ss$ decay.  If the interaction in Eq.~(\ref{eq:HtoS}) generates the
  $s$ mass, the result is the orange curve; the other curves are for
  fixed and independent values of $\zeta$ and $m_s$.  {\bf Right:}
  Exotic Higgs branching fraction to a new fermion $\psi$ interacting
  with the Higgs as in Eq.~(\ref{eq:HtoPsis}) to illustrate the
  sensitivity of exotic Higgs decay searches to high scales, here
  $\Lambda$.  We take here $\mu = m_\psi$.
  \label{fig:intro_BR}}
\end{center}
\end{figure}

In fact, we have very good reasons to expect that new physics may
couple preferentially to the Higgs boson.  The brief survey in
\S\ref{subsec:theory} of simplified models and theories that produce
exotic Higgs decays will provide ample examples that corroborate this
statement.  More generally, the Higgs provides one of only a few
``portals'' that allow SM matter to interact with hidden-sector matter
that is not charged under the SM forces (e.g.~\cite{Silveira:1985rk,Binoth:1996au,Schabinger:2005ei,Strassler:2006im,Patt:2006fw}), and where the
leading interaction can be (super-)renormalizable.\footnote{The other
  two portals are the ``vector portal'' at mass dimension 2, namely
  the hypercharge field strength $B^{\mu\nu}$, and the ``neutrino
  portal'', given by the product of the Higgs and a lepton doublet, $H
  L$, with mass dimension $5/2$.  The vector portal can mediate, e.g.,
  kinetic mixing between hypercharge and a new $U(1)$ gauge field with
  the renormalizable interaction $F'_{\mu\nu} B^{\mu\nu}$; the
  neutrino portal operator can mediate the renormalizable coupling
  $HLN$, with $N$ a sterile neutrino.}  Since the operator $|H|^2$ is
a SM singlet, we can couple it to a singlet scalar field $s$ through
the Higgs portal as
\beq
\label{eq:HtoS}
\Delta\mathcal{L}=\frac{\zeta}{2} s^2 |H|^2\,,
\eeq
where we have assumed for simplicity that $s$ has a conserved $Z_2$
parity.  This kind of interaction is a very common building block in
models of extended Higgs sectors.  If $m_s<m_h/2$, this interaction
allows $h\to ss$ after electroweak symmetry breaking (EWSB), and {\em
  even a coupling as small as $\zeta = 10^{-2}$} yields
$\mathrm{Br}(h\to{\rm BSM}) = 10\%$.  In Fig.~\ref{fig:intro_BR} ({\bf
  left}), we plot $\mathrm{Br}(h\to ss)$ for various couplings $\zeta$
as a function of the singlet mass $m_s$.  (The orange line shows the expected branching fraction if the interaction in Eq.~(\ref{eq:HtoS}) generates the $s$ mass. Achieving larger branching fractions requires a cancellation between the Higgs contribution and another contribution to the $s$ mass.) 
Even very small couplings of
the Higgs boson to new states beyond the SM can lead to potential
signals at the LHC. 

There are many possible interactions through the Higgs portal. One
striking and generic feature of these interactions is that searches
for exotic Higgs decays can easily be sensitive to new physics scales
$\gtrsim 1$~TeV.  As one example, consider the (effective)
dimension-six Higgs-portal interaction
\beq
\label{eq:HtoPsis}
\Delta \mathcal{L}=
    \frac{\mu}{\Lambda^2}|H|^2 \bar \psi \psi \,,
\eeq
where $\psi$ is some new singlet fermion and $\mu$ is a chiral
symmetry breaking parameter with dimensions of mass.  Taking $\mu\sim
m_\psi$ for simplicity, we show the resulting
$\mathrm{Br}(h\to\bar\psi\psi)$ versus $m_\psi$ for various $\Lambda$
in Fig.~\ref{fig:intro_BR} ({\bf right}).  Even
$\mathrm{Br}(h\to\bar\psi\psi)\sim\mathcal{O}(10^{-2})$ induced by the
higher-dimensional operator of Eq.~(\ref{eq:HtoPsis}) is sensitive to
scales $\Lambda\gtrsim 1$ TeV.  The scaling $\mu\sim m_\psi$ is
conservative --- some models can yield $\mu\sim v$ or greater,
allowing even further reach (see, e.g.~, Fig.~\ref{fig:BRexSMF}).  Thus exotic Higgs decays can indirectly
probe new physics scales beyond the kinematic reach of the LHC, and
may provide the {\em only} evidence of a new sector that is accessible
to the LHC.
%

Given the large Higgs sample that is being collected, it may at first
glance seem surprising that the majority of possible exotic Higgs
decay modes are poorly constrained, if at all, by existing searches.
A major reason for this is that the dominant Higgs production process,
gluon fusion, creates Higgs bosons largely at rest, without any
associated objects.  In a four-body exotic cascade decay of such a
Higgs boson, for example, the characteristic transverse momenta of the
daughter particles is not large, $p_T \lesssim 30$ GeV.  Typical
exotica searches at the LHC place much higher analysis cuts on object
energies, leaving such decays largely unconstrained.  In addition, the
SM backgrounds are larger at lower energies, so that dedicated
analyses are required to find a new physics signal.  In many cases,
exotic Higgs decay signals are thus {\em not} seen or constrained by
existing non-targeted analyses. It is necessary to perform
\emph{dedicated} searches for exotic Higgs decays.
Since there are dozens of possible exotic decay modes, dozens of new
searches are needed to discover or constrain a broad and generic class
of theories beyond the SM.

In some cases, particularly if the exotic decay produces only jets
with or without $\met$, it may be difficult to trigger on Higgs events
produced in the (dominant) gluon-gluon-fusion channel.  However, even
under these pessimistic assumptions, a few hundred events should still
be on tape in the existing 7 and 8 TeV datasets, since the associated
production of the Higgs boson with a leptonically-decaying $Z$- or
$W$-boson will usually be recorded due to the presence of one or two
leptons.  Moreover, additional events may have triggered in the vector
boson fusion (VBF) channel due to the rapidity gap of two of the jets
in these events (see next paragraph).  In some cases, more
sophisticated triggers on combinations of objects, possibly with low
thresholds, may be required to write a larger fraction of events to
tape.

In addition to the ``standard'' LHC7 and LHC8 datasets, an additional
300--500 Hz of data was collected and ``parked'' during the LHC8
running.  This parked dataset was not reconstructed immediately, but
may present additional opportunities for exotic Higgs analyses.  For
example, at CMS, it included a trigger on Higgs VBF production
($M_{jj}>650$~GeV and $|\Delta\eta_{jj}|>3.5$)~\cite{parked-CMS}.  In
ATLAS \cite{Petersen:1495760}, the applications for Higgs physics are
less direct but the lowered object $p_T$ thresholds in the ATLAS
delayed data stream may present opportunities.  More generally, it is
important for the LHC14 run to be aware of cases in which simple
changes in the trigger could appreciably increase or decrease the
number of recorded exotic decays.

The subject of exotic Higgs decays is not a new one.  There is an
extensive literature on exotic Higgs decays, much of it driven by the
past desire to hide a light Higgs from LEP searches, both to preserve
electroweak naturalness and to maximize agreement with precision
electroweak fits that yielded a best-fit Higgs mass below the LEP
bound of $\sim 114$ GeV (see e.g.~\cite{Chang:2008cw} for a review).
Now that the Higgs boson has been discovered, however, the questions
have changed.  We know the mass of (at least one) Higgs boson, and we
also know that its branching fraction into exotic states cannot exceed
$\approx 60\%$.  The relevant question is now: \emph{for various
  exotic final states, what branching fractions can be probed at the
  LHC, and how can the sensitivity to these final states be
  maximized?}

The search for exotic Higgs decays is a program which deserves to be
pursued in a systematic fashion.  Our aim in this work is to make such
a physics program easier by providing a centralized assessment of
models, signatures, and limits.
	\subsection{Exotic Decay Modes of the 125 GeV Higgs  Boson}\label{subsec:list}

In this section, we list the exotic decay modes that are the focus of
this paper.  We organize them by decay topology.  While this is not
the only possible way to make a systematic list of possible exotic
decays, it has the advantage that it is well-adapted to a large number
of specific models in the literature, allowing a relatively simple
mapping between these models and our list; however, since any number
of final state particles can be invisible, different topologies can
yield the same experimental signature.  We also focus on topologies
that arise in models commonly found in the literature, many of which 
we review in \S\ref{subsec:theory}. 

In our discussion of exotic decays we will make {\bf three simplifying
  assumptions}:
\begin{itemize}
\item[{\bf 1)}] {\bf The observed Higgs at 125 GeV is principally
    responsible for breaking the electroweak symmetry.}  This means
  that in models with additional physical scalars, the theory is
  usually close to a decoupling limit in which the 125 GeV state is
  SM-like.  The production cross sections for this particle are then
  close to those predicted for the SM Higgs.  The decay modes are also
  SM-like, but modifications of $\mathcal{O}(10-50\%)$ are
  theoretically easily obtained and consistent with current data (see
  discussion in \S\ref{subsec:general-motivation}).  We note that this
  is not the only scenario allowed by current LHC data, as some
  non-decoupling limits are still viable for BSM models (see
  e.g.~\cite{Belanger:2012sd,
    Bechtle:2012jw,Barbieri:2013nka,Han:2013mga}), but the assumption
  of a decoupling-like limit is generic and minimal.  We emphasize
  that any exotic-decay search that targets a 125 GeV Higgs should
  also scan over a much wider Higgs mass range, looking for additional
  Higgs bosons that may appear in a more complex Higgs sector and may
  often decay to a final state not found for an SM Higgs.
\item[{\bf 2)}] {\bf The observed Higgs at 125 GeV decays to new
    particles beyond the SM.}  We consider scenarios in which the
  newly-discovered Higgs boson enables the discovery of new,
  weakly-coupled particles, which in many cases have exotic Higgs
  decays as their primary or only production mode at the LHC.  We do
  not consider rare Higgs decays to SM particles, which can be very
  sensitive to new physics, whether through its effects in loops (such
  as in $\gamma\gamma$ or $Z\gamma$), through its modifications of the
  $V$-$V$-$H$ couplings \cite{gunion2000higgs} or its nonstandard
  flavor structures (as in lepton family number-violating decays $h\to
  \tau\mu$, see \cite{Blankenburg:2012ex,Harnik:2012pb} and references therein).
\item[{\bf 3)}] {\bf The initial exotic 125 GeV Higgs decay is to two neutral BSM particles.}  Generally, to compete with the SM decay modes, the
  Higgs decay to exotic particles needs to begin as a two-body decay, and LEP limits place stringent constraints on light charged particles \cite{L3:2003ex, Batell:2013psa}.
  Three-body or higher-body exotic decays typically require new states
  with masses $m\lesssim m_h$ that have substantial couplings to the
  Higgs boson, in order to induce any appreciable BSM branching
  fraction after the phase space suppression~\cite{Giddings:2013gh}.
  In some cases, these light particles can appear in loops and change
  the Higgs decay rates to $\gamma\gamma$ and/or $Z\gamma$ final
  states.  While this is certainly worthy of further study we will not
  do so here.
  \end{itemize}

\begin{figure}
\begin{center}
\begin{tabular}{m{2cm}m{4mm}m{3cm}m{4mm}m{4cm}m{4mm}m{4cm}}
\includegraphics[width=2cm]{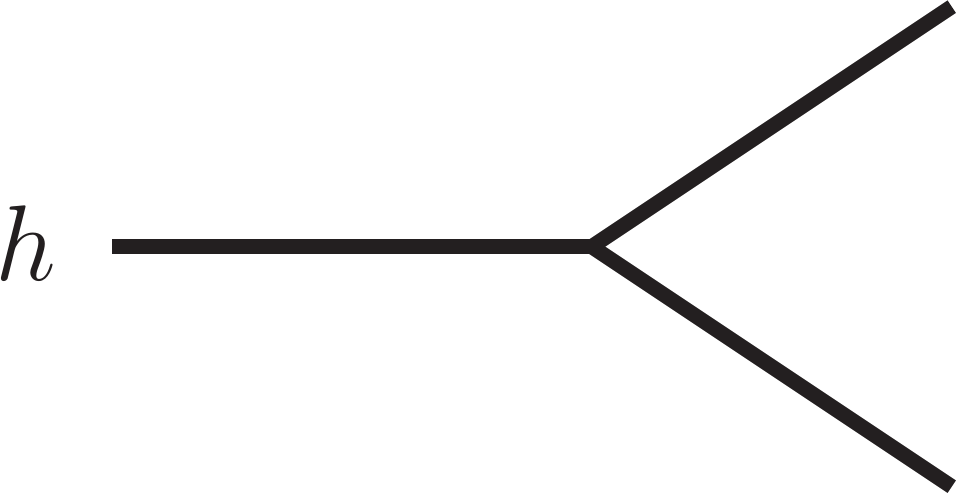}&&
\includegraphics[width=3cm]{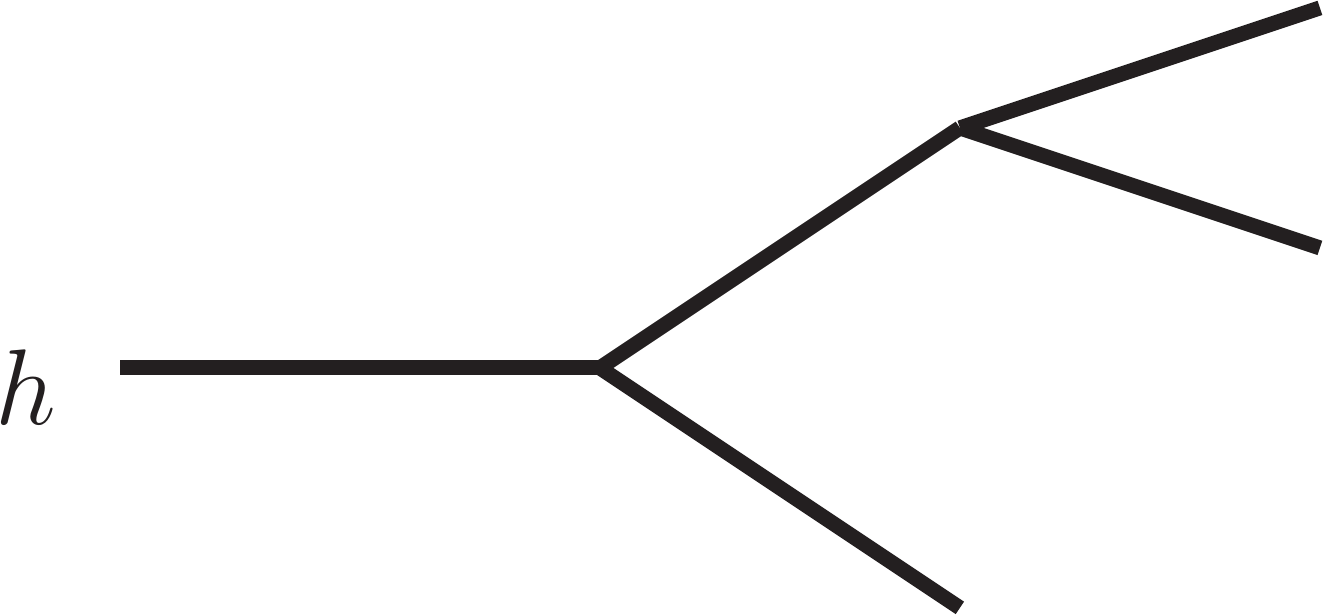}&&
\includegraphics[width=4cm]{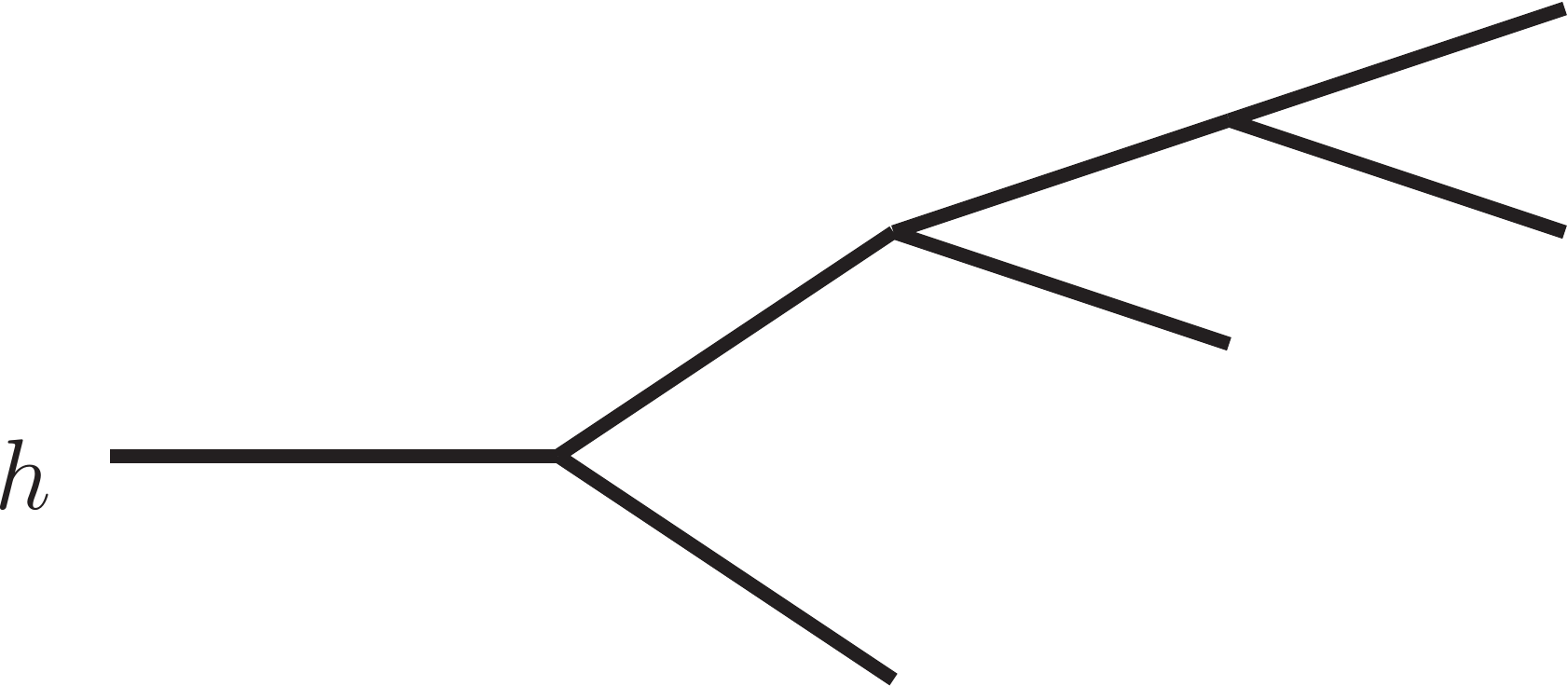}&&
\includegraphics[width=4cm]{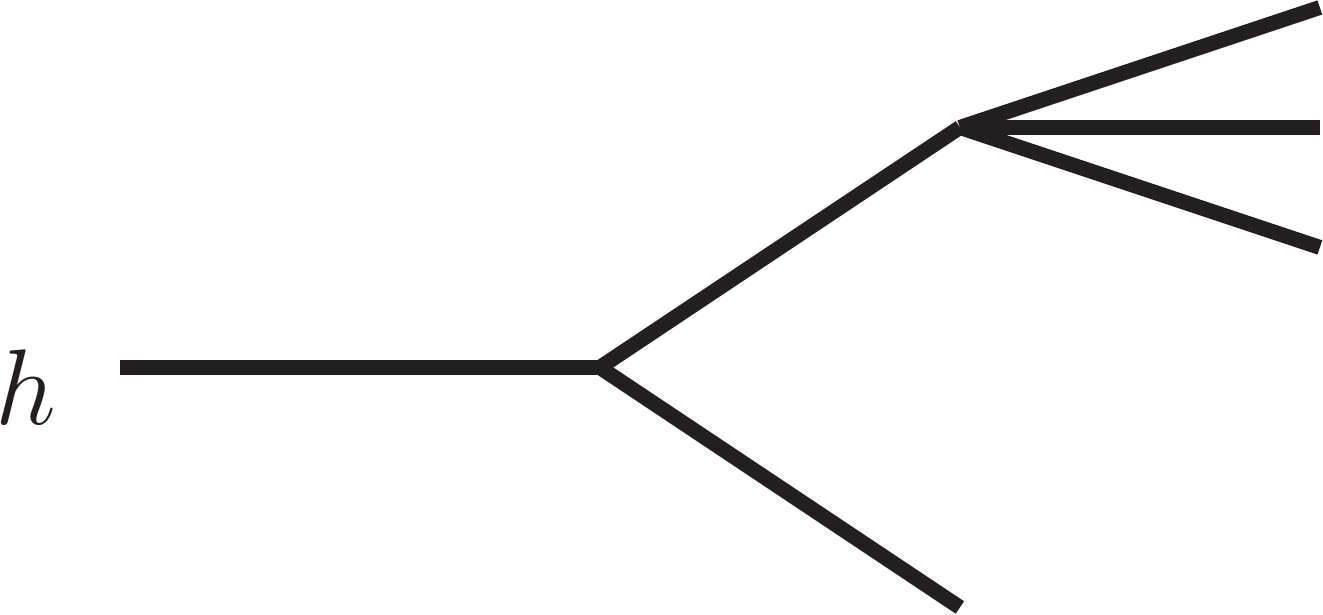}
\\
\hspace{4mm} $h\rightarrow2$ &&
\hspace{3mm} $h\rightarrow2\rightarrow3$ 
&& 
\hspace{3mm} $h\rightarrow2\rightarrow3\rightarrow4$
&&
\hspace{2mm} $h\rightarrow2\rightarrow(1+3)$
\end{tabular}
\begin{tabular}{m{4cm}m{4mm}m{4cm}m{4mm}m{4cm}}
\\
\vspace{8mm}
\\
\includegraphics[width=4cm]{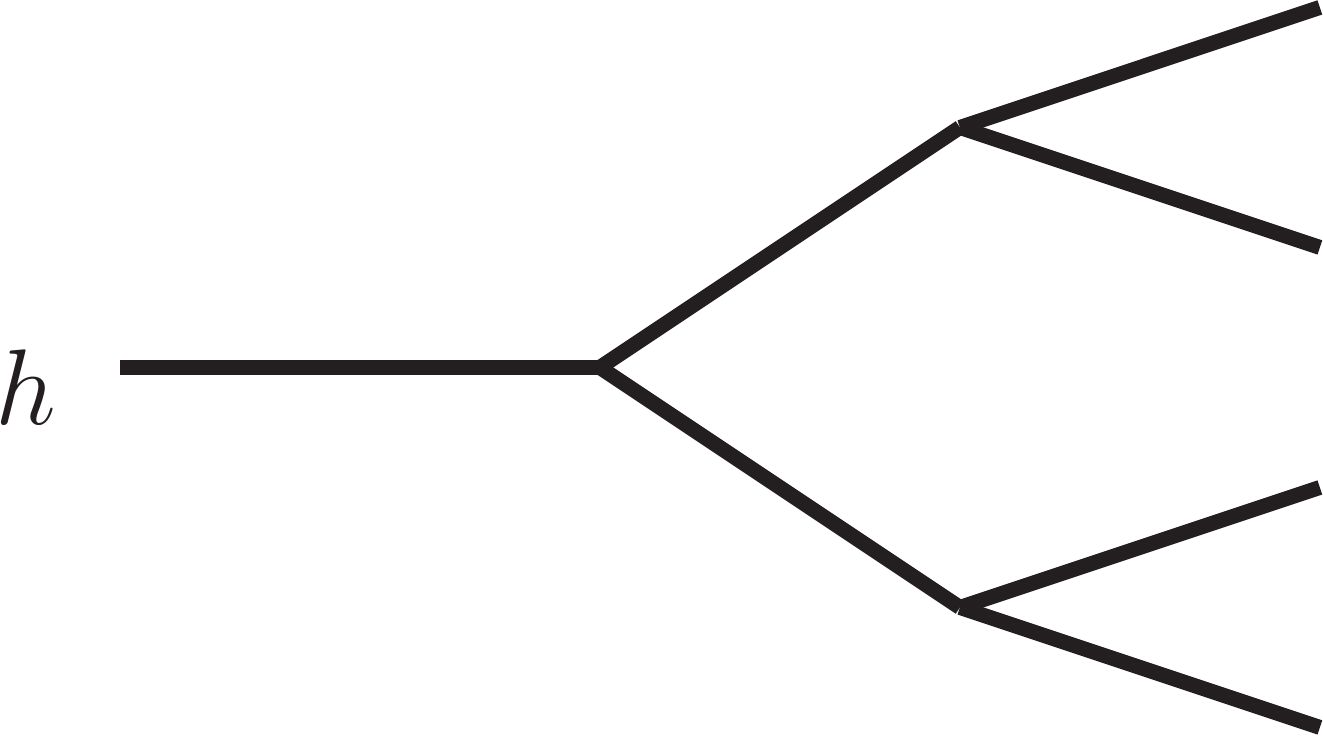}&&
\includegraphics[width=4cm]{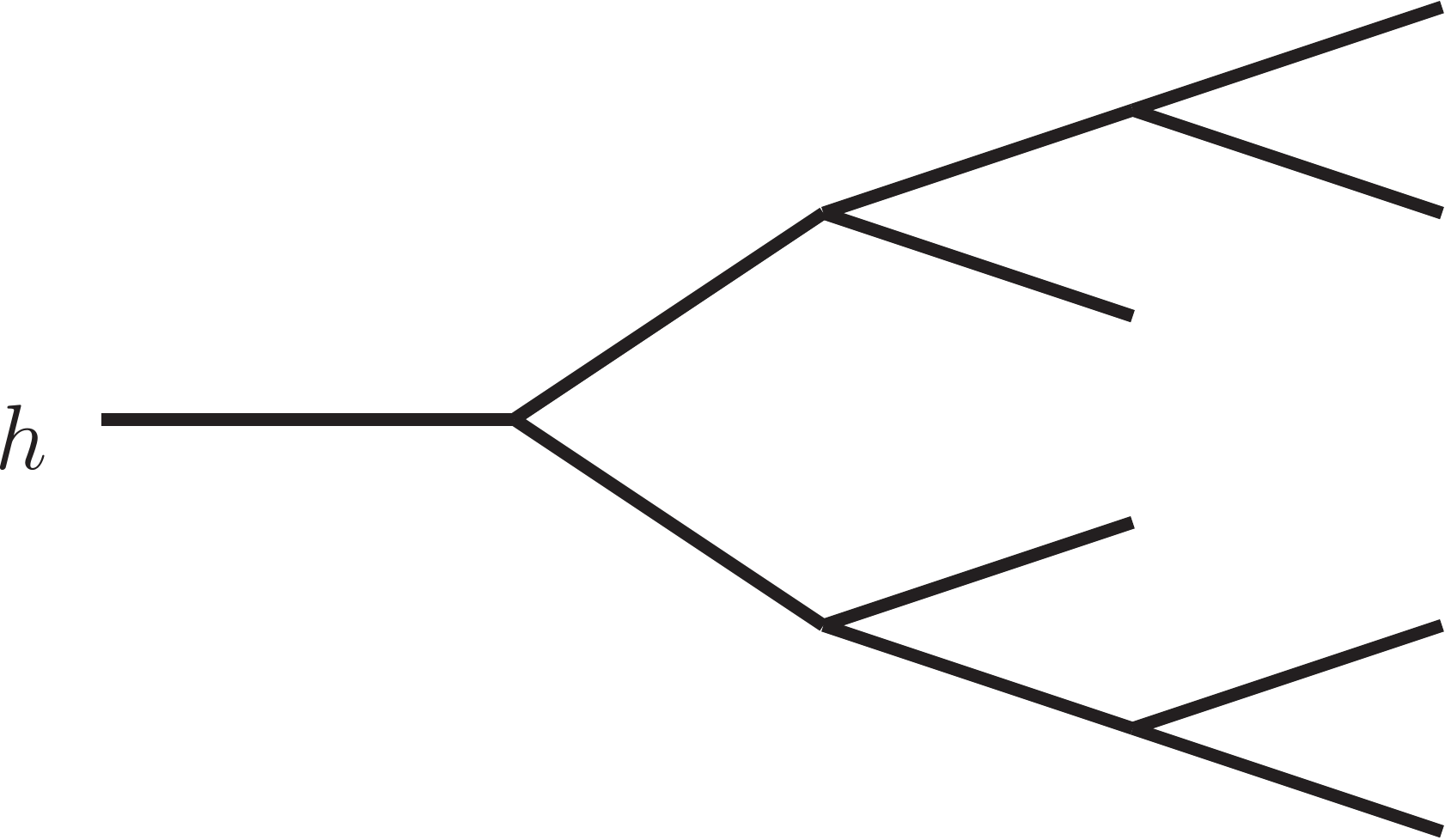}&&
\includegraphics[width=4cm]{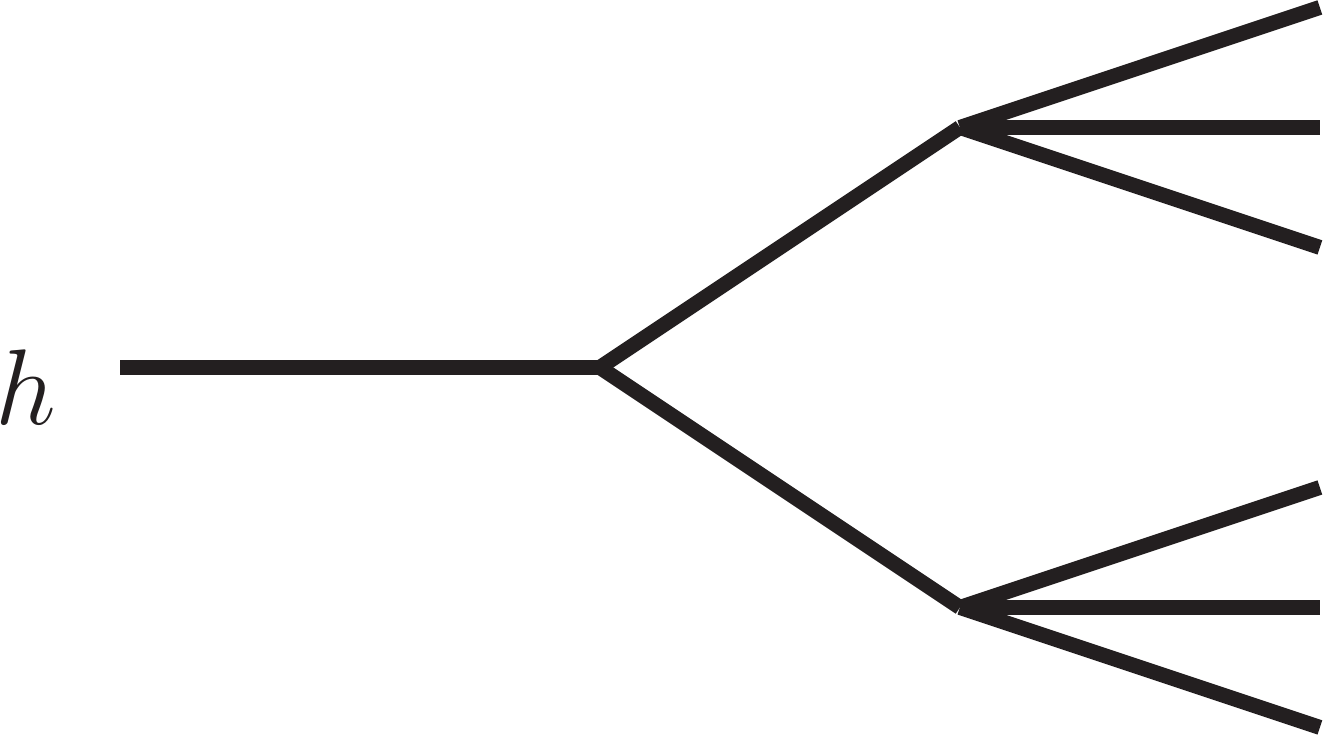}
\\
\hspace{7mm} $h\rightarrow2\rightarrow4$
&& 
\hspace{7mm} $h\rightarrow2\rightarrow4\rightarrow6$
&& 
\hspace{10mm} $h\rightarrow2\rightarrow6$
\end{tabular}
\end{center}
\caption{ The exotic Higgs decay topologies we consider in this
  document, along with the labels we use to refer to them. Every
  intermediate line in these diagrams represents an \emph{on-shell,
    neutral} particle, which is either a $Z$-boson or a BSM
  particle.}
\label{f.higgsdecaytopologies}
\end{figure}

Our focus is thus on decays that begin via the two-body process $h\to
X_1 X_2$, where $X_{1,2}$ are BSM states (possibly identical).
Depending on the properties of $X_1$ and $X_2$, a large number of
distinct exotic Higgs decay modes are possible.  The topologies we
consider are shown in Fig.~\ref{f.higgsdecaytopologies}. Our choice is
guided by existing models in the literature, but of course there are
other possibilities as well. The specific modes we consider (as well
as some modes that fall into the same category but that we do not discuss
further) are listed below.  In parentheses we list the section numbers
where a particular decay mode will be discussed in more detail.
A pair of particles in parentheses denotes that they form a resonance.

\begin{itemize}

\item {\bf $h\to 2$} \\
  This topology occurs for Higgs decays into BSM particles with a
  lifetime longer than detector scales.  It includes \emph{$h\to $
    invisible}
  decays~\cite{Shrock:1982kd,Eboli:2000ze,Bai:2011wz,Riva:2012hz} and,
  in principle, {\em $h\to$~$R$-hadrons}, although the latter scenario
  is strongly constrained.  In this paper, we consider only:
\begin{enumerate}
\item {\em $h\to$~invisible ($\met$)}   (\S\ref{sec:MET})
\end{enumerate}

\item $h\to 2\to 3$ \\
  Here the Higgs decays to one final-state particle that is
  detector-stable and another one that decays promptly or with a
  displaced vertex. Possibilities include
\begin{enumerate}
\item {\em $h\to \gamma+\met$}  (\S\ref{sec:1gaMET}).

\item {\em $h \rightarrow (bb) + \met$}  (\S\ref{bbmet}).

\item {\em $h \rightarrow (\tau\tau) + \met$}  (\S\ref{tatamet}).

\item {\em $h \rightarrow (\gamma\gamma) + \met$}  (\S\ref{sec:2gaMET}).

\item {\em $h \rightarrow (\ell\ell) + \met$}  (collimated leptons \S\ref{Clepton}).

\end{enumerate}
One might also consider $h\to \gamma + Z$ or $\gamma + Z'$, where the $Z'$ decays to two SM particles and may have different decay modes
than the $Z$; for instance, the $Z'$ could be leptophilic.  In the SM,
$\mathrm{Br}(h\to \gamma Z) \sim 10^{-3}$, but this can be enhanced in
BSM models, e.g.~\cite{Azatov:2013ura}.  The \emph{semi-invisible}
$h\to \gamma+\met$ signature arises in the SM ($h\to\gamma Z\to
\gamma \nu\bar\nu$), but can also be enhanced in BSM theories, e.g.,
$h\to \tilde B \tilde G\to \gamma + 2\tilde G$, where $\tilde{B}$ and
$\tilde G$ are a bino and gravitino respectively
\cite{Petersson:2012dp}.

\item {\bf $h\to 2\to 3\to 4$}

  For this topology, we only consider signatures that contain $\met$.
  In particular, we consider Higgs decays to neutral fermions $h\to
  \chi_1 \chi_2$, where $\chi_2 \to a \chi_1$ or $\chi_2 \to V \chi_1$
  and $\chi_1$ is invisible.  Similar decays can occur in more general
  hidden sectors where the roles of $\chi_{1,2}$ may be played either by
  fermionic or bosonic fields \cite{Strassler:2006im,Strassler:2006qa}. Such single-resonance topologies give rise to
  semi-invisible decays, and appear in (for example) the PQ-symmetry
  limit of the Next-to-Minimal Supersymmetric Standard Model
  (NMSSM)~\cite{draper2011dark,HLWY}, where the resonance is exotic,
  or the SM extended with a neutrino sector like the
  $\nu$SM~\cite{0706.1732,0705.2190,0704.0438}, where the resonance is
  the $W$ or $Z$. Discussion in a simplified model context can be
  found in \cite{Chang:2007de}. We consider in more detail:
\begin{enumerate}
\item ($b\bar b$) + $\met$ (\S\ref{bbmet})
\item ($\tau\tau$) + $\met$ (\S\ref{tatamet})
\item ($\gamma \gamma$) + $\met$ (\S\ref{sec:2gaMET})
\item ($\ell^+ \ell^-$) + $\met$ (isolated \S\ref{sec:2lMET}, collimated \S\ref{Clepton})
\end{enumerate}

\item {\bf $h\to 2\to (1+3)$}

  This topology occurs when the resonant cascade decays of the $h\to
  2\to 3 \to 4$ topology go off-shell.  Here again we only consider
  semi-invisible signatures, and focus on leptonic signatures.

\begin{enumerate}
\item $\ell^+ \ell^-$ + $\met$ (isolated \S\ref{sec:2lMET})
\end{enumerate}

\item {\bf $h\to 2\to 4$}

  In this topology the Higgs decays as $h\to a a', s s', V_1 V_2, a
  V_1 \to (x x) (y y)$, where $a$ and $a'$ ($s$ and $s'$, $V_1$ and
  $V_2$) are not necessarily distinct pseudo-scalars (scalars,
  vectors).  In most cases we can reconstruct two resonances.  The
  scalars and pseudo-scalars can typically decay to $x, y =
  \textrm{quarks, leptons, photons, or gluons}$, while the vectors can
  typically decay to $x, y = \textrm{quarks or leptons}$.  This
  topology occurs in well-known BSM theories like the R-symmetric limit
  of the NMSSM~\cite{Ellwanger:2003jt,
    Ellwanger:2005uu,Almarashi:2011bf, Ellwanger:2004gz}, Little Higgs
  Models~\cite{Kilian:2004pp, Cheung:2006nk, Csaki:2003si}, or any
  theory that features additional SM singlet scalars, such
  as~\cite{Cheng:1999bg, Cvetic:1997ky,
    Erler:2002pr,Panagiotakopoulos:1999ah, Panagiotakopoulos:2000wp,
    Dobrescu:2000yn, Strassler:2006im, Barger:2006rd}.  Also possible is the fermionic
  decay $h\to \chi_2\chi_2\to 2(\gamma \chi_1)$, which occurs in,
  e.g., the MSSM with gauge-mediated
  SUSY-breaking~\cite{Mason:2009qh} (see also~\cite{Englert:2012wf} for discussions 
    of 1 to 3 light jets $+\met$ in simplified models with this topology).  In this paper, we consider in
  more detail:
\begin{enumerate}
\item ($b\bar b$)($b\bar b$) (\S\ref{sec:hto4b})
\item ($b\bar b$) ($\tau^+ \tau^-$)  (\S\ref{sec:h2b2tau})
\item ($b\bar b$) ($\mu^+ \mu^-$)  (\S\ref{sec:h2b2mu})
\item ($\tau^+ \tau^-$) ($\tau^+ \tau^-$)   (\S\ref{sec:hto4tau})
\item ($\tau^+ \tau^-$) ($\mu^+\mu^-$)  (\S\ref{sec:hto4tau})
\item ($jj$)($jj$)  (\S\ref{sec:h4j})
\item ($jj$) ($\gamma\gamma$)  (\S\ref{sec:2gamma2jet})
\item ($\ell^+\ell^-$) ($\ell^+\ell^-$)   (\S\ref{sec:htoZa} for $h \rightarrow Z  Z_D$,  \S\ref{sec:hto4l} for $h \rightarrow Z_D Z_D$, \S\ref{sec:collep_nomet} for collimated leptons)  
\item ($\gamma\gamma$) ($\gamma\gamma$)  (\S\ref{sec:4gamma})
\item  {\em $\gamma\gamma + \met$} (no $\gamma\gamma$-resonance, \S\ref{sec:2gaMET})
\end{enumerate}

\item $h \to 2 \to 4 \to 6$

  Here both the Higgs' daughters undergo on-shell cascade decays.  As
  for the single-cascade topology $h\to 2\to (1+3)$, examples of such
  cascades include NMSSM neutralinos, decaying via $\chi_2 \to \chi_1
  a$, $a\to f\bar f$, or right-handed neutrinos, decaying via $N_R\to
  \nu Z, \ell W$.  More elaborate hidden sectors allow for many
  possibilities, such as $\phi_2 \to a \phi_1$, $a\to gg
  (\gamma\gamma)$, or $\phi_1\to Z_D \phi_2$, $Z_D\to \ell\ell, q\bar
  q$ (here $\phi_{1,2}$ are BSM states that may be either fermions or
  scalars).
 
 We only consider final states with leptons for this topology:
\begin{enumerate}
\item {\em $h \rightarrow 2(\ell\ell) + \met$} (isolated \S\ref{sec:IsoLeptonMET}, collimated \S\ref{sec:collep_nomet}).
\item {\em $h\rightarrow (\ell\ell) +\met +X$} (isolated \S\ref{sec:2lMET}, collimated \S\ref{Clepton})
\end{enumerate}

\item $h\to 2 \to 6$

  There are various possibilities here.  Examples include Higgs decays
  to R-parity violating neutralinos, which can yield $h
  \to \chi_1 \chi_1 \rightarrow 6 j,\; 4j+ 2\ell, \;4\ell
  +2\nu$.  In addition, any of the resonant cascade decays discussed
  above may become three-body. Another example is flavored dark matter, 
  where the Higgs can decay to two heavy dark flavors first and then into light 
  quarks and the dark matter candidate via higher dimensional operator, 
  resulting in $h \to 4j +\met$~\cite{Batell:2011tc}.
    
  We only consider final states with isolated leptons for this
  topology:
\begin{enumerate}
\item{\em $h \rightarrow 2\ell + \met + X$} (\S\ref{sec:2lMET}).
\item {\em $h \rightarrow 4\ell + \met$} (\S\ref{sec:IsoLeptonMET}).
\end{enumerate}

\item $h\to 2 \to$ {\em many}, where ``many'' refers to many SM
  particles, including ``weird jets''.  This occurs \cite{Strassler:2006im} in Higgs decays to
  hidden-sector particles that undergo a long series of cascade decays
  or a hidden-sector parton shower to (many) SM particles and possibly
  detector-stable hidden-sector particles that appear as $\met$.  The
  SM particles produced could be dominated by leptons, photons, or
  hadrons, leading to \emph{lepton-jets, photon-jets, or ``weird''
    high-multiplicity jets}.  We do not consider
  any of these final states in more detail.

\item Finally, in all of the decay topologies listed above, {\it
    displaced vertices are possible and should be considered in the
    LHC analyses}.  A simple example \cite{Strassler:2006im, Strassler:2006ri}  
 is $h\to 2 \to 4$, where the two particles produced in
  the Higgs decay are long-lived and decay far out in the detector; a similar signature arises in
R-parity violating supersymmetry \cite{Carpenter:2007zz}.  These signatures
offer opportunities for LHCb \cite{Strassler:2006ri,Carpenter:2007zz} as well as ATLAS and CMS, but we do not cover them here.  A number of relevant experimental searches have already been performed \cite{Abazov:2006as,Abulencia:2007ut,Aaltonen:2008dm,Abazov:2008zm,Abazov:2009ik,CMS:2011iwa,CMS:2011zva,Aad:2012kw,ATLAS:2012av,Chatrchyan:2012ir,LHCb:2012gja,CMS:2012rza,CMS:2012uza,Chatrchyan:2012jwg,Chatrchyan:2012jna,Aad:2012zx,ATLAS:2012ioa,CMS:2013oea}.

\end{itemize}

In the following sections we examine most of the above decay modes in
detail, outline their theoretical motivations, and review existing
collider studies and relevant experimental searches. For some channels
with significant discovery potential we also define benchmark models
that can be used to design future searches, obtain limits from already
performed searches, and/or perform collider studies to demonstrate how
much exclusion can be achieved with the extant LHC dataset.

\subsection{Theoretical Models for Exotic Higgs Decays}\label{subsec:theory}

In this section, we describe and review theoretical models that give rise to 
exotic Higgs decays.  
We begin with several ``simplified models'' (in the spirit of e.g.~\cite{Alves:2011wf}), which 
capture the essential ingredients that are involved in more complicated BSM models.  
It often makes sense to present experimental results in a simplified model framework, 
as only a few parameters are needed to capture the relevant details; for example,  
non-SM four-body decays of the Higgs of the form $h\to \phi \phi \to (f\bar f)(f'\bar f')$ (where $\phi$ is a
singlet particle and $f, f'$ are SM fermions) can be
parametrized merely by $m_h=125\,\gev$, $m_{\phi}$, ${\Br} (h\to\phi
\phi)$, and ${\Br} (\phi \to f\bar f)$. More parameters can be added if
the decays are displaced or involve multi-step cascades.

We discuss adding to the SM a scalar, one or two fermions, or a vector.  We also describe 
various two-Higgs-doublet (2HDM) models with the addition of a scalar.  
We then turn our attention to more complicated models that have ingredients similar to 
the simplified models, namely the MSSM, NMSSM, and Little Higgs models. 
Finally, we summarize the rich phenomenology possible in Hidden-Valley models. 

\subsubsection{SM + Scalar}
\label{SMS}
A particularly simple extension of the SM is to add to it one real scalar singlet $\sing$.
This model can easily produce non-trivial exotic Higgs decays, since 
1.) the Higgs can decay to  pair of singlets; and 2.) the singlet decays to SM particles
(by virtue of mixing with the Higgs).  Singlet scalars coupled to the Higgs also provide a well-known avenue for enhancing the electroweak phase transition in the early universe, which is a necessary ingredient for electroweak baryogenesis (see e.g. \cite{Morrissey:2012db}).
We describe this simple model below, as well as two small variations (one with more 
symmetry, one with a complex scalar), but all three models, as well as other variations, can yield essentially identical phenomenology.  
In \S\ref{2HDMS}, this will be generalized to two-Higgs-doublet models with a singlet. 

\vskip 8mm
\noindent {\bf Three Examples}

At the renormalizable level, gauge invariance allows the singlet $\sing$ to couple only to itself and to $H^\dagger H \equiv |H|^2$.  
The resulting potential is given by
\begin{equation}
   V(H,\sing)=V(H)+\hat V(\sing)+\acplg \, \sing \, |H|^2 + \frac{1}{2}\,
   \zeta \, \sing^2 \, |H|^2\,, \label{eq:H+real}
\end{equation}
where $\hat V(\sing)$ is a general quartic polynomial that may give $\sing$ a vacuum expectation value. The couplings $\acplg$ and $\zeta$ generate mixings between $H$ and $S$. Assuming those mixings are small, we identify the uneaten doublet degree of freedom to be the SM-like Higgs with $m_h = 125$~GeV and take the singlet field to have a mass below $m_h/2$. The small mixings give mass eigenstates $h$ and $s$, which are mostly doublet- and singlet-like, respectively. The decays $h \rightarrow s s$ are generated by an effective cubic term, and $s$ decays to SM particles via its doublet admixture. 

Imposing a $Z_2$ symmetry $\sing\to -\sing$, we can obtain a simpler version of this model with similar phenomenology.  
In this case, $\hat V(\sing)$ contains only quadratic and quartic terms and $\acplg=0$, e.g.
\begin{equation}\label{VSH}
   V(H,\sing)=-\mu^2\, |H|^2 -\half\,{\mu'}^2 \, \sing^2 + \lambda \, |H|^4 + \frac{1}{4} \, \kappa \, S^4 + \half\, 
   \zeta \, \sing^2\, |H|^2. 
\end{equation}
Depending on the choice of couplings, the potential may have a minimum at $\sing=0$, in which case the $Z_2$ is unbroken, there is no mixing between $H$ and $\sing$, and the $\sing$ does not decay; the coupling $\zeta $ induces the invisible decay $h\to s s$.  If the minimum instead has $\sing\neq 0$, then the $Z_2$ is broken, and the coupling $\zeta $ now not only produces a cubic term but also a quadratic term that allows $H$ and $\sing$ to mix.  In this case, the phenomenology is just as described in the previous paragraph, i.e.~$h\to ss$ for $m_s<m_h/2$,  
with $s$ decaying to SM particles. 

A third model, with essentially identical phenomenology, involves a theory with a {\it complex} scalar and an 
\emph{approximate} $U(1)$ global symmetry.\footnote{An exact $U(1)$ symmetry leads to invisible decays, while a spontaneously broken $U(1)$ gives 
rise to an unacceptable massless Nambu-Goldstone boson; a gauged $U(1)$ will be discussed in \S\ref{2HDMS}.}  
Here the scalar potential is as above, with $\sing$ now complex, and with a small $U(1)$ breaking part:
\begin{equation}\label{VSH01}
   V(H,\sing)=V_0(|H|^2,|\sing|^2) + V_1(|H|^2, \sing, \sing^\dagger)
\end{equation}
\begin{equation}
\label{VSH0}
V_0 =   -\mu^2\, |H|^2 -\mu'^2 \, |\sing|^2 +   \lambda \, |H|^4 +  \kappa \, |\sing|^4 +
   \zeta \,  |\sing|^2|H|^2 
\end{equation}
\begin{equation}
V_1 = ( \rho + \xi_\sing \, |\sing|^2 +\xi_H \, |H|^2)\, \sing + {\rm \ hermitean \ conjugate }\ +\ {\mathrm{other \ terms}} 
\end{equation}
where we have chosen not to consider the most general $V_1$ for illustration purposes. If the potential is such that $\sing$ develops 
a non-zero vacuum expectation value, the spectrum consists of a massive scalar $\sing$ and a light pseudo-Nambu-Goldstone boson $a$ with mass $m_a$.  If $m_s > \half m_h > m_a$, then $h\to aa$ is possible, which is an invisible decay 
unless the $U(1)$-violating terms also violate charge conjugation.  In that case, $a$ can mix with the massive state $s$, which in turn mixes with $H$ as in previous examples, allowing the $a$ to decay to SM particles, with couplings inherited from $H$.  

\vskip 2mm
\noindent {\bf Phenomenology}

After electroweak symmetry breaking there are two relevant \emph{mass-eigenstates}: the SM-like scalar $h$ at 125 GeV containing a small admixture of $S$, and the mostly-singlet scalar $s$ containing a small admixture of $H$. The phenomenology of all three variants above is the same, as far as decays of the form $h \rightarrow s s \rightarrow \mathrm{SM}$ are concerned. It can be captured in terms of three parameters:
\ben
\item
The effective Lagrangian contains a term of the form $\mu_v\, h\, s\, s$, which gives $h \rightarrow s s$ with $\BR(h\to\,{\rm exotic})$ determined by $\mu_v$.  
\item 
The singlet's mass $m_s$ affects $\BR(h\to\,{\rm exotic})$ and the type of SM final states available for $s \rightarrow \mathrm{SM}$.  
\item
The mixing angle between $S$ and $H$, denoted here by $\theta_{\sing}$, determines the overall width of $s\to{\rm SM}$.
If $s$ cannot decay to other non-SM fields, $\theta_{\sing}$ controls its lifetime.
\een
Apart from these continuous parameters, the parity of $s$ also
affects the partial widths to different final states, mostly near thresholds. Note that the total width of $s$ is usually not important 
for phenomenology if it decays promptly.  However, the lifetime of $s$ is
macroscopic ($c\tau\sim$ meters) if $\theta \lesssim 10^{-6}$.
This possibility is technically natural and thus the experimental search for displaced vertices deserves serious consideration \cite{Strassler:2006ri}; 
however, we do not discuss this further here.  
Therefore, for a large part of parameter space, only $\mu_v$ and $m_s$ is relevant for collider phenomenology as this 
fixes $\BR(h \rightarrow s s)$ and $\BR(s \rightarrow \mathrm{SM})$.  

The partial width for exotic Higgs decays is given by
\begin{equation}
\label{eq:GammahssSMS}
\Gamma(h\rightarrow ss)\; =\; \frac{1}{8 \pi} \frac{\mu_v^2}{m_h} \sqrt{1- \frac{4 m_s^2}{m_h^2}} 
\; \approx\;
\left(\frac{\mu_v/v}{0.015}\right)^2 \Gamma(h \rightarrow \mathrm{SM})\,,
\end{equation}
where the last step assumes $m_{s}\ll m_h/2$. 
Therefore, the new branching ratio is $\ord(1)$ even for small values of $\mu_v/v$.
This is not surprising, if we recall that in the SM the bottom quark takes up almost
$60\%$ of the total width although its Yukawa coupling is only  $\sim 0.017$. 
\begin{figure}[t]
\begin{center}
\includegraphics[width=7cm]{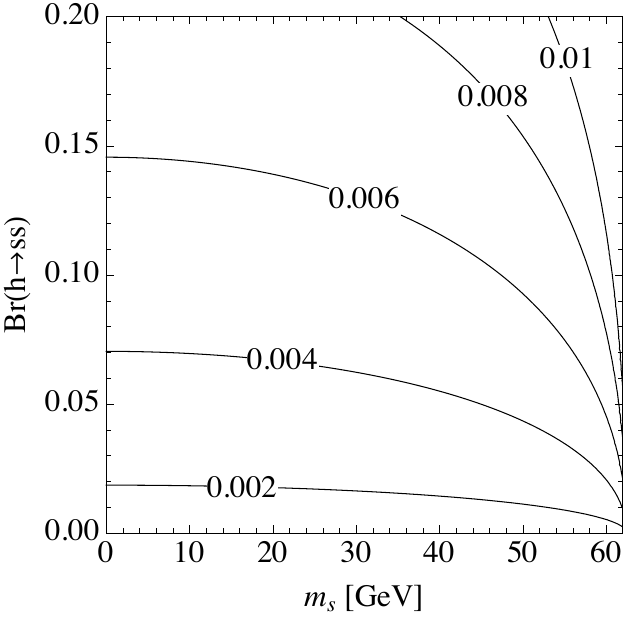}
\caption{\small Size of the cubic coupling $\mu_v$ in units of Higgs expectation value $v$ to yield the indicated $h \rightarrow s s$ branching fraction as a function of singlet mass, as given by~\Eqref{GammahssSMS}.
}
\label{fig:BRexSMS}
\end{center}
\end{figure}
In~\figref{BRexSMS}, we show contours of $\mu_v/v$ in the $\BR(h\to ss)$ versus $m_s$ plane.  

\begin{figure}
\begin{center}
\hspace*{-12mm}
\begin{tabular}{ccc}
\includegraphics[width=0.465\textwidth]{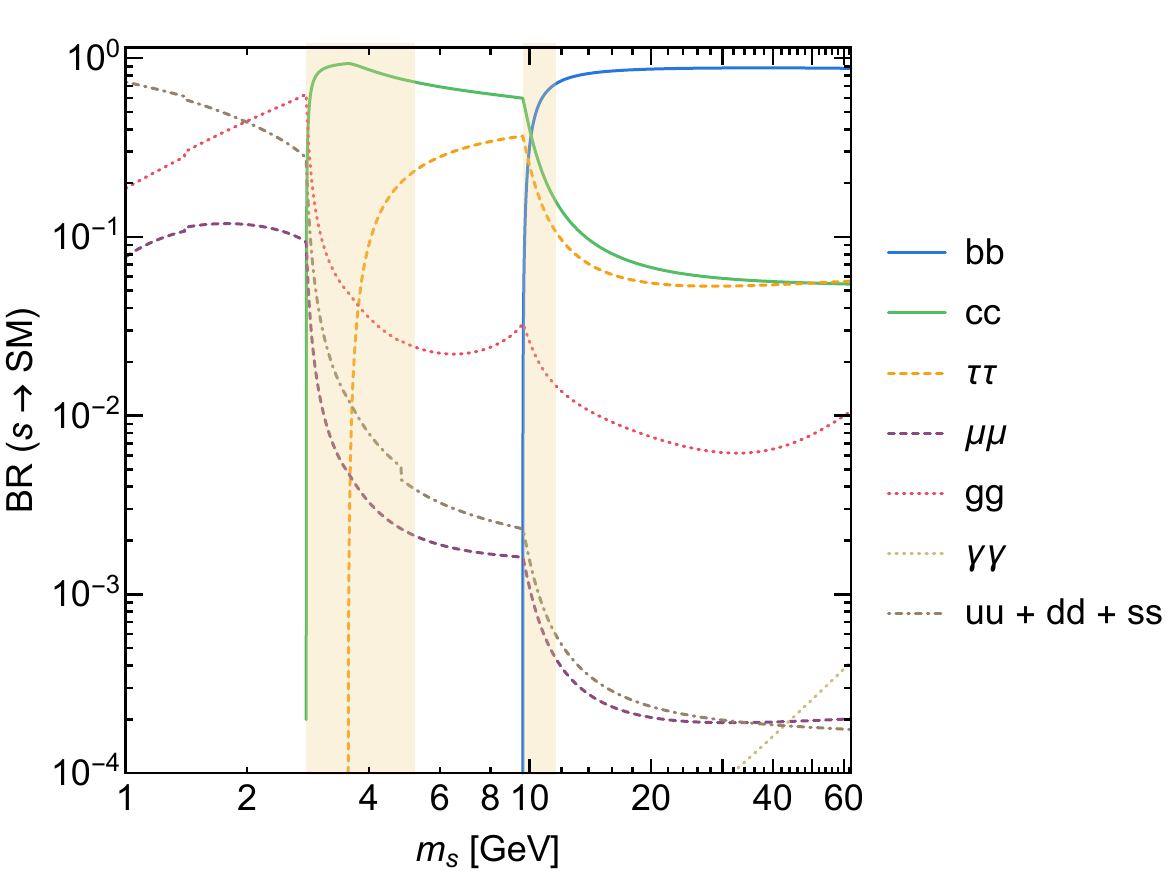}
\includegraphics[width=0.42\textwidth]{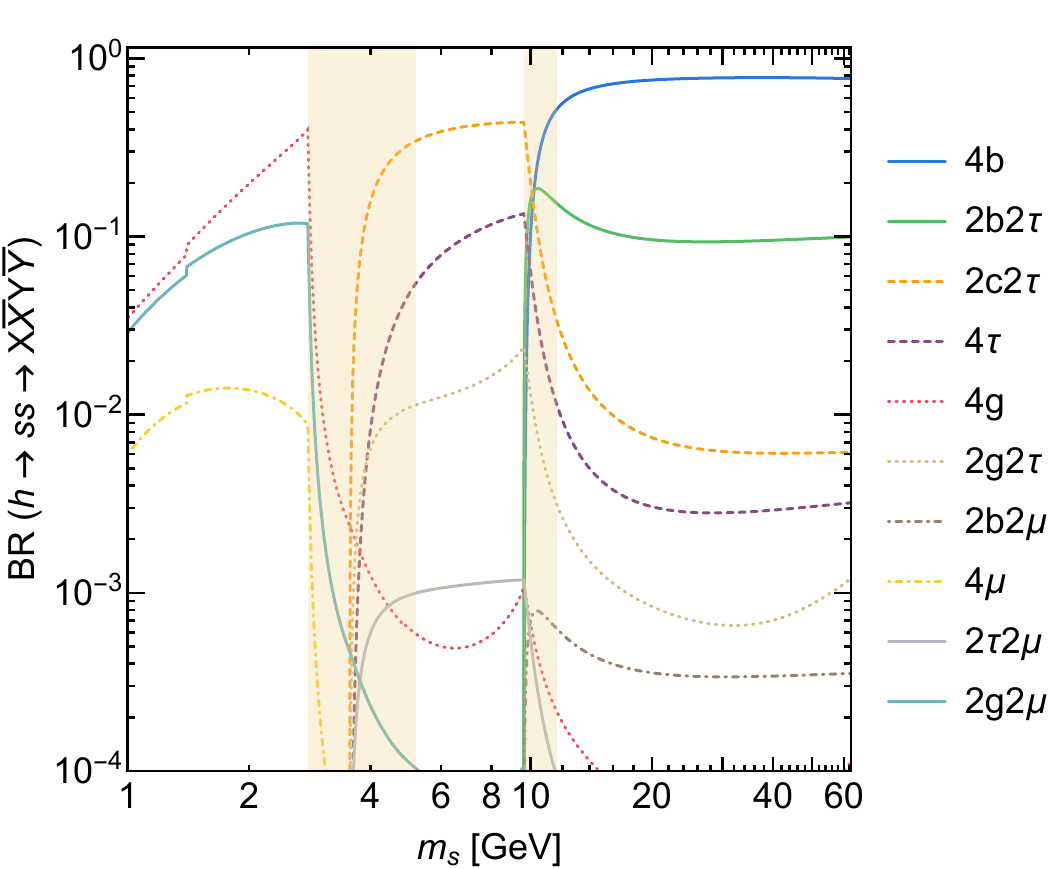}
\end{tabular}
\caption{\small {\bf Left}: Branching ratios of a CP-even scalar singlet to SM particles, as function
of $m_{s}$.
{\bf Right}: Branching ratios of exotic decays of the 125~GeV Higgs boson as function of $m_s$, in the \emph{SM + Scalar} model described in the text, 
scaled to $\BR(h\to ss) = 1$. Hadronization effects likely invalidate our simple calculation in the shaded regions. 
\label{fig:BRsSMS}}
\end{center}
\end{figure}

The individual partial widths of the singlet $s$ to SM particles are readily computed using existing calculations for Higgs decays, e.g.~\cite{Djouadi:2005p17109, Djouadi:2005p17157}. Decays into $W^*W^*$ and $Z^*Z^*$ are negligible for $m_{s}<m_h/2$. At lowest order, the partial decay width to fermions is given by 
\begin{equation}
\Gamma(s \rightarrow f \bar f) = \sin^2 \theta_S \frac{N_c}{8 \pi} \frac{m_s m_f^2}{v^2}  \beta_f^3\,,
\end{equation}
where $\beta_f = \sqrt{1-4m_f^2/m_s^2}$ and $N_c$ is the number of colors, equaling $3\;(1)$ for quarks (leptons). For the pseudoscalar singlet state $a$, $\beta_f^3$ is replaced by $\beta_f$. The mixing suppression $\sin^2 \theta_S$ is common to all partial widths, including those to gluons and photons, and thus does not affect branching ratios if $s$ only decays to SM particles. 
$\BR(s\to {\rm SM})$ and $\BR(h\to ss \to {\rm SM})$ are shown for $m_s > 1 \gev$ in \figref{BRsSMS} on the {\bf left} and {\bf right}, respectively. 
It is clear that a simple singlet extension of the SM generically implies significant branching ratios of exotic Higgs decays to 4 SM objects.

The theoretical calculations become increasingly inaccurate as $m_s$ is lowered to $\sim 1$~GeV, where perturbative QCD breaks down, or when $m_s$ is close to a hadronic resonance, 
which can enhance the decay rates~\cite{gunion2000higgs}. 
Decays to quarkonium states are suppressed for $s$ but may be important for $a$.
For $m_s < 1 \gev$ and above the pion threshold, partial widths have to be computed within a low energy effective theory of QCD, such as soft-pion theory or the chiral Lagrangian method. 
Nevertheless, it is clear that the dominant decay of the singlet is to some combination of hadrons, which are boosted due to the large mass difference between the singlet and $h$. 
The resulting two-track jet may look like a low-quality hadronic $\tau$-decay. 
Between the muon and pion thresholds ($210 \mev \lesssim m_s \lesssim 270 \mev$), the dominant decay is to $\mu^+\mu^-$, while for $m_s \lesssim 210~\mev$, the 
dominant decay is to $e^+e^-$. Photons are the only possible final state for $m_s < 2\, m_e$, in which case the scalar is detector-stable.

Further details of the branching ratio calculation can be found in \S\ref{2HDMS} and Appendix~\ref{sec:2HDM+Sapp}, which also includes a more detailed discussion of pseudoscalar decays.

For $m_s \lesssim 2m_b$, the $s\bar bb$ coupling can in principle be probed by bottomonium decay~\cite{Dermisek:2010mg, Echenard:2012hq}. The strongest limits are $\Br(\Upsilon(1S) \to \gamma \tau^+ \tau^-) \lesssim 10^{-5}$ by BaBar \cite{Lees:2012te}, which constraints the Yukawa coupling to satisfy $y_{sbb} \lesssim 0.4$ for $\Br(s\rightarrow \tau^+ \tau^-) = 1$ \cite{McKeen:2008gd, Lisanti:2009uy}. In the SM+S scenario, $y_{sbb} = \sin \theta_S~y_{hbb}$ with $y_{hbb} \approx 0.02$ in the SM. Clearly the Upsilon decay measurement provides no meaningful bounds on singlet extensions. Similar arguments apply to pseudoscalars, and hence the 2HDM+S and NMSSM in the next sections.

%
\subsubsection{2HDM (+ Scalar)}
\label{2HDMS}

The SM Higgs sector is made up of a single $SU(2)_L$ doublet $H$ with hypercharge $Y = +\frac{1}{2}$, denoted by $H \sim 2_{+1/2}$. 
Adding a doublet to this minimal picture is one of the simplest extensions of the Higgs sector compatible with a $\rho$-parameter close to 1. Such extensions are found in several well-motivated theories, such as supersymmetry~\cite{Haber:1984rc} and axion models~\cite{Kim:1986ax, Peccei:1977hh}, where holomorphy and the Peccei-Quinn symmetry, respectively, necessitate an additional doublet; theories of electroweak baryogenesis, which might be made viable with additional doublets~\cite{Trodden:1998qg}; and grand unified models~\cite{gunion2000higgs}. For this reason, it makes sense to define the most general Two-Higgs Doublet Model (2HDM) and study it in detail (for a comprehensive review, see e.g.~\cite{Branco:2011iw}; for a discussion on the impact of recent SM-like Higgs boson discovery, see e.g.~\cite{Chen:2013jvg}).
Below we will then add a light scalar to the 2HDM to obtain a rich set of exotic Higgs decays.  

The most general 2HDM Higgs potential 
is given by~\cite{gunion2000higgs}
\begin{eqnarray}
V & = & m_1^2 |H_1|^2 + m_2^2 |H_2|^2 + \frac{\lambda_1}{2} |H_1|^2 + \frac{\lambda_2}{2} |H_2|^2 + \lambda_3
|H_1|^2|H_2|^2 + \lambda_4 |H_1^\dagger H_2|^2 + \\
\nonumber 
&&\frac{\lambda_5}{2} \left( (H_1 H_2)^2 + \cc \right) + m_{12}^2 \left( H_1 H_2 + \cc \right) + \\  \nonumber 
&& \left( \lambda_6 |H_1|^2 (H_1 H_2) + \cc \right) +  \left(\lambda_7 |H_2|^2 (H_1 H_2) + \cc \right)~.    
\end{eqnarray}
We choose the charges of the Higgs fields such that $H_1 \sim 2_{-1/2}$ and $H_2 \sim 2_{+1/2}$. 
Note that we choose conventions that differ slightly from the ``standard'' 
conventions of~\cite{gunion2000higgs,Branco:2011iw}; this will simplify the transition to supersymmetry  
models below.\footnote {To recover the conventions of~\cite{gunion2000higgs} set $\Phi_2 = H_2, \ \Phi_1 = i \sigma^2 H_1^*$.}
The scalar doublets $H_{1,2}$ acquire vacuum expectation values $v_{1,2}$, which we assume here are real and aligned. Expanding around the minima yields two complex and four real degrees of freedom 
\begin{equation}
H_1 = \frac{1}{\sqrt{2}}\left( \begin{array}{c}
v_1 + H_{1,R}^0 + i H_{1,I}^0 \\
H_{1, R}^- + i H_{1,I}^-
\end{array}
\right), \ \ \ 
H_2 =  \frac{1}{\sqrt{2}}\left( \begin{array}{c}
H^+_{2,R} + i H_{2,I}^+ \\
v_2 + H_{2,R}^0 +  i H_{2, I}^0
\end{array}
\right)~. 
\end{equation}
The charged scalar and pseudoscalar mass matrices are diagonalized by a rotation angle $\beta$, defined as $\tan \beta = v_2/v_1$. 
One charged (complex) field and one neutral pseudoscalar combination of $H^0_{1,2,\, I}$ are eaten by the SM gauge bosons after electroweak symmetry breaking. 
The other complex field yields two charged mass eigenstates, $H^\pm$, which we assume are heavy and will thus play no further role in our discussions. 
The surviving three real degrees of freedom yield one neutral pseudoscalar mass eigenstate,  
\begin{equation}
A = H^0_{1,I} \sin \beta - H^0_{2,I} \cos \beta\,,
\end{equation}
and two neutral scalar mass eigenstates,   
\begin{equation}
\label{eq:hH2HDM}
\left(
\begin{array}{c}
h \\ H^0
\end{array}\right)
=
\left(
\begin{array}{cc}
- \sin \alpha & \cos \alpha \\
\cos \alpha & \sin \alpha 
\end{array}
\right)
\left(
\begin{array}{c}
H^0_{1,R} \\ H^0_{2,R}
\end{array}\right)\;,
\end{equation}
where\footnote{Contrast this to the MSSM Higgs potential, where $-\pi/2\leq\alpha\leq 0$.} $-\pi/2 \leq \alpha \leq \pi/2$.  
Our notation anticipates the assumption below that the model is in a decoupling limit, so that $h$ is the SM-like Higgs and $H^0$ is the other, heavier, scalar.  

Allowing the most general Yukawa couplings to fermions would result in large Flavor-Changing Neutral Currents (FCNCs). This can be avoided by imposing $\mathbb{Z}_2$ symmetries to ensure that fermions with the same quantum numbers all couple to only one Higgs field. This results in four ``standard" types of fermion couplings commonly discussed in the literature: Type I (all fermions couple 
to $H_2$), Type II (MSSM-like, $d_R$ and $e_R$  couple to $H_1$, $u_R$ to $H_2$), Type III (lepton-specific, leptons/quarks couple to $H_1$/$H_2$ respectively) and Type IV (flipped, with $u_R, e_R$ coupling to $H_2$ and $d_R$ to $H_1$). 
The couplings of the $h,\ H^0$, and $A$ mass eigenstates to fermions and gauge fields relative to the SM Higgs couplings are summarized in \tabref{2HDMcoupling}.\footnote{More general fermion couplings are possible within the framework of Minimal Flavor 
Violation~\cite{Dery:2013aba,  Altmannshofer:2012ar}. We do not discuss this case here since we use the 2HDM to illustrate a range of possible exotic Higgs decay signatures, which would not be qualitatively different in the MFV scenarios.} 

\begin{table}
   \centering
\begin{tabular}{ cccccc }
\hline
  & Couplings & I & II  & III (Lepton specific) & IV (Flipped) \\ \hline
\multirow{5}{*}{$h$} 
& $g_{h VV}$ & $\sin (\beta-\alpha)$ &$ \sin (\beta-\alpha)$ &$ \sin (\beta-\alpha)$ &$ \sin (\beta-\alpha)$ \\
 & $g_{h t \bar t}$ & $ \cos\alpha/\sin\beta$& $ \cos\alpha/\sin\beta$ &$ \cos\alpha/\sin\beta$&$ \cos\alpha/\sin\beta$ \\
 &  $g_{h b \bar b}$ & $ \cos\alpha/\sin\beta$&$- \sin\alpha/\cos\beta$&$ \cos\alpha/\sin\beta$&$- \sin\alpha/\cos\beta$ \\
 &  $g_{h \tau \bar \tau}$ & $ \cos\alpha/\sin\beta$&$- \sin\alpha/\cos\beta$&$- \sin\alpha/\cos\beta$&$ \cos\alpha/\sin\beta$ \\  
  \hline
 \multirow{5}{*}{$H^0$} 
& $g_{H^0 VV}$ & $\cos (\beta-\alpha)$ &$ \cos (\beta-\alpha)$ &$ \cos (\beta-\alpha)$ &$ \cos (\beta-\alpha)$ \\
 & $g_{H^0 t \bar t}$ & $ \sin\alpha/\sin\beta$& $ \sin\alpha/\sin\beta$ &$ \sin\alpha/\sin\beta$&
$ \sin\alpha/\sin\beta$ \\
 &  $g_{H^0 b \bar b}$ & $ \sin\alpha/\sin\beta$&$ \cos\alpha/\cos\beta$&$ \sin\alpha/\sin\beta$&$ 
\cos\alpha/\cos\beta$ \\
 &  $g_{H^0 \tau \bar \tau}$ & $ \sin\alpha/\sin\beta$&$\cos\alpha/\cos\beta$&$ \cos\alpha/\cos\beta$&
$ \sin\alpha/\sin\beta$ \\  
 \hline

\multirow{4}{*}{$A$} 
 & $g_{A VV}$ & 0&0&0&0 \\
 & $g_{A t \bar t}$  & $ \cot \beta$ &  $ \cot \beta$ &$ \cot \beta$ &$ \cot \beta$ \\
 & $g_{A b \bar b}$ & $-  \cot \beta$ & $ \tan \beta$ &$- \cot \beta$ & $ \tan \beta$ \\
 & $g_{A \tau \bar \tau}$ & $-  \cot \beta$ & $ \tan \beta$& $ \tan \beta$ &$- \cot \beta$\\ \hline
\end{tabular}
   \caption{Couplings of the neutral scalar and pseudoscalar mass eigenstates in the four types of 2HDM with a $\mathbb{Z}_2$ symmetry, following the notation of \cite{Craig:2012pu}. The couplings are normalized to those of the SM Higgs.}
   \label{tab:2HDMcoupling}
\end{table}

In general, 2HDMs could allow for exotic decays of the 125~GeV state of the form $h \rightarrow A A$, $H^0 \rightarrow h h, A A$ or 
$h \rightarrow Z A$ (where we temporarily identified the 125~GeV state with either $h$ or $H^0$), where the daughter (pseudo)scalars decay to SM fermions or gauge bosons. 
However, while this possibility can be realized in certain corners of parameter space, 2HDMs are by now too constrained from existing data~\cite{Gunion:1997aq, Belanger:2013xza} to allow for a 
wide variety of exotic Higgs decay phenomenology.

These restrictions are easily avoided as follows.  First, we assume the 2HDM is near or in the decoupling limit, 
\begin{equation}
\alpha \rightarrow \beta - \pi/2 \, , 
\end{equation}
where the lightest state in the 2HDM is $h$, which we identify with the observed 125 GeV state.  
In this limit, the fermion couplings of $h$ also become identical to the SM Higgs, while the gauge boson couplings are very close to SM-like for $\tan \beta \gtrsim 5$. 
All of the properties of $h$ are determined by just two parameters, $\tan \beta$ and $\alpha$, and the type of fermion couplings. 
The remaining parameters, which control the rest of the Higgs spectrum and its phenomenology, are in general constrained by the measured production and decays of 
$h$~\cite{Ferreira:2012my, Alves:2012ez, Ferreira:2011p17387, Craig:2012vn, Craig:2012pu, Craig:2012bs,Bai:2012ex, Azatov:2012qz, Dobrescu:2012td, Chen:2013p17161}, 
but plenty of viable parameter space exists in the decoupling limit.  

Second, we add to the 2HDM one complex scalar singlet,  
$$S = \frac{1}{\sqrt{2}}(S_R + i S_I)\,,$$ 
which may attain a vacuum expectation value that we implicitly expand around. 
This singlet only couples to $H_{1,2}$ in the potential and has no direct Yukawa couplings, acquiring all of its couplings to SM fermions through its mixing with $H_{1,2}$.  This mixing needs to be small to avoid spoiling the SM-like nature of $h$.  

Under these two simple assumptions, exotic Higgs decays of the form
\begin{equation}
h \rightarrow s s  \rightarrow X \bar X Y \bar Y \ \ \ \ \mathrm{or} \ \ \ \ \ h \rightarrow  a a \rightarrow X \bar X Y \bar Y
\end{equation}
as well as 
\begin{equation}
h \rightarrow a Z \rightarrow X \bar X Y \bar Y
\end{equation}
are possible, where $s(a)$ is a (pseudo)scalar mass eigenstates mostly composed of $S_R(S_I)$ and $X, Y$ are SM fermions or gauge bosons. We refer to this setup as the 2HDM+S.
For Type II 2HDM+S, a light $a$ corresponds roughly to the R-symmetry limit of the NMSSM (see section~\ref{NMSSMscalar}). However, the more general 2HDM framework allows for exotic Higgs decay phenomenologies that are much more diverse than those usually considered in an NMSSM-type setup. 

To incorporate the already analyzed constraints on 2HDMs into the 2HDM+S (e.g.~\cite{Chen:2013p17161}), one can imagine adding a decoupled singlet sector to a 2HDM with $\alpha, \beta$ chosen so as to not yet be excluded.\footnote{As we have pointed out in \S\ref{SMS}, bottomonium decays provide no meaningful constraint on the 2HDM+S scenario.} The real and imaginary components of $S$ can be given separate masses, and small mixings to the 2HDM sector can then be introduced as a perturbation. Approximately the same constraints on $\alpha, \beta$ apply to this 2HDM+S, as long as $\mathrm{Br}(h \rightarrow ss/aa/Za) \lesssim 10\%$. This allows for a wide range of possible exotic Higgs decays. There are some important differences depending on whether the lightest singlet state with a mass below $m_h/2$ is scalar or pseudoscalar. We will discuss them in turn.

\begin{figure}
\begin{center}
\includegraphics[width=0.52\textwidth]{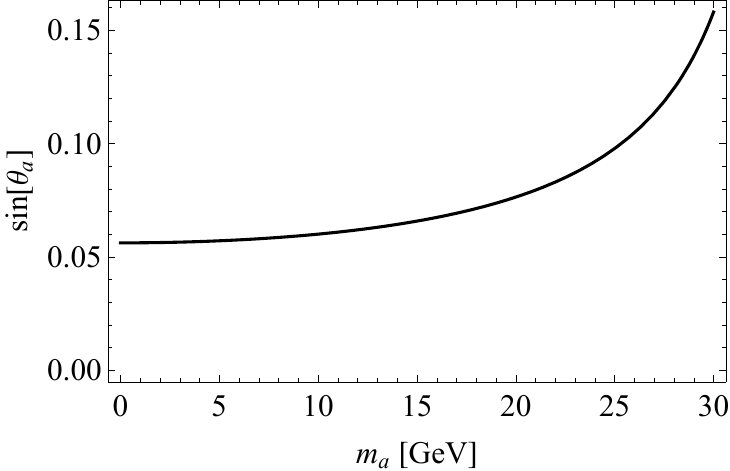}
\end{center}
\caption{
Required mixing angle between the doublet and singlet-sector pseudoscalar for $\mathrm{Br}(h \rightarrow a Z) = 10\%$, assuming no other exotic Higgs decays and $\alpha = \pi/2 - \beta$ (decoupling limit).
}
\label{fig:2HDMShzareqmix}
\end{figure}

\vskip 2mm
\paragraph*{\bf Light Pseudoscalar ($a$)}
There are two pseudoscalar states in the 2HDM+S, one that is mostly $A$ and one that is mostly $S_I$. One can choose the mostly-singlet-like pseudoscalar
\begin{equation}
\label{eq:2HDMSpseudoscalar}
a = \cos \theta_a S_I + \sin \theta_a A \ \ , \ \ \ \ \theta_a \ll 1,
\end{equation}
to be lighter than the SM-like Higgs. There are two possible exotic Higgs decays: $h \rightarrow Z a$ for $m_a < m_h - m_Z \approx 35 \gev$ and $h \rightarrow a a$ for $m_a < m_h/2 \approx 63 \gev$. 

The partial width $\Gamma(h\rightarrow Z a)$ is entirely fixed by the 2HDM parameters $\alpha, \beta$ and the mixing angle $\theta_a$. 
The relevant interaction term in the effective Lagrangian is 
\beq
\label{eq:2HDMShzalagrangian}
\mathcal{L}_\mathrm{eff}\supset g_\mathrm{eff} (a\partial^\mu h - h \partial^\mu a) Z_\mu  \,,  \ \ \ \mathrm{where} \ \ \ \ \ 
g_\mathrm{eff}=\sqrt {\frac {g^2+g'^2} {2}}\, \sin(\alpha-\beta) \, \sin{\theta_a},
\eeq
which gives 
\begin{equation}
\Gamma(h\rightarrow Z a) = \frac{g_\mathrm{eff}^2}{16 \pi} \frac{[ 
(m_h + m_Z + m_a) (m_h - m_Z + m_a) (m_h + m_Z - m_a) (m_h - m_Z - m_a)]^{3/2}}{m_h^3 m_Z^2}.
\end{equation}
\figref{2HDMShzareqmix} shows that $\theta_a \sim 0.1$ gives $\mathrm{Br}(h \rightarrow Za) \sim 10\%$ in the absence of other exotic decays.

Two terms in the effective Lagrangian give rise to $h\rightarrow aa$ decays:
\begin{equation}
\mathcal{L}_\mathrm{eff} \ \supset \  g_{hAA} \ h A A  \ + \  \lambda_S |S^2|^2 \  .
\end{equation}
In terms of mass eigenstates, this contains
\begin{equation}
\mathcal{L}_\mathrm{eff} \  \ \supset \ \  g_{hAA} \ \sin^2 \theta_a \ h a a \ \ + \ \  4 \, \lambda_S \, v_s \, \sin{\zeta_1} \, \cos^2 \theta_a \  h a a \ ,
\end{equation}
where $\langle S \rangle = v_s$ is the singlet vacuum expectation value, and the (presumably small) mixing angle $\zeta_1$ determines the singlet scalar content of the SM-like Higgs, see \Eqref{2HDMSscalarorig}. The first term by itself can easily give rise to $\mathrm{Br}(h \rightarrow a a) \sim 10\%$ if $g_{hAA}\sim v$ and $\theta_s \sim 0.1$, see~\figref{BRexSMS}. 
(\figref{BRexSMS} shows the results for Higgs partial widths to scalars, but these are almost identical to pseudoscalars, except near threshold.) The additional contribution from the second term (even without a singlet scalar below the Higgs mass) means that $\mathrm{Br}(h \rightarrow a a)$ and $\mathrm{Br}(h \rightarrow Z a)$ can be independently adjusted.

The decay of $a$ to SM fermions proceeds via the $A$ couplings in~\tabref{2HDMcoupling}, multiplied by $\sin \theta_a$.  Therefore, once the type of 2HDM model has been specified, 
the exotic Higgs decay phenomenology is entirely dictated by the two exotic branching ratios $\mathrm{Br}(h \rightarrow a a)$ and $\mathrm{Br}(h \rightarrow Z a)$,  as well as $\tan \beta$, which determines $a$'s fermion couplings.  Perturbative unitarity of the Yukawa couplings sets a lower bound of $\tan \beta > 0.28$~\cite{Chen:2013p17161}; we will show results for 
$\tan \beta$ as low as $\sim 0.5$. 

\begin{figure}
\begin{center}
\includegraphics[width=9cm]{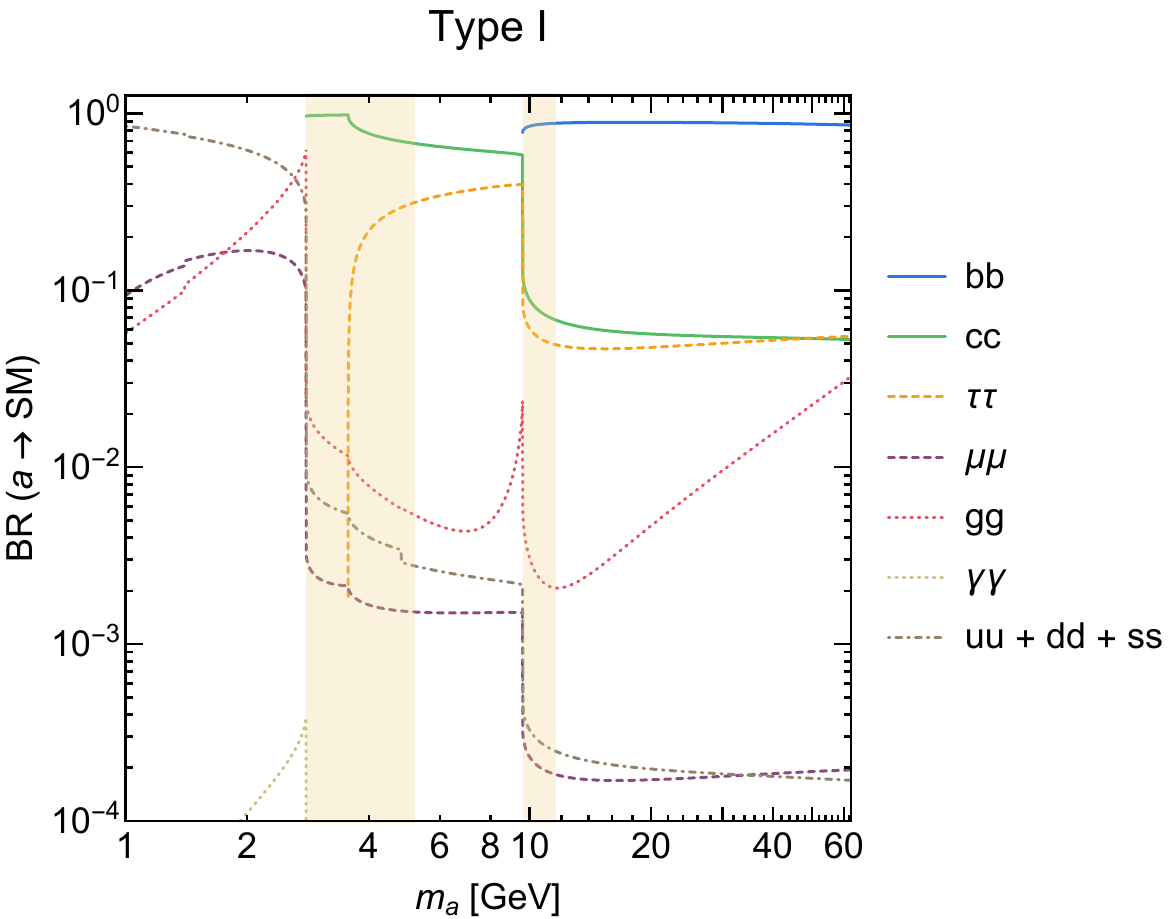}
\end{center}
\caption{
Branching ratios of a singlet-like pseudoscalar in the 2HDM+S for Type I Yukawa couplings. Decays to quarkonia likely invalidate our simple calculations in the shaded regions. 
}
\label{fig:fBRs2HDMatype1}
\end{figure}

\begin{figure}
\begin{center}
\includegraphics[width=8cm]{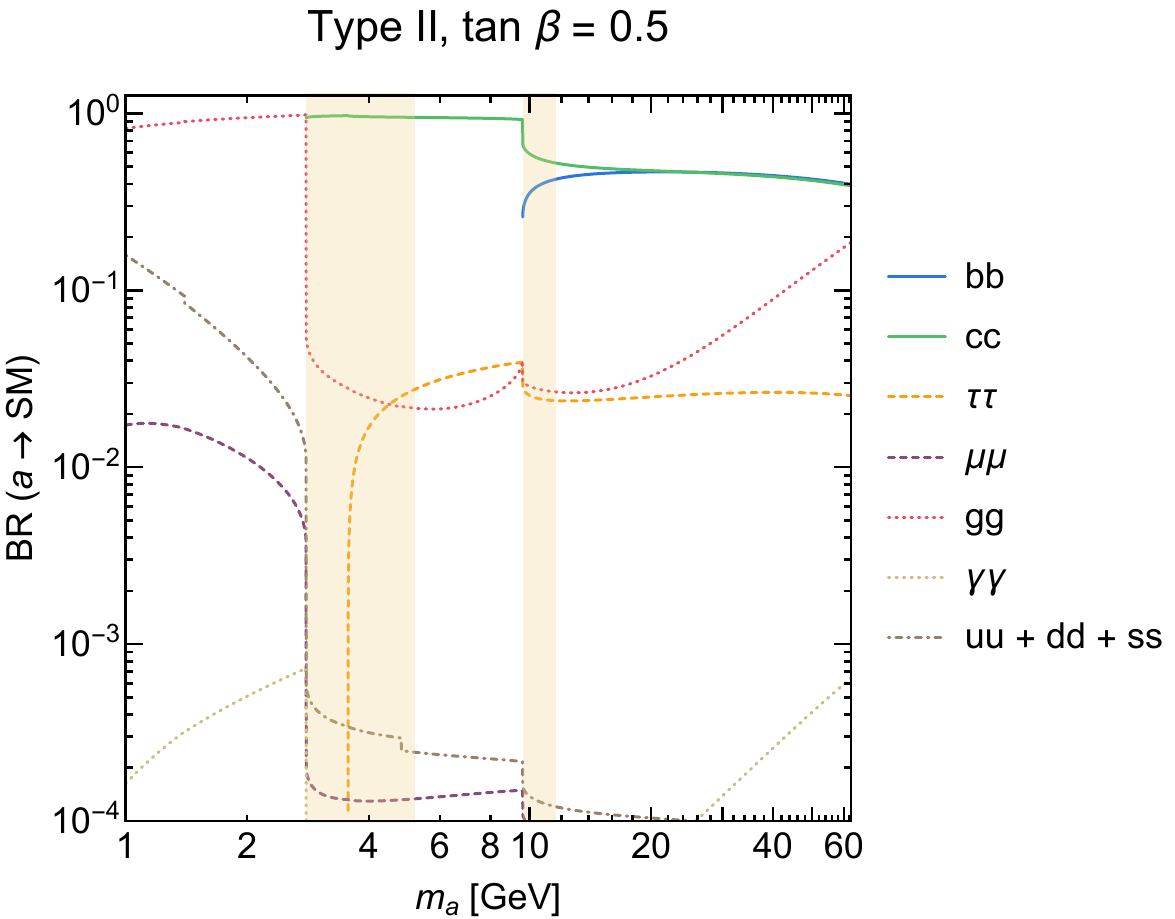}
\includegraphics[width=8cm]{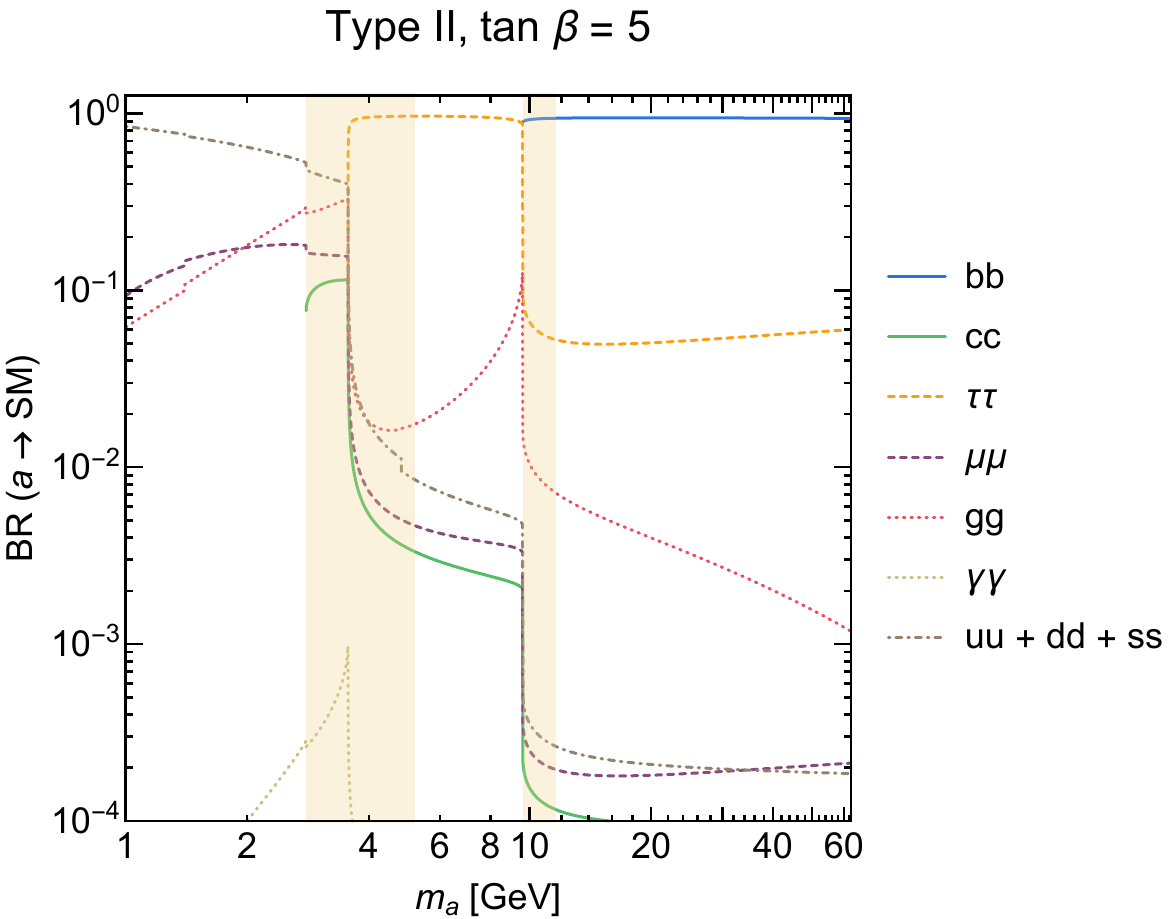}
\end{center}
\caption{
Branching ratios of a singlet-like pseudoscalar in the 2HDM+S for Type II Yukawa couplings. Decays to quarkonia likely invalidate our simple calculations in the shaded regions. 
}
\label{fig:fBRs2HDMatype2}
\end{figure}

\begin{figure}
\begin{center}
\includegraphics[width=8cm]{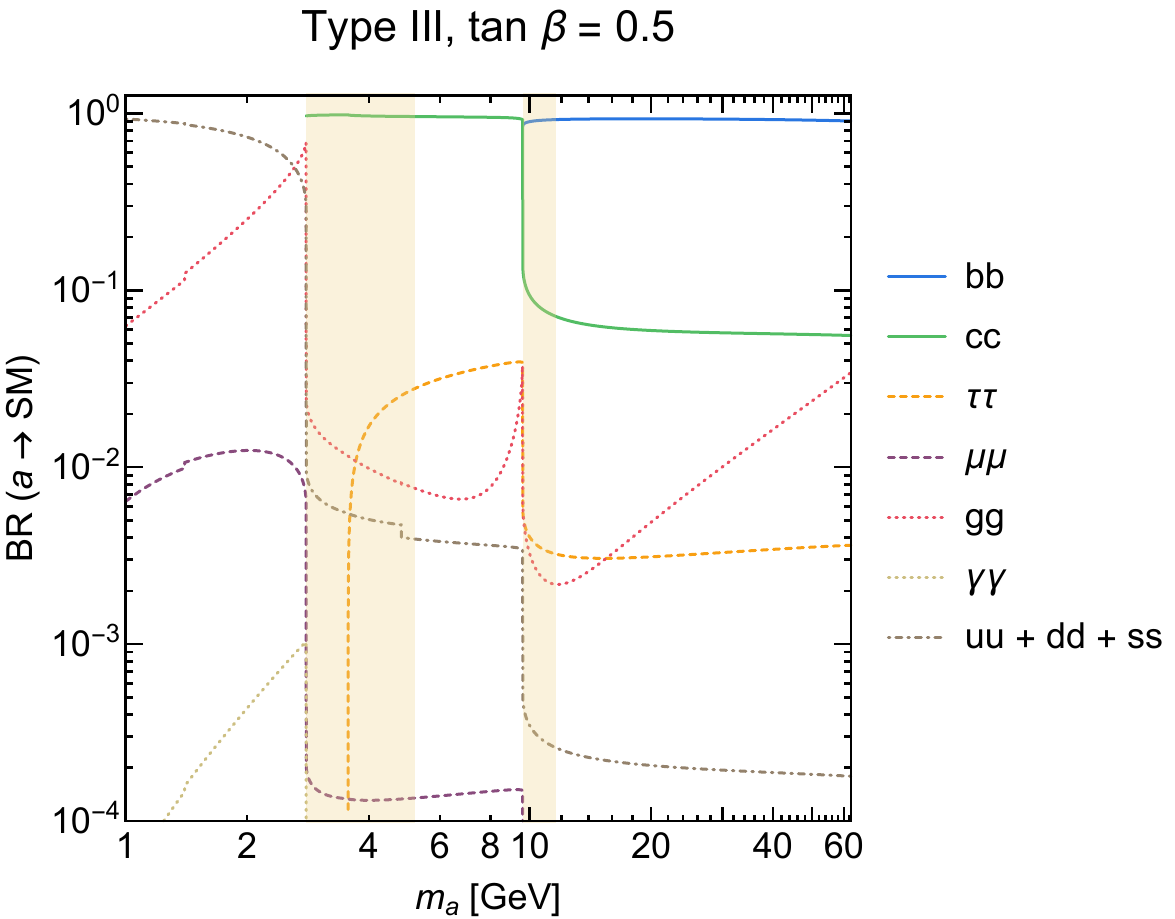}
\includegraphics[width=8cm]{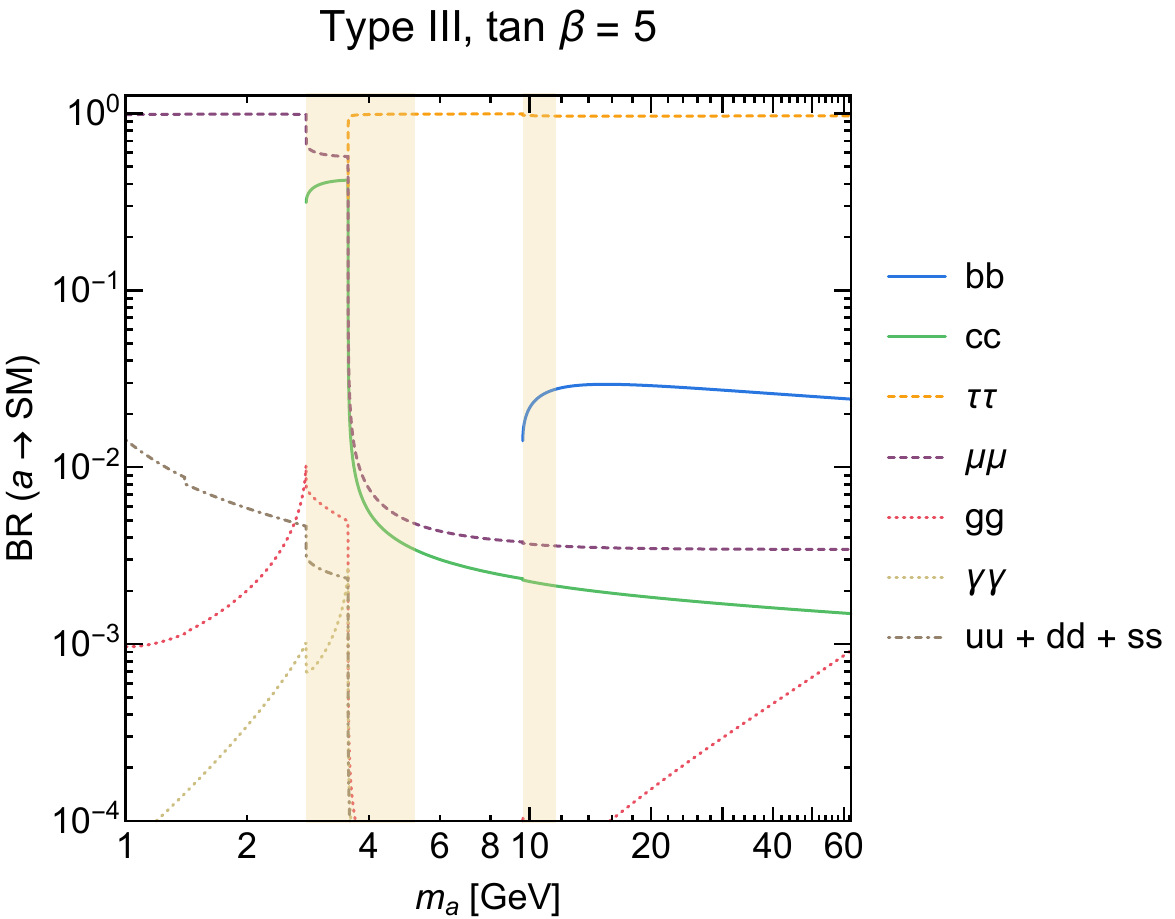}
\end{center}
\caption{
Branching ratios of a singlet-like pseudoscalar in the 2HDM+S for Type III Yukawa couplings. Decays to quarkonia likely invalidate our simple calculations in the shaded regions. 
}
\label{fig:fBRs2HDMatype3}
\end{figure}

\begin{figure}
\begin{center}
\includegraphics[width=8cm]{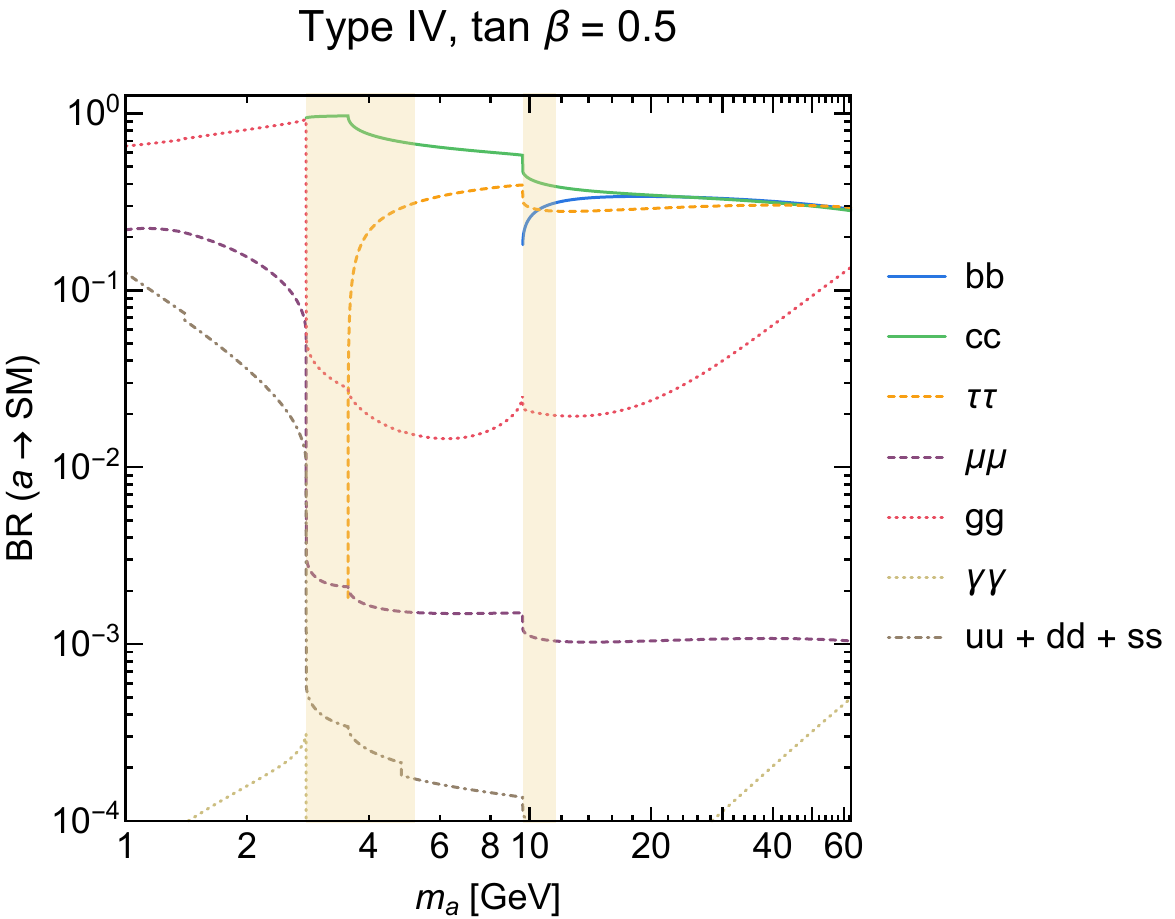}
\includegraphics[width=8cm]{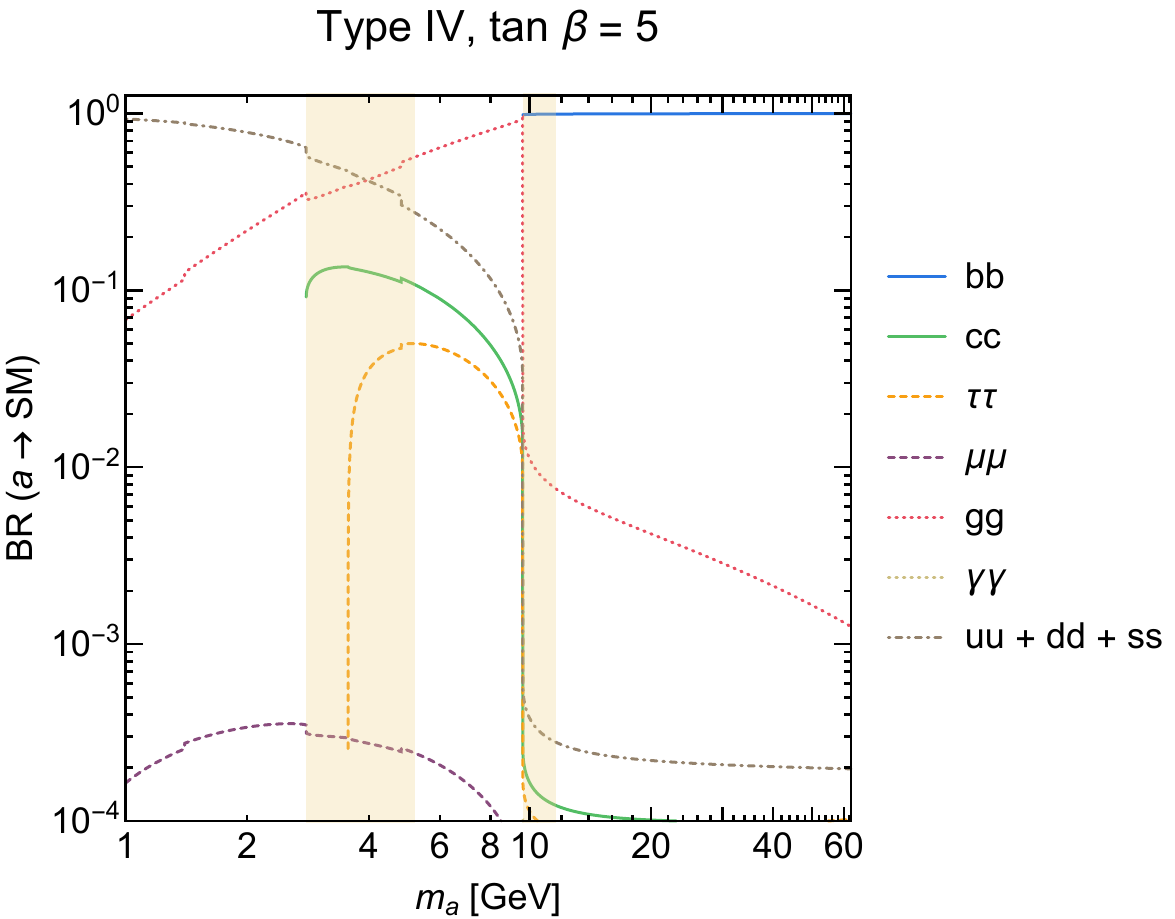}
\end{center}
\caption{
Branching ratios of a singlet-like pseudoscalar in the 2HDM+S for Type IV Yukawa couplings. Decays to quarkonia likely invalidate our simple calculations in the shaded regions. 
}
\label{fig:fBRs2HDMatype4}
\end{figure}

In Figs.~\ref{fig:fBRs2HDMatype2}--\ref{fig:fBRs2HDMatype4}, we show $\BR(a \rightarrow X \bar X)$, where $X$ is a SM particle.  These include $\mathcal{O}(\alpha_s^2, \alpha_s^3)$ radiative corrections for decays to quarks, which can be readily computed~\cite{Djouadi:2005p17157, Djouadi:2005p17109} (for details see Appendix~\ref{sec:2HDM+Sapp}). As mentioned in Section \ref{SMS}, perturbative QCD can be used for pseudoscalar masses above $\sim 1 \gev$, though the calculation breaks down near quarkonium states~\cite{Baumgart:2012pj}. A detailed investigation of this is beyond the scope of this paper. The results can be summarized as follows:
\begin{itemize}
\item Type I (\figref{fBRs2HDMatype1}): Since all fermions couple only to $H_2$, the branching ratios are independent of $\tan\beta$.  
The pseudoscalar couplings to all fermions are proportional to those of the SM Higgs, all with the same proportionality constant, and the branching ratios are thus very similar to those of the 
SM+S model with a complex $S$ and a light pseudo-scalar $a$ (i.e., for example, proportional to the mass of the final state fermions).  
\item Type II (\figref{fBRs2HDMatype2}): The exotic decay branching ratios are those of NMSSM models.  Unlike Type I models, they now depend on $\tan\beta$, 
with decays to down-type fermions suppressed (enhanced) for down-type fermions for $\tan \beta < 1$ ($\tan \beta > 1$).  
\item Type III (\figref{fBRs2HDMatype3}): The branching ratios are $\tan\beta$ dependent.  
For $\tan \beta > 1$, pseudoscalar-decays to leptons are enhanced over decays to quarks.  For example, unlike the NMSSM above the $b\bar{b}$-threshold, 
decays to $\tau^+ \tau^-$ can dominate over decays to $b\bar{b}$; similarly, above the $\mu^+ \mu^-$ threshold, decays to $\mu^+\mu^-$ can dominate over decays to heavier, kinematically 
accessible quark-pairs.  This justifies extending, for example, NMSSM-driven $4\tau$ searches over the entire mass range above the $b\bar{b}$-threshold.  For $\tan\beta<1$, decays to quarks are 
enhanced over decays to leptons.  
\item Type IV (\figref{fBRs2HDMatype4}): The branching ratios are $\tan\beta$ dependent.  For $\tan \beta < 1$ and compared to the NMSSM, the pseudoscalar-decays to up-type quarks and leptons 
can be enhanced with respect to down-type quarks, so that branching ratios to $b \bar b$, $c \bar c$ and $\tau^+ \tau^-$ can be similar.  This opens up the possibility of detecting this model in the $2b2\tau$ or $2c2\tau$ final state. 
\end{itemize}
Note that the branching ratios are only independent of $\tan \beta$ for Type I, and all types reduce to Type I for $\tan \beta = 1$.  

A sizable $\mathrm{Br}(h \rightarrow Z a)$ would open up additional exciting search channels with leptons that reconstruct the $Z$-boson. This is discussed in~\secref{htoZa}.

For $3 m_\pi < m_a < 1 \gev$ the decay rate calculations suffer large theoretical uncertainties 
but the dominant decay channels will likely be muons and hadrons. Below the pion, muon, and electron thresholds, the pseudoscalar decays dominantly to muons, electrons, and photons, 
respectively, except for $\tan \beta < 1$ in Type II, III and $\tan \beta > 1$ in Type IV, where the suppressed lepton couplings can also cause decays to photons to dominate below the pion threshold.  
If the pseudoscalar couples to both quarks and leptons, then requiring its mixing angle to be small enough to not conflict with constraints from e.g.~meson decays and the muon anomalous magnetic moment implies that any allowed decay to two muons 
(for $2m_\mu < m_a < 3m_\pi$) is likely to have at least a displaced vertex (or be detector-stable), while any allowed decay to two electrons 
(for $2m_e< m_a < 2 m_\mu$) will be detector stable~\cite{Essig:2010gu}. 
For pseudoscalars that couple preferentially to leptons, the meson-decay constraints are absent and prompt decays to muons are allowed; however, allowed decays to electrons will likely have at least a displaced vertex, and need to be detector-stable as $m_a$ is decreased well below the muon threshold~\cite{Essig:2010gu}.

\vskip 2mm
\paragraph*{\bf Light Scalar ($s$)}
We now assume that the mass of the real singlet $S_R$ is below $m_h/2$. 
The scalar Higgs spectrum, \Eqref{hH2HDM}, gets extended by the additional real singlet, which mixes with the doublet sector 
\begin{eqnarray}
\nonumber \left(
\begin{array}{c}
h \\ H^0 \\ s
\end{array}\right)
&=&
\left(
\begin{array}{ccc}
1&0&0\\
0&\cos \zeta_2 & \sin \zeta_2\\
0&- \sin \zeta_2 & \cos \zeta_2
\end{array}
\right)
\left(
\begin{array}{ccc}
\cos \zeta_1 & 0 & \sin \zeta_1\\
0 & 1 & 0 \\
- \sin \zeta_1 & 0 & \cos \zeta_1
\end{array}
\right)
\left(
\begin{array}{ccc}
- \sin \alpha & \cos \alpha  & 0\\
\cos \alpha & \sin \alpha  & 0\\
0 & 0 & 1
\end{array}
\right)
\left(
\begin{array}{c}
H^0_{1,R} \\ H^0_{2,R} \\ S_R
\end{array}\right).
\end{eqnarray}
If we assume that the mixing angles $\zeta_{1,2}$ are small, this simplifies to
\begin{eqnarray}
\label{eq:2HDMSscalarorig}
\left(
\begin{array}{c}
h \\ H ^0\\ s
\end{array}\right)
&=&
\left(
\begin{array}{ccc}
- \sin \alpha & \cos \alpha  & \zeta_1\\
\cos \alpha & \sin \alpha  & \zeta_2\\
(-\zeta_2 \cos \alpha + \zeta_1 \sin \alpha) & (-\zeta_1 \cos \alpha - \zeta_2 \sin \alpha) & 1
\end{array}
\right)
\left(
\begin{array}{c}
H^0_{1,R} \\ H^0_{2,R} \\ S_R
\end{array}\right).
\end{eqnarray}
In this approximation, $h$ and $H$ have the same Yukawa couplings as in the regular 2HDM but now contain a small $S_R$ component that allows the decay $h\rightarrow ss$. 
The mostly-singlet state $s$ on the other hand mixes with some admixture of $H^0_{1,R}$ and $H^0_{2,R}$. This can be expressed in more familiar notation by adopting 
the following parameterization for the small singlet-doublet mixing angles 
\begin{equation}
\zeta_1 = - \zeta \cos(\alpha - \alpha')  \ \ \ \ \ , \ \ \ \ \ \ \zeta_2 = - \zeta \sin(\alpha - \alpha') \ ,
\end{equation}
\begin{eqnarray}
\label{eq:2HDMSscalar}
\Longrightarrow \ \ 
\left(
\begin{array}{c}
h \\ H^0 \\ s
\end{array}\right)
&=&
\left(
\begin{array}{ccc}
- \sin \alpha & \cos \alpha  & - \zeta \cos(\alpha - \alpha')\\
\cos \alpha & \sin \alpha  & - \zeta \sin(\alpha - \alpha')\\
- \zeta \sin \alpha' & \zeta \cos \alpha'& 1
\end{array}
\right)
\left(
\begin{array}{c}
H^0_{1,R} \\ H^0_{2,R} \\ S_R
\end{array}\right). \ \ \ \ \ 
\end{eqnarray}
The \emph{arbitrary} angle $\alpha'$ determines the $H^0_{1R,2R}$ admixture contained within $s$, while the \emph{small} mixing parameter $\zeta$ gives its overall normalization. The couplings of $s$ to SM fields are now identical to those of the SM-like Higgs $h$ in \tabref{2HDMcoupling}, scaled down by $\zeta$ and with the replacement $\alpha \rightarrow \alpha'$. Since $\alpha$ and $\alpha'$ can be independently chosen, $s$ can have an even broader range of branching fractions than $a$ and mirrors the range of possible $h$-decays in the regular 2HDM, but without a mass restriction beyond $m_s < m_h/2$. Just as for $h$, choosing $\alpha' \rightarrow \frac{\pi}{2} - \beta$ amounts to giving $s$ fermion couplings that are SM-Higgs-like (up to the overall mixing 
factor $\zeta$). In this limit, the 2HDM+S theory reduces to the SM+S case discussed in \S\ref{SMS}. On the other hand, choosing $\alpha' = \beta$ gives the same couplings as the pseudoscalar case. 

\begin{figure}
\begin{center}
\includegraphics[width=8cm]{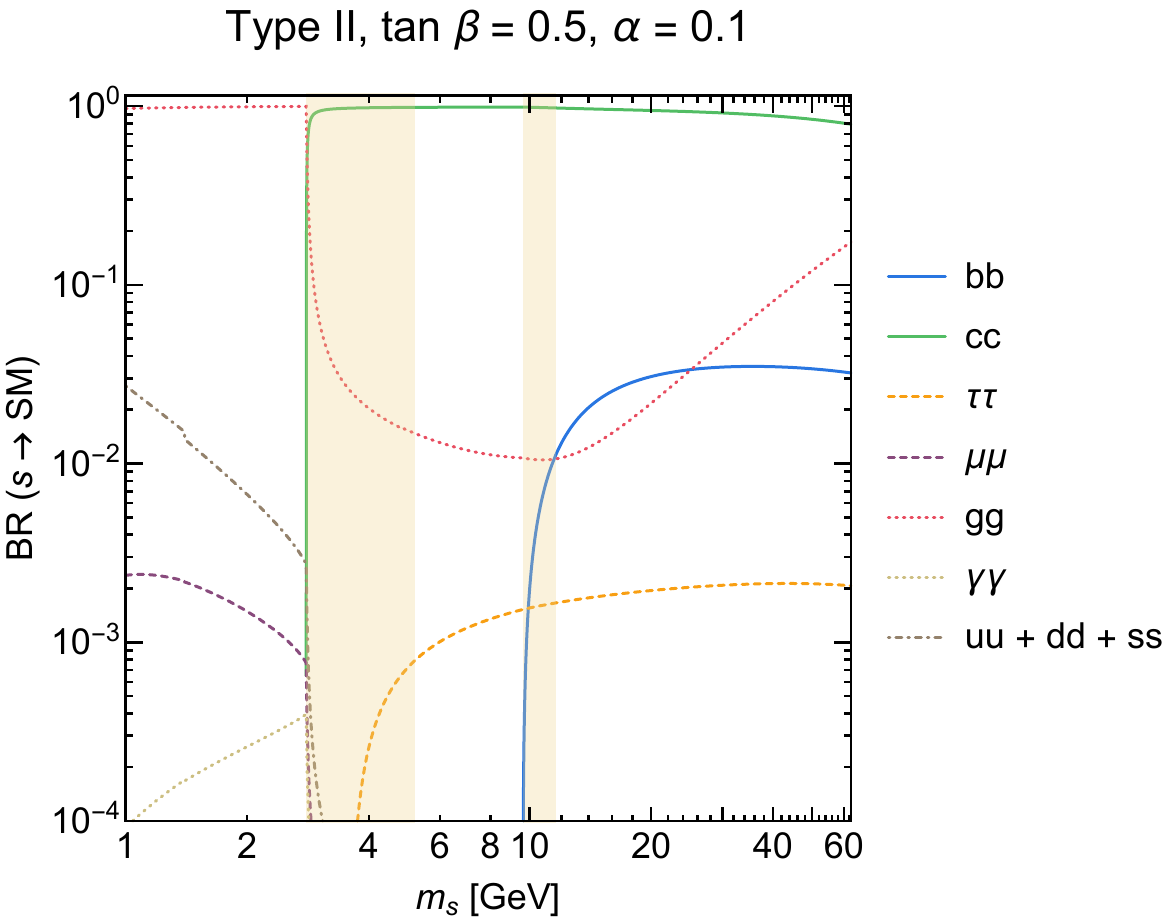}
\includegraphics[width=8cm]{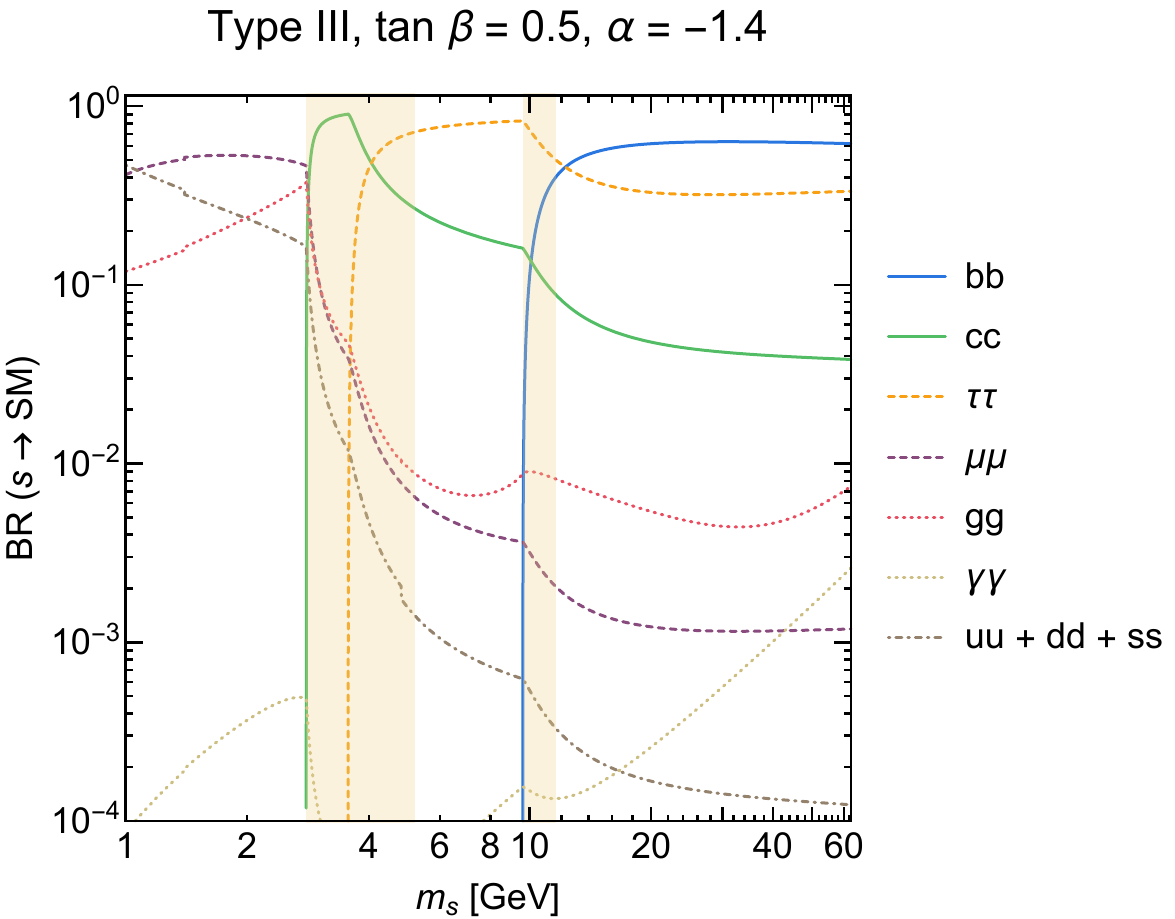}\\
\includegraphics[width=8cm]{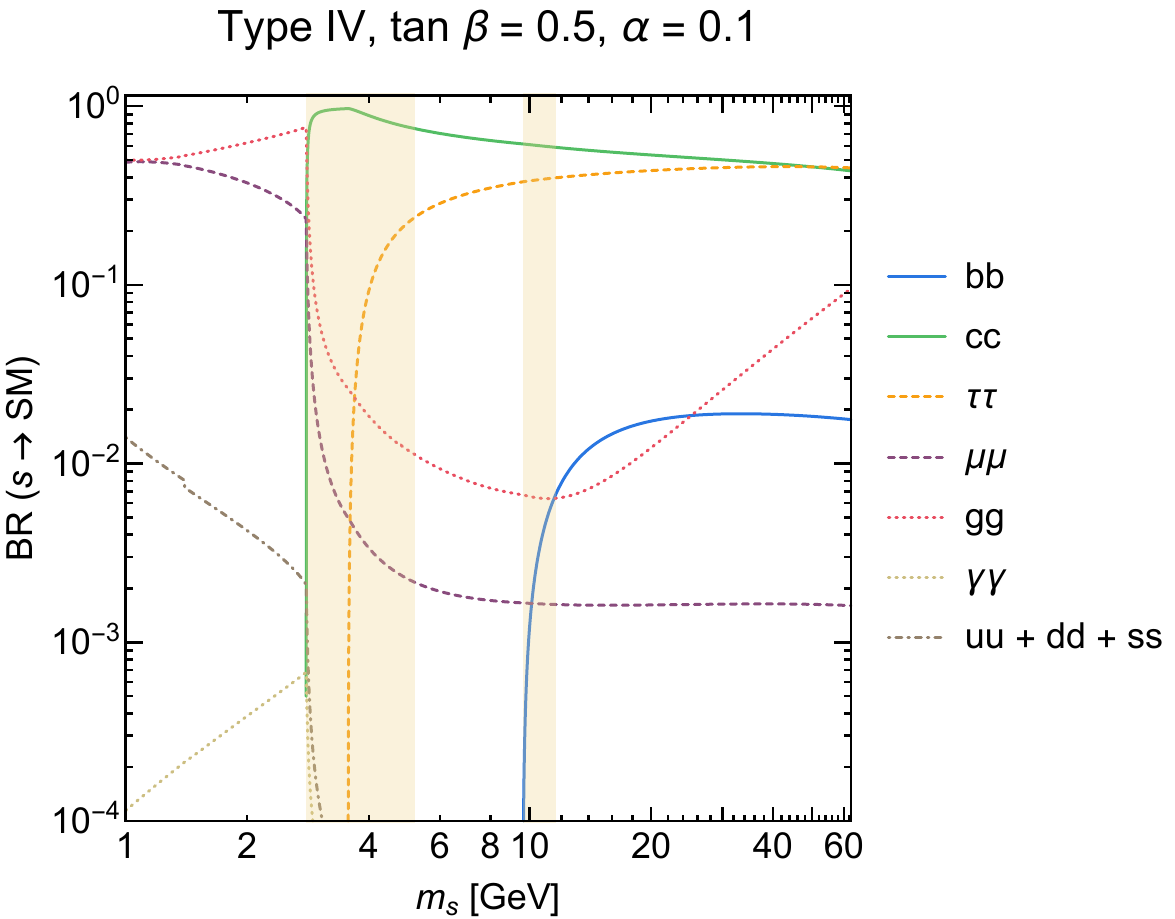}
\end{center}
\caption{
Singlet scalar branching ratios in the 2HDM+S for different $\tan \beta, \alpha'$ and Yukawa coupling type. These examples illustrate the possible qualitative differences to the pseudoscalar case, such as dominance of $s \rightarrow c \bar c$ decay above $b\bar b$-threshold; democratic decay to $b \bar b$ and $\tau^+\tau^-$; and democratic decay to $c \bar c$ and $\tau^+ \tau^-$. Hadronization effects likely invalidate our simple calculations in the shaded regions. 
}
\label{fig:fBRs2HDMs}
\end{figure}

The $s \rightarrow X \bar X$ branching ratios are computed analogously to the pseudoscalar case, with further details again given in Appendix~\ref{sec:2HDM+Sapp}. There is a large range of possible decay phenomenologies. \figref{fBRs2HDMs} illustrates some examples that have qualitatively new features compared to the pseudoscalar case, namely the possible dominance of $s \rightarrow c \bar c$ decays above the $b\bar b$-threshold; similar decay rates to $b \bar b$ and $\tau^+\tau^-$; and similar decay rates to $c \bar c$ and $\tau^+ \tau^-$.

\vskip 2mm
\paragraph*{\bf Summary}
The 2HDM+S allows for a large variety of Higgs decay phenomenologies  $h \rightarrow aa \rightarrow X \bar X Y \bar Y$, $h \rightarrow ss \rightarrow X \bar X Y \bar Y$, and $h \rightarrow a Z \rightarrow X \bar X Y \bar Y$ by coupling the SM-like Higgs $h$ to a singlet-like scalar $s$ or pseudoscalar $a$. While the singlet's couplings within each fermion ``family" (down-type quarks, up-type quarks, or leptons) are ranked by their Yukawa couplings, the relative coupling strength to each family can be adjusted, and arbitrarily so in the scalar case. 

A simple illustration of the rich decay phenomenology is to consider, for example, the dominant decay mode(s) above the $b \bar b$ threshold. With the three largest Yukawa couplings in each family being to the bottom, charm, or tau, we demonstrated every possible combination of dominant decays: similar decays widths to $b \bar b$, $c \bar c$, and $\tau^+ \tau^-$, dominant decay widths to 
any two out of those three, or just one dominant mode. This motivates searches for a large variety of non-standard four-body final states of exotic Higgs decays. 

In \S\ref{subsec:SMvector}, we motivate additional four-body Higgs decay channels, ranked by gauge coupling instead of Yukawa coupling.  We will see that even 
decays to $\mu^+\mu^-$ and $e^+e^-$ can dominate above the $b\bar{b}$-threshold.


\subsubsection{SM + Fermion} \label{subsub:SMF}

We here discuss exotic Higgs decays that can arise by the addition of
a light fermion to the SM.  We focus on two possibilities,
\emph{neutrino portal-mediated} and \emph{Higgs portal-mediated} Higgs
decays.

The leading interaction of a single Majorana fermion $\chi$ with the
SM fields is given by the renormalizable but lepton-number violating
``neutrino portal'' operator,
\beq\label{eq:YukawaNeutrino}
\mc{L}_{N} = y \chi H L.
\eeq
If this lepton-number violating coupling is forbidden, the leading
coupling between $\chi$ and the SM is through the dimension five Higgs
portal operator\footnote{The dipole operator
  $\chi^\dag\sigma^{\mu\nu}\chi F_{\mu\nu}$ is also dimension five,
  but vanishes for a Majorana $\chi$.},
\begin{equation}
\label{eq:1fH2}
\mathcal{L}_{\chi H}=\frac{\kappa}{2M}( \chi \chi+\chi^\dag \chi^\dag) |H|^2.
\end{equation}
This kind of coupling occurs, for instance, in the MSSM when all BSM
degrees of freedom except a bino-like neutralino are integrated out at
a high scale.  In the MSSM, the states integrated out to generate this
operator are fermionic, with electroweak quantum numbers.  In UV
completions where the state being integrated out is bosonic, the
operator of Eq.~(\ref{eq:1fH2}) has effective coupling
$\frac{\mu}{2M^2}$, where $\mu$ is some hidden sector mass scale. This
is a consequence of chiral symmetry, and, as we frequently may have
$\mu\ll M$, may result in the Higgs portal interaction becoming
effective dimension six.  As an example of this kind of UV completion,
consider a simple hidden sector consisting of a singlet scalar $S$
together with the fermion $\chi$,
\beq
\mathcal{L} =  (c\, S+ m_0) ( \chi \chi + \chi^\dag \chi^\dag) + V(S) + \zeta S^2 |H|^2,
\eeq 
and let $V(S)$ allow $S$ to develop a vacuum expectation value, $\langle S \rangle \equiv
\mu$.\footnote{For simplicity, we do not consider the possible
  interaction $S|H|^2$.  This operator could be forbidden in the
  presence of a global symmetry taking $S\to -S$, $\chi\to i\chi$,
  which would also forbid the mass term $m_0 (\chi\chi+\chi^\dag
  \chi^\dag)$. } Then integrating out the excitations of $S$ around
this $\langle S \rangle$, with mass $m_s$, we obtain the operator
\beq
\mathcal{L}_{\chi H}= \frac{c\, \zeta \mu}{m_s^2}( \chi \chi+\chi^\dag \chi^\dag) |H|^2.
\eeq
The mass of the fermion is $m_\chi= m_0+c\mu$, so either there are
large cancellations or $c\mu\sim m_0\sim m_\chi\ll m_s$, and the
operator is effective dimension-six.

\smallskip
\noindent {\bf Neutrino portal-mediated Higgs decays}  

We first consider exotic Higgs decays mediated by the neutrino portal operator,
Eq.~(\ref{eq:YukawaNeutrino}).  The renormalizable neutrino portal
coupling occurs in the so-called $\nu$SM, the minimal model that can
give mass to the SM neutrinos. Here the SM is extended by sterile
neutrinos, allowing the SM neutrinos to get a mass from a see-saw type
mechanism triggered by a Majorana mass term $(M/2) \chi\chi$.  The
operator of Eq.~(\ref{eq:YukawaNeutrino}) mixes the sterile neutrino
$\chi$ with the active SM neutrino $\nu$ arising from the $SU(2)$
doublet $L$. In the absence of large cancellations in the neutrino
mass matrix, sterile neutrinos must be extremely heavy, $M\gg v$, or
extremely decoupled, $y\ll y_e \ll 1$.  In this limit, the decay $h\to
\chi \nu$ is negligible, even if kinematically allowed.  However, the
authors of~\cite{Kersten:2007vk,0706.1732} show that active-sterile
mixing angles as large as several percent are possible, with
(accidental) cancellations among the Yukawa couplings still allowing
for small active neutrino masses.  Mixing angles of the order of a few
percent may imply a sizable partial width for $h\to\nu \chi$,
 \begin{equation}
\Gamma(h\to \nu \chi)=\frac{|y|^2}{8\pi} m_h\left(1-\frac{m_\chi^2}{m_h^2}\right)^{3/2}\,,
\end{equation}
where $m_\chi$ is the mass of the sterile neutrino $\chi$.  For
$m_h<130$ GeV, neutrino data and pion decay constraints on $W$-lepton coupling universality 
still allow the partial width into
$h\to\nu \chi$ to exceed that into $h\to b\bar b$, see \cite{0706.1732} for a detailed discussion
(see also \cite{Chang:2007de}).

The mass mixing between sterile (right-handed (RH)) neutrinos and
active (left-handed (LH)) neutrinos introduces couplings of the RH
neutrinos to $W$ and $Z$ gauge bosons. Therefore, in the region of
parameter space for which the active-sterile mixing angle $\Theta$ is
close to its phenomenological upper bound, the RH neutrinos decay
promptly into $\chi\to \ell W^*\to \ell f f^\prime$ and $\chi\to \nu
Z^*\to \nu f \bar f$, where $f$ and $f^\prime$ are either a lepton or
a quark of the SM, and with all branching ratios fixed by the
electroweak quantum numbers of the SM fermions.  In general $\chi$ may
have non-zero mixings with one, two, or all three SM neutrinos.

\smallskip
\noindent {\bf Higgs portal-mediated Higgs decays}  

\begin{figure}[t]
\begin{center}
\includegraphics[width=0.48\textwidth]{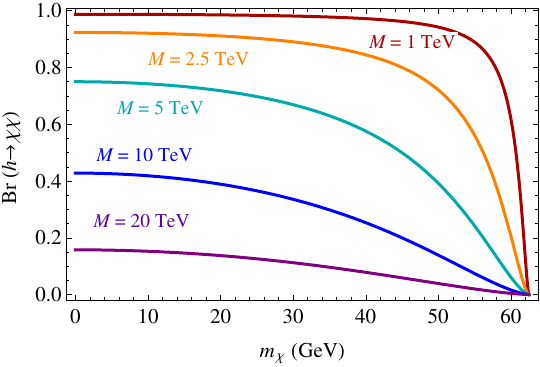}~~~~
\caption{\small Higgs branching fraction into Majorana fermions $\chi$
  resulting from the partial width of \Eqref{GammahffSMF}, as a
  function of the Higgs portal scale $M$ and the mass of the fermion
  $m_\chi$. We fix the coupling $\kappa$ to be equal to 1. }
\label{fig:BRexSMF}
\end{center}
\end{figure}

We next turn to the higher-dimension decays, mediated by the higher-dimension operator
of Eq.~(\ref{eq:1fH2}).  After electroweak symmetry breaking, this
operator yields a coupling $\lambda h(\chi\chi + \chi^\dag
\chi^\dag)$, with effective Yukawa coupling given by $\lambda ={\kappa
  v/2M}$.  The resulting partial width into $\chi$ is then
\beq 
\label{eq:GammahffSMF}
\Gamma(h\to\chi\chi) = \frac{m_h}{8\pi}\left(
  \frac{\kappa v}{M}\right)^2 \left(1- \frac{4 m_\chi^2}{m_h^2}\right)^{3/2}.
\eeq
As the effective Yukawa coupling $\lambda$ is only competing with the small
$b$-quark Yukawa, substantial branching fractions Br$(h\to\chi\chi)$
can be obtained even for Higgs portal scales $M$ significantly above a
TeV, as shown in Fig.~\ref{fig:BRexSMF}, where we fix $\kappa=1$ for
simplicity.

The kinds of signatures that are realized depends on how $\chi$
decays.  If the Higgs portal coupling of Eq.~(\ref{eq:1fH2}) is the only
interaction that the new fermion $\chi$ possesses, then $\chi$ is
absolutely stable, and the resulting Higgs decay is invisible.  In
general, however, $\chi$ will possess additional interactions. If
these interactions preserve the $\mathbb{Z}_2$ symmetry taking
$\chi\to-\chi$, then $\chi$ will remain stable.  On the other hand, if
the $\mathbb{Z}_2$ is violated by a dimension-six operator of the form
\begin{equation}
\mathcal{L}_{f}=\frac{1}{\Lambda^2} \chi f_1 f_2 f_3
\end{equation}
where $f_1f_2f_3$ is a gauge-invariant combination of quarks and
leptons, then $\chi$ will undergo the three-body decay $\chi\to
f_1f_2f_3$.  Some of these decays are familiar from previous study of
R-parity violating neutralino decays in the MSSM, namely those
involving holomorphic combinations of SM fermion fields (we suppress
spinor structures for simplicity),
\beq
\lambda_{ijk} L_i L_j e^c_k,\phantom {spacer}
\lambda'_{ijk} L_i Q_j d^c_k,\phantom {spacer}
\lambda''_{ijk} u^c_i d^c_j d^c_k .
\eeq
One may also consider the non-holomorphic operators~\cite{Csaki:2013jza}
\beq
\kappa_{ijk} Q_i Q_j {d^c}^\dag_k,\phantom {spacer}
\kappa'_{ijk} L^\dag_i Q_j u^c_k ,\phantom {spacer}
\kappa''_{ijk} u_i {d^c}^\dag_j e^c_k.
\eeq
Another flavor-violating possibility appearing at dimension six is the
radiative decay $\chi\to \gamma\nu$, mediated by
\beq
\mathcal{O}_{\gamma\nu} = \chi H L_i \sigma^{\mu\nu} B_{\mu\nu}.
\eeq
While this operator can yield two-body final states, it naturally
scales with a loop factor.  All of these lepton and/or baryon-number
violating decays necessarily have nontrivial flavor structure, and the
combinations of operators that appear depends on the flavor structure
of the UV theory.  Unlike the SM plus scalar interactions considered
in \S\ref{SMS} and \S\ref{2HDMS} or the neutrino-portal decays discussed
earlier, the possible decays of $\chi$ are not determined by the Higgs
coupling to the fermion, but require additional interactions,
involving the flavor structure of the theory.

\smallskip

To summarize, the exotic Higgs signatures from a single additional
(Majorana) fermion species are then Higgs decays to either invisible
particles, or to one or more four- or six-body final states, where the
six bodies form two three-body resonances of equal mass.  When
neutrinos are among the final state partons, the final states will
include missing energy, and the resonances will not be
reconstructable.  This is always the case in the possible four-body
final states where neutrinos are always involved, and is sometimes the
case in the six-body final states.

\subsubsection{SM + 2 Fermions}

It is worth generalizing the previous discussion to the case with two
new singlet fermions $\chi_1$ and $\chi_2$.  The Majorana mass matrix
for these two fermions has three parameters, and the dimension-five
Higgs portal operators form a matrix
\begin{equation}
\label {eq:multiHf}
\mathcal {L}_{\chi}=\frac{c_{ij}}{\Lambda}\chi_i\chi_j|H|^2 .
\end{equation}  
After electroweak symmetry breaking, the BSM fermions form two mass
eigenstates $\chi_1$ and $\chi_2$, with mass $m_2 > m_1$.  If we take
relatively light fermions $m_h>2 m_2$, the decays $h\to \chi_2\chi_2$,
$h\to\chi_1\chi_2$ and $h\to\chi_1\chi_1$ are all possible.
This kind of  interaction appears in, for instance, the NMSSM (see \S\ref{NMSSMfermion}.), where
$\chi_2$ and $\chi_1$ are mostly bino- and singlino-like, respectively,
and the higher-dimension Higgs portal coupling of Eq.~(\ref{eq:multiHf})
results after integrating out the charged Higgsinos.  It can also arise in (possibly supersymmetric) Hidden Valleys; see \S\ref{sec:hiddenvalley}.

Let us first consider the case where there is a $\mathbb{Z}_2$
symmetry which takes $\chi_i\to -\chi_i$.  In
this case, $\chi_1$ is stable, but the heavier new state decays as $\chi_2 \to \chi_1+X$.  If the Higgs portal coupling of
Eq.~(\ref{eq:multiHf}) is the only coupling of the $\chi_i$, then the
decay will proceed through an off-shell Higgs, $\chi_2\to h^*
\chi_1\to (f\bar f, gg, \gamma\gamma) \chi_1$.  In this case, branching
fractions into different SM partons will be determined by the Higgs
couplings, and will typically result in Higgs decays to $\met$ plus
one or two non-resonant quark-antiquark, lepton-anti-lepton, or gluon
pairs, depending on the available phase space.

If the $\chi_i$ have additional interactions besides their coupling to
the Higgs, such as a dipole coupling to the hypercharge field
strength,
\begin{equation}
\label {eq:SMplusfdipole}
\mathcal {L}_{\chi}=\frac{1}{\mu}\chi_1^\dag\sigma_{\mu\nu}\chi_2 B^{\mu\nu}
\end{equation}  
or a coupling to the $Z$ boson induced by mixing with states
transforming under $SU(2)_L$,
\begin{equation}
\label {eq:SMplusfZ}
\mathcal {L}_{\chi}= h_{ij} \chi_i^\dag\sigma^{\mu}\chi_jZ_\mu ,
\end{equation}  
then other decay patterns are possible. The dipole operator allows the
decays $\chi_2\to\gamma\chi_1$, as well as $\chi_2\to\chi_1 Z$ if
$m_2-m_1 > m_Z$ (phase space suppression renders decays through an
off-shell $Z$ largely irrelevant when $m_2-m_1 < m_Z$).  The 
operator of Eq.~(\ref{eq:SMplusfZ}) also yields $\chi_2\to \chi_1 Z$
when phase space allows, or if $m_2-m_1< m_Z$, will mediate the
three-body decays $\chi_2\to f\bar f \chi_1$ with branching ratios set
by the $Z$ branching fractions.  

Note that a common feature of all these decays is that the pairs of
SM partons have a kinematic endpoint at $m_{f\bar f, gg,\gamma\gamma}< m_2-m_1$,
and that the transverse mass of the visible partons and the $\met$ is
bounded from above.

The $Z$ boson coupling can arise in NMSSM-like models, see
e.g.~\S\ref{NMSSMscalar}, or in models with additional RH
neutrinos~\cite{0705.2190,0704.0438} that mix with the SM
neutrinos. In the latter case, the couplings $h_{ij}$ in
(\ref{eq:SMplusfZ}) are sufficiently small that the neutrino decay
lengths are macroscopic. In the former case, the couplings can instead
be larger, and the Majorana fermions can have a prompt decay into SM
fermions.
Additional examples are models with a fourth generation of fermions where
the two fourth generation neutrinos do not mix with the SM
neutrinos~\cite{Carpenter:2010sm,1103.3765,1107.2123}. In these models, 
the mass range $M_1\gtrsim 30$ GeV, $M_2-M_1\lesssim 20$ GeV is
allowed by LEP measurements of the $Z$ width and LEP bounds on
$e^+e^-\to \chi_1 \chi_2, \chi_2 \chi_2$~\cite{Carpenter:2010sm}. In
this region of parameter space, $h\to \chi_2 \chi_2$, as well as $h\to
\chi_1 \chi_1$, can have a sizable branching
ratio~\cite{1103.3765}. Furthermore, the heavier neutrino $\chi_2$ can
decay promptly via $\chi_2\to Z^* \chi_1$, while the lighter neutrino
$\chi_1$ is long-lived.

If the $\mathbb{Z}_2$ parity is violated, allowing $\chi_1$ to decay,
Higgs decays to as many as ten partons may result.  We will not
consider such complex decays in this work, but one should bear in mind
that they can occur.

Many models with new fermion species also contain new bosonic degrees
of freedom, which, if light, open new possibilities for the decays of
the $\chi_i$.  We will see examples of this in
\S\ref{NMSSMfermion}. 
\subsubsection{SM + Vector}
\label{subsec:SMvector}

\noindent {\bf Preliminaries}

An additional $U(1)_D$ gauge symmetry added to the SM is theoretically
well-motivated and occurs in many top-down and bottom-up extensions of
the SM.  The $U(1)_D$ vector boson (the ``dark photon'' or the
``dark-$Z$'') is usually referred to as $A'$, $Z'$, $\gamma_D$, or
$Z_D$ in the literature and various possibilities exist to connect the
additional $U(1)_D$ to the SM (see
e.g.~\cite{Langacker:2008yv,Jaeckel:2010ni,
  Hewett:2012ns,Essig:2013lka} for reviews).  In
\S\ref{sec:hiddenvalley}, we will discuss more complicated
hidden-valley phenomenology, involving non-abelian gauge symmetries
and/or composite states~\cite{Strassler:2006im,Strassler:2008bv}.
Here we focus on Higgs decays that involve an $A'$, with the $A'$ mass
between $\sim$MeV--63 GeV.  A sub-GeV $A'$ has generated a lot of
interest in the last few years due to anomalies related to dark
matter~\cite{ArkaniHamed:2008qn,Pospelov:2008jd,Finkbeiner:2007kk,Fayet:2004bw}
and as an explanation of the discrepancy between the calculated and
measured muon anomalous magnetic moment~\cite{Pospelov:2008zw}.

\begin{figure}
\begin{center}
\includegraphics[width=0.6\textwidth]{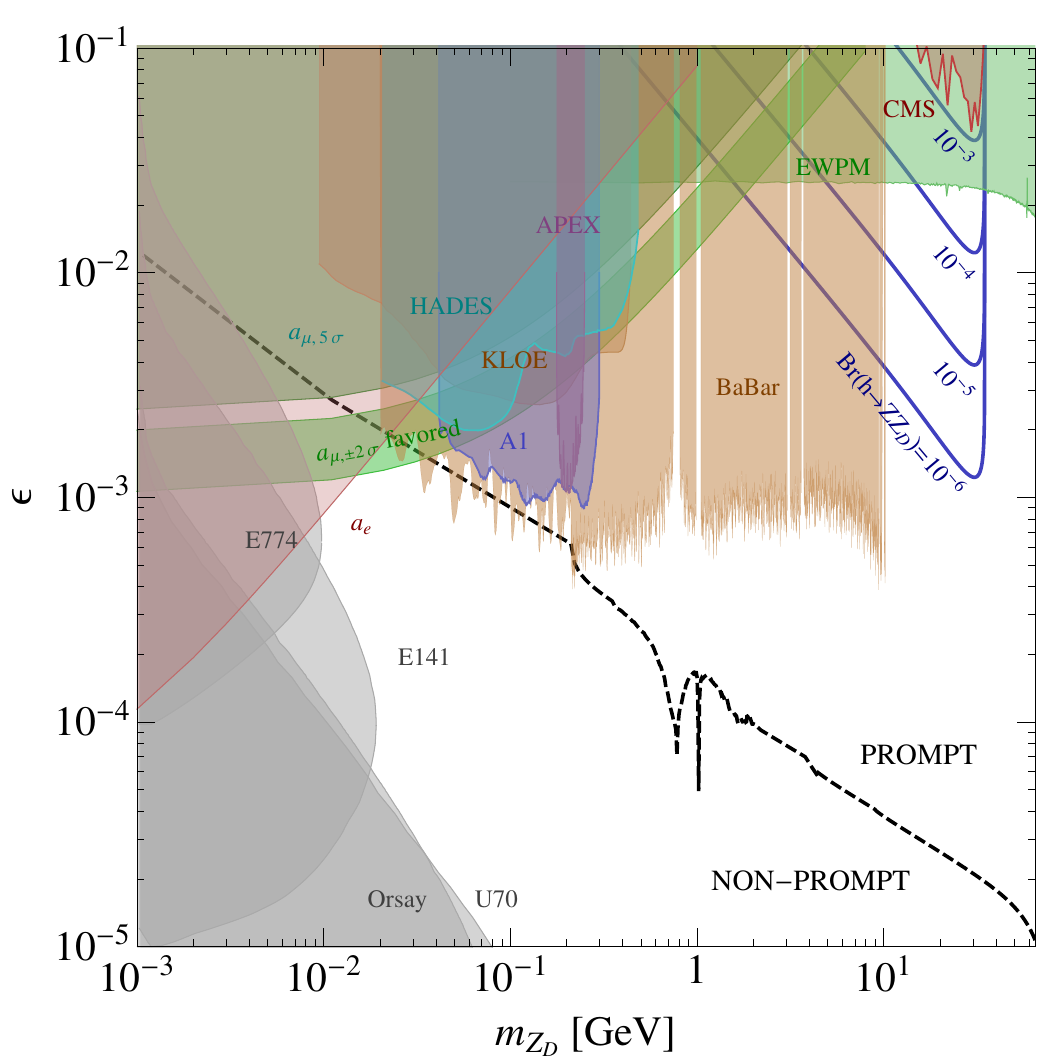}
\end{center}
\caption{ Constraints on $\epsilon, m_{Z_D}$ for pure kinetic mixing
  (no additional source of $Z$-$Z_D$ mass mixing) for $m_{Z_D} \sim
  $MeV--10~GeV.  The black dashed line separates prompt ($c\tau < 1
  \mu m$) from non-prompt $Z_D$ decays.  The three blue lines are contours of
  $\mathrm{Br}(h \rightarrow Z Z_D)$ of $10^{-4}, 10^{-5}, 10^{-6}$
  respectively. Shaded regions are existing experimental
  constraints~\cite{Bjorken:2009mm,Bjorken:1988as,Riordan:1987aw,Bross:1989mp,Batell:2009yf,Blumlein:2011mv,Andreas:2012mt,Pospelov:2008zw,Reece:2009un,Aubert:2009cp,Hook:2010tw,Babusci:2012cr,Archilli:2011zc,Abrahamyan:2011gv,Merkel:2011ze,Dent:2012mx,Davoudiasl:2012ig,Endo:2012hp,Adlarson:2013eza,Agakishiev:2013fwl,Merkel:2014avp,Lees:2014xha},
  see e.g.~\cite{Essig:2013lka} for a recent review.  
 The red shaded region ``CMS'' is a new limit we derived by recasting the CMS 20+5$\ifb$ $h\to Z Z^*$ analysis \cite{CMS-PAS-HIG-13-002}, as described in \S\ref{sec:htoZa}. (We obtain a similar bound from the corresponding ATLAS analysis  \cite{ATLAS:2013nma}.)  This new bound can be optimized with a dedicated LHC measurement, likely improving upon the Electroweak Precision Measurement Bounds (green region labelled ``EWPM'' \cite{Hook:2010tw}) for some masses.
  }
\label{fig:SMVZdbounds}
\end{figure}

The $U(1)_D$ can couple to the SM sector via a small gauge kinetic
mixing term $\frac{1}{2} \epsilon F'_{\mu \nu} B^{\mu
  \nu}$~\cite{Holdom:1985ag,Galison:1983pa,Dienes:1996zr} between the
dark photon and the hypercharge gauge boson. This renormalizable
interaction can be generated at a high scale in a grand unified theory
or in the context of string theory with a wide range of $\epsilon\sim
10^{-17} - 10^{-2}$~\cite{Holdom:1985ag, Abel:2003ue,Abel:2008ai,
  Goodsell:2009xc,Cicoli:2011yh,Goodsell:2011wn,Baumgart:2009tn,ArkaniHamed:2008qp,Essig:2009nc}.
This term effectively gives SM matter a dark milli-charge, made more
obvious by a $\mathrm{GL}(2,R)$ field redefinition $B_\mu \rightarrow
B_\mu - \epsilon A'_\mu$ which yields canonical kinetic terms, and
allows for dark photon decay to SM particles and possible experimental
detection.  To avoid the tight constraints on new long-range forces, a
`dark Higgs' $S$ with a non-zero vacuum expectation value can give a
non-zero mass to the $A'$.  An $A'$ with a sub-GeV mass can be probed
at beam dumps and colliders, and with measurements of the muon
anomalous magnetic moment, supernova cooling, and rare meson
decays~\cite{Bjorken:2009mm,Bjorken:1988as,Riordan:1987aw,Bross:1989mp,Batell:2009yf,Essig:2009nc,Blumlein:2011mv,Andreas:2012mt,Pospelov:2008zw,Reece:2009un,Aubert:2009cp,Archilli:2011zc,Abrahamyan:2011gv,Merkel:2011ze,Dent:2012mx,
  Davoudiasl:2012ag, Davoudiasl:2013aya}, see
Fig.~\ref{fig:SMVZdbounds} and e.g.~\cite{Essig:2013lka} for a recent
review.
 
A broken $U(1)_D$ can also lead to exotic Higgs decays, especially if
there is mixing between the two Higgs sectors.  In this context we
refer to the corresponding vector field as $Z_D$.

The possibility of $h \rightarrow Z_D Z_D$ through Higgs-to-dark-Higgs
mixing or $h \rightarrow Z Z_D$ through $Z$-$Z_D$ mass mixing (which
is also induced by the above-mentioned kinetic mixing) was discussed
in~\cite{Gopalakrishna:2008dv} and~\cite{Davoudiasl:2012ag,
  Davoudiasl:2013aya}, respectively, with both occurring, for example,
in hidden valley models \cite{Strassler:2008bv, Strassler:2006im}.

To examine the range of possible exotic Higgs phenomena due to a
$U(1)_D$ sector we examine the model of \cite{Gopalakrishna:2008dv},
but with $m_h$ set to 125 GeV and allowing for the full range of dark
Higgs and dark-$Z$ masses relevant to exotic Higgs decay
phenomenology.\footnote{Ref. \cite{Chang:2013lfa} appeared while this
  work was being completed, performing a similar analysis with a
  different focus on constraining the couplings of the extended Higgs
  potential for relatively low $m_{Z_D} < 5 \gev$.} This includes
Higgs-to-dark-Higgs mixing and kinetic mixing between the $B$ boson
and the dark vector $Z_{D}$, but no explicit mass mixing between the
$Z$ and $Z_D$.\footnote{The constraints shown in
  Fig.~\ref{fig:SMVZdbounds} are altered in the presence of such pure
  mass mixing, which requires additional Higgs doublets that also
  carry dark charge. The resulting $Z_D \rightarrow \mathrm{SM}$
  decays would be more $Z$-like and lead to additional constraints
  from rare meson decays as well as new parity-violating interactions
  \cite{Davoudiasl:2012ag}. However, we stress that the exotic Higgs
  phenomenology would not be qualitatively different.}  We will assume
prompt $Z_D$ decays, which requires $m_{Z_D} \gtrsim 10 \mev$ given
the current constraints shown in~Fig.~\ref{fig:SMVZdbounds}.

For $m_{Z_D}>10$ GeV, the most stringent constraints
come from precision electroweak measurements;\footnote{We thank Adam Falkowski for useful correspondence on the electroweak precision bounds shown in the green ``EWPM'' region in Fig.~\ref{fig:SMVZdbounds}.} we have verified the results in \cite{Hook:2010tw}. These constraints are largely driven by the tree-level shift to the
$Z$ mass,\footnote{Additional and more model-dependent constraints
arise when $m_{Z_D}$ is approximately equal to the center-of-mass
energy of $e^+$-$e^-$ experiments \cite{Hook:2010tw}.} and limit
$\epsilon \lesssim 0.02$ for $m_{Z_D}< m_h/2$.

\label{pageref.ZZdlimits}
Also shown in Fig.~\ref{fig:SMVZdbounds} is a new constraint we derived by recasting the CMS 20+5$\ifb$ $h\to Z Z^*$ analysis \cite{CMS-PAS-HIG-13-002}, as described in \S\ref{sec:htoZa}. (We obtain a similar bound from the corresponding ATLAS analysis  \cite{ATLAS:2013nma}.)  This new bound is already almost competitive with the Electroweak Precision Measurement Bounds (green region labelled ``EWPM'') for some masses, and can  be optimized further with a dedicated search. We expect LHC14 with $300 \ifb$ to be sensitive to $\mathrm{Br}(h\to Z Z_D)$ as low as $\sim 10^{-4}$ or $10^{-5}$. This would make the LHC the best probe of dark vector kinetic mixing for $10 \gev \lesssim m_\mathrm{Z_D} \lesssim m_h/2$ in the foreseeable future.

\vskip 2mm
\noindent {\bf Model Details}

The model is defined by a $U(1)_D$ gauge sector and a SM singlet $S$
that has unit charge under the $U(1)_D$. The kinetic terms of the
hypercharge and $U(1)_D$ gauge bosons (adopting mostly the notation of
\cite{Davoudiasl:2012ag}) are
\begin{equation}
\label{eq:SMVkinmix}
\mathcal{L}_\mathrm{gauge} = -\frac{1}{4} \hat B_{\mu \nu} \hat B^{\mu \nu} - \frac{1}{4} \hat Z_{D\mu\nu} \hat Z_D^{\mu \nu} + \frac{1}{2} \frac{\epsilon}{\cos \theta_W} \hat B_{\mu \nu} \hat Z_D^{\mu \nu},
\end{equation}
with $\hat B_{\mu \nu} = \partial_\mu \hat B_\nu - \partial_\nu \hat
B_\mu$, $\hat Z_{D\mu \nu} = \partial_\mu \hat Z_{D\nu} - \partial_\nu
\hat Z_{D\mu}$, and $\cos \theta_W = g/\sqrt{g^2 + g'^2}$ is the usual
Weinberg mixing angle. The hatted quantities are fields before
diagonalizing the kinetic term. The Higgs potential is
\begin{equation}
V_0 =   -\mu^2|H|^2 +  \lambda |H|^4 -\mu_D^2 |\sing|^2 + \lambda_D |\sing|^4 +
   \zeta  |\sing|^2|H|^2.
\end{equation}
The dark Higgs $S$ acquires a vacuum expectation value and gives
$Z_D$, which `eats' the pseudoscalar component of $S$, some mass
$m_{Z_D}$. There are two connections between the dark and the SM
sectors: the gauge kinetic mixing $\epsilon$ and the Higgs mixing
$\zeta$. The phenomenology depends on which one dominates.

The gauge kinetic term is diagonalized by transforming the gauge
fields
\begin{equation}
\left(\begin{array}{c} Z_{D} \\ B\end{array}\right) = 
\left( \begin{array}{cc} 1 & 0 \\ \displaystyle- \frac{\epsilon}{\cos \theta_W} & 1 \end{array}\right) \left(\begin{array}{c} \hat Z_{D} \\ \hat B\end{array}\right)\ ,
\end{equation}
where we always work to lowest order in the small $\epsilon$. $\hat B$
therefore gets replaced by $B + \frac{\epsilon}{\cos \theta_W} Z_D$,
giving all SM fermions a dark milli-charge proportional to their
hypercharge, while particle-couplings to $\hat B$ remain unchanged
when transforming to $B$.

The $Z_D$ and $Z$ gauge boson mass terms are
\begin{equation}
\mathcal{L}_\mathrm{mass} = \frac{1}{8} w^2 g_D^2 ( \hat Z_{D\mu})^2 + \frac{1}{8} v^2 (- g \hat W^3_\mu + g' \hat B_\mu)^2\ ,
\end{equation}
where $g_D$ is the gauge coupling of $U(1)_D$ and $w$ is the vacuum
expectation value of $S$. Writing in terms of canonically normalized
gauge fields this becomes
\begin{equation}
\mathcal{L}_\mathrm{mass} = \frac{1}{8} w^2 g_D^2 ( Z_{D\mu})^2 + \frac{1}{8} v^2 (- g W^3_\mu + g' B_\mu + g' \frac{\epsilon}{\cos \theta_W} Z_{D\mu})^2 .
\end{equation}
The SM gauge boson $Z_\mu = - \sin \theta_W B_\mu + \cos \theta_W
W^3_\mu$ is no longer a mass eigenstate:
\begin{equation}
\mathcal{L}_\mathrm{mass} = \frac{1}{2} m_{Z_D}^2  ( Z_{D\mu})^2 + \frac{1}{2} m_Z^2  (Z_\mu - \epsilon \tan\theta_W Z_{D\mu})^2 .
\end{equation}
To leading order in $\epsilon$ the mass eigenstates with masses $m_Z,
m_{Z_D} + \mathcal{O}(\epsilon^2)$ are
\begin{eqnarray}
\label{eq:SMVmassbasisZZD}
\nonumber \tilde Z &=& Z + \epsilon_Z Z_D \\
\tilde Z_D &=& Z_D - \epsilon_Z Z , \ \ \ \ \mathrm{where} \ \ \ \epsilon_Z = \frac{ \epsilon \tan\theta_W  m_Z^2}{m_Z^2 - m_{Z_D}^2}.
\end{eqnarray}
(Henceforth, we omit the tildes and will refer to the mass eigenstates
unless otherwise noted.)  Therefore, there are interaction terms of
the form $2 \epsilon_Z \frac{m_{Z_D}^2}{v} h Z_\mu Z_D^\mu$ and
$\epsilon_Z^2 \frac{m_{Z_D}^4}{ m_Z^2 v} h Z_{D\mu} Z_D^\mu$ which
lead to $h \rightarrow Z_D Z$ and $h \rightarrow Z_D Z_D$ decays
(though the latter is strongly suppressed), see \Figref{SMVvertices}.

If $Z_D$ is the lightest state in the dark sector it will decay to SM
particles. This is entirely due to the kinetic mixing in
\Eqref{SMVkinmix}, but in the basis of \Eqref{SMVmassbasisZZD} it is
due to the dark milli-charge of SM fermions and the accompanying mass
mixing with the $Z$. Explicitly, the coupling of $Z_D$ to SM fermions
is
\begin{equation}
\mathcal{L} \  \supset  \ g_{Z_Dff}  \ Z_D^\mu  \ \bar f \gamma_\mu f ,
\end{equation}
where
\begin{equation}
g_{Z_D ff} = - g' \frac{\epsilon}{\cos \theta_W} Y - \epsilon \tan \theta_W \frac{m_Z^2}{m_Z^2-m_{Z_D}^2} \frac{1}{\sqrt{g'^2 + g^2}} \left( g^2 T_3 - g'^2 Y \right).
\end{equation}
The first and second term come from dark milli-charge and
$Z$-$Z_D$ mass mixing, respectively. This coupling is dominantly
photon-like, up to deviations $\sim \mathcal{O}(m_{Z_D}^2/m_Z^2)$:
\begin{equation}
\label{eq:SMVcoupling}
g_{Z_D f  f} =
\epsilon g' \left\{
 - (T_3 + Y) \cos \theta_W \left( 1 +\frac{m_{Z_D}^2}{m_Z^2}\right)   + \frac{Y}{\cos \theta_W} \frac{m_{Z_D}^2}{m_Z^2} + \mathcal{O}\left( \frac{m_{Z_D}^4}{m_Z^4}\right)
\right\}
\end{equation}
For $m_{Z_D} \gtrsim \gev$ the $Z_D$, branching ratios are easily
computed to lowest order and without QCD corrections, and are shown in
\Figref{SMVBrZdark} (a). For $m_{Z_D} \lesssim \gev$, non-perturbative
QCD effects are important. They can be computed from the QCD
contribution to the imaginary part of the electromagnetic two-point
function, which in turn is determined from cross-section measurements
of $e^+ e^- \rightarrow \
\mathrm{hadrons}$~\cite{Beringer:1900zz}. The resulting branching
ratios are shown in \Figref{SMVBrZdark} (b).

The most important qualitative difference to the scalar decays
considered in \S\ref{SMS} and \ref{2HDMS} is that branching ratios are
ordered by gauge coupling instead of Yukawa coupling, meaning decays
to $e^+ e^-$ and $\mu^+ \mu^-$ remain large above the $\tau$
thresholds. Prompt $Z_D$ decay requires $\epsilon \gtrsim 10^{-5} -
10^{-3}$, as indicated in \Figref{SMVZdbounds}, which summarizes the
constraints on $Z_D$ kinetic mixing for our regime of interest.

\begin{figure}
\begin{center}
\begin{tabular}{ccc}
\includegraphics[height=0.48\textwidth]{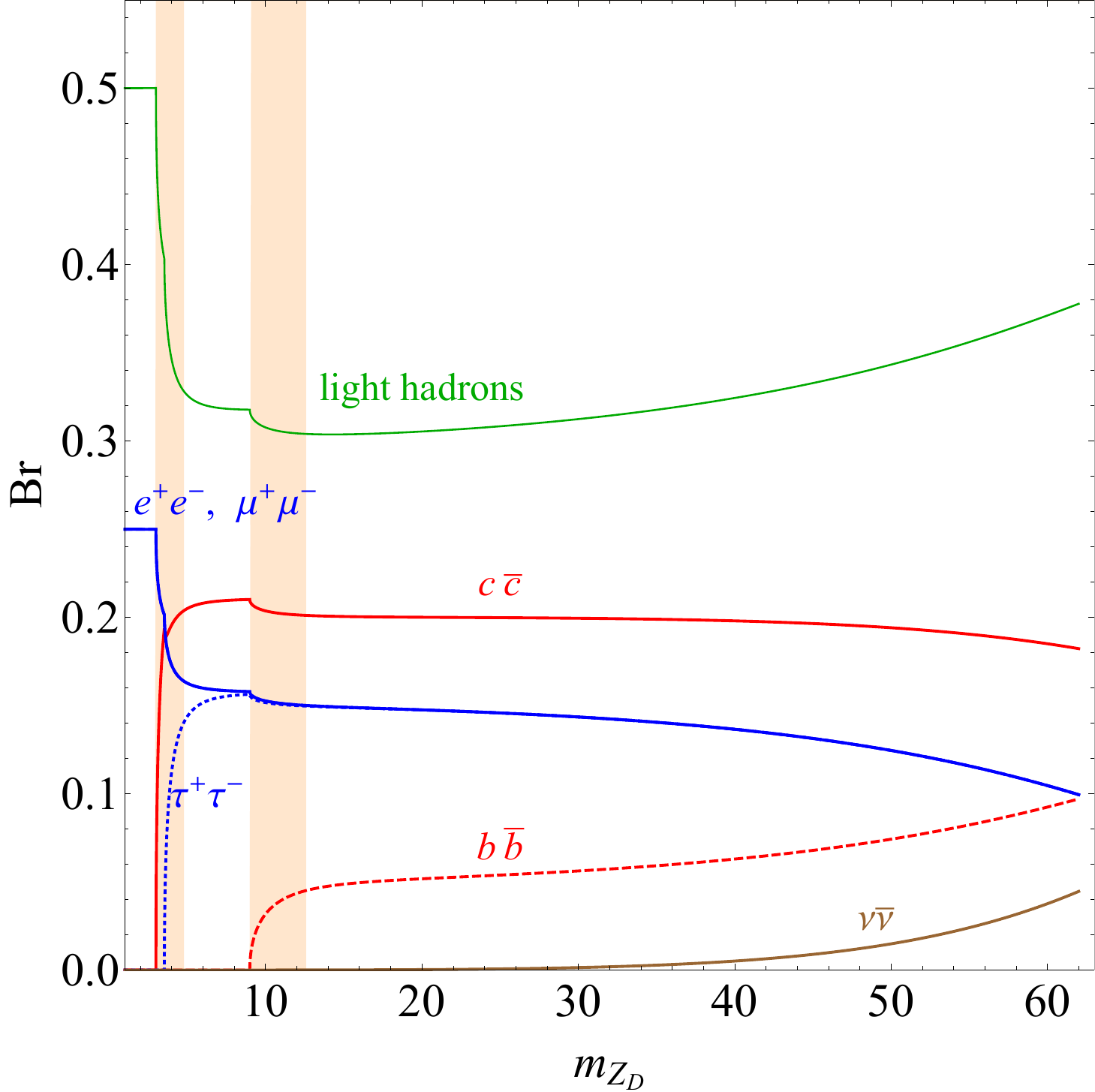} & \hspace{2mm} &
\includegraphics[height=0.48\textwidth]{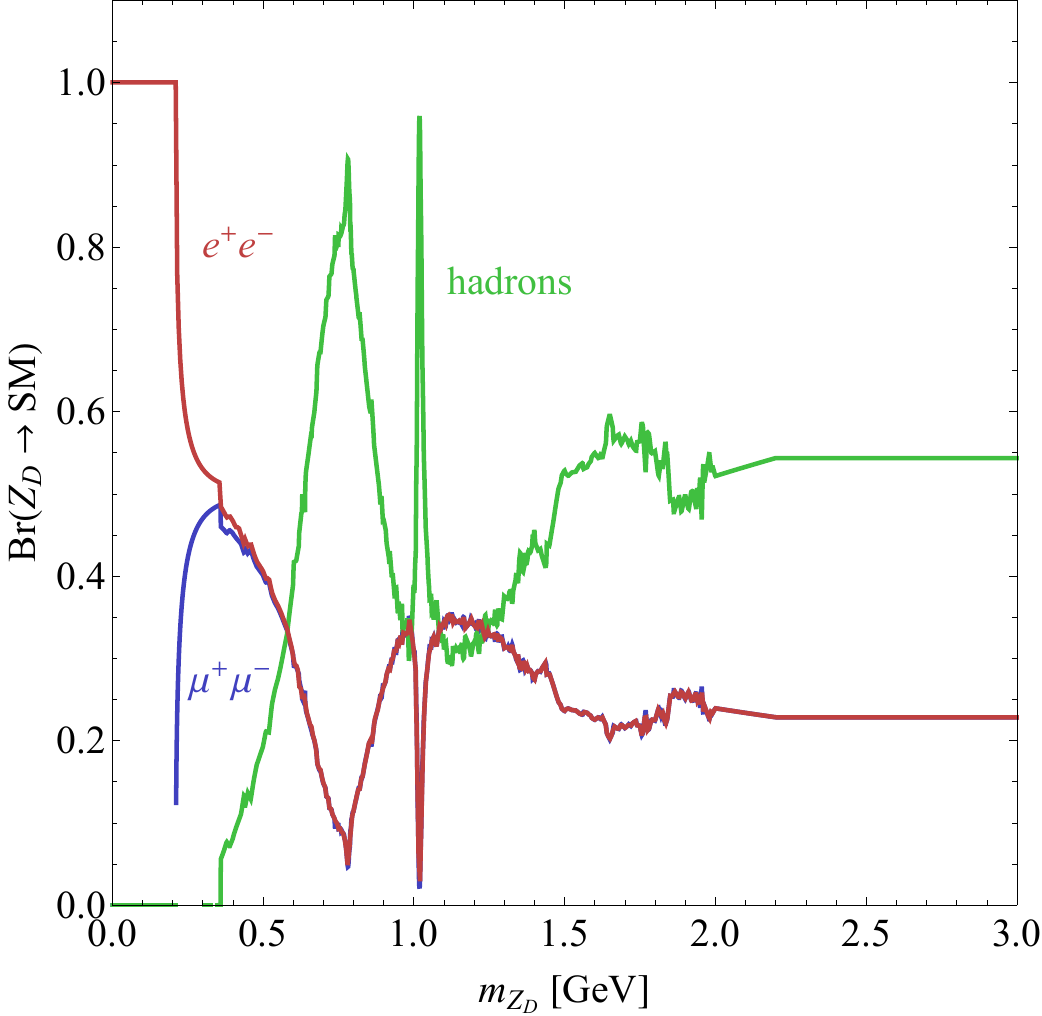}
\\
(a) && \hspace{8mm} (b)
\end{tabular}
\end{center}
\caption{
(a)~Branching ratios for $Z_D$ decay, to lowest order and without QCD corrections, assuming decays to the dark sector are kinematically forbidden. Hadronization effects likely invalidate our simple calculation in the shaded region.  (b)~Branching ratios for $Z_D$ decay for $m_{Z_D} \lesssim 3 \gev$, including non-perturbative QCD effects.}
\label{fig:SMVBrZdark}
\end{figure}

The Higgs potential is minimized by vacuum expectation values of $H^0$ and $S$
\begin{equation}
H^0 = \frac{1}{\sqrt{2}}(h + v) \ \ \ , \ \ \ \  S = \frac{1}{\sqrt{2}} (s + w) \ ,
\end{equation}
where to leading order in the small Higgs mixing $\zeta$,
\begin{equation}
v = \frac{\mu}{\sqrt \lambda} - \zeta \frac{\mu_D^2 }{4  \lambda_D \sqrt{\lambda} \mu} \approx 246 \gev
\ \ \ \mathrm{and} \ \ \ \ 
w = \frac{\mu_D}{\sqrt \lambda_D} - \zeta \frac{\mu^2 }{4  \lambda \sqrt{\lambda_D}  \mu_D }.
\end{equation}
The mass eigenstates 
\begin{eqnarray}
\label{eq:SMVmassbasishs}
\nonumber \tilde h &=& h - \epsilon_h s \\
\tilde s  &=& s + \epsilon_h h , \ \ \ \ \mathrm{where} \ \ \ \epsilon_h = 
\zeta \frac{\mu \mu_D}{2 \sqrt{\lambda \lambda_D} | \mu^2 - \mu_D^2|},
\end{eqnarray}
have masses 
\begin{equation}
m_h^2 = 2 \mu^2 - \zeta \frac{\mu_D^2}{\lambda_D} \  \ \ \mathrm{and} \ \ \ \ 
m_s^2 = 2 \mu_D^2 - \zeta \frac{\mu^2}{\lambda}.
\end{equation}
(Again we drop the tildes from now on and always refer to the mass
eigenstates.) The effective Lagrangian contains terms of the form
$\kappa h s s$ where $\kappa = \zeta (m_h^3 + 2 m_h m_s^2)/(\sqrt{16
  \lambda} (m_h^2 - m_s^2))$, and $2 \epsilon_h \frac{m_{Z_D}^2}{w} h Z_{D
  \mu} Z_D^\mu$, which lead to exotic Higgs decays $h \rightarrow s s$
and $h \rightarrow Z_D Z_D$, see \Figref{SMVvertices}.  The vertex $h
s Z_D$ is present but is suppressed by both mixings.

\begin{figure}
\begin{center}
\hspace*{-5mm}
\begin{tabular}{m{4cm} m{30mm} m{4cm} m{4mm} m{4cm} m{20mm}}
\includegraphics[width=4cm]{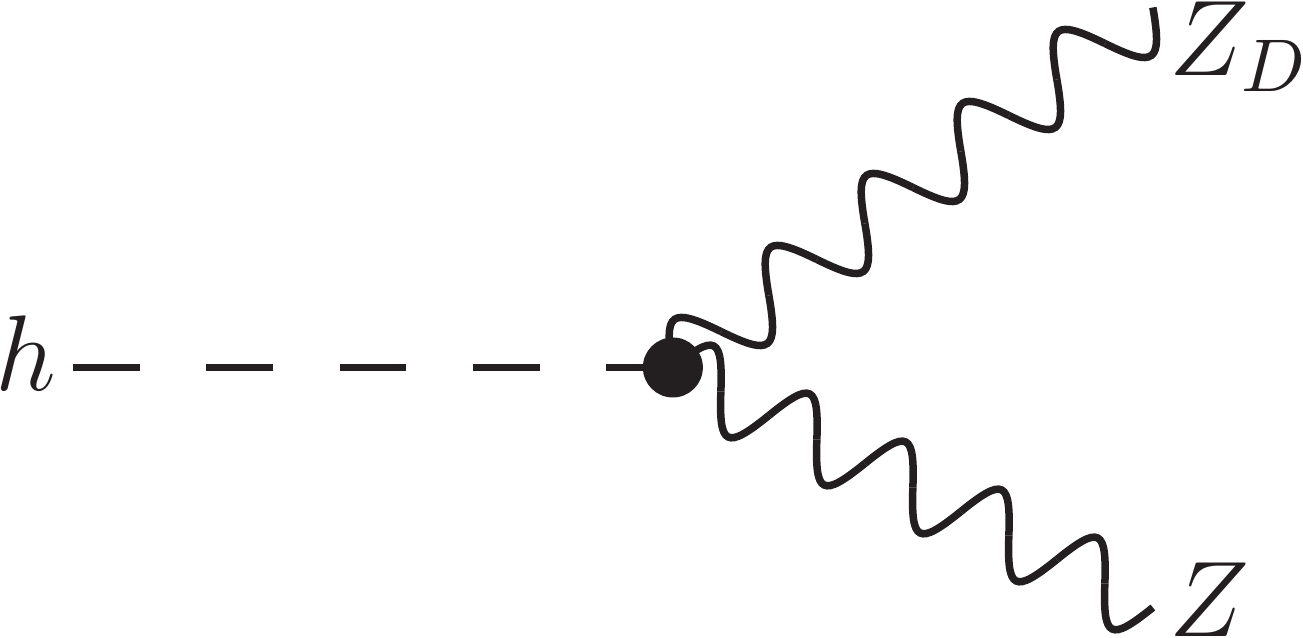}& $\mathcal{M} \propto \epsilon$ &
\includegraphics[width=4cm]{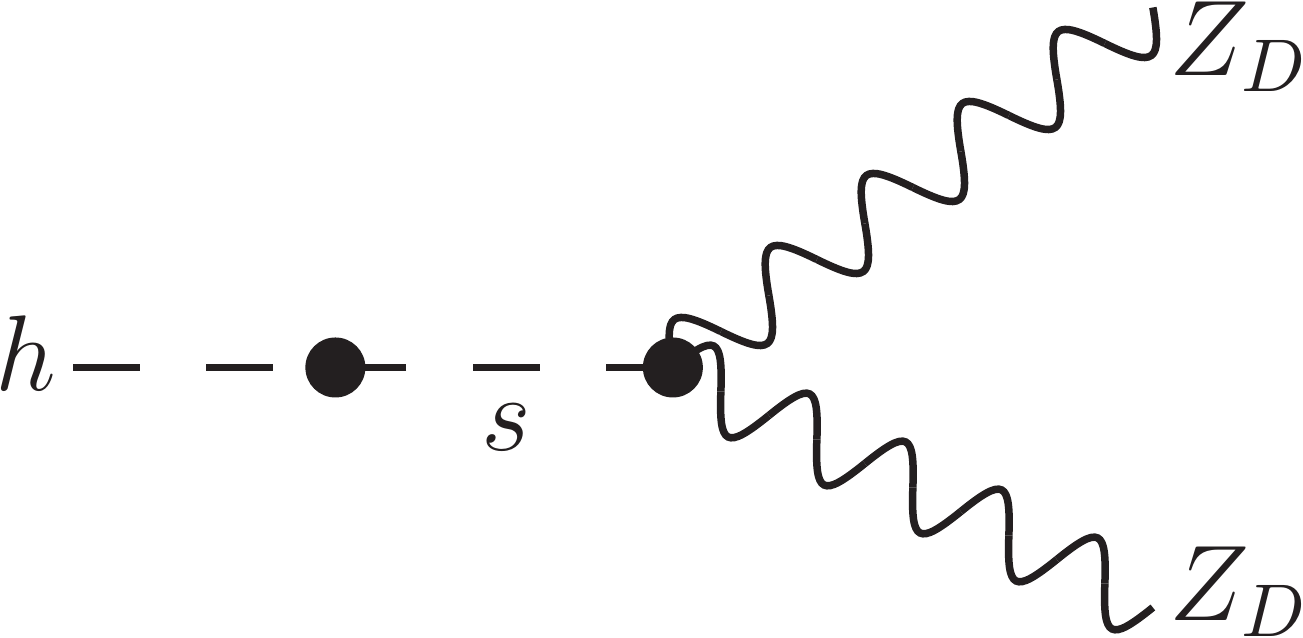}& ,  &
\includegraphics[width=4cm]{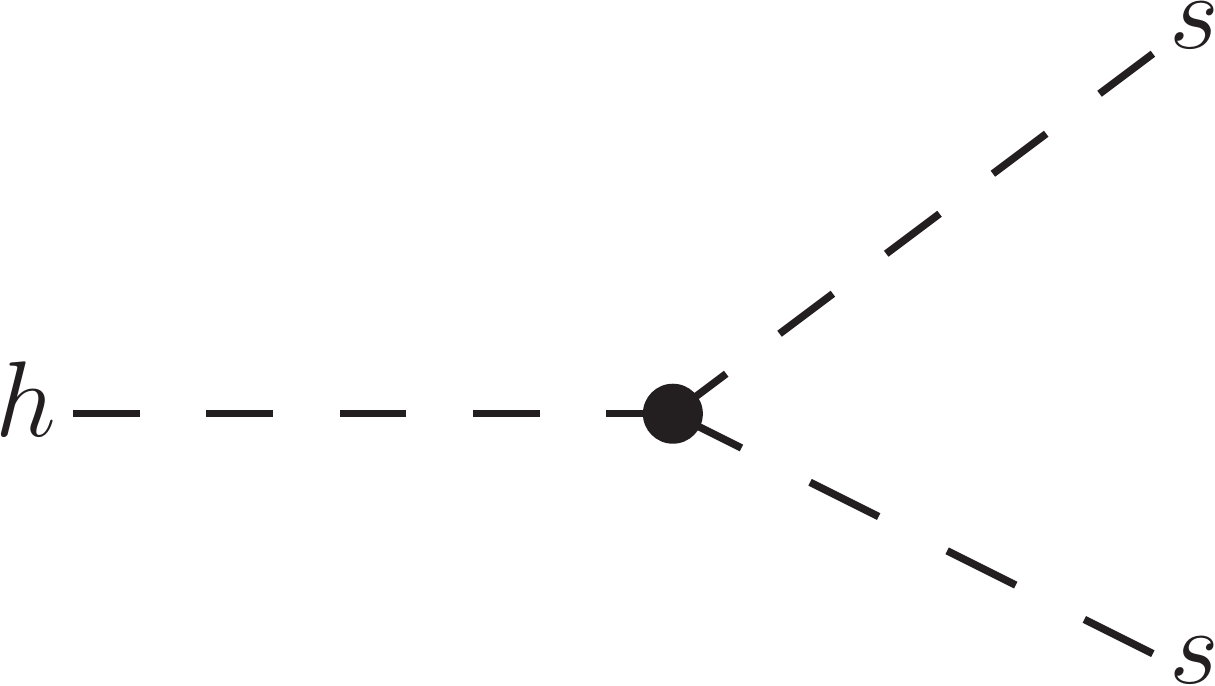} & $\mathcal{M}  \propto \zeta$
\end{tabular}
\end{center}
\caption{ The dominant exotic Higgs decays in the SM+V model. The $h
  \rightarrow Z Z_D$ matrix element is proportional to the gauge
  kinetic mixing $\epsilon$, while $h \rightarrow Z_D Z_D$ and $h
  \rightarrow s s$ are controlled by the Higgs mixing parameter
  $\zeta$. The vertex $h s Z_D$ is present but suppressed by both
  mixings.  }
\label{fig:SMVvertices}
\end{figure}

We can now discuss the relevant limits of this theory for exotic Higgs phenomenology:
\begin{itemize}
\item Gauge mixing dominates:\\
  For $\epsilon \gg \zeta$ the dominant exotic Higgs decay is $h
  \rightarrow Z Z_D$. To leading order in $m_{Z_D}^2/m_Z^2$ the
  partial width is
\begin{equation}
\Gamma(h \rightarrow Z Z_D) = \frac{\epsilon^2 \tan^2 \theta_W}{16 \pi} \frac{m_{Z_D}^2 (m_h^2 - m_Z^2)^3}{m_h^3 m_{Z}^2 v^2}.
\end{equation}
This agrees with the full analytical expression to $\sim 10\%$ for
$m_h - m_Z - m_{Z_D} > 1 \gev$.  \Figref{SMVZdbounds} shows contours
of $\mathrm{Br}(h \rightarrow Z Z_D) = 10^{-4}, 10^{-5}, 10^{-6}$. The
largest $\Br$ allowed by indirect electroweak precision constraints is
$\sim 3 \times 10^{-4}$.

In this regime, the SM+V theory leads to the $f \bar f + Z$ exotic
Higgs signatures discussed in \S \ref{sec:htoZa}. 
As outlined on page \pageref{pageref.ZZdlimits}, dedicated LHC searches for this signal at Run I and II can improve upon the electroweak precision limit. 
For very light $Z_D$ above the electron threshold this would
also lead to lepton-jets + $Z$ signatures, see
\S\ref{Clepton}~\cite{ArkaniHamed:2008qp}.

Note that $\Gamma(h \rightarrow Z Z_D) \propto \epsilon^2$. 
In addition, the dark vector will also contribute \emph{at the same order} to the $\Gamma(h \to Z \ell^+ \ell^-)$ partial width (in the non-resonant region)  via its interference with $Z^*$ in $h \to Z Z^* \to Z \ell^+ \ell^-$. Since kinetic mixing shows up in both $Z_D$ production and decay, this will lead to $\mathcal{O}(\epsilon^2)$ deviations in the dilepton spectrum and may represent a discovery opportunity, particularly for $m_{Z_D} > m_h - m_Z$. We leave this for future investigation.

\item Higgs mixing dominates:\\
  When $\zeta \gg \epsilon$ and Higgs mixing dominates then $h
  \rightarrow Z_D Z_D, s s$ are both possible, depending on the
  spectrum of the dark sector. (We still assume that $\epsilon$ is
  large enough for $Z_D$ to decay promptly.) The partial decay widths
  to leading order in $\zeta$ are
\begin{eqnarray}
\nonumber \Gamma(h \rightarrow Z_D Z_D) &=& \frac{\zeta^2}{32 \pi}  \ \frac{v^2}{m_h} \sqrt{1 - \frac{4 m_{Z_D}^2}{m_h^2}}  \  \frac{(m_h^2 + 2 m_{Z_D}^2)^2 - 8 (m_h^2 - m_{Z_D}^2)m_{Z_D}^2}{(m_h^2 - m_{s}^2)^2},
\\
\Gamma(h \rightarrow s s) &=& \frac{\zeta^2}{32 \pi} \frac{v^2}{m_h} \sqrt{1- \frac{4 m_s^2}{m_h^2}}    \ \ \ \ \frac{(m_h^2 + 2 m_s^2)^2}{(m_h^2 - m_s^2)^2}.
\end{eqnarray}
Different regions of of the $(m_{Z_D}, m_s)$ mass plane are shown in
\Figref{SMVdelta}, along with the size of the Higgs mixing $\zeta \sim
10^{-3} - 10^{-2}$ required for $\mathrm{Br}(h \rightarrow Z_DZ_D , s
s) = 10\%$ and the relative rates of $h \rightarrow ss$ vs $h
\rightarrow Z_D Z_D$ decays when both are allowed.

In Region $A$ ($m_s > m_h/2, m_{Z_D} < m_h/2$) the only relevant
exotic Higgs decay is $h \rightarrow Z_D Z_D$. This allows for
spectacular $h \rightarrow 2\ell 2\ell'$ decays ($\ell, \ell' = e$ or
$\mu$) with a reconstructed $Z_D$ resonance above the $\tau$- or
$b$-thresholds.

Region $B$ allows exotic Higgs decays both to $Z_D Z_D$ and $ss$. The
presence of two resonances below half the Higgs mass gives a rich
exotic decay phenomenology. $h \to s s \to 4 Z_D$ occurs with roughly equal probability as  $h \to Z_D Z_D$ and can result in spectacular final states with as
many as 8 leptons. Note that, in this simplified model, there is no corresponding $Z_D \to s s$ decay in the lower right corner of that mass plane. However, a (pseudo)scalar pair could be produced from dark vector decay in e.g. a 2HDM+V framework, resulting in final states with as many as 8 $b$-quarks.

Already with current data, limits of $\Br(h\rightarrow Z_D Z_D)
\lesssim 10^{-4}$ can be achieved, see~\S \ref{sec:hto4l}. Each of the
above cases may, for suitable masses, also lead to interesting
`lepton-jet' signatures, see \S\ref{Clepton}.

\item Intermediate Regime:\\
  Here the decays induced by kinetic and Higgs mixing are
  comparable. For example, \Figref{SMVZdbounds} shows that $\epsilon
  \sim 10^{-2}$ is not excluded for some values of $m_{Z_D}$, allowing
  $\Br(h\rightarrow Z Z_D) \sim 10^{-4}$. The branching ratios for
  $h\rightarrow Z_D Z_D, s s$ will be similar if $\zeta \sim 10^{-4}$.
\end{itemize}

\begin{figure}
\begin{center}
\hspace*{-8mm}
\begin{tabular}{ccc}
\begin{tabular}{c}
\includegraphics[width=8.5cm]{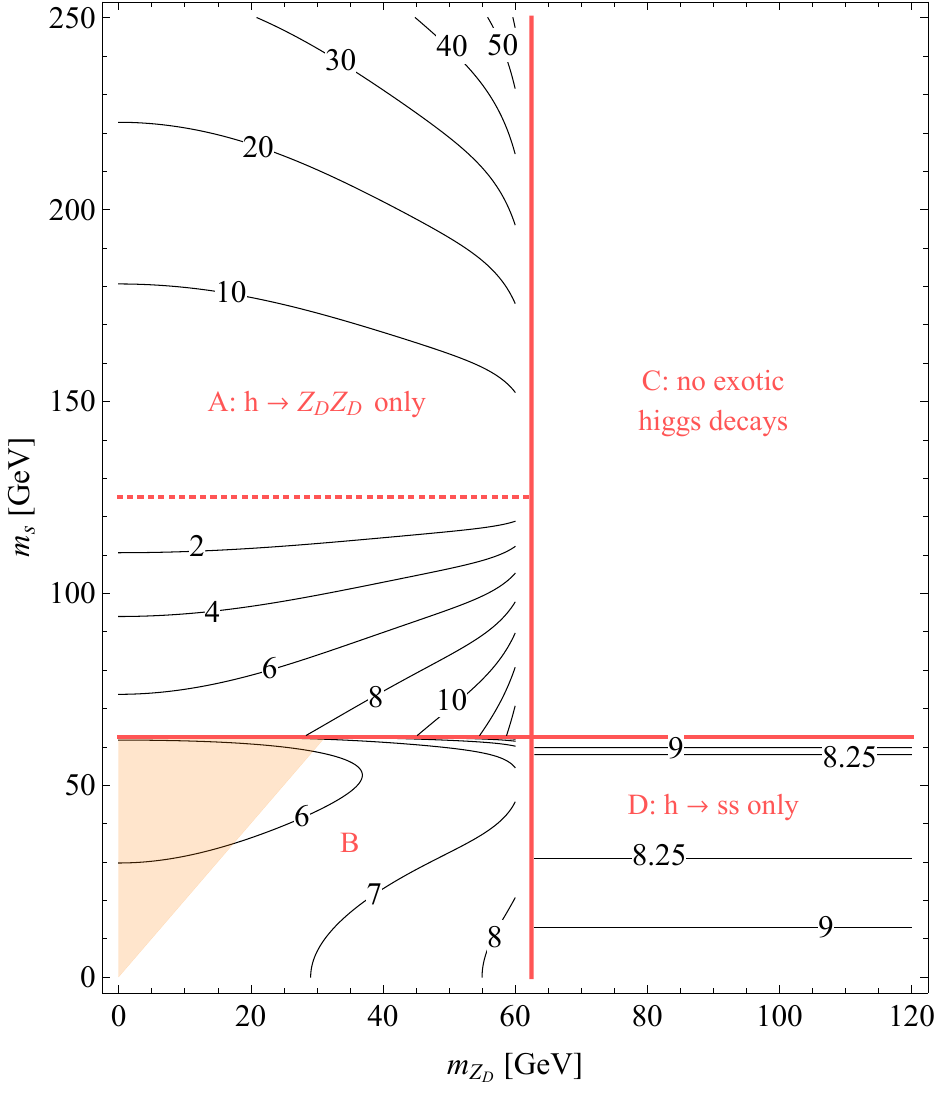}
\end{tabular}
 & \hspace{2mm} &
\begin{tabular}{c}
Region $B$: $\frac{\Gamma(h \rightarrow s s)}{\Gamma(h \rightarrow s s)+ \Gamma(h \rightarrow Z_D Z_D)}$ \vspace{2mm} \\
\includegraphics[width=8.5cm]{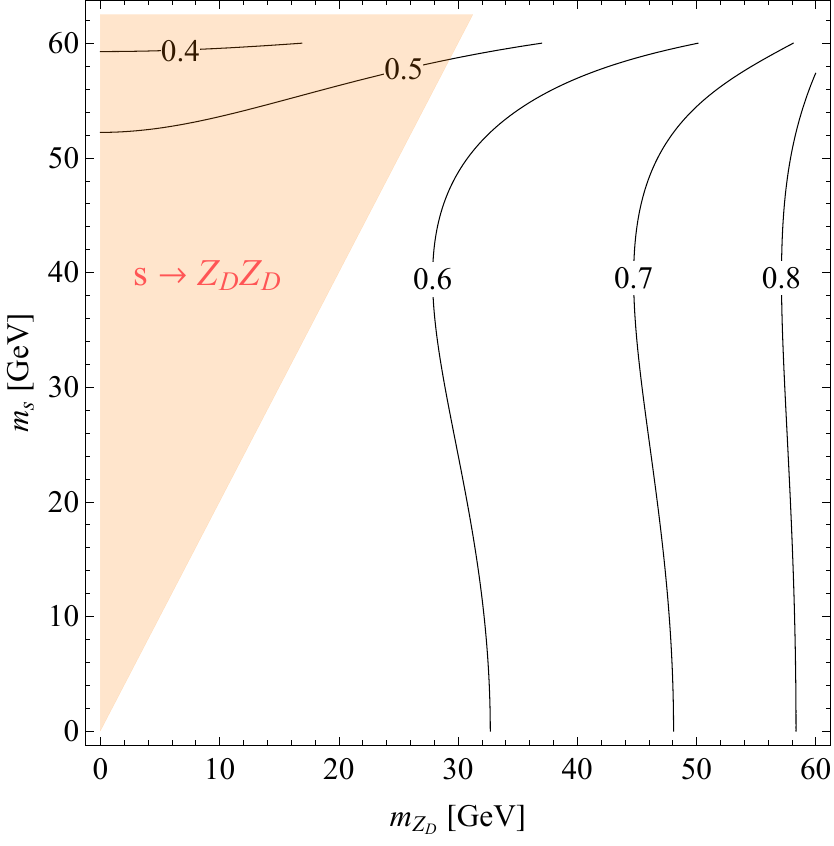}
\end{tabular}
\end{tabular}
\end{center}
\caption{
{\bf Left:} mass plane in the SM+V model with different exotic Higgs decays for $\zeta \gg \epsilon$
(i.e.~when the mixing between the Higgs and dark-Higgs dominates over the kinetic mixing). 
The black contours are the values of $\zeta  \times 10^3$ required for $\mathrm{Br}(h \rightarrow Z_D Z_D, s s) = 10\%$. 
Region $A$ is the case examined by \cite{Gopalakrishna:2008dv} (the dotted red line indicates $m_h = m_s$). Region $C$ has no exotic Higgs decays. Region $D$ reproduces the SM+S model of \S\ref{SMS}. 
Region B has both $h \rightarrow s s$ and $h \rightarrow Z_D Z_D$ decays, with the $h \rightarrow ss$ fraction of exotic decays  shown on the {\bf right}.
In the upper left shaded region, $s \rightarrow Z_D Z_D$ is the dominant decay mode of the dark scalar. This allows the Higgs to decay to up to 8 SM fermions. 
}
\label{fig:SMVdelta}
\end{figure}

\vskip 2mm
\noindent {\bf Summary}

In summary, the SM+V setup allows for many different kinds of exotic Higgs decays, including 
$h \rightarrow Z Z_D$, $h\to Z_D Z_D$, and $h\to ss$, with $Z_D \to f\bar{f}$, and $s \to f\bar{f}$ or 
$s \to Z_D Z_D \to (f\bar{f})(f\bar{f})$.  This leads to final states of 
$Z + (f\bar{f})$, $(f\bar{f})(f\bar{f})$, and $((f\bar{f})(f\bar{f}))((f\bar{f})(f\bar{f}))$, where parentheses 
around a set of particles denotes a resonance (all final-state particles combined will form the Higgs resonance). Since the $Z_D$ (although not the $s$) couples to the fermions' gauge charges, final states with several light leptons have sizable branching fractions over the entire kinematically permitted mass range. 
Certain spectra can produce interesting  lepton-jet signatures. 
		
\subsubsection{MSSM}
\label{MSSM}

In this section, we study the possible Higgs exotic decays in the
framework of the Minimal Supersymmetric Standard Model (MSSM) with
R-symmetry.

The Higgs sector of the MSSM has been extensively studied in the light
of the recent Higgs discovery. In particular a Higgs at around 125 GeV
with SM-like properties can be realized in the decoupling limit where
the additional scalars and pseudoscalars are heavy
($m_{a,H,H^\pm}\gtrsim 300$ GeV).  In this regime, exotic decays of
the type $h\to A^0 Z,\, h\to HH,\, h\to A^0A^0,\, h\to H^\pm W$ are
kinematically forbidden (here $A^0$ denotes the CP-odd
scalar).\footnote{SM-like Higgs bosons can also be achieved in a
  corner of parameter space where the additional scalar and
  pseudoscalars are lighter than $m_h$ (see for
  example~\cite{Hagiwara:2012mga,Drees:2012fb,Bechtle:2012jw}). Low
  energy flavor observables like $b\to s \gamma$, however, set
  important constraints on this region of parameter
  space~\cite{Han:2013mga,Barenboim:2013bla}.  Furthermore, the decays
  of the SM-like Higgs into lighter scalars are still not
  kinematically accessible.}  In general, the regime $m_A\leq m_h/2$
is highly constrained. This is due to the fact that the masses of the
$H$, $A^0$, and $H^\pm$ scalars of the MSSM are closely tied to one
another. In particular, at the tree level $m_{H^\pm}^2=m_A^2+m_W^2$,
leading to a charged Higgs boson already excluded by LEP searches, for
$m_A\lesssim 60$ GeV.

Additional Higgs exotic decays could be realized if some of the
sparticles are lighter than the Higgs boson. This possibility is
however very constrained by LEP and LHC searches. In particular,
assuming a LEP bound at around 100 GeV for electrically charged
sparticles, the only possible Higgs exotic decays, in the framework of
the MSSM, are to sneutrinos or to neutralinos.\footnote{Light sbottoms
  are another possibility, but this is now almost entirely ruled out
  \cite{Batell:2013psa}.}  However, in view of the LEP lower bound on
the masses of the left handed sleptons, which are related through
$SU(2)$ symmetry to the sneutrino masses, the decay to sneutrinos are
generically kinematically closed.

 \begin{figure}[t]
\begin{center}
\begin{tabular}{cc c}
\includegraphics[width=0.47\textwidth]{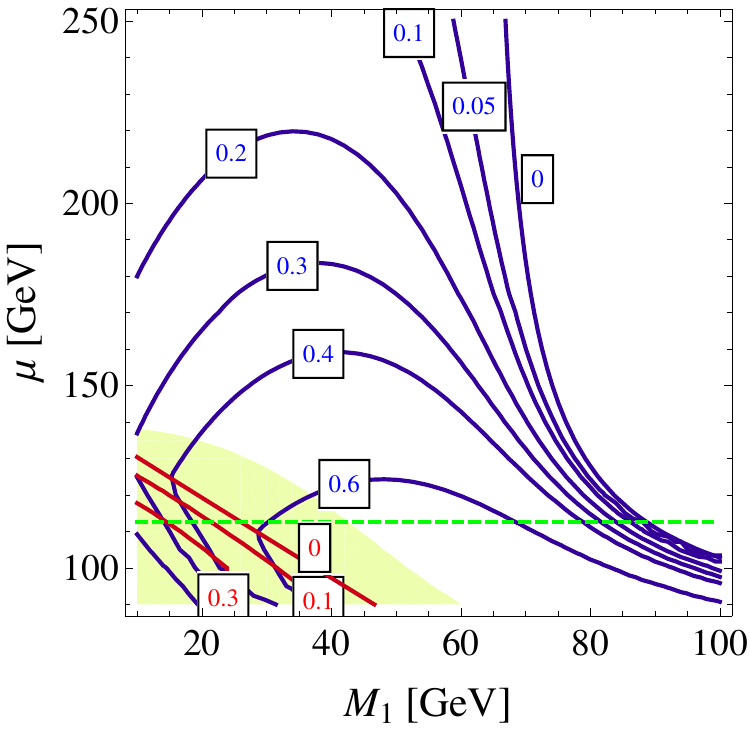}& \hspace{5mm} &
\includegraphics[width=0.47\textwidth]{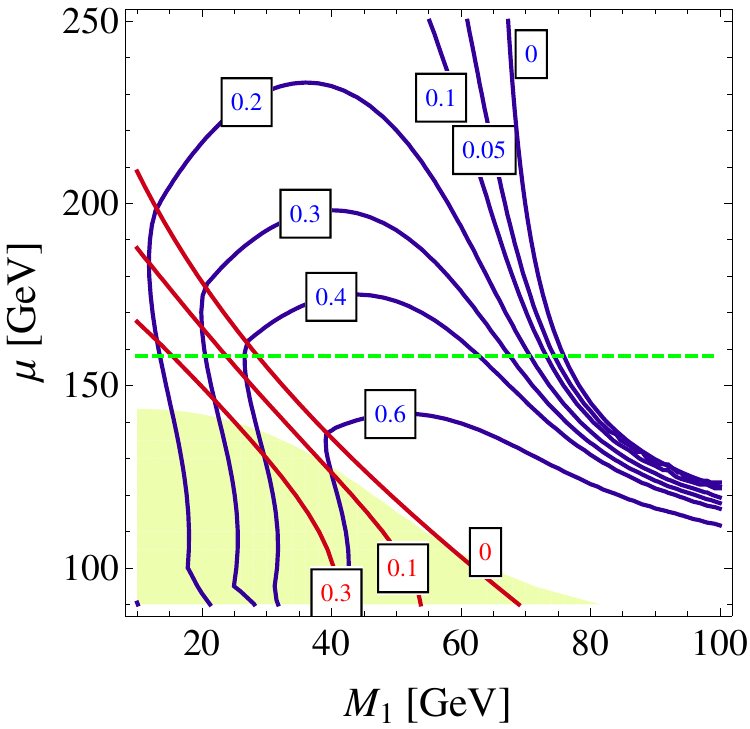}\\
(a) & & (b)
\end{tabular}\vspace*{-6mm}
\end{center}
\caption{Branching ratios of the Higgs into neutralinos:
  Br$(h\to\chi_1\chi_1)$ and Br$(h\to\chi_1\chi_2)$ are shown
  in blue and red, respectively. The yellow region is the region
  excluded by the LEP bound on the $Z$ invisible width. The region
  below the dashed green line is the region with a lightest chargino
  below the LEP bound of $\sim 100$ GeV. The input parameters are
  $\tan\beta=10$ and $M_2=300$ GeV ({\bf left}), $M_2=150$ GeV ({\bf
    right}).}
\label{Fig:MSSM300}
\end{figure}

The decay of the Higgs into neutralinos $h\to \chi_i\chi_j$
\cite{Gunion:1988yc} is therefore typically the only accessible decay
(here, as elsewhere, we suppress the superscript ``$0$'' on
neutralinos to streamline notation). This decay mode is most easily
realized in models with non-universal gaugino masses, for which the
universality relation $ M_1\sim \frac{M_2}{2}\sim \frac{M_3}{7}$ at
the electroweak scale is relaxed, allowing light LSPs while still
satisfying the LEP and LHC bounds on chargino and gluino masses. As
neutralinos which couple to the Higgs boson also typically couple to
the $Z$, the main constraint on Higgs decays to neutralinos comes from
the precise LEP measurements of the invisible and total widths of the
$Z$ boson, for $m_{\chi_i}+m_{\chi_j}<m_Z$. However, as
Fig. \ref{Fig:MSSM300} shows, for mainly bino LSPs, it is possible to
accommodate a sizable branching ratio for the decay $h\to
\chi_1\chi_1$ while still maintaining compatibility with the LEP $Z$
measurements (see
also~\cite{AlbornozVasquez:2011aa,Desai:2012qy,Dreiner:2012ex,Han:2013gba,Ananthanarayan:2013fga}
for recent studies). The parameter space for which $h\to \chi_1\chi_2$
is open is strongly constrained by both LEP $Z$ measurements (the
yellow region in Fig.~\ref{Fig:MSSM300} is the region excluded by the
LEP measurement of the $Z$ invisible width) and chargino searches.

In summary, the MSSM generally can now only provide for Higgs decays
into neutralinos.  These neutralinos may either be detector-stable, in
which case the Higgs decay is invisible (as discussed in
\S\ref{sec:MET}), or, in models with gauge-mediated supersymmetry
breaking, they may decay within the detector to photon-gravitino pairs
\cite{Mason:2009qh} (as studied in \S\ref{sec:2gaMET}).  Higgs decays
to other sparticles or to other (pseudo-)scalars in the extended MSSM
Higgs sector are now strongly constrained by the LEP and LHC
experiments.

In the following, we will investigate the possible Higgs exotic decays
in the framework of the Next to Minimal Supersymmetric Standard Model
(NMSSM). In this model, both the Higgs as well as the neutralino
sectors are significantly richer, which provides us with a larger set
of possibilities.
		
\subsubsection{NMSSM with exotic Higgs decay to scalars}
\label{NMSSMscalar}

The field content of the NMSSM is very similar to the MSSM; it differs
merely by the addition of a singlet superfield $S$, which is
introduced to address the $\mu$-problem of the MSSM (for an exhaustive
review of the NMSSM see e.g.~\cite{Ellwanger:2009dp}).  The
superpotential and soft supersymmetry-breaking terms of the Higgs
sector are given by
\begin{eqnarray}
W & = & \lambda S H_u H_d + \frac{\kappa}{3} S^3~, \\
 V_{\it soft} &= &{m^2_{H_d}} |H_d|^2 + {m^2_{H_u}} |H_u|^2 +{m^2_S}|S|^2  + (- \lambda A_{\lambda} H_u H_d S + \frac{1}{3} A_\kappa \kappa S^3 + h.c.).
\end{eqnarray}
The phenomenology of this model can be easily connected to the
simplified models that we have reviewed in previous sections.  If we
disregard the Higgsinos and singlino (which if heavy are largely
irrelevant for Higgs phenomenology) the Higgs sector of the NMSSM is
essentially that of a Type II `2HDM + Scalar' model (see
\S\ref{2HDMS}), where we can immediately identify $H_d, H_u$ as $H_1,
H_2$.

The singlet scalar $S = \frac{1}{\sqrt{2}}(S_R + i S_I)$ can obtain a
vacuum expectation value $\langle S \rangle = v_s$, generating an
effective $\mu$ parameter $\mu_{\rm eff} = \lambda v_s$.  The presence
of additional light singlet scalars, pseudoscalars, and fermions
allows for exotic Higgs decays within the NMSSM.  In this section we
discuss decays to light CP-even scalars $s$ or pseudoscalars $a$ of
the form
\begin{equation}
h \rightarrow s s \ \ , \ \ \ \ 
h \rightarrow a a \ \ , \ \ \ \ 
h \rightarrow a Z.
\end{equation}
Decays to fermions are covered in the next section,
\S\ref{NMSSMfermion}.

There are three ways of realizing the above decays within the
NMSSM. In each case, the exotic Higgs decay phenomenology is a subset
of the Type~II 2HDM+S discussed in \S\ref{2HDMS}, with some additional
restrictions (like $-\pi/2 < \alpha < 0$).

The first is an accidental cancellation resulting in a light
singlet-like $s$ or $a$. Recent examples of such models have been
found in a parameter scan~\cite{Christensen:2013dra} (for recent studies on the 
constraint on $\mathrm{Br}(h\to ss, aa)$, e.g., see~\cite{Cao:2013gba}).  By choosing
$\lambda, \kappa \sim 0.5$, $|A_\lambda| \lesssim 150 \gev$ and
$A_\kappa \sim 0$ the lightest pseudoscalar can satisfy $m_a < m_h/2$
for a SM-like Higgs $h$, with $\mathrm{Br}(h \rightarrow a a)$ or
$\mathrm{Br}(h \rightarrow Z a) \sim \mathcal{O}(0.1)$. On the other
hand, $\lambda, \kappa \sim 0.5$, $A_\lambda \sim 0-200 \gev$ and
$A_\kappa \sim - 500 \gev$ can result in a singlet-like light Higgs
satisfying $m_s < m_h/2$ with $\mathrm{Br}(h \rightarrow s s) \sim
\mathcal{O}(0.1)$. 
 
There are also two symmetry limits resulting in light pseudoscalars,
namely the R-limit and the PQ-limit of the NMSSM. The R-symmetry limit
is realized for $A_\lambda ,\ A_\kappa \to 0$~\cite{Dobrescu:2000yn,
  Dermisek:2006wr, Morrissey:2008gm}, defined by the scalar field
transformations
\beq
\label{eqn:NMSSMR} 
H_u \to H_u\, e^{i\varphi_{R}},\quad H_d \to H_d\, e^{i\varphi_{R}}, \quad S \to S\,
e^{i\varphi_{R}}\;.  
\eeq 
This global symmetry is spontaneously broken by the Higgs vacuum
expectation values $v_u, v_d, v_s$, which results in a massless
Nambu-Goldstone boson (the R-axion) appearing in the spectrum:
\begin{eqnarray}
A_R &\propto& v \sin 2 \beta \ A + v_s \ S_I,
 \end{eqnarray}
where
\begin{equation*}
A = \cos \beta \ H_{uI} + \sin \beta \ H_{dI} \ \ , \ \  v = \sqrt{v_u^2 + v_d^2}\ .
\end{equation*}
In most of the parameter space $v_s = \frac{\mu_{\rm eff}}{\lambda} \gg v
\sin 2 \beta$, making $A_{R}$ mostly singlet-like. To avoid
cosmological constraints on a massless axion and to help stabilize the
vacuum, the R-symmetry is usually taken to be approximate. This leads
to a light, mostly singlet-like pseudo-goldstone boson, and depending
on the exact parameters chosen opens up the possibility of $h
\rightarrow aa$ for $a = A_{R}$. Through
its $A$ component, $a$ then decays to SM fermions, dominantly $b\bar
b$ and $\tau^+ \tau^-$ above the respective thresholds (see
Fig.~\ref{fig:fBRs2HDMatype2}).

For $\kappa ,\ A_\kappa \to 0$ ~\cite{Peccei:1977ur, Peccei:1977hh,
  Chun:1994ct, Ciafaloni:1997gy, Hall:2004qd, Feldstein:2004xi,
  Miller:2003hm, Barbieri:2007tu, Lebedev:2009ag, Dobrescu:2000jt},
there is an approximate PQ-symmetry: 
\beq\label{eqn:NMSSMPQ} 
H_u \to H_u\, e^{i\varphi_{PQ}},\quad H_d \to H_d\, e^{i\varphi_{PQ}}, \quad S
\to S\, e^{-2i\varphi_{PQ}}\; .  
\eeq 
The PQ-symmetry limit is also shared by some other singlet-extensions
of the MSSM, including the nearly-MSSM (nMSSM) \cite{nMSSM} and the
general NMSSM (e.g., see~\cite{Ellwanger:2009dp}). Analogously to the
R-limit there is a PQ-axion,
\begin{eqnarray}
A_{PQ} &\propto& v \sin 2 \beta \ A - 2 \ v_s \ S_I.
\end{eqnarray}
Exotic Higgs decays to this pseudoscalar, and even the singlet-like
scalar, are in principle possible. However, for $m_h = 125 \gev$,
exotic Higgs decays to (pseudo-)scalars are generically not dominant in
the PQ-limit. Instead, decays to binos and singlinos can
dominate. This will be discussed in the next subsection.

\subsubsection{NMSSM with exotic Higgs decay to fermions}
\label{NMSSMfermion}

While both the R- and the PQ-limit lead to a light pseudoscalar as
discussed in \S\ref{NMSSMscalar}, the PQ-limit with $m_h = 125 \gev$
typically leads to different exotic Higgs decay phenomenology, in
which decays to fermions can be as or more important than decays to
scalars~\cite{draper2011dark,HLWY}.

When $v_s \gg v_u, v_d$, the dominant tree-level contributions to the
masses of the singlet-like scalars and singlino-like fermion $\tilde
S$ are \cite{Ciafaloni:1997gy, Miller:2005qua, draper2011dark}
\beq 
m^2_{s} \sim \kappa v_S \left(A_\kappa + 4 \kappa v_S\right)\; ,
\ \ m^2_{a} \sim -3\kappa v_S A_\kappa\; , \ \ m_{\tilde S} \sim 2
\kappa v_S\; .  
\eeq 
The pseudoscalar $a$ is light in both the R- and PQ-limits, but in the
PQ-limit $s$ and $\tilde S$ must be light as well.  This cannot be
realized in the R-limit, since vacuum stability for small $\kappa$
requires $A_\lambda \sim \mu \tan\beta$, strongly breaking R-symmetry.

This abundance of possible light singlet-like states opens up many
different exotic Higgs decays, giving phenomenology that is
qualitatively unlike the decays in the R-limit. In the R-limit, the
coupling of the SM-like Higgs to the R-axion eigenstate is $g_{h a a}
\sim \mathcal O \left({m_h^2/v_S^2}\right) \times v$
~\cite{Dobrescu:2000yn,Dermisek:2006wr}.  The trilinear coupling
$g_{haa}$ is equivalent to the mass parameter $\mu_v$ of
\figref{BRexSMS}, and as can be seen from that figure, $v_s$ as large
as $10 m_h$ can still yield a sizeable branching fraction Br$(h\to
aa)\sim 0.1$.  

The corresponding couplings in the PQ-limit instead
scale as \cite{draper2011dark,HLWY}
\begin{equation}
g_{haa}, g_{hss} \sim \mathcal{O}(\lambda^2 \epsilon' v), 
\end{equation}
where 
\beq
\label {eq:varepsdef}
\epsilon' = \left| \frac{A_\lambda}{\mu_{\rm eff} \tan \beta} -
  1\right| < \frac{m_Z}{\mu_{\rm eff} \tan \beta}
\eeq
is required by vacuum stability (avoiding a runaway in the
$S$-direction).  For a given $\mu_{\rm eff}$, small $\lambda$
corresponds to small singlet-doublet mixing and mostly SM-like Higgs
phenomenology.  Correspondingly, parameter scans using NMSSMTools
\cite{NMSSMTools, Ellwanger:2006rn, Muhlleitner:2003vg, Das:2011dg}
indicate that $\lambda \lesssim 0.2$ dominates the surviving parameter
space in the PQ-limit ($\kappa \ll \lambda$) (see
App.~\ref{sec:nmssmplots}).
It is thus common in the PQ-limit to obtain
$g_{haa}, g_{hss}\ll v$, suppressing exotic Higgs decays to
\mbox{(pseudo-)scalars}.  However, the PQ-limit allows the SM-like Higgs
boson to decay into a pair of light neutralinos $h \to \chi_i
\chi_j$~\cite{draper2011dark,HLWY,Cao:2011re}. The relevant vertex couplings for a singlino-like $\chi_1$
and a bino-like $\chi_2$ are~\cite{draper2011dark,HLWY}
\begin{eqnarray}
C_{h \chi_1 \chi_2} \sim \mathcal{O} \left(\frac{g_1 v}{v_s}
\right) \ ,  \  \ 
C_{h \chi_1 \chi_1} \sim \mathcal{O} \left(
\frac{\lambda v}{v_s \tan \beta} \right) \ .
\end{eqnarray}
For $m_{\chi_2} \lesssim 100$ and $m_{\chi_1} \sim \mathcal{O}(1-10
\gev)$ the off-diagonal decay $h \to \chi_1 \chi_2$ can be
kinematically accessible with an $\mathcal{O}(0.1)$ branching
fraction. The purely invisible decay $h \to \chi_1 \chi_1$ is
suppressed by a factor of $\sim \lambda/(g_1 \tan \beta)$ relative to
the off-diagonal decay, ignoring phase space factors.  Meanwhile,
Higgs decay to a pair of bino-like $\chi_2$ also scales as a single
factor of the bino-Higgsino mixing angle, $C_{h\chi_2\chi_2}\sim
\mathcal{O} (g_1 /\lambda) C_{h\chi_1\chi_2} $ and if $h\to
\chi_2\chi_2$ is kinematically available, this branching fraction can
be important.

For $m_{\chi_2}-m_{\chi_1} > {\rm min} \{m_{s}, m_{a}\}$, the heavier
neutralino can decay via $\chi_2\to\chi_1 a$ or $\chi_2\to\chi_1
s$~\cite{draper2011dark,HLWY}. This leads to a plethora of possible $h
\rightarrow (xx) + \met$ or $h \rightarrow (xx)(yy) + \met$ decays,
where $x, y$ are SM partons (most likely $b$, $\tau$, or light jets,
see \S\ref{2HDMS}) that reconstruct the singlet boson mass $a$ or $s$.
If $m_{\chi_2} - m_{\chi_1} < {\rm min} \{m_{s}, m_{a} \}$, the
principal decay mode of $\chi_2$ is the three-body decay $\chi_2\to
(a,s)^*\chi_1 \to (xx) \chi_1$, while the radiative mode $\chi_2 \to
\chi_1 \gamma$ may become significant, with Br$(h \to \chi_1 \chi_1
\gamma)$ as high as $\mathcal O(0.1)$.  On-shell $\chi_2 \to \chi_1
Z$ does not occur until $m_{\chi_2} - m_{\chi_1} > m_Z$.  Given that
we require $m_{\chi_2} - m_{\chi_1} < m_{h}-2 m_{\chi_1}$, these
points are sparse.  Fig.~\ref{fig:NMSSMdecay_topo} shows the
corresponding exotic decay topologies.  Further discussion can be
found in Appendix \ref{sec:nmssmplots}, together with some example model
points which illustrate the main exotic Higgs decay modes in the
PQ-symmetry limit of the NMSSM in Table~\ref{table:NMSSMbenchmark}.

 \begin{figure}[t]
\begin{center}
\begin{tabular}{cc c}
\includegraphics[width=0.45\textwidth]{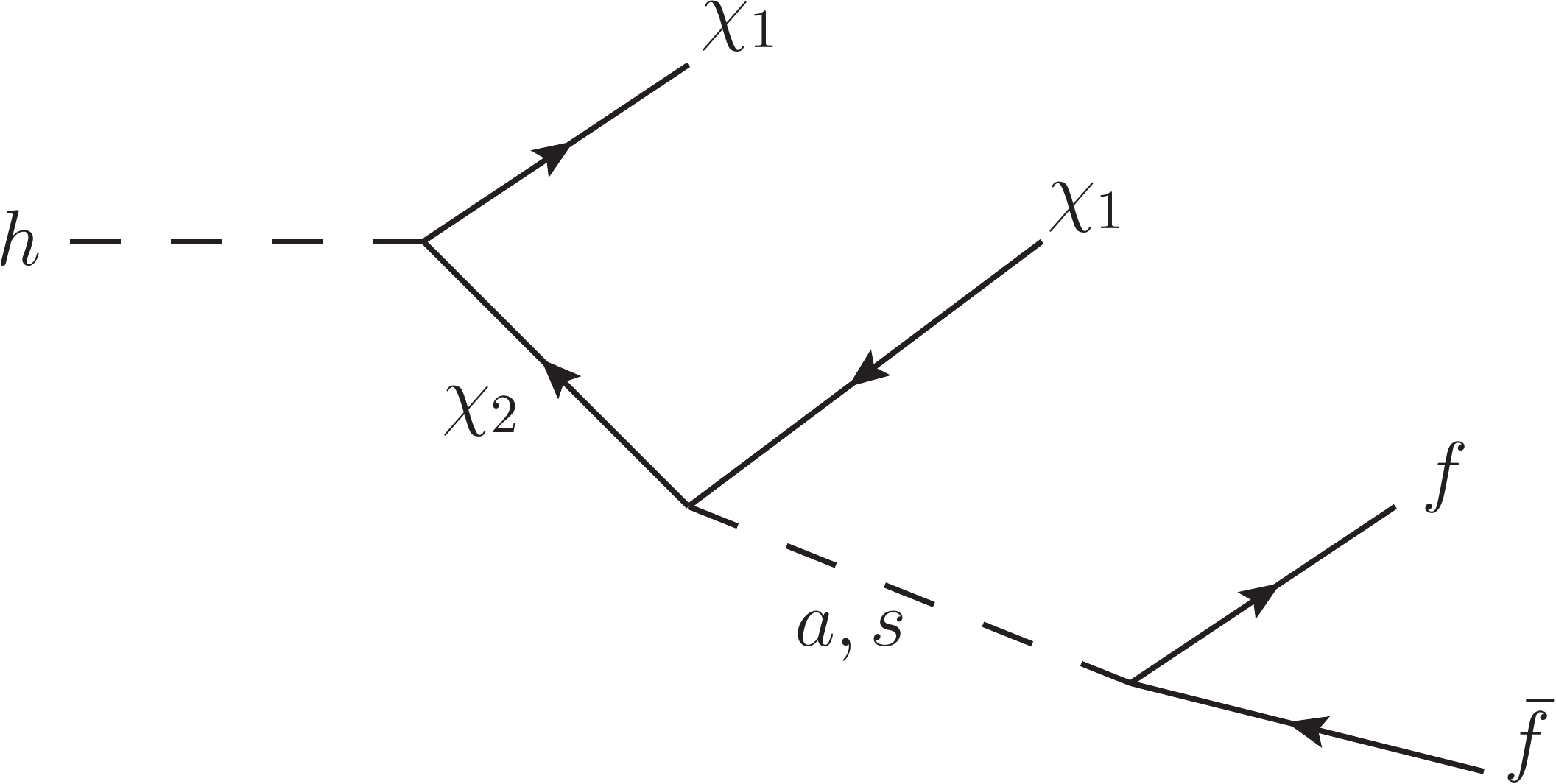}& \hspace{5mm} &
\includegraphics[width=0.38\textwidth]{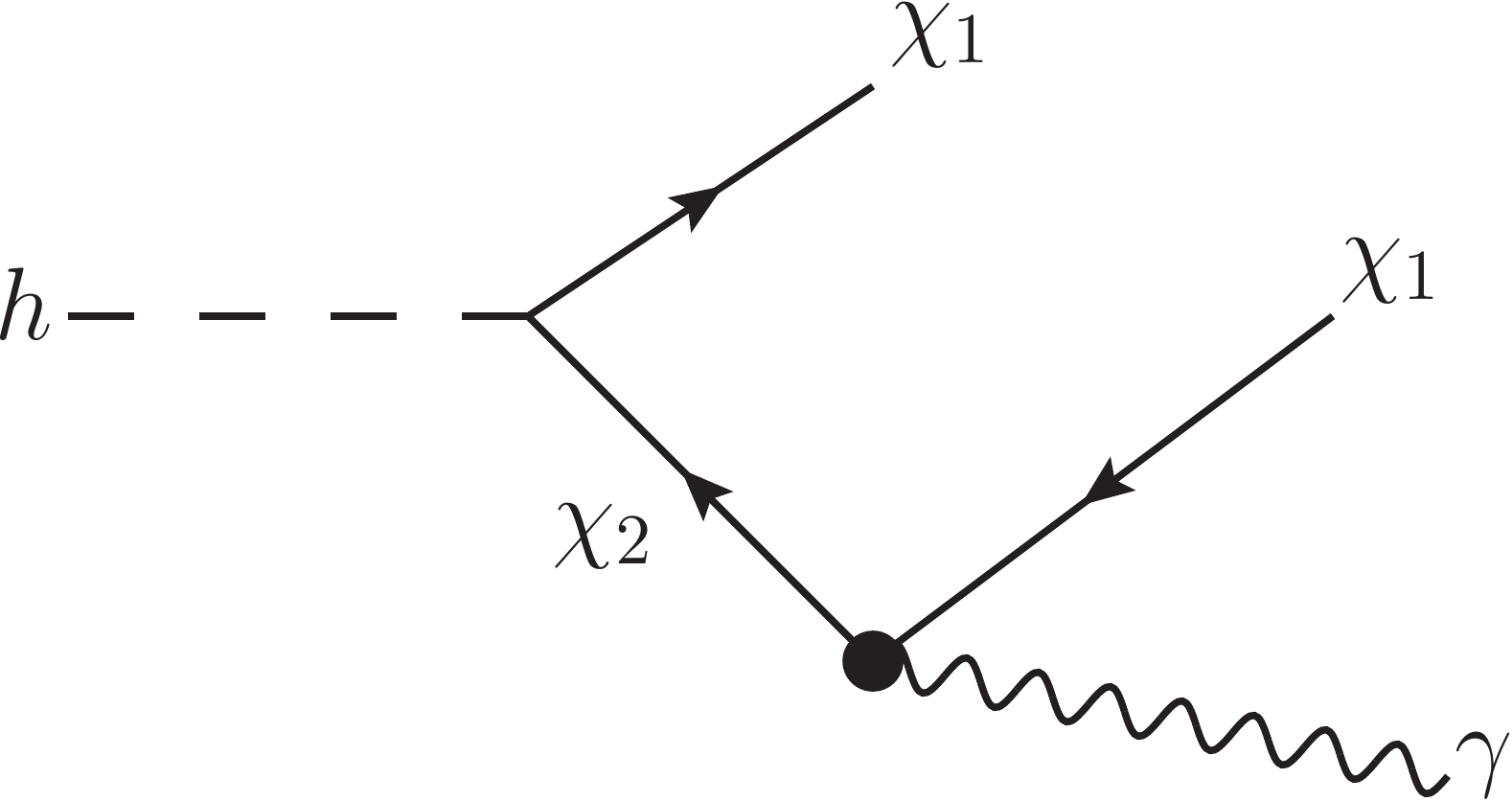}\\
(a) & & (b)
\end{tabular}\vspace*{-6mm}
\end{center}
\caption{Two significant fermionic decay topologies of the SM-like Higgs boson in the
  PQ symmetry limit.  {\bf Left (a):}
  depending on whether $\mathrm{min}\{m_s, m_a\}$ exceeds $m_{\chi_2}
  - m_{\chi_1}$, $a(s)$ may or may not be on shell. {\bf Right (b):}
  to be non-negligible, the radiative $\chi_2$ decay requires
  $\mathrm{min}\{m_s, m_a\} > m_{\chi_2} - m_{\chi_1}$.)}
\label{fig:NMSSMdecay_topo}
\end{figure}

\vskip 2mm
\noindent {\bf Summary:}

The PQ-limit of the NMSSM yields semi-invisible exotic Higgs decays
into pairs of light neutralinos, most typically $h\to \chi_2\chi_1$ or
$h\to \chi_2\chi_2$, with $\chi_2 \to \chi_1 a$, $\chi_1 s$, and $a,
s\to (f\bar f, gg, \gamma \gamma)$~\cite{draper2011dark,HLWY}.  This yields final states of the
form $(b\bar b)+\met$, $(\tau \tau)+\met$, $(b\bar b) (b\bar b)+\met$,
$(\tau \tau)(\tau\tau)+\met$, $(b\bar b) (\tau\tau)+\met$, and the
rarer but cleaner $\gamma+\met$, $(2,4)\mu+\met$, $(\mu\mu)(b\bar
b)+\met$.  Depending on the spectrum, the visible particles may be
collimated or isolated.  Current experimental constraints and future
prospects for a subset of these decays are discussed in
\S\ref{sec:1gaMET} ($\gamma+\met$), \S\ref{sec:2gaMET}
($2\gamma+\met$), \S\ref{Clepton} (collimated $2\ell+X$),
\S\ref{sec:collep_nomet} (collimated $4\ell + X$), \S\ref{bbmet}
($bb+\met$), and \S\ref{tatamet} ($\tau\tau+\met$).

		
\subsubsection{Little Higgs}
\label{LHiggs}

Another class of models with additional potentially light spin-0
fields is Little Higgs~ \cite{ArkaniHamed:2001nc, ArkaniHamed:2002qx,
Perelstein:2005ka}.
In these models, the SM Higgs doublet serves as a pseudo Nambu Goldstone boson (PNGB) of multiple approximate global symmetries. 
Explicit breaking of this set of symmetries is {\it collective}, namely, apparent only in the presence of at least two terms in the Lagrangian. This ensures that quadratically divergent diagrams
contributing to the Higgs mass parameter require two loops, thereby allowing to push the
cutoff scale to $\Lambda\sim(4\pi)^2v\sim10\tev$ instead of the usual $4\pi v\sim1\tev$.

In order to implement collective symmetry breaking, the electroweak gauge group is extended to
a larger global symmetry, which is partially gauged.
The partial gauging introduces the explicit breaking, which is crucial for having a nonzero
Higgs mass as well as Yukawa couplings.
In most Little Higgs models, all the spontaneously broken global generators are explicitly broken
by the partial gauging, thereby giving mass to the associated Goldstone bosons.
However, in some models, not all global generators are explicitly
broken at leading order, either because they are collectively broken like the ones related
to the Higgs doublet, or because that would interfere with collective symmetry
breaking~\cite{Perelstein:2003wd,Csaki:2003si}.
A consequence of this is the presence of light (pseudo-)scalars  $a$ with direct couplings
to the SM Higgs, which potentially leads to exotic Higgs
decays~\cite{Kilian:2004pp,Surujon:2010ed}.

If one imposes Minimal Flavor Violation
(MFV)\cite{D'Ambrosio:2002ex,Chivukula:1987py,Hall:1990ac,Buras:2000dm}
in order to avoid large flavor changing neutral currents, the
couplings of $a$ to SM fermions are proportional to the SM Yukawas, and thus the coupling
to the $b$ quark is typically enhanced.

However, an enhanced decay rate of $a$ to gluons is possible in some cases,
as well as an enhanced rate to charm quarks -
which arises for models with enhanced up-Yukawa couplings compared to down-Yukawa.
The former possibility results in a ``buried Higgs''~\cite{Bellazzini:2009xt,Falkowski:2010hi}
scenario, 
with the Higgs decaying to four gluon-originated jets, while the latter implies $h\to4c$
decays, also known as ``charming Higgs''~\cite{Bellazzini:2009kw}
(see also~\cite{Lewis:2012pf} for a more recent jet substructure study), where $a$
may decay to $c\bar c$ even if $m_a>2m_b$. 
Although the original version of the charming Higgs is
excluded by the observed Higgs mass, other versions may exist (and in
any case the same final state arises in other models, such as the type
IV 2HDM+Scalar models mentioned in \S~\ref{2HDMS}.)

As a final comment, note that in models with multiple light particles, cascade decays among
these particles, and more complex final states, such as
$h\to a'a'\to (aaa)(aaa)$, could result.
		
\subsubsection{Hidden Valleys}
\label{sec:hiddenvalley}

In the hidden valley scenario~\cite{Strassler:2006im,
  Strassler:2006ri, Strassler:2006qa, Strassler:2008bv,Juknevich:2009ji,Juknevich:2009gg}, a sector of SM-singlet
particles, interacting amongst themselves, is appended to the SM.
These are then coupled to the SM through irrelevant operators at the
TeV scale, or through marginal operators with weak couplings.  An
important additional feature of a hidden valley, distinct from a
general hidden sector, is that a mass gap (or a symmetry) forbids one
or more of the valley particles from decaying entirely to
hidden-sector particles; instead, these particles decay to SM
particles.  Interactions between the SM and hidden valley may also
allow the 125 GeV Higgs to decay to valley particles, which in turn
decay to SM particles.

The phenomenology of Higgs decays to hidden valleys can sometimes be
captured by ``simplified'' models, including the ones studied earlier
in this section, but much more complex patterns of decays may easily
arise.  This is especially true if hidden valleys have strong and
perhaps confining interactions.  For instance, if hidden-valley
confinement generates hidden ``hadrons'', then, just as QCD has a
variety of hadrons that decay to non-hadronic final states, often with
long lifetimes, and with masses that are spread widely around 1 GeV,
the hidden valley may have multiple particles of comparable masses
that decay to SM particles, sometimes with very long lifetimes.

More generally, common features that arise in hidden valleys,
generally as a result of self-interactions of one sort or another,
include the following.
\begin{itemize}
\item Multiple types of neutral particles with narrow widths arise, decaying
  to the SM particles via very weak interactions.
\item Because their decays are mediated by very weak interactions,
  their lifetimes may be long, though they are sensitive to unknown
  parameters; decays may occur promptly, at a displaced vertex, or far
  outside the detector, giving a $\met$ signal.
\item As they interact so weakly with the SM, they are rarely produced
  directly; instead, they are dominantly produced in the decays of
  heavy particles, including the Higgs, neutralinos, etc.
\item When created in the decays of heavy particles, the new
  particles, if sufficiently light, may commonly be highly boosted.
\item Because of their self-interactions, the new particles are often
  produced in clusters, just as QCD hadrons (and their parent gluons)
  are produced in the showering and hadronization that forms QCD jets.
\end{itemize}

Hidden valleys arise in several theoretical contexts.  Dark matter may
well be from a hidden sector; for instance, the ``WIMP miracle'' can
apply to particles that are not WIMPs at all~\cite{Feng:2008ya}.  Many
of the models that have attempted to explain recent hints of indirect
and direct dark matter detection have involved hidden valleys, the
most famous being \cite{ArkaniHamed:2008qn,Pospelov:2008jd}.
Supersymmetry breaking models typically have a hidden sector, within
which some particles (often just a single spin-one or spin-zero
particle) occasionally survives to low energy.  And model building
that attempts to generate the SM from string theory generally leads to
additional non-SM gauge groups under which no SM particles are
charged.  Hidden valleys have also appeared in certain attempts to
address the hierarchy problem (cf. Twin Higgs~\cite{Chacko:2005pe}, in
which the top quark and $W$ loops that correct the Higgs mass are
cancelled by particles in a hidden valley).

Entry to the hidden valley may occur through a wide variety of
``portals''; any neutral particle, or particle/anti-particle pair, may
couple to operators made from valley fields, and consequently may
itself decay to such particles, and may mediate transitions between SM
and valley fields.  The $Z$ boson can be a portal; rare $Z$ decays,
and rare $Z$-mediated processes, can be used to put significant bounds
on certain types of hidden valleys.  However, explicit calculation
shows these bounds are not sufficient to rule out the
possibility~\cite{Strassler:2006im,Strassler:2006ri} that the Higgs
itself has decays to a hidden valley that could be discovered in
current or future LHC data.  This is because of the Higgs' narrow
width, which makes it far more sensitive to very small couplings than
is the $Z$, which is nearly 3 orders of magnitude wider.

Aside from direct limits from $Z$ decays, rare $B$ and other meson
decays, and direct production limits, constraints on hidden valleys
can arise from precision tests of the SM.  However, these are
generally rather weak \cite{Strassler:2006im}, since the hidden valley
sector is weakly coupled to the SM.  Cosmological constraints are
sometimes important, but very large classes of models evade them
easily \cite{Strassler:2006im}.

The hidden valley scenario is relevant for our current purposes
because new Higgs decays commonly arise in hidden valley models.  What
makes hidden valleys an experimental challenge is that the range of
theoretical possibilities are very large.  None of the potential
motivations --- dark matter, supersymmetry breaking, naturalness, or
string theory --- point us toward any particular type of hidden
valley, nor is there a strong reason for it to be minimal.  The
diversity of phenomena in quantum field theory in its various
manifestations (e.g.~extra dimensions) is enormous, and any of these
phenomena might appear in a hidden sector. Fortunately, many models
produce similar experimental signals.  Indeed, in many hidden valleys,
the dominant discoverable process is the same as one that occurs in
one of the models that we have already discussed.

We first give a few examples of phenomena that can arise in hidden
valleys that, though very different in their origin from theories we
have already discussed, give signals that we have already discussed.
We then give some examples of phenomena that we have not discussed
that can arise in these models.

\vskip 2mm
\noindent {\bf SM + Scalar, 2HDM + Scalar} (\S\ref{SMS} and \S\ref{2HDMS}): 

Consider a confining hidden valley, with its own gauge group $G$ and
quarks $Q_i$, and a Higgs-like scalar $S$ that gives mass to the $Q_i$
via a $S Q_i \bar Q_i$ coupling, but does not break $G$.  We imagine
that $S$ mixes with one of the SM Higgs doublets; for example, this
model could be an extension of the NMSSM.  If the gauge group confines
and breaks chiral smmetry, with PNGBs $K_v$, then a $S K_v K_v$
coupling and the mixing of $S$ and the Higgs allows the decay $h\to
K_v K_v$.  The $K_v$ may then decay to SM fermions, with the heaviest
fermions available typically most common; this can occur for instance
via mixing with a heavy $Z'$ or with a SM pseudoscalar Higgs.  An
example (not at all unique) is given in the model of
\cite{Strassler:2006im}, which shows decays may be prompt for
$m_{K_v}$ above about 20 GeV.

\vskip 2mm
\noindent {\bf SM + 2 Fermions (and similar)} (\S\ref{subsub:SMF}): 

The same signal that arises in a simplified model with fermions may
arise in hidden valleys, for much the same reasons.  But it may arise
even when there are no fermions at all.  Consider the same model just
mentioned, but with two flavors of PNGBs (as with pions and kaons in
the SM), $\pi_v$ and $K_v$.  It may be that the $\pi_v$ are stable or
very long-lived, and produce only \MET, while $K_v$ cannot decay to
two or more $\pi_v$. This could be due to kinematic constraints (like
Kaons in QCD if $m_K$ were less than $2 m_\pi$), or symmetries. In
that case $K_v$ may decay via a small coupling to a scalar field $S$
that mixes with $h$, or via a spin-one vector $V$ that mixes with $Z$.
This opens up the possibility of $K_v\to \pi_v h^*$ or $K_v\to \pi_v
Z^*$, which would produce a non-resonant pair of SM fermions, or
resonant decays such as $K_v\to \pi_v S$ or $K_v\to \pi_v V$.

In other hidden valleys, it can happen that there are two states, the
heavier of which can only decay to the lighter via a loop of heavy
particles, which allows for a radiative (i.e. photon emission) decay.
If the lighter state is stable or decays invisibly, then the signal of
two photons + $\met$ can arise.

The lesson here is that these signals can arise whenever we have two
states, the lighter of which is invisible and the heavier of which can
only decay to the lighter via emission of an on- or off-shell particle
that decays to SM fermions or gauge bosons.

\vskip 2mm
\noindent {\bf SM + Vector} (\S\ref{subsec:SMvector}): 

There are several ways for spin-one particles to arise naturally in a
hidden valley, and for these to mix with the photon and/or $Z$ to
allow them to decay to SM fermions.  There could be a broken $U(1)$
symmetry, giving what is often called a ``dark photon''.  Mixing with
the hypercharge boson is through renormalizable kinetic mixing. There
could be a broken non-abelian gauge symmetry; in this case, there
could be several spin-one particles, with the heavier ones decaying to
the lighter ones via a cascade.  Such a scenario only permits mixing
with hypercharge through a dimension-five version of kinetic
mixing. Finally, the spin-one particles could be stable bound states
$\rho_v$, like a $\rho$ meson in a theory with no chiral symmetry
breaking and no pions.  (An example with a stable vector and a stable
pseudovector was given in \cite{Strassler:2006im}.)

Decays of the Higgs to such particles can be induced using any of the
mechanisms mentioned above or in the simplified model discussion.  For
instance, decay of a Higgs to two $\rho_v$ (or, if there are two
vectors $\rho_1, \rho_2$, the decay $h\to \rho_1 \rho_2$) can occur
along the same lines as the decay $h\to K_v K_v$ mentioned earlier.

A particularly well-known example of this type of hidden valley is
\cite{ArkaniHamed:2008qp}, in which an elementary ``dark photon'' of
low mass preferentially creates light leptons with very few photons or
neutral pions.  Dark matter annihilation can create these dark photons
and thus provide leptonic final states potentially consistent with
certain astrophysical observations.  Because the dark photon must be
lightweight, it tends to be produced with a high boost, giving the
now-famous phenomenon of a ``lepton-jet''.  A simple lepton-jet
contains two nearby leptons, isolated from other particles but not
from one another.  (More complex lepton jets will be addressed below.)

\vskip 3mm 

In this paper, we have limited ourselves to relatively simple final
states to which the Higgs might decay.  However, the complex final
states that are common in hidden valleys are important to keep in
mind, as they can pose considerable (though interesting) experimental
challenges.  For instance, even limited complexity can lead to 8 or
more visible partons, from four hidden valley scalars, pseudoscalars,
or vectors (possibly plus \MET) in a Higgs decay. The kinematics are
then dependent on the hidden sector's mass spectrum and internal
dynamics, giving rise to a wide array of signals.

This direction of research lies beyond our scope and should be
returned to in the future.  However, a couple of relatively simple
experimental cases deserve note.  First, any of the final states
mentioned above may be accompanied by valley particles that are
long-lived on detector time-scales and therefore invisible.  This
motivates searches for similar final states accompanied by $\met$,
which we address in \S \ref{sec:1gaMET}, \S\ref{sec:2gaMET},
\S\ref{sec:IsoLeptonMET}, \S\ref{sec:2lMET}, \S\ref{Clepton},
\S\ref{sec:collep_nomet},
\S\ref{bbmet}, and \S\ref{tatamet}.

Second, many models produce ``complex'' lepton-jets, in which multiple
``dark-photons'' (or dark non-abelian bosons or $\rho$ mesons) are
created near one another, clustered either by the kinematics of a
cascade decay or by the physics of hidden-valley showering and
hadronization.  Some efforts have been made to find such objects
\cite{Aad:2013yqp}.  Another interesting possibility would give
several such dark photons created with low momentum along with \MET,
leading to many unclustered very soft leptons. An attempt to search
for such final states was made by CDF
\cite{Aaltonen:2012pu}. Unfortunately, in models where the vector
bosons can decay also to pions, the leptons are fewer and hadrons
often take their place, making the challenges much greater.  One
important signature, which is useful for particles of mass up to
several GeV, is a di-pion resonance with the same mass as a di-lepton
resonance.  In models where the light particles are pseudo-scalars,
and often produce taus and rarely muons, it is not clear whether a
good search strategy exists, unless rates are sufficient for a di-muon
resonance search.

Another issue that commonly arises in hidden valleys is long-lived
neutral particles~\cite{Strassler:2006im}.  Valley particles, by
definition, are neutral under all SM gauge groups.  The case of
hadrons in QCD offers a useful analogy.  Most hadrons in QCD are
highly unstable, but a few are stable, and others are metastable, for
a diversity of reasons (exact and approximate symmetries, weak forces,
kinematic constraints, etc.)  Their decays are often very slow on QCD
time-scales, and their lifetimes are spread across many orders of
magnitude, from the neutron at fifteen minutes to the $\pi^0$ at a
hundredth of a femtosecond.  The same could be true of a sector of
hidden valley particles.  The particles that are stable on detector
time-scales will give us nothing but \MET.  The shorter-lived
particles will give us prompt decays, of the sort that we discuss in
this article.  But it is quite common, given a rich spectrum of
particles with a variety of lifetimes, that one or more will decay
typically with a displaced vertex.  An example of a natural theory
where such particles may arise in Higgs decays \cite{Strassler:2006ri}
is the Twin Higgs \cite{Chacko:2005pe}, though the details are still
to be worked out. This issue takes us beyond our current purposes, but
this possibility has already received some amount of experimental
study, as in
\cite{Abe:1998ee,
Scott:2004wz,
Abazov:2006as,
Abulencia:2007ut,
Aaltonen:2008dm,
Abazov:2008zm,
Abazov:2009ik,
CMS:2011iwa,
CMS:2011zva,
Aad:2012kw,
ATLAS:2012av,
Chatrchyan:2012ir,
LHCb:2012gja,
CMS:2012rza,
CMS:2012uza,
Chatrchyan:2012jwg,
Chatrchyan:2012jna,
Aad:2012zx,
ATLAS:2012ioa,
CMS:2013oea}


\section{\secsize $\secbold{h \rightarrow \met}$}\label{sec:MET}

\subsection{Theoretical Motivation}

Higgs decays into a new stable, neutral particle have a venerable
history, going back to the pioneering work of Suzuki and Shrock
\cite{Shrock:1982kd}.  Since the astrophysical evidence for particle
dark matter strongly suggests the existence of new neutral degrees of
freedom, potential Higgs decays to dark matter (DM) are a topic of
particular interest \cite{Silveira:1985rk,McDonald:1993ex,Burgess:2000yq}.  While the
most minimal models of Higgs-coupled DM with $2 m_{DM}< 125$ GeV have
been excluded by LHC observations of the Higgs boson alone (direct
detection, particularly from XENON100, also constrains these models;
see, e.g.~\cite{Mambrini:2011ik}), non-minimal models can easily still
allow for light thermal DM coupling to the SM predominantly through
the Higgs \cite{Batell:2010bp,Pospelov:2011yp,Weinberg:2013kea}.  Dark matter
therefore constitutes one of the most robust motivations for the
invisible decay mode.

The possibility that the Higgs might dominantly decay to neutralinos
in models with weak-scale supersymmetry \cite{Gunion:1988yc} has
received comparatively less attention due to the difficulty of
achieving this signature in traditional CMSSM-type models of
supersymmetry breaking \cite{Martin:1999qf}.  With less restricted
spectra, or in non-minimal models such as the NMSSM, it is easier to
realize Higgs decays to neutralinos \cite{Han:2004yd,Barger:2006dh,Hall:2011aa,
  Ananthanarayan:2013fga,Kozaczuk:2013spa} and/or goldstini
\cite{Bertolini:2011tw, Riva:2012hz}.

Beyond supersymmetry and DM, many theoretical frameworks predict one
or more new neutral particles, often naturally light, which can
furnish an invisible BSM decay mode for the Higgs boson.  Frequently
considered examples are majorons \cite{Shrock:1982kd,Li:1985hy,
  Joshipura:1992ua} as well as more general PNGBs \cite{Dedes:2008bf};
hidden sectors \cite{Binoth:1996au,Schabinger:2005ei,Strassler:2006im,Delgado:2007dx,Craig:2013fga};
fourth-generation neutrinos \cite{Rozanov:2010xi,Keung:2011zc}; and
right-handed neutrinos \cite{Graesser:2007yj} and their K-K
excitations \cite{Martin:1999qf} or superpartners
\cite{Banerjee:2013fga}.

\subsection{Existing Collider Studies}    

The Higgs decay to missing energy is a difficult experimental
signature due to the lack of kinematic information in the final state
and the irreducible background from SM $Z\to\nu\bar \nu$ production.
Nevertheless, the excellent theoretical motivation for this signal has
made it a focus of study for many years.  A Higgs decaying invisibly
must be produced in association with another object in order to be
observed. In order of production cross-section, the reasonable
candidates are then:
\begin{itemize}
\item $gg\to h + $jets
\item VBF production of $h+2j$
\item $Wh$, $W\to \ell\nu$
\item $Zh$, $Z\to \ell^+\ell^-, (b\bar b)$.
\end{itemize}  
While $t\bar t h$ associated production initially appeared promising
\cite{Gunion:1993jf,Kersevan:2002zj}, the small cross-section and
complex final state make this mode challenging.

The monojet$+\met$ signal, sensitive to gluon fusion production with
ISR,\footnote{There is a potentially significant contribution from VBF
  to monojet $+\met$ searches, depending on the jet criteria adopted
  in the search \cite{Djouadi:2012zc}.} has a large rate, but its
reach is limited by the lack of kinematic handles to separate an
invisible Higgs from the nearby background $Z+j$ \cite{Bai:2011wz}.
Similarly, production in association with a leptonic $W$ is not useful
for an invisibly-decaying Higgs boson, due to the lack of kinematic
information in the final state that could separate the signal from the
large Drell-Yan background $qq\to W^*\to \ell\nu$
\cite{Cavalli:2002goa,Godbole:2003it,ATL-PHYS-PUB-2009-061}.

The VBF production mode offers the best combination of cross-section
and signal-to-background discrimination at the LHC, both for 14 TeV
\cite{Eboli:2000ze, Davoudiasl:2004aj} and 7 and 8 TeV
\cite{Bai:2011wz}.\footnote{Note that searches targeting the VBF
  production mode also see a secondary signal contribution from $gg\to
  h + 2j$, which is relatively more important at 7 and 8 TeV than at
  14 TeV.}  Ref.~\cite{Bai:2011wz} estimates that 20 fb$^{-1}$ at 8
TeV can allow limits to be placed for $\mbox{Br}(h\to \met)\gtrsim
0.4$, while Ref.~\cite{Ghosh:2012ep} estimates the sensitivity
$\mbox{Br}(h\to \met)\gtrsim 0.25$ with 300 fb$^{-1}$ at 14 TeV.
Meanwhile Ref.~\cite{ATL-PHYS-PUB-2009-061} estimates sensitivity for
$\mbox{Br}(h\to \met)\gtrsim 0.50$ with 30 fb$^{-1}$ at 14 TeV.
Assumptions about systematic errors are critical in obtaining these
estimates.

Associated production with a leptonically decaying $Z$ boson has
significantly smaller LHC cross-section than any of the above
production modes, but on the other hand the final state contains more
kinematic information
\cite{Frederiksen:1994me,Cavalli:2002goa,Godbole:2003it}.  For a $125$
GeV Higgs, $Zh$, $Z\to \ell\ell$ can nearly approach the reach of VBF
at the 14 TeV LHC \cite{ATL-PHYS-PUB-2009-061}, though its utility at
7 and 8 TeV is more limited \cite{Bai:2011wz}.  Including $Z\to b\bar
b$ as well as $Z\to \ell^+\ell^-$ decays can incrementally improve the
reach, at both the Tevatron \cite{Martin:1999qf} and the LHC
\cite{Ghosh:2012ep}.

\subsection{Existing Experimental Searches and Limits}

The best existing constraints come from ATLAS measurements targeting
$Zh$ associated production with $Z\to\ell\ell$, which limit the
invisible branching fraction to be
\beq
\mbox{Br} (h\to\mbox{invisible}) <0.65\: (0.84 \mbox{ expected})
\eeq
at $95\%$ CL \cite{ATLAS:2013pma} with 4.7 fb$ ^ {-1} $ at 7 TeV and
13.0 fb$^{-1}$ at 8 TeV.  The measurement by CMS in the same channel
with the full 7 and 8 TeV data sets places a $95\%$ CL upper bound on
the invisible branching fraction of Br$(h\to\mbox{ invisible}) <0.75
(0.91)$ \cite{CMS-PAS-HIG-13-018}. CMS also has a measurement in the
VBF channel, with a 95\% CL upper limit \cite{CMS-PAS-HIG-13-013}
\beq
\mbox{Br} (h\to\mbox{invisible}) <0.69 \mbox{ observed}\: (0.53 \mbox{ expected})
\eeq
with 19.6 fb$^{-1}$ of 8 TeV data.  Much weaker limits come from
reinterpretation of monojet $+\met$ measurements
\cite{Djouadi:2012zc}.


\section{\secsize $\secbold{h \rightarrow 4b}$}
\label{sec:hto4b}
One possible exotic Higgs decays is to four $b$ quarks via a light
resonance $X$: $\hsm\to XX \to b\bar b b\bar b$.  Below, we outline
the theoretical motivation to consider such decays, and discuss their
LHC phenomenology.

\subsection{Theoretical Motivation}
In the SM, a 125 GeV Higgs can decay to four $b$ quarks via $ZZ^*$.
This branching ratio is small: $\BR(\hsm\to ZZ^*)\times\BR(Z\to b\bar
b)^2\sim10^{-4}$.  The $b\bar b$ pair associated with the on-shell $Z$
boson is relatively uncollimated because of the large $Z$ mass, and
the resulting signature has a large irreducible QCD background. A more
experimentally viable situation occurs in models where the Higgs
decays to new particles ``$X$'' which further decay to a pair of
$b$-quarks.  Such a decay topology can arise in several new physics
scenarios, such the general 2HDM+S (\S\ref{2HDMS}), extensions of the
SM with hidden light gauge bosons (\S\ref{subsec:SMvector}), the
(R-symmetry limit of the) NMSSM (\S\ref{NMSSMscalar}), the Little
Higgs model (\S\ref{LHiggs}), and commonly in the Hidden Valley scenario (\S\ref{sec:hiddenvalley}.  In all of these models, $X \rightarrow
b \bar b$ can be the dominant decay mode in certain regions of
parameter space, therefore strongly motivating the study of the $h
\rightarrow 4b$ decay channel.

\begin{itemize}

\item 2HDM+S: In two-Higgs-doublet models with an additional light
  singlet, the decay $h\to ss$ or $h\to aa$, where $s$ ($a$) is the
  mostly-singlet (pseudo)scalar is generic.  Depending on $\tan\beta$,
  the decays $s\to b\bar b$ or $a\to b\bar b$ are also generic
  (although not guaranteed) in all four 2HDM Types as long as
  $m_a,~m_s > 2 m_b$.

\item R-symmetry limit in the NMSSM: The additional two degrees of
  freedom in the NMSSM Higgs sector (which corresponds to a Type II
  2HDM+S model) make a light pseudoscalar $a$ with sizable coupling to
  the SM-like Higgs and SM fermions possible. In the case of an
  approximate R-symmetry, the imaginary component of the new singlet
  is naturally light, since it serves as a pseudo-Goldstone boson of
  the spontaneously broken $U(1)_{\rm R}$, once the singlet acquires a
  vacuum expectation value.
For $m_a \leq m_h/2$, the decay $h \rightarrow a a$ opens up.  (Note, however, that while $a$ is light in the PQ limit of the NMSSM, the decay $h\to a a$ is generically suppressed compared to other decays; see \cite{draper2011dark} or \S\ref{NMSSMfermion}.)  The pseudoscalar $a$ couples to fermions proportional to  the Yukawa matrices, which are enhanced by $\sin\beta/\sin\alpha$. This makes large decay branching ratios for $a \rightarrow b \bar b$ natural in large regions of parameter space.

\item Little Higgs models: Another class of models with potentially light pseudo-scalars is the Little Higgs model. The couplings of $a$ to SM fermions are again proportional to the SM Yukawas if one imposes Minimal Flavor Violation (MFV)~\cite{D'Ambrosio:2002ex,Chivukula:1987py,Hall:1990ac,Buras:2000dm} in order to get rid of large flavor violation; thus the coupling to the $b$-quark is typically enhanced.

\end{itemize}

\subsection{Existing Collider Studies}    
Most of the existing collider studies are performed within the NMSSM framework (the Little Higgs model was considered in \cite{Cheung:2007sva}) under the assumption that $\mathrm{Br}(h\rightarrow a a) \simeq 1$. Those studies that have been performed at the LHC were done for $\sqrt s = 14 \ \tev$. The case with $\sqrt s = 8 \ \tev$ has not explicitly been studied, but insight can still be gained from previous work. 

\vspace{3mm}
\noindent\textbf{LEP and Tevatron}

Much of the earlier literature on exotic Higgs decays was framed in the context of trying to evade the LEP limit of $m_h > 114 \gev$ for a Higgs produced with SM-like strength, allowing for a lighter and more natural Higgs. 
For example,~\cite{Chang:2005ht} presented constraints from LEP on NMSSM cascade decays; for $h \rightarrow 4b$, the Higgs mass constraint is around 110 GeV, only slightly weaker than the LEP constraint on a SM Higgs.  
The 125 GeV Higgs is not constrained by LEP, as it is above LEP's kinematic limit.  
The Tevatron also does not have any exclusion power for $h\rightarrow 4b$ with SM-strength production~\cite{Dobrescu:2000jt,Stelzer:2006sp, Carena:2007jk,Cheung:2007sva}.

\vspace{3mm}
\noindent\textbf{LHC}

The literature contains several collider studies examining $h \rightarrow 4b$ decay at the 14 TeV LHC.  Refs.~\cite{Ellwanger:2003jt, Ellwanger:2005uu} considered the $4b$ final state in the context of VBF Higgs production, but this signature is very difficult to distinguish from QCD background. More recently the focus has been on the $Wh$ production mode~\cite{
Carena:2007jk, 
Cheung:2007sva,Cao:2013gba, 
Kaplan:2011vf}, where the tagged lepton greatly reduces backgrounds and enhances discovery potential.

Ref. \cite{Cao:2013gba} is the most recent study demonstrating how a very simple $4b$ search could constrain $h\rightarrow a a \rightarrow b \bar b b \bar b$ at LHC14. It makes use of the known Higgs mass and utilizes full showering and fast detector simulation. The total signal cross section is parameterized in terms of the  associated Higgs production cross section $\sigma_{Wh}$,
\begin{equation}
\sigma_{4b} = C^2_{4b} \sigma_{Wh},
\end{equation}
where 
\begin{equation}
\label{e.C4b}
C^2_{4b} = \kappa_{hVV}^2 \mathrm{Br}(h\rightarrow a a) \mathrm{Br}(a \rightarrow b \bar b),
\end{equation}
and $\kappa_{hVV}$ is the $WWh$ coupling strength relative to the SM. Within the assumptions we make in this survey, $C^2_{4b} = \Br(h\rightarrow 2a \rightarrow 4b)$. The selection requirements are exactly 4 $b$-tags (with assumed $70, 5, 1\%$ efficiency for $b$, $c$, light flavor jet), one isolated lepton, and a reconstructed $m_{4b}$ in the Higgs mass window. This greatly reduces the main backgrounds ($t \bar t$ +  jets and V + jets). At the 14  TeV LHC with 300 $\ifb$ of data, this gives signal significance $S/\sqrt{B} = 2$ for $\Br(h\rightarrow 2a \rightarrow 4b) \approx 0.1$ if $m_a > 30 \gev$.

Searching for $h\to2a\to4b$ decay if $m_a < 30 \gev$ requires the use of jet substructure. This case was addressed by \cite{Kaplan:2011vf}, which primarily deals with the much more difficult signature $h \rightarrow a a \rightarrow 4g$ (also considered in \cite{Bellazzini:2010uk, Chen:2010wk, Luty:2010vd, Falkowski:2010hi}), with $h \rightarrow 4b$ considered as a special case that can also make use of heavy flavor tagging. 
They focus on boosted Higgs production in association with a $W$ or $Z$ (with $C^2_{4b} = 1$ in the above notation) by requiring a reconstructed vector boson to have $p^T_V > 200 \gev$. A range of pseudoscalar masses is considered for a 120 GeV Higgs.

For $m_a \lesssim 30 \gev$, a boosted Higgs decaying as $h \rightarrow a a \rightarrow 4j$ can produce a 2-, 3-, or 4-pronged fat jet. Pseudoscalar candidates are constructed to minimize their mass difference,  requiring the lighter pseudoscalar candidate to have at least 75\% of the mass of the heavier one, and by selecting events with a fat jet mass close to the hypothesized Higgs mass and looking for a pseudoscalar mass resonance. 

\emph{Assuming Br$(h\to 4b)=1$}, without heavy flavor tagging the $h\rightarrow 4j$ signature can be observed at $3\sigma$ with $100 \ifb$ of LHC14 luminosity; adding 1 (2) $b$-tags improves the $h\rightarrow 4b$ discovery signal to $\sim 6 \sigma$ ($\gtrsim 10 \sigma$). Naively scaling this sensitivity to $300 \ifb$ we obtain a signal significance $S/\sqrt{B} \approx 2$ for $\Br(h\rightarrow 2a \rightarrow 4b) \approx 0.1$. This is comparable to the result for $m_a > 30 \gev$ by \cite{Cao:2013gba}.

It therefore seems reasonable to expect the LHC14 to have $2 \sigma$ sensitivity to $\Br(h\rightarrow 2a \rightarrow 4b) = 0.1$ $(0.2)$ with $300\ifb$ ($100 \ifb$) of data across the kinematically allowed mass range for the pseudoscalar $a$.

\subsection{Existing Experimental Searches and Limits}

Due to large QCD backgrounds to the $4b$ final state, the only realistic discovery mode for $h\rightarrow b \bar b b\bar b$ at the LHC is 
$Wh$ associated production. The produced lepton allows for the event to be triggered on, which is difficult for the relatively soft all-hadronic final state resulting from gluon-fusion or vector-boson fusion Higgs production. Therefore, the relevant final state for experimental searches is  $4b + \ell + \met$ (or some variety with fewer $b$-tags).  

To the best of our knowledge, no such search has been performed. $V (h \rightarrow b \bar b)$ searches \cite{cmshbb, atlashbb} have not yet reached SM sensitivity and are even less likely to find the softer signal from 4 $b$'s. Searches for $b (h \rightarrow b \bar b)$ production \cite{cmsbhbb, atlasbhbb} do not look for an isolated lepton or large amounts of $\met$, which results in large backgrounds, and SUSY searches for final states containing several $b$-jets like \cite{ATLAS-CONF-2012-161} also typically do not require a lepton while requiring an amount of missing energy that is much too high for $Vh$ production.

The $h \rightarrow aa \rightarrow 4b$ process will contribute to the signal region of SM  $h \rightarrow 2b$ searches. The recent CMS analysis \cite{Chatrchyan:2013zna} observes a $2\sigma$ excess consistent with a SM-like 125 GeV Higgs, constituting the first indication of $h\rightarrow\bar b b$ decay at the LHC. The signal strength corresponding to this excess is 
\begin{equation}
\mu_{2b} \equiv \frac{\sigma_{h}  \ \Br{h\rightarrow b \bar b} \ \ \ \ }{[\sigma_{h}  \ \Br{h\rightarrow b \bar b}]_\mathrm{SM}} = 1.0 \pm 0.5.
\end{equation}
We can, in principle, use this to derive a limit on $\mathrm{Br}(h\to4b)$. Define the $m_a$-dependent efficiency ratio
\begin{equation}
r_{4b}(m_a) = \frac{\epsilon_{h\to2a\to4b}}{\epsilon_{h\to2b}}
\end{equation}
for a $h\to2a\to4b$ event to end up in the signal region of the $h\to2b$ search, relative to a SM-like $h\to2b$ event. Assuming a SM-like partial width $\Gamma_{h\to2b}^\mathrm{SM}$ as well as SM-like Higgs production, and defining the total Higgs with in the SM to be $\Gamma_h^\mathrm{SM}$, the \emph{expected} signal strength observed in a $h\to2b$ search will be
\begin{eqnarray}
\nonumber \mu_{2b} &=& \frac{\Gamma_{h\to2b}^\mathrm{SM} + r_{4b} \Gamma_{h\to2a\to4b}}{\Gamma_h^\mathrm{SM} + \Gamma_{h\to2a\to4b}}  \frac{\Gamma_h^\mathrm{SM}}{\Gamma_{h\to2b}^\mathrm{SM}} \\
&=& 1 + \mathrm{Br}(h\to2a\to4b)\left[\frac{r_{4b}}{\mathrm{Br}(h\to2b)^\mathrm{SM}} - 1 \right]
\end{eqnarray}
For $\mathrm{Br}(h\to2b)^\mathrm{SM} \approx 0.6$, this expected signal strength is shown in Fig.~\ref{fig:4bsigstrengthexpectation__38}.

To estimate $r_{4b}$ for the analysis in \cite{Chatrchyan:2013zna} we  simulated $h\to2b$ and $h\to2a\to4b$ events in MadGraph and Pythia.  Applying the analysis cuts from  \cite{Chatrchyan:2013zna} we find that $0.5 \lesssim r_{4b} \lesssim 1.5$, with higher efficiency for lighter pseudoscalar masses $m_a \sim 15 \gev$, since the resulting collimated 2b-jets are tagged as single $b$-jets from $h\to2b$ decay. 
Given the $2\sigma$ limit of $\mu_{2b} < 1.9$ by \cite{Chatrchyan:2013zna} we can then read off a limit on $\mathrm{Br}(h\to2a\to4b)$ from Fig.~\ref{fig:4bsigstrengthexpectation__38}. For $m_a \sim 15 \gev$, the limit\footnote{The assumption of SM-like $\Gamma_{h\to2b}$ in our interpretation does not take into account the reduced $h b \bar b$ coupling when $\mathrm{Br}(h\to2a\to4b)$ is high due to large higgs-singlet mixing in a model like SM+S or 2HDM+S. In such a case, consistently taking the reduced $\Gamma_{h\to2b}$ into account would make this limit slightly weaker.} is $\mathrm{Br}(h\to2a\to4b) \lesssim 0.7$, while no meaningful limits are derived for heavier pseudoscalars.

 \begin{figure}[t]
\begin{center}
\includegraphics[width=8cm]{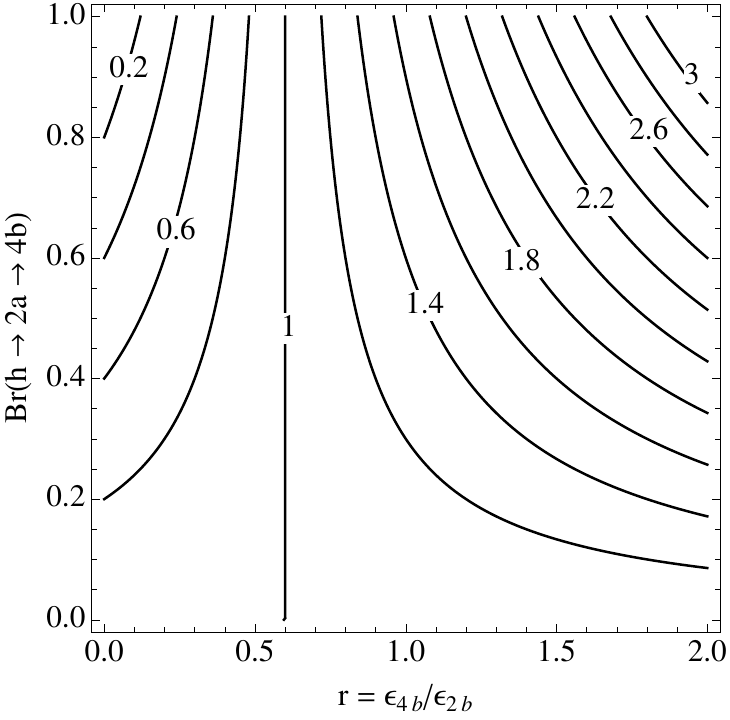}
\end{center}
\caption{
Expected signal strength observed in a $h\to2b$ search, assuming SM-like higgs production and couplings with the exception of a new $h\to2a\to4b$ decay mode with selection efficiency $r_{4b}$ relative to the efficiency of SM-like $h\to2b$ events.
}
\label{fig:4bsigstrengthexpectation__38}
\end{figure}

Clearly there exists motivation for a dedicated experimental search, which could easily be performed by triggering on leptons and missing energy from associated Higgs production, and performing a $4b$ search similar to the studies by \cite{Cao:2013gba, Kaplan:2011vf}.

\subsection{Proposals for New Searches at the LHC}

The  LHC14 studies~\cite{Carena:2007jk, 
Cheung:2007sva,Cao:2013gba, 
Kaplan:2011vf} as well as the above-mentioned limit from the $h\rightarrow 2b$ search make it plausible that a $2\sigma$ sensitivity for the $\Br(h \rightarrow 4b) \lesssim 0.5$ could be obtained using $25\ifb$ of LHC8 data (this is 
based on a naive scaling of cross sections and luminosity).  More study would be needed to investigate the sensitivity in more detail.  
 The boosted regime is also worth exploring at LHC8, either by looking for explicitly boosted pseudoscalars from Higgs decay giving two-pronged double-$b$-jets (depending on $m_a$) or for fully boosted Higgses as in~\cite{Kaplan:2011vf}, or by looking indirectly via a diagonal cut in the  $(p_{T,2b}, m_{2b})$ plane and requiring low $\Delta R_{2b}$. These analyses can be easily parameterized in a simplified model with a single pseudoscalar $a$ of mass $m_a$ and a 125 GeV Higgs with SM-like production modes. The signature-space then only has two parameters, $m_a$ and $C^2_{4b}$ as defined in Eq.~(\ref{e.C4b}).

	\section{\secsize $\secbold{h \rightarrow 2b 2\lowerGreekBold\tau}$}
\label{sec:h2b2tau}

\subsection{Theoretical Motivation}
This channel can become very important in the case that the Higgs decays into a pair of light (pseudo)-scalars, $h \to a a $,
with $a$ further mostly
decaying into the third generation fermions $b \bar b$ or $\tau^+ \tau^-$. In the mass range $2m_{b} < m_a < m_h/2$
the Higgs can have a relatively large branching ratio into $aa$, while both decays into $b \bar b$ and $\tau^+ \tau^-$ are
allowed by phase space. In many models, e.g.~the NMSSM (see \S\ref{NMSSMscalar}), Little Higgs models (see \S\ref{LHiggs}) and certain Hidden Valley models (see \S\ref{sec:hiddenvalley}), the couplings of $a$ to SM fermions will be roughly proportional to
the SM Yukawa couplings (with some corrections that depend on $\tan \beta$), leading to
$\BR(a \to b \bar b) \approx 94\%$ and $\BR(a \to \tau^+ \tau^-) \approx 6\%$.  In this case $\sim 90\%$ of all the
$aa$ decays will end up in $b \bar b b \bar b$, $\sim10\%$ in $b \bar b \tau^+ \tau^-$ and less than $1\%$ in
$\tau^+ \tau^-\tau^+ \tau^-$.  The first mode was discussed in \S\ref{sec:hto4b} and is very challenging, especially
in the range of $m_a \lesssim 30$~GeV, where the $b$-jets start merging.  The last channel, $h\to 4\tau$, is discussed in \S\ref{sec:hto4tau} for general models.  However, in the class of models considered here, where $\BR(a\to b\bar b)/\BR(a\to\tau^+\tau^-) \simeq 3m_b^2/m_\tau^2$, the $4\tau$ rate is likely too small to be exploited.
In this case, $b \bar b \tau^+ \tau^- $ can be a reasonable compromise between branching fraction and visibility of the signal.  In particular, more than
50\% of the ditau decays include at least one isolated lepton.

\subsection{Existing Collider Studies}
This channel has attracted the attention of several research groups both in the context of the Tevatron and of the LHC.
Most of the studies assumed a ${\cal O}(1)$ branching fraction for the decay $h \to aa$.
\begin{itemize}
\item Refs.~\cite{Aglietti:2006ne,Carena:2007jk} performed a feasibility study for this mode at the Tevatron. This study used
associated production of the Higgs with a leptonic $W$. The study found very few sources of irreducible
backgrounds, but also very small $\sigma (Wh)\times \BR(h \to b \bar b \tau^+ \tau^-)$. For example, for
$\BR(h \to a a ) = 1$, which is bigger than what we can realistically assume today,
effective production rates after the acceptance cuts
$\sigma (Wh)\times \BR(h \to b \bar b \tau^+ \tau^-) = 0.55$~fb have been found for a Higgs with mass $m_h= 120$~GeV
and with a very optimistic assumption on the branching ratios of the pseudoscalar $a$: $\BR(a \to b \bar b)=0.7$ and $\BR(a \to \tau^+ \tau^-)=0.3$~\cite{Aglietti:2006ne}\footnote{Note that these branching ratios can only be obtained in a small region of parameter space of the NMSSM that  predicts very large radiative corrections to the $a\tau^+\tau^-$ and $ab\bar b$ couplings.}.
This can probably be improved by $\sim 40\%$ if this channel is combined with $Z(\to \ell^+ \ell^-)h$ associated
production.  But probably little more can be gained at the Tevatron, and one cannot hope for more than just
a few signal events in the realistic case.
\item This study was performed in Ref~\cite{Adam:2008aa} for the LHC at 14~TeV. Motivated by the SM
$  h \to \tau^+  \tau^-$ channel, the authors concentrated on the VBF Higgs production mode. 
This study largely relies on a precise reconstruction of $m_{b \bar b} $ for rejection of the
dominant $t \bar t$ background, while $m_{\tau \tau}$ and $m_{b\bar b\tau \tau}$ are not considered. 
The study is rather preliminary, and it claims that with 100~fb$^{-1}$ data, a significance of $S/\sqrt{B} \sim 2$ is possible after $b$-tagging.  

It is also worth noticing that this study only considered channels with both $\tau$s decaying leptonically (denoted $\tau_\ell$), 
and the situation can probably be significantly improved by including $\tau$'s decaying hadronically 
(denoted $\tau_h$), e.g.~$\tau_\ell \tau_h$ and maybe even $\tau_h \tau_h$ final states. 
Unfortunately we have not found any other dedicated studies along these lines.
\item Ref.~\cite{Carena:2007jk} also very briefly discussed this search for 14 TeV LHC, 
considering only associated production with a $W$ or $Z$, decaying leptonically. This study found this mode largely unfeasible at 
100~fb$^{-1}$ due to very small $S/B$ ratio.
\end{itemize}

\subsection{Discussion of  Future Searches at the LHC}
We are not aware of any current experimental searches in this channel. 
Searches for $h+2b$ with $h\to 2\tau$~\cite{CMS-PAS-HIG-12-050,Aad:2012cfr} are not sensitive to the $2b2\tau$ 
decay mode, as they did not search for $2\tau$-resonances below 90~GeV. 
Nonetheless, this channel might be a very important direction for studies of the LHC at 14~TeV. 
Probably, in order to have optimal reach, all three major productions modes (gluon
fusion, VBF, and $W/Z$ associated production) should be combined together.  Different production modes
may be dominated by different backgrounds.  While $t \bar t$ looks indeed like a formidable background for VBF, it
is possible that $\gamma^*/Z^* + $~jets dominates the two other channels.

It is also worth noticing additional complications for very small values of $m_a$.  First, as the mass of $a$ is getting close
to the $\Upsilon$ mass, the branching ratio $a\to b \bar b$ can be significantly reduced in favor of $a \to \eta + X$,
leading effectively to a $\tau^+ \tau^- j$ event topology and opening up additional possible backgrounds from bottomonium
decays~\cite{Englert:2011us} (see~\cite{Baumgart:2012pj} for a detailed discussion and calculation of the branching ratios).  
In addition, the $\tau$'s tend to merge in this region of parameter space, failing
isolated reconstruction criteria and yielding effectively a single $\tau$-like jet instead of two.
Finally, triggering on these events may be an issue.  
In particular, one can only be confident that associated production events are triggered with a reasonable efficiency. At LHC8, one
can also probably use parked data at CMS gathered via the (low-efficiency) VBF trigger.  It is not clear, though, whether a search in this channel is feasible.
At the 14 TeV LHC, the trigger thresholds may be too high for this type of decay, and therefore one probably has to focus on
associated production.

We conclude that more dedicated feasibility studies for the LHC are needed in this particular channel.
	
\section{\secsize $\secbold{h \rightarrow 2b2\lowerGreekBold\mu}$}
\label{sec:h2b2mu}

The possibility of the Higgs boson decaying to $(b\bar b)(\mu^+\mu^-)$ is intriguing. In the context of NMSSM and 2HDM+S models it represents a compromise between the very difficult but often dominant $4b$ mode (see \secref{hto4b}) and the spectacular but rare $4\mu$ signature.
Below we present the theoretical motivation to consider this decay mode and demonstrate the reach of a dedicated search at both Run I and II of the LHC.
A detailed study will appear in~\cite{bbmumu:2013yz}.

\subsection{Theoretical Motivation}

The $h\to(b\bar b)(\mu^+\mu^-)$ decay mode occurs when the Higgs field couples to one or more
bosons $a^{(i)}$ that couple to $b$ quarks and muons, with at least one $a^{(i)}$ heavy enough to decay to $b \bar b$. As discussed in \S\ref{SMS}, the simplest realization of such a scenario is given by extending the SM to include an additional real singlet scalar. However, searching for this mode is motivated in any model with additional singlets that couple to quarks in proportion to their masses.\footnote{If the coupling is through gauge interactions, fully leptonic final states are generally the preferred discovery channel, see \S\ref{sec:htoZa} and \S\ref{sec:hto4l}.} This includes the  2HDM+S (\S\ref{2HDMS}) and the well-known NMSSM (\S\ref{NMSSMscalar}), as well as many hidden valleys (\S\ref{sec:hiddenvalley}).

The small coupling to muons leads to very hierarchical branching ratios,
\beq
   \Br(h\to4\mu) = \frac{\varepsilon}{2}\Br(h\to2b2\mu) = \varepsilon^2\Br(h\to4b),
   \label{eq:BRmu}
\eeq
with
$\varepsilon\equiv\Br(a\to\mu^+\mu^-)/\Br(a\to b\bar b)\sim m_\mu^2/3m_b^2\approx 2\times 10^{-4}$ in the SM+S. (Non-minimal scalar models can modify this ratio, but the ratio is in general very small.) Assuming SM Higgs production and $\Br(h\rightarrow aa) = 10\%$ leads to zero $h\to 2a \to 4\mu$ events from gluon fusion at LHC Run I, while about twenty $h \to 2a \to 2b2\mu$ events are expected to occur. Even though this is much less than the few hundred $h\to 2a \to 4b$ events expected from associated production, the backgrounds for the $4b$ search are so challenging (see \S\ref{sec:hto4b}) that the $2b2\mu$ channel may provide much better sensitivity. This is even more attractive in non-minimal models, where e.g.~$\tan \beta$ can enhance the leptonic pseudoscalar branching fraction significantly. It is also possible that the Higgs decays to two pseudo scalars, $h\to a_1 a_2$, which have large branching fractions to $2b$ and $2\mu$, respectively. 
The presence of a clean dimuon resonance makes the $2b2\mu$ decay mode very attractive for discovering SM extensions with extra 
singlets. 

\subsection{Existing Collider Studies and Experimental Searches}
To the best of our knowledge there have been no theoretical collider studies of this final state, and there are no limits on this 
decay channel from existing searches.
A similar topology is searched for in $h\to b\bar b$ from associated production with a $Z$ boson,
where the $Z$ decays to $\mu^+\mu^-$.
However, this search is not relevant for $(2b)(2\mu)$, since the required $b\bar b$ invariant mass was $\mathcal{O}(125~{\rm GeV})$,  
and the two muons were required to reconstruct the $Z$-boson.  
A dedicated search is therefore needed for this channel.

\subsection{Proposals for New Searches at the LHC}

We estimate the discovery potential of a very simple search for $h\to2a \to 2b 2\mu$ with Run I LHC data as well as $100~\ifb$ at 14 TeV. This preliminary study is simulated at parton-level for signal and backgrounds (see~\cite{bbmumu:2013yz} for a more complete study).

\vspace{5mm}

\noindent \textbf{LHC 7 and 8 TeV}

We assume the Higgs is produced through gluon fusion and has a non-zero branching ratio as $h\to aa \to (b \bar b)(\mu^+ \mu^-)$. 
We do not include Higgs bosons produced through VBF in our analysis, although this would slightly increase the sensitivity to this channel.  
The final state consists of two opposite-sign muons and two $b$-tagged jets and is simulated for $m_h = 125 \gev$ and $m_a \in (15, 60) \gev$. (Lower masses involve complicated decays to quarkonia~\cite{Baumgart:2012pj}, which are beyond the scope of this study.) The main background is Drell-Yan (DY) production with associated jets, $Z/\gamma^* + 2j/2c/2b$, where the $Z$-decay/$\gamma^*$ produces two muons. In this preliminary estimate, we neglect backgrounds arising from lepton-misidentification of jets, diboson production $VV$, and $t \bar t$ production, which are expected to be subdominant to DY.  (The $t\bar{t}$ background has a total cross section comparable 
to DY + jets but does not contribute significantly in the low dimuon invariant mass region~\cite{Chatrchyan:2011cm, Gonzalez:2002fxa}, and 
also typically produces a sizable amount of $\met$ that is not present for the signal.)

\begin{table}[t!]
\resizebox{\textwidth}{!}{%
   \centering
   \begin{tabular}{@{} |l|cccccccc| @{}} 
      \hline

      Selection Criteria    & $S$ (rel.) & $S$ (cum.) & $bb$ (rel.)& $bb$ (cum.) & $cc$ (rel.) & $cc$ (cum.)& $j'j'$ (rel.) & $j'j'$ (cum.)\\
\hline
\hline

$N_{\rm ev,~initial}$ ($25\ifb$) & & 80.8  && $1.9\times 10^{5}$ && $4.3\times10^5$ && $1.1\times 10^{7}$ \\
 \hline
\hline
  Two opposite sign $\mu$'s      & 100\% & 100\%& 100\% & 100\% & 100\% & 100\% & 100\% & 100\%\\
  $ |\eta(\mu_1)|, |\eta(\mu_2)|<2.5$ &&&&&&&&\\
  $p_{T_{\mu_1,\mu_2}}> 17$~GeV, 8~GeV       &  58\%  & 58\% &  69\% & 69\% & 41\% & 41\% & 63\% & 63\%\\

       At least  two jets    & 100\%  &  58\% & 100\% & 69\% & 100\% & 41\% & 100\% & 63\%\\
      $|\eta(j_1)|, |\eta(j_2)|<2.5$,   &&&&&&&&    \\
      $p_T(j_1), p_T(j_2)>25~\text{GeV}$    & 6.6\%  & 3.8\% &  18\%& 12\% & 16\% & 6.4\% & 18\% & 11\%\\
      $\Delta R_{j_1 j_2, j\mu, \mu_1\mu_2}>0.7, 0.4, 0.4$ & 100\%& 3.8\%&96\%&12\%&97\%&6.2\%&95\%&11\%\\ 
      \hline

$|m (j_1, j_2)-m_a| < 15~\text{GeV}$ & 100\% & 3.8\% & 5.3\% & $6.4\times10^{-3}$ & 5.5\% & $3.4\times 10^{-3}$ & 5.3\% & $5.7\times 10^{-3}$\\

     $|m (\mu_1, \mu_2, j_1, j_2)-m_h|< 15$ GeV& 100\% & 3.8\% & 2.7\% & $1.7\times 10^{-3}$ & 8.6\% & $2.9\times 10^{-4}$ & 4.3\% & $2.4\times 10^{-4}$ \\

    $|m(\mu_1,\mu_2)-m_a|< 1$ GeV      &  100\%  & 3.8\% &  4.1\% & $7 \times10^{-6}$ & 2.8\%  & $8 \times10^{-6}$ & 3.6\% & $8.7\times 10^{-6}$\\
   \hline
\hline

$N_{\rm ev,~final}$ ($25 \ifb$, no $b$-tag) & & 3.1  && 1.3 && 3.4 && 97.8 \\
 \hline
 \hline
   & & $S=3.1$ && $B_{\text{total}}=102.5$ && $S/B=0.03$ && $S/\sqrt{B}=0.31$  \\

      \hline

   \end{tabular}}
   \caption{Relative and cumulative efficiencies of the signal ``$S$'' ($h\to aa \to b\bar{b}\mu^+\mu^-$) 
   and backgrounds  for $m_a=30$ GeV (without $b$-tagging) at 8 TeV LHC. 
   The labels $bb$, $cc$, and $jj$ indicate SM Drell-Yan ($Z/\gamma^*$) productions with final states $b \bar b \mu^+ \mu^-$, $c \bar c \mu^+ \mu^-$, and $ j  j \mu^+ \mu^-$, respectively. For the signal normalization, we assume $\Br(h\to aa)=10\%$ and a 2HDM-Type III (leptonic-specific) + S model with 
   $\tan\beta=2$. The latter assumption leads to $2\times \Br(a\to b\bar b) \Br(a\to \mu^+\mu^-)=1.7\times 10^{-3}$ (see \S\ref{2HDMS}).}
   \label{tab:2b2mu_cuts}
   \end{table}

Both signal and background are simulated to lowest order at parton-level in MadGraph~5~\cite{Alwall:2011uj}. The signal is renormalized by the NLO gluon-fusion cross section $\sigma_{ggF}\simeq 19.3~\text{pb}$~\cite{Dittmaier:2011ti}. The obtained leading-order cross sections for backgrounds\footnote{We impose generator-level cuts $p_T(j) > 10 \gev, p_T(l) > 5 \gev, \eta(j)<5, \eta(l)<2.5, \Delta R_{jj, \mu\mu, j\mu}>0.4$. Here $j$ includes heavy flavor. 
} are 
$\sigma_{b \bar b \mu^+ \mu^-}\simeq 3.7 \pb$, $\sigma_{c \bar c \mu^+ \mu^-}\simeq 8.6 \pb$, and $\sigma_{j j \mu^+ \mu^-}\simeq 226 \pb$. 
These samples are scaled up by a representative $K$-factor of 2. We approximate the total Run I data with $25 \ifb$ at $\sqrt{s} = 8$~TeV.

To approximate trigger threshold and detector reconstruction requirements, we impose the following preselection cuts: only use partons with $|\eta| < 2.5$; require $\Delta R$ between any two jets to be $ > 0.7$, and between two muons or between a muon and a jet $ > 0.4$; two leading jets with $p_{Tj_{1,2}} > 25 \gev$; two muons with $p_{T\mu_{1,2}} > 17\gev,~8 \gev$, respectively. To roughly simulate 
$b$-(mis)tagging we reweight events according to constant tagging probabilities of 65\%, 10\% and $0.5\%$ for $b, c,$ and light jets,  respectively~\cite{ATLAS-CONF-2012-097}. 
Following this preselection, we require either 0, 1, or 2 $b$-tags and use mass reconstruction cuts to focus in on the signal for each pseudoscalar mass:
\begin{equation}
|m_{\mu\mu} - m_a| < 1 \gev \ \ , \ \ \ |m_{jj} - m_a| < 15 \gev\ \ , \ \ \ |m_{jj\mu\mu} - m_h| < 15 \gev.
\end{equation}

 \begin{figure}[t!]
   \begin{center} $
   \begin{array}{lll}
   \includegraphics[width=0.32\textwidth]{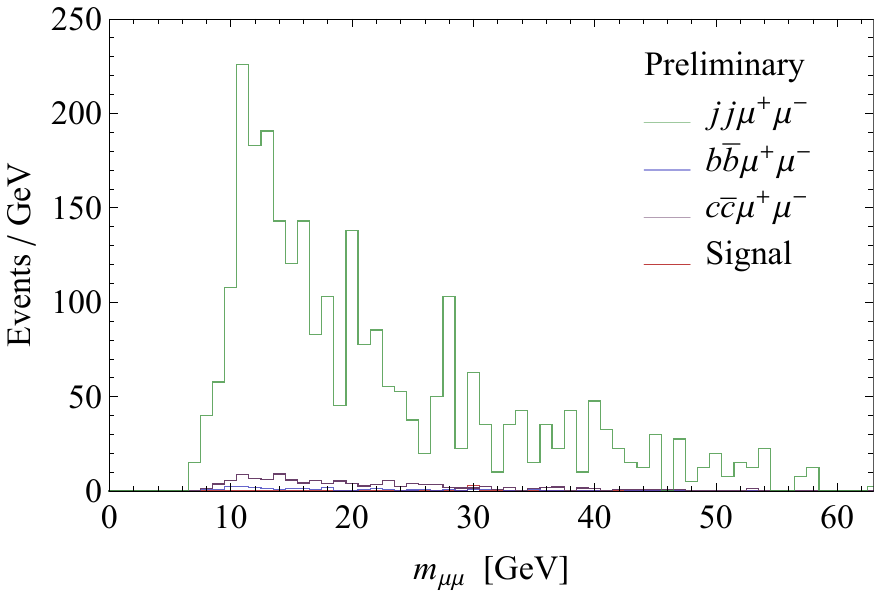} &  
   \includegraphics[width=0.32\textwidth]{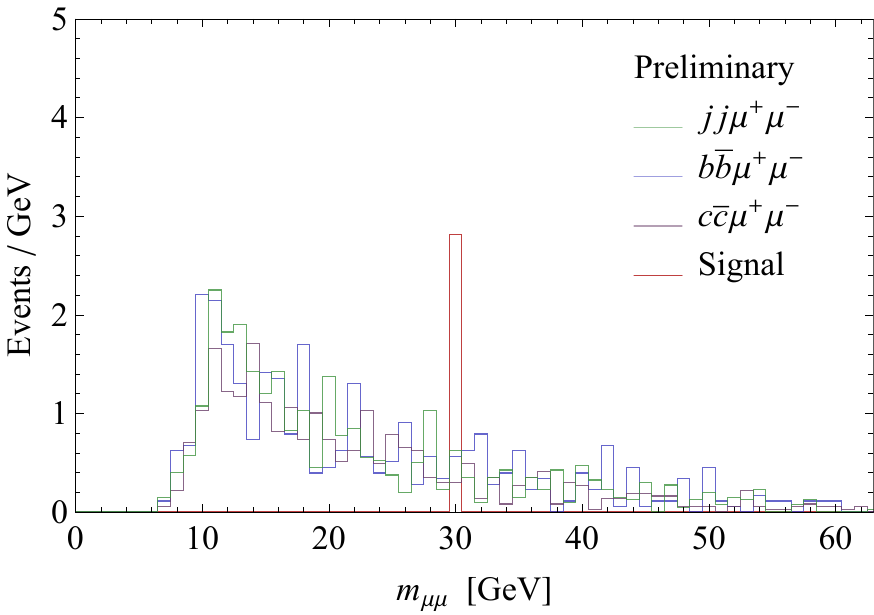} &  
   \includegraphics[width=0.32\textwidth]{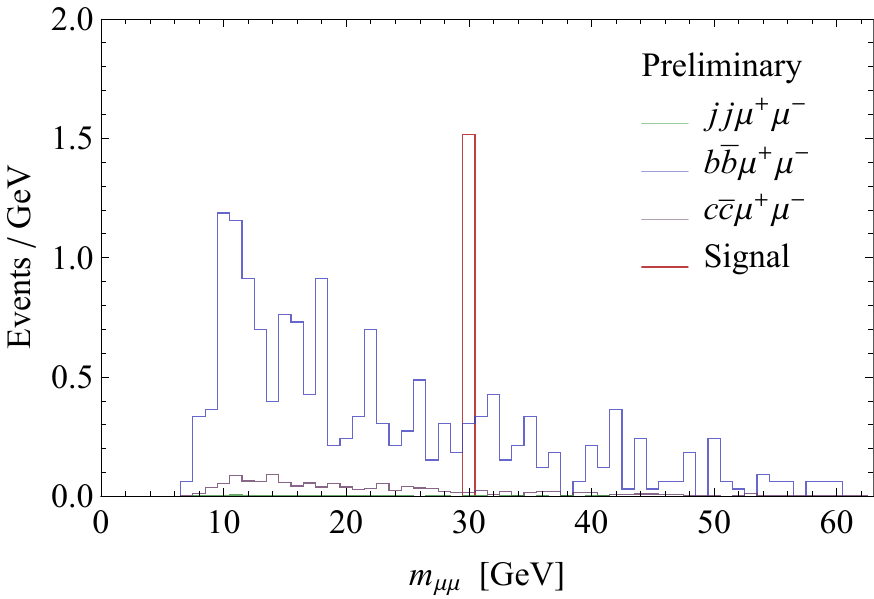} 
      \end{array}$
   \end{center}
   \caption{\small Dimuon invariant mass spectrum, $m_{\mu\mu}$, for signal ($m_a=30 \gev$) and backgrounds for $25\ifb$ at 8 TeV LHC after all kinematic cuts (except for $m_{\mu\mu}$ cuts) with ({\bf left}) no $b$-tag, ({\bf middle}) at least one $b$-tag, and ({\bf right}) two $b$-tags. 
   For the signal normalization, we assume $\Br(h\to aa)=10\%$ and $2\times \Br(a\to b\bar b) \Br(a\to \mu^+\mu^-)=1.7\times 10^{-3}$ 
   as in Table \ref{tab:2b2mu_cuts}. 
   }
   \label{fig:2b2mubtag}
\end{figure}

Table \ref{tab:2b2mu_cuts} shows the relative and cumulative efficiencies for the signal and background.  
Fig.~\ref{fig:2b2mubtag} shows an example of distributions of the signal with $m_a=30 \gev$ and backgrounds after applying the kinematic cuts and tagging probabilities above. As expected, $Z/\gamma^*$ production clearly dominates the signal if no $b$-tag is applied. The signal is visible only in the $b$-tagged cases.

We demonstrate 95\% C.L. sensitivity of $\Br(h\to aa \to b\bar b \mu^+ \mu^-)$ with respect to $m_a$ in Fig.~\ref{fig:2b2mu95}. 
For $m_a\leq 25 \gev$, the $\bar b b$ from $a$-decay are collimated enough to fail our simple reconstruction cuts. 
A more sophisticated substructure analysis is required in this regime~\cite{bbmumu:2013yz}.

The upper limits on $\Br(h\to aa \to b\bar b \mu^+ \mu^-)$ can be further translated into upper bounds for $\Br(h\to aa)$ for a fixed $m_a$ by noticing
\beq
\Br (h\to aa)=\frac{\Br(h\to aa \to b\bar b \mu^+ \mu^-)}{2\Br(a\to b \bar b)\Br(a\to \mu^+\mu^-)}=\frac{\Br(h\to aa \to b\bar b \mu^+ \mu^-)}{2\Br(a\to b \bar b)\Br(a\to \tau^+\tau^-)}\frac{m_\tau^2 \beta_\tau}{m_\mu^2 \beta_\mu},
\eeq
where $\beta_f\equiv(1-4m_f^2/m_a^2)^{1/2}$. This allows us to show $\Br (h\to aa)$ limits in the plane of $a$ branching ratios to $\bar b b$ and $\tau \tau$, which can be free parameters relative to each other (see e.g. 2HDM+S, \S\ref{2HDMS}), while the ratio between $\tau \tau$ and $\mu \mu$ is fixed by their masses. From Fig.~\ref{fig:2b2mu95} the corresponding upper limits on  $\Br(h\to aa \to b\bar b \mu^+ \mu^-)$ are $4.6\times10^{-4}$ ($m_a=30 \gev$, at least one $b$-tag), $5.2\times10^{-4}$ ($m_a=30 \gev$, two $b$-tags), $1.3\times10^{-4}$ ($m_a=60 \gev$, at least one $b$-tag), and $1.4\times10^{-4}$ ($m_a=60 \gev$, two $b$-tags).

\begin{figure}[t!]
   \begin{center}
   \includegraphics[width=0.6\textwidth]{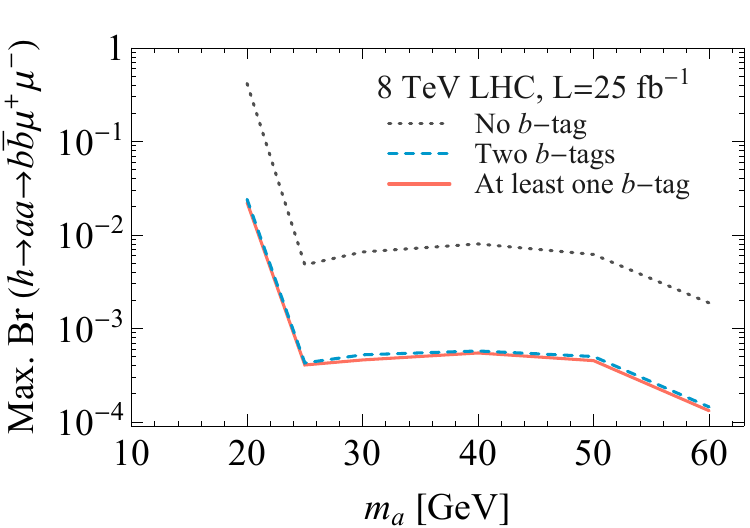}
      \end{center}
   \caption{
Expected 95\% C.L.~sensitivity to $\Br(h\to aa \to b\bar b \mu^+ \mu^-)$ for $25\fb$ data at 8 TeV LHC.  The solid, dashed, and dotted lines show the limits for at least one $b$-tag, two $b$-tags, and no $b$-tag, respectively. 
   }
   \label{fig:2b2mu95}
\end{figure}

\begin{figure}[htbp]
   \begin{center} $
   \begin{array}{ll}
   \includegraphics[width=0.45\textwidth]{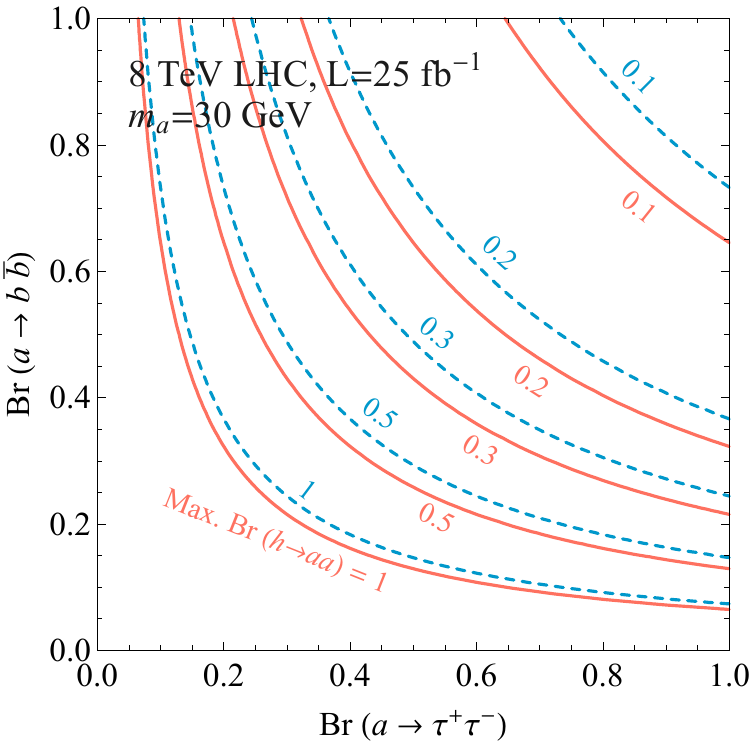} &  \includegraphics[width=0.45\textwidth]{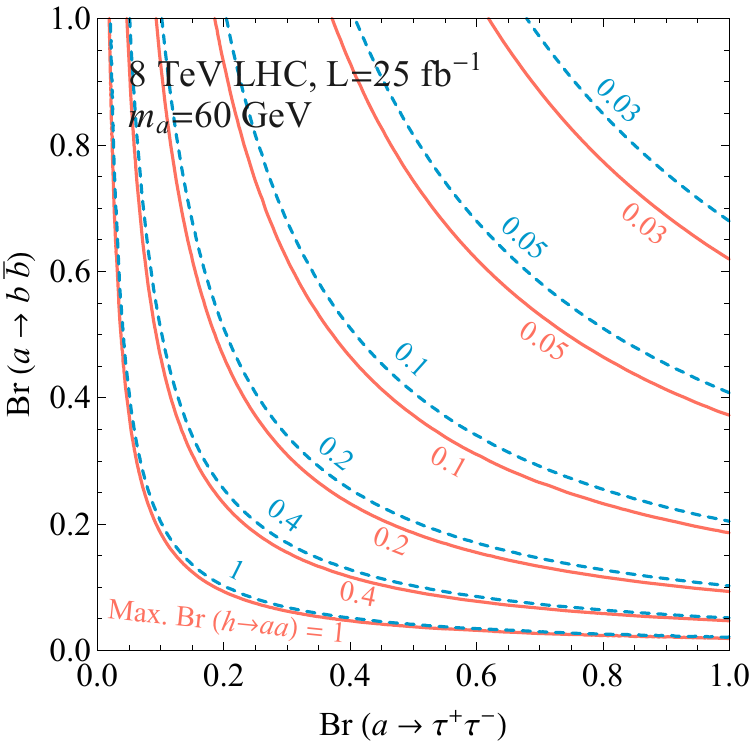}       \end{array}$
   \end{center}
   \caption{\small  
   Expected 95\% C.L.~sensitivity to $\Br(h\to aa)$ from a $h\to b\bar b\mu^+\mu^-$ search as a function of $\Br(a\to b\bar b)$ and $\Br(a\to \tau^+\tau^-)$, assuming that the pseudoscalar coupling to leptons is proportional to the lepton masses.  We show $m_a=30 \gev$ ({\bf left}) and $m_a=60\gev$ ({\bf right}) with $25\fb$ of data at the 8 TeV LHC (see text for further details). The red solid lines and blue dashed lines present the limits for at least one $b$-tag and two $b$-tags, respectively. The corresponding sensitivities to $\Br(h\to aa \to b\bar b \mu^+ \mu^-)$ are given in Fig.~\ref{fig:2b2mu95}.}
   \label{fig:2b2mubrbr}
\end{figure}

\vspace{5mm}

\noindent \textbf{LHC 14 TeV}

We repeat the study with identical cuts for $100\ifb$ of data at the 14 TeV LHC. The gluon fusion NLO Higgs production cross section is $\sigma_{ggF}=49.85 \pb$~\cite{Dittmaier:2011ti}. Drell-Yan background cross sections at LO from MadGraph with identical generator level cuts are $\sigma_{b \bar b \mu^+ \mu^-}=9.68 \pb$, $\sigma_{c \bar c \mu^+ \mu^-}=20.5 \pb$, and $\sigma_{j j \mu^+ \mu^-}=452.5 \pb$, again upscaled by a $K$-factor of 2.

The expected 95\% C.L. sensitivity of the 14 TeV LHC is shown in Fig.~\ref{fig:LHC142b2mu95}. 
We then translate this sensitivity to the expected 95\% C.L. sensitivity to Br$(h\to aa)$ as a function of the branching ratios of $a$ to $b \bar b$ and $\tau^+ \tau^-$, assuming that the pseudoscalar coupling to $\tau$'s and $\mu$'s is proportional to $m_\tau$ and $m_\mu$, respectively. Fig.~\ref{fig:LHC142b2mubrbr} demonstrates the expected sensitivity to $m_a=30 \gev$ and $m_a=60 \gev$. The corresponding expected sensitivities to $\text{Br}(h\to aa \to b\bar b \mu^+ \mu^-)$ are $1.8\times10^{-4}$ ($m_a=30 \gev$, at least one $b$-tag), $1.5\times10^{-4}$ ($m_a=30 \gev$, two $b$-tags), $6.2\times10^{-5}$ ($m_a=60 \gev$, at least one $b$-tag), and $5.3\times10^{-5}$ ($m_a=60 \gev$, two $b$-tags).

\begin{figure}[tp]
   \begin{center}
   \includegraphics[width=0.6\textwidth]{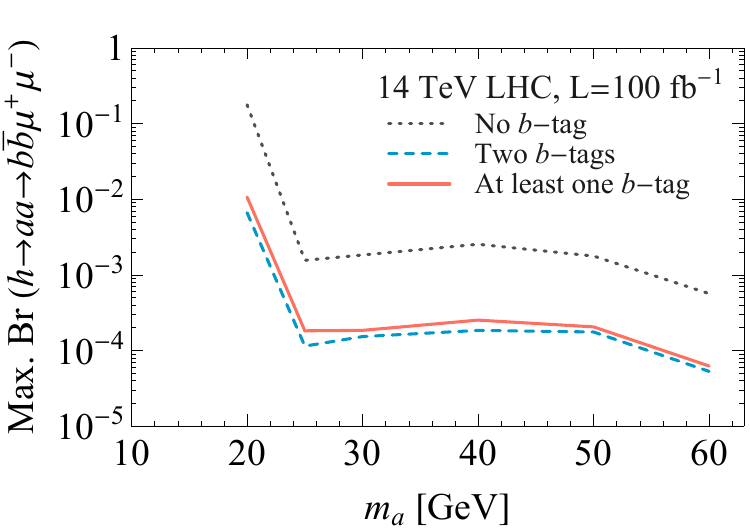}
      \end{center}
   \caption{Expected 95\% C.L.~sensitivity to $\Br(h\to aa \to b\bar b \mu^+ \mu^-)$ for $100\fb$ of data at 14 TeV LHC.  The solid, dashed, and dotted lines show the limits for at least one $b$-tag, two $b$-tags, and no $b$-tag respectively.}
   \label{fig:LHC142b2mu95}
\end{figure}

\begin{figure}[tp]
   \begin{center} $
   \begin{array}{ll}
   \includegraphics[width=0.45\textwidth]{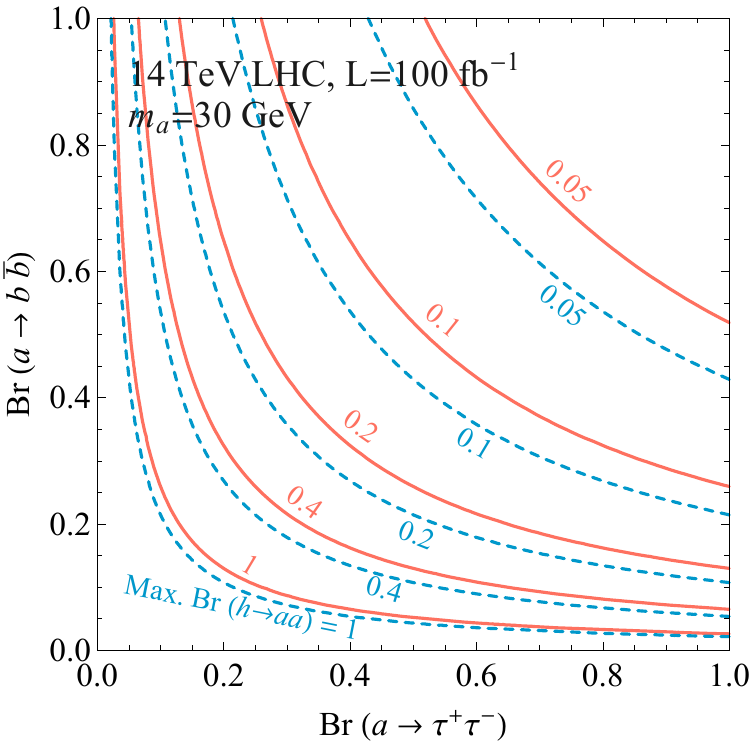} &  \includegraphics[width=0.45\textwidth]{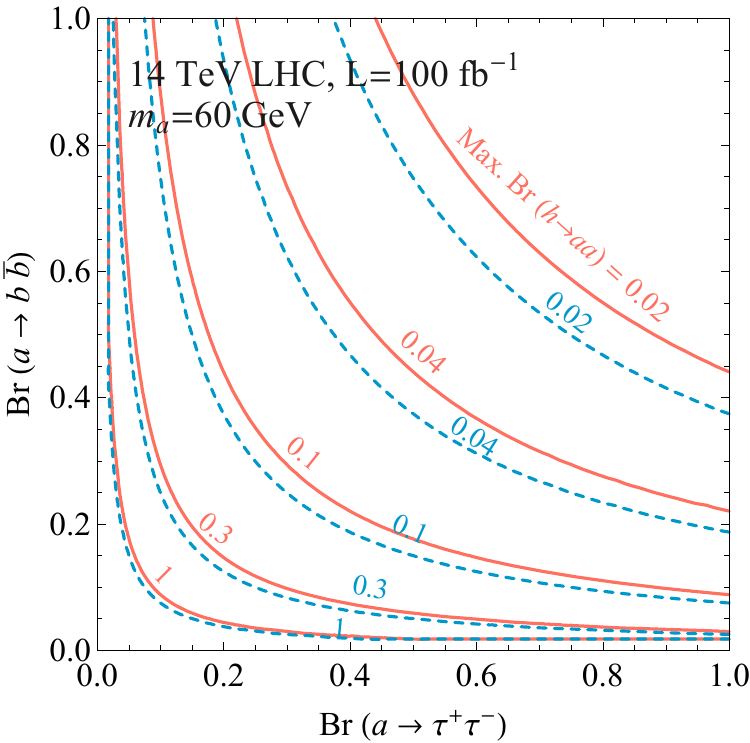}       \end{array}$
   \end{center}
   \caption{\small 
   Expected 95\% C.L.~sensitivity to $\Br(h\to aa)$ from a $h\to b\bar b\mu^+\mu^-$ search as a function of $\Br(a\to b\bar b)$ and $\Br(a\to \tau^+\tau^-)$, assuming that the pseudoscalar coupling to leptons is proportional to the lepton masses.  We show $m_a=30 \gev$ ({\bf left}) and $m_a=60\gev$ ({\bf right}) with $100\fb$ of data at the 14 TeV LHC (see text for further details). The red solid lines and blue dashed lines present the limits for at least one $b$-tag and two $b$-tags, respectively. The corresponding expected sensitivities to $\Br(h\to aa \to b\bar b \mu^+ \mu^-)$ are given in Fig.~\ref{fig:2b2mu95}.}

   \label{fig:LHC142b2mubrbr}
\end{figure}


\vskip 2mm
{\bf Summary}

Our simple parton-level study demonstrates that $\sim 10^{-4} - 10^{-3}$ sensitivity to $\Br(h\to 2a \to 2b2\mu)$ is possible at the LHC. We will investigate this channel more closely in \cite{bbmumu:2013yz}, but these preliminary results already strongly suggest conducting a corresponding search with available Run I data.

	
\section{\secsize $\secbold{h \rightarrow 4\lowerGreekBold\tau,\, 2\lowerGreekBold\tau2\lowerGreekBold\mu}$}
\label{sec:hto4tau}
\def\lsim{\mathrel{\rlap{\lower4pt\hbox{\hskip1pt$\sim$}}
    \raise1pt\hbox{$<$}}}
\def\gsim{\mathrel{\rlap{\lower4pt\hbox{\hskip1pt$\sim$}}
    \raise1pt\hbox{$>$}}}

\subsection{Theoretical Motivation}
\label{sec:h4tauintro}

In this section, we consider scenarios where the Higgs can decay into a pair of scalar or pseudoscalar bosons ``$a$'', with a mass between $2m_\tau$ and $m_h/2$, and with a sizable decay rate to tau pairs.  As discussed in \S\ref{2HDMS}, such a state can arise in 2HDM models supplemented with a singlet scalar field, especially if $m_a$ is below the bottomonium region.  A well-known example is the
NMSSM with an approximately-conserved R-symmetry~(\ref{NMSSMscalar}), which is a class of Type-II models with a very light pseudo-Goldstone boson; see also hidden valleys, \S\ref{sec:hiddenvalley}.  Another simple example is the set of Type-III (lepton specific) 2HDM models with modestly large $\tan\beta$, with or without extra singlet fields~(\ref{2HDMS}).  There, leptonic decays can dominate for new scalar or pseudoscalar states of almost any mass.

Besides focusing on the mass range $m_a = [2m_\tau,m_h/2]$, the main assumption that we will employ is that the couplings of $a$ are in direct proportion to the lepton masses.  For $a$ above the tau pair threshold, this means that the branching fractions to lepton pairs are in proportion $\tau^+\tau^-:\mu^+\mu^-:e^+e^- \; \simeq \; m_\tau^2:m_\mu^2:m_e^2 \; \simeq \; 1:3.5\times 10^{-3}:8\times 10^{-8}$.  By far the dominant $2\to 4$ fully leptonic branching fraction is then $4\tau$, though there is also a nearly 1\% relative $\Br$ to $2\tau2\mu$, which contains a tight $2\mu$ resonance~\cite{Lisanti:2009uy}.\footnote{Lighter states, between the muon and tau pair thresholds, can decay dominantly to muons and lead to a $4\mu$ final state with multiple resonant features.  For dedicates
searches see~\cite{Abazov:2009yi,Chatrchyan:2012cg}. Note that in this particular regime the leptons are highly collimated, such that
searches for ``lepton-jets'' can also place non-trivial bounds (see e.g.~\cite{Aad:2012qua})}  We do not need to make any explicit assumptions about the branching fractions to non-leptonic states, though here we will not consider possible signal contributions from decays with these states.  For example, if $a$ is above the $b$-quark pair threshold, $a \to 2b$ can dominate, and the $2a \to 4b$ and mixed $2b2\tau$ decay modes can be much larger than $4\tau$.  We discuss these in detail in \S\ref{sec:hto4b} and \S\ref{sec:h2b2tau}, respectively.

Taus can decay either leptonically (35\%) or hadronically (65\%).  These further subdivide into electron/muon leptonic decays, and one- and three-prong (and very rarely five-prong) hadronic decays.  In cases where $m_a \ll m_h$, the two taus or prompt muons from an individual $a$ decay can merge according to standard isolation criteria. (Generally, $\Delta R \sim 4m_a/m_h$.  E.g., roughly 0.3 for $m_a = 9$~GeV.)  We therefore are presented with a large number of final-state channels containing various combinations of isolated or non-isolated leptons, in association with a number of tau-like jets.  The number of options is further multiplied when we consider the various Higgs production modes.  To get a sense of orientation, we show in Table~\ref{tab:rawNumbers2012} the expected raw number of events in several non-exclusive $4\tau$ final-state channels for the 2012 LHC data set, taking as a benchmark $\Br(h\to 2a) = 10\%$ and $\Br(a\to 2\tau) \simeq 1$.  We pay special attention to muons, which are easier to identify than electrons, especially with nearby hadrons or other electrons.  In Table~\ref{tab:rawNumbers2012_2tau2mu} we show an analogous set of numbers for the $2\tau2\mu$ final-state channels.

\begin{table}[tp]
 \centering
\begin{tabular}{l||r||r|r|r|r|r|r|r|r}
   2012  $4\tau$       & Total & $\ge1\mu$ & $\ge2\mu$ & $\ge3\mu$ & $\ge2\ell$ & $\ge3\ell$ & $4\ell$ & $2\times(\ge1\mu)$ & $(\mu\mu/\mu e)+(0\mu)$  \\ \hline\hline
ggF                    & 38000 &  20200    &   10100   &    700    &   28600    &   4600     &   580   &        3800        &    4700       \\
VBF                    &  3200 &   1700    &     850   &     60    &    2400    &    400     &    50   &         320        &     400       \\
$W(\to \ell\nu)h$      &   300 &    160    &      80   &      5    &     220    &     40     &     5   &          30        &      40       \\
$Z(\to\nu\bar\nu)h$    &   150 &     80    &      40   &      3    &     110    &     20     &     2   &          15        &      20       \\
$Z(\to \ell^+\ell^-)h$ &    55 &     30    &      15   &      1    &      40    &      7     &     1   &           5        &       7       \\
\end{tabular}
\caption{\small Approximate raw numbers of events for a selection of $h \to 2a \to 4\tau$ decay channels, assuming $\Br(h\to 2a) = 10\%$ and $\Br(a\to \tau^+\tau^-) \simeq 1$, with the 2012 LHC data set (8~TeV, 20~fb$^{-1}$).  No trigger or reconstruction cuts have been applied.  (Categories are not all mutually exclusive, and leptons from $W/Z$ decay are not being counted.)}
\label{tab:rawNumbers2012}
\end{table}

\begin{table}[tp]
 \centering
\begin{tabular}{l||r||r|r|r}
   2012  $2\tau2\mu$          &  Total & $\ge1\mu$ & $\ge1\ell$ & $2\ell$ \\ \hline \hline
ggF                           &  266   &    75     &    120     &   33              \\
VBF                           &   22   &     6.3   &     10     &    2.7              \\
$W/Z(\to \ell$'s$/\nu$'s$)h$  &    3.5 &     1.0   &      1.6   &    0.4              \\
\end{tabular}
\caption{\small Approximate raw numbers of events for a selection of $2\tau$ decay channels within $h \to 2a \to 2\tau2\mu$, assuming $\Br(h\to 2a) = 10\%$, $\Br(a\to \tau^+\tau^-) \simeq 1$, and $\Br(a\to \mu^+\mu^-) = 0.35\%$, with the 2012 LHC data set (8~TeV, 20~fb$^{-1}$).  No trigger or reconstruction cuts have been applied.  (Categories are not all mutually exclusive, and leptons from $W/Z$ and $a\to\mu^+\mu^-$ decay are not being counted.)}
\label{tab:rawNumbers2012_2tau2mu}
\end{table}

While these raw numbers start at the tens of thousands, the various decay channels all have tradeoffs.  One of the primary concerns is that the mass-energy of the Higgs must be distributed between a large number of final-state particles, many of which are invisible neutrinos.  A typical $\tau$ receives
${\cal O}(1/4)$ of the energy, suggesting $p_T(\tau) \sim 30$~GeV.  However, when the $\tau$ decays, the visible $p_T$ frequently falls below normal reconstruction thresholds.  The leptonic decays, which are naively cleaner than the hadronic decays, have more neutrinos and less visible energy.  Therefore, while we appear to be presented with many opportunities for clean leptonic tags, the leptons are often too soft to either trigger or reconstruct.  The fact that these leptons can be non-isolated from each other or from a nearby hadronic tau further complicates matters.  If non-isolated leptons and/or hadronic taus are considered, backgrounds from QCD must be carefully accounted for.  In particular the signal can be faked by $\Upsilon(1S$--$3S)$ leptonic decays, for which the $\Br$'s are a few percent, and by events with $\gamma^*/Z^*$ emissions.

Another handle is the kinematics of the decay.  In principle, each event is triply-resonant, reconstructing to two $a$'s and the 125~GeV Higgs.  However, the neutrinos in the tau decays present a complication.  In the $4\tau$ mode, assuming that every visible $\tau$
decay can even be identified, typically the best that we can do is to attempt to reconstruct the Higgs's visible mass or variants of its transverse mass folding in the $\met$.  There is therefore no sharp resonance peak.  Reconstruction of the $a$ mass further suffers from the fact that the $\met$ contributed by each individual $a$ is a priori unknown.  The $a$ mass's utility as a discriminating variable against backgrounds is also highly reduced if $m_a$ is at or below the bottomonium region.  These difficulties highlight the major advantage of the $2\tau2\mu$ mode.  Though the overall rate is much smaller than $4\tau$, every event is tightly localized around the same value of $m(\mu^+\mu^-)$.  The prompt muons also tend to be much more energetic than the leptons produced in tau decays, significantly enhancing the relative rate once realistic momentum cuts are applied.

The complications associated with $h \to 2a \to 4\tau$ and the low rates for $2\tau2\mu$ means that at present these decays are difficult to constrain, and no significant limits exist from dedicated searches.  Nonetheless, the signals are distinct enough that they can ultimately be observed or constrained, even for $\Br(h \to 2a \to 4\tau) \lsim 10\%$.  This will especially be true over the lifetime of the LHC, as the higher statistics will allow better exploitation of the cleaner subleading final-state channels.  In the following subsections, we discuss ways in which theorists and experimentalists have sought to construct viable search strategies, review existing dedicated and non-dedicated searches, and quantify to what extent the non-dedicated searches might place meaningful constraints.  In particular, we estimate that a combination of recent CMS 3-lepton and 4-lepton searches 
at 8~TeV may already constrain $\Br(h\to 2a \to 4\tau) \lsim 20$--40\% for $m_a \gsim 15$~GeV.  We further estimate that a dedicated $\mu^+\mu^-$ resonance search in 3/4-lepton events could indirectly probe down to $\Br(h\to 2a \to 4\tau) \lsim 10\%$ with the 2012 data, even for $m_a < 10$~GeV.

\subsection{Existing Collider Studies}
\label{sec:4tauProposals}

Recent interest in $h \to 2a \to 4\tau$ searches was in part spurred by the observation~\cite{Dermisek:2005gg} of a ``blind spot'' between the direct OPAL bound of 86~GeV~\cite{Abbiendi:2002in} (limited only by an unfortunate choice of signal simulation range) and the LEP kinematic reach of approximately 115~GeV.  In particular, this would have allowed a lighter SM-like Higgs, requiring a less fine-tuned NMSSM.  However, as we now know, the SM-like Higgs was beyond LEP's reach.

Subsequent search proposals at the Tevatron and LHC have exploited the fact that the majority of the $2a$ decay channels contain one or more leptons.  The chance of producing a fully-hadronic final state is only about $(0.65)^4 = 18\%$.  It has also been pointed out that closeby hadronic taus (or a hadronic tau and an electron) still constitute a jet-like object with unusually low track activity and a distinctive calorimeter pattern, leaving various options for tagging it as a ``ditau-jet''.

Below, we briefly review several recent proposals using a variety of strategies.  Note that these all typically assume $\Br(h\to 2a\to 4\tau) \simeq 1$ and masses in the range $m_a \simeq [2m_\tau,m_\Upsilon]$, so that the $a \to 2\tau$ decays are highly collimated.

{\bf Trilepton and collinear ${\mathbf e\mu}$:}  In Ref.~\cite{Graham:2006tr}, the $h \to 2a \to 4\tau$ decay mode is studied in the context of the Tevatron.  For ggF, they consider trilepton channels and channels where one of the tau pairs decayed to a roughly collinear $e\mu$ pair (to reduce $\gamma^*$ and hadronic decay backgrounds).  The starting efficiency for trilepton from its $\Br$ is roughly 10\%, but after accounting for cuts on lepton $p_T$ (3~GeV), $\eta$ (2.0), and isolation, the final efficiency becomes only 0.5\%.  The estimated cross section times acceptance for ggF is then 4~fb, or ${\cal O}$(40~events) for Run~II.  The collinear $e\mu$ case, assumed to recoil against a low-track ditau-jet, could have higher efficiency but also faces higher backgrounds that are much more difficult to model.  No attempt is made to estimate these.  Utilizing the associated $Wh$ and $Zh$ production modes is also suggested, though the rates tend to be even smaller.  While the rate limitations at the Tevatron make all of these searches unlikely to yield a signal, especially since recent LHC results imply that exotic Higgs decays cannot dominate, most of these ideas can readily be adapted to the LHC.

{\bf Two $\mathbf \mu\tau_h$-jets:}  In~\cite{Belyaev:2008gj}, the $4\tau$ decay is studied for VBF and $Wh$ production at LHC14, exploiting a pair of decays $a \to \mu\tau_h$(1-prong).  For VBF, the events are assumed to be selected with a same-sign dimuon trigger allowing an offline selection of $p_T > 7$~GeV, while the $Wh$ channel is triggered with the leptonic $W$ decay.  The specific requirements of the two channels are not identical, but each demands two muons (same-sign for VBF) and two one-prong hadronic taus, forming two approximately collinear $\mu\tau$ systems.  For LHC14 and $m_h = 125$~GeV, VBF is predicted to have $\sigma\times A \sim 20$--70~fb and $Wh$ 4--10~fb, increasing for lighter pseudoscalars.  Scaling to LHC8 with 20~fb$^{-1}$, and multiplying by a reference $\Br(h\to 2a) = 10\%$, we estimate 15--55~events (VBF) and 3.7--9~events ($Wh$).  The upper ranges of these numbers are close to the raw counts expected from $\Br$ alone, suggesting very high estimated reconstruction efficiency and/or other exclusive final-states being picked up by the analysis.  VBF is more promising in terms of raw event counts, but backgrounds are not assessed.  The $Wh$ search is expected to be ``almost background free.''  No search of this type has been performed yet.

{\bf Dimuon resonance:}  Ref.~\cite{Lisanti:2009uy} considers the subleading decay sequence $h \to 2a \to 2\tau2\mu$, with a focus on identifying the sharp $2\mu$ resonance at $m_a$.  The taus are assumed to decay hadronically, and are simply treated as a jet with aligned $\met$.  The Higgs resonance is also shown to be approximately reconstructable, though this is not used for discrimination, as $S/B$ is already $\gg 1$.  For a 125~GeV Higgs and 7~GeV pseudoscalar $a$, 5~fb$^{-1}$ at LHC14 is estimated to give 2$\sigma$ sensitivity to $\Br(h\to 2a) < 10\%$ via ggF production.  Note that the statistics from the 2012 run corresponds to about 8~fb$^{-1}$ of LHC14, so this strategy may already be capable of rather stringent limits.  D0 has performed a search of this type, which we describe in the next subsection.

{\bf Ditau-jets:} In~\cite{Englert:2011iz}, a calorimeter based ``ditau-jet tag'' is assessed in the context of $Zh \to (\ell^+\ell^-)(4\tau)$.  (See also~\cite{Katz:2010iq} for tracker-based techniques tailored to boosted $h \to 2\tau$ ``jets.'')  For this purpose no lepton identification is used.  The main ditau-jet discriminating variables considered are the N-subjettiness ratio $\tau_3/\tau_1$ operating on ECAL cells and the $m/p_T$ ratio.  (A more powerful likelihood-based tag is also studied.) $\met$ and $p_T(Z)$ are also applied to purify the signal.  For LHC14, $\sigma\times A \gtrsim 1$~fb is achieved with $S/B \simeq 0.5$.  Scaling $\Br(h\to 2a) \to 10\%$, and $\sigma$ and luminosity to a 2012-like dataset, this would yield only ${\cal O}$(1~event) with $S/B \ll 1$.  However, the ditau-jet tag can also be considered for searches in channels with higher cross sections.

\subsection{Existing Experimental Searches  and Limits}

Dedicated searches for prompt $4\tau$ and $2\tau2\mu$ final states of the Higgs have been performed at LEP~\cite{Schael:2010aw,Abbiendi:2002in} and at the Tevatron~\cite{Abazov:2009yi}, respectively, but no significant constraints have yet been established for $m_h = 125$~GeV.  No dedicated search has yet been performed at the LHC.  We briefly discuss the Tevatron search, and also some non-dedicated searches at the LHC that may have sensitivity to our signal, or can serve as starting points for new dedicated searches.  We then recast a subset of the non-dedicated searches to derive new, nontrivial limits.

{\bf Tevatron $\mathbf 2\tau2\mu$:}  With 4 fb$^{-1}$, D0 searched for $2\tau2\mu$ (and $4\mu$) in ggF events~\cite{Abazov:2009yi}, based on the strategy presented in~\cite{Lisanti:2009uy}.  Most accepted events pass a 4--6~GeV dimuon trigger.  Muon ID is relaxed for one of the muons in the $a \to 2\mu$ candidate, but its inner track can still be reconstructed.  The search is a bump-hunt in the muon-pair mass spectrum over the range $m_a = [3.6,19]$~GeV.  The $a \to 2\tau$ ditau-jet is minimally identified by requiring significant $\met$, possibly near a jet with low track multiplicity.  Assuming unit branching fractions for a 125~GeV Higgs, the limit is approximately a factor of 4 above the SM production cross section at the low range of $m_a$, and steadily weakens for larger $m_a$.

{\bf LHC high-multiplicity leptons:}  A variety of high-multiplicity lepton ($\ge 3\ell$) searches have now been completed at the LHC, mainly motivated by supersymmetry, including scenarios with R-parity violation.  Several searches are focused on tau signals.  Typical SUSY multilepton searches demand large amounts of $\met$, hadronic activity, and/or one or more $b$-tags, any one of which can very efficiently eliminate the $4\tau$ and $2\tau2\mu$ Higgs signals.  Still, relatively more inclusive 3- and 4-lepton searches have been performed by
CMS~\cite{Chatrchyan:2012mea,CMS026,CMS027,CMS13010} (most recently 9.2~fb$^{-1}$ 3/4-lepton and 19.5~fb$^{-1}$ 4-lepton at LHC8) and ATLAS~\cite{Aad:2012xsa} (4.7~fb$^{-1}$ at LHC7).  While these largely utilize standard lepton and tau isolation requirements, they use quite low $p_T$ thresholds.  The analysis of~\cite{CMS13010} uses particle-flow isolation, and does not count nearby leptons against each other.  The multilepton searches are especially interesting to consider for $m_a \gsim 15$~GeV, where the isolation issues are less severe and experimental vetoes on low-mass dilepton pairs are avoided.

{\bf LHC same-sign dilepton:}  Same-sign dileptons are also a standard signal of supersymmetry, and we expect that the usual searches are similarly unconstraining.  However, ATLAS has performed an inclusive search for new physics in same-sign dileptons
using the full 2011 data set~\cite{ATLAS:2012mn}.  While this again relies on lepton isolation, it is nonetheless useful to understand what kind
of limit might apply to our scenarios.

While the existing dedicated searches are not constraining, we can explore the power of the non-dedicated searches.  We keep our study as model-independent as possible by scanning across the full kinematic range $m_a = [2m_\tau,m_h/2]$, and leaving $\Br(h \to 2a)$ and $\Br(a\to\tau^+\tau^-)$ as free parameters.  We express our results as a function of the limits on total branching fraction $\Br(h\to2a\to4\tau) = \Br(h\to2a) \times \Br(a\to\tau^+\tau^-)^2$ versus $m_a$.  Note that while masses above $m_\Upsilon$ are not usually considered in conjunction with an appreciable $\Br$ to leptons, we again emphasize that they can arise easily if $a$ is mostly composed of (or mixed into) the leptonic Higgs field in the Type-III 2HDM.  Depending on the $a$'s coupling to $b$-quarks, there can also be nontrivial effects from decays and mixings into the bottomonium sector when $m_a \simeq m_\Upsilon$, which we neglect (see~\cite{Englert:2011us, Baumgart:2012pj} for more details).

A remaining free parameter is the CP phase of the $a$'s Yukawa couplings.  Assuming CP conservation, $a$ may be a CP-odd pseudoscalar or a CP-even scalar.  We fix $a$ to be the former.  There are two consequences of favoring CP-odd over CP-even.  First, this choice can affect the relative $\Br$'s to $2\tau$ and $2\mu$, but only for $m_a$ very close to $2m_\tau$.  (E.g., for $m_a = 5$~GeV, the ratio $\Br(a\to\mu^+\mu^-)/\Br(a\to\tau^+\tau^-)$ is approximately twice as large in the CP-even case.)  Second, there is an imprint of the $a$'s CP on the azimuthal decay angle correlations of the two taus in the $a$ rest frame.  We expect this to be a minor effect, but it can in principle affect isolation rates.

We simulate ggF, VBF, and $(W/Z)h$ production of a 125~GeV Higgs decaying to $2a$ in Pythia 8.176~\cite{Sjostrand:2007gs}, which includes a full treatment of tau spin correlations.\footnote{We thank Philip Ilten for help tracking down and fixing a bug in Pythia's $2\tau$ spin correlation code.}$^{,}$\footnote{We have also checked $t\bar t h$.  This production channel is rare, but it gives many opportunities for lepton production.  We estimate that this represents up to a 10\% contribution to the signal in the 4-lepton and same-sign dilepton searches below, but do not explicitly incorporate it into the derivation of constraints.}  We set the cross sections to the values recommended by the LHC Higgs Cross Section Working Group~\cite{LHCHiggsCrossSectionWorkingGroup:2011ti}.  For ggF, we reweight the $p_T$ spectrum after showering to the NLO+NLL predictions of HqT~2.0~\cite{Bozzi:2005wk,deFlorian:2011xf}.

We do not apply a detector model nor simulate pileup.  For the leptons, particle-level should still furnish an adequate zeroth-order approximation of the full detector, including isolation.  However, lepton identification efficiencies can be important, especially for soft leptons.  CMS provides a detailed discussion and parametrizations of these efficiencies in the appendix of~\cite{CMS13010}, and we apply these for our CMS analyses.  For ATLAS, which uses harder lepton $p_T$ cuts for the analysis that we study, we coarsely assume flat efficiencies of 90\% for muons and 75\% for electrons.  Lepton isolation requirements vary by analysis, and we have adjusted them on a case-by-case basis.

The hadronic taus are much more difficult to reliably model.  For these, we take a minimalistic approach, simply ``rebuilding'' each hadronic tau out of its visible decay products and applying a flat 50\% identification efficiency if its visible $p_T$ exceeds 15~GeV.  However, two hadronic taus within $\Delta R < 0.45$ (averaging between ATLAS and CMS radii) are assumed to be unidentifiable, as are hadronic taus with a lepton with $p_T > 2$~GeV within the same radius.  This mimics the isolation failures that would occur in these cases.

For the jets and missing energy, we reconstruct the former with the anti-$k_T$ algorithm with $R=0.45$, and the latter from the 2-vector sum of all neutrinos.  Jets that overlap with identified hadronic taus are removed.

We consider constraints from three recent LHC multilepton analyses\footnote{We do not consider the related but
superceded analyses~\cite{Chatrchyan:2012mea,Aad:2012xsa}.  We also do not consider~\cite{CMS027}, which is very closely related to (1) and uses the same data set, but divides the analysis bins by $S_T$ instead of by $\met$.  This division tends to give lower $S/B$ in the 3-lepton bins.}:
\begin{enumerate}
\item {\bf CMS PAS SUS-12-026:}  3- and 4-leptons in many exclusive bins, 9.2~fb$^{-1}$ at 8~TeV~\cite{CMS026}.
\item {\bf CMS PAS SUS-13-010:}  4-leptons with at least one OSSF pair, 19.5~fb$^{-1}$ at 8~TeV~\cite{CMS13010}.
\item {\bf ATLAS 1210.4538:}  Same-sign dileptons, 4.7~fb$^{-1}$ at 7~TeV~\cite{ATLAS:2012mn}.
\end{enumerate}

As a first step, we use the reported background rates to verify our treatment of the reconstructions.  We generate diboson events in Pythia, and $W^\pm W^\pm$ and $t\bar t(W/Z)$ in MadGraph, normalizing each to NLO.  For (1) and (2), we compare 4-lepton analysis channels to our $ZZ$ simulation.  For (1), we use the channel ``OSSF2, on-$Z$, $H_T < 200$~GeV, $\met < 50$~GeV, $0\tau$, $0b$.''  We predict 56~events, and CMS predicts $73\pm16$.  For (2), we compare to the bin ``$M_1 = [75,110]$~GeV, $M_2 = [75,110]$~GeV.''  It is normalized to the central CMS $ZZ$ cross section measurement, which is about 10\% higher than the NLO prediction.  Weighting our sample accordingly, we predict 130~events, and CMS predicts 150.  For (3), we compare our simulations to the ``Prompt'' same-sign dilepton background estimated by ATLAS.  In the ($e^\pm e^\pm$, $e^\pm\mu^\pm$, $\mu^\pm\mu^\pm$) channels we obtain (78, 275, 165) events, and ATLAS predicts ($101\pm13$, $346\pm43$, $205\pm26$).  In all of the comparisons there is a systematic tendency for our predictions to underestimate the experiments by about 20\%.  This may be related to our idealized treatment of isolation, and suggests that our Higgs signal estimates may be slightly conservative.

\begin{table}[tp]
 \centering
\begin{tabular}{l}
  CMS PAS SUS-12-026 (9.2~fb$^{-1}$, 8~TeV)  \\ \hline
1a) 3-lepton, OSSF0, $H_T < 200$~GeV, $\met < 50$~GeV, $0\tau$, $0b$     \\
1b) 3-lepton, OSSF0, $H_T < 200$~GeV, $\met = [50,100]$~GeV, $0\tau$, $0b$   \\
1c) 3-lepton, OSSF0, $H_T < 200$~GeV, $\met > 100$~GeV, $0\tau$, $0b$               \\
1d) 3-lepton, OSSF0, $H_T > 200$~GeV, $\met > 100$~GeV, $0\tau$, $0b$               \\
1e) 3-lepton, OSSF1, below-$Z$, $H_T < 200$~GeV, $\met < 50$~GeV, $0\tau$, $0b$     \\
1f) 3-lepton, OSSF1, below-$Z$, $H_T > 200$~GeV, $\met = [50,100]$~GeV, $0\tau$, $0b$   \\
1g) 3-lepton, OSSF1, below-$Z$, $H_T > 200$~GeV, $\met > 100$~GeV, $0\tau$, $0b$   \\
  \\
 CMS PAS SUS-13-010 (19.5~fb$^{-1}$, 8~TeV)  \\ \hline
2a) $M_1 < 75$~GeV, $M_2 < 75$~GeV  \\
2b) $M_1 = [75,110]$~GeV,  $M_2 < 75$~GeV  \\
  \\
 ATLAS 1210.4548 (4.7~fb$^{-1}$, 7~TeV)  \\ \hline
3a) $e^\pm e^\pm$,  $m(\ell^\pm \ell^\pm) > 15$~GeV  \\
3b) $e^\pm \mu^\pm$,  $m(\ell^\pm \ell^\pm) > 15$~GeV  \\
3c) $\mu^\pm \mu^\pm$,  $m(\ell^\pm \ell^\pm) > 15$~GeV  \\
\end{tabular}
\caption{\small Analysis bins used in setting our $h\to2a\to4\tau$ limits.}
\label{tab:CMSbins}
\end{table}

\begin{table}[tp]
\centering
\begin{tabular}{c|c |c |c |c}
Channel & $m_a=12$ GeV &  $m_a=50$ GeV  & Background & Observed\\ \hline
1a) & 2.57 & 3.31 & $27 \pm 6.7$ & 23 \\
1b) & 0.19 & 1.1 & $17.75 \pm 7.5 $ & 16 \\
1c) & 0.01 & 0.18 & $4.5\pm 2.3$ & 3 \\
1d) & 0 & 0.3 & $1.9\pm 1.2 $ & 1 \\
1e) & 2.5 & 9.5 & $282\pm 29$ & 258 \\
1f) & 0 & 0.29 & $4.5 \pm 0.9$ & 4 \\
1g) & 0.02 & 0.68 & $3.5 \pm 0.8$ & 2 \\ \hline
2a) & 1.48 & 0.2 & $10.4 \pm 2$ & 14 \\
2b) & 0.97 & 0.22 & $35\pm 8$ & 30 \\ \hline
3a) & 2.8 & 3.7 & $346 \pm 44$ & 329 \\
3b) & 7.2 & 9.2 & $639 \pm 71$ & 658 \\
3c) & 3.7 & 5.5 & $247 \pm 30$ & 264 \\
\end{tabular}
\caption{\small Signal predictions and SM backgrounds in all of the analysis bins considered for exclusions in this subsection. See Table~\ref{tab:CMSbins} for descriptions. The signal prediction here is given fixing $\Br(h\to2a\to4\tau) = 10\%$ for reference, though it is a free parameter in setting the exclusions.}
\label{tab:allsignals}

\end{table}

We run the search using a number of preselected bins from the different analyses.  From the CMS multilepton searches (1) and (2), we focused on bins with high $S/B$.  The selected bins are listed in Table~\ref{tab:CMSbins}.  From the ATLAS same-sign dilepton search (3), we have added positive-charge and negative-charge counts for the $m(\ell^\pm \ell^\pm) > 15$~GeV bins, but maintained the binning in flavor.  In Table~\ref{tab:allsignals} we display the expected number of signal events for two example mass points ($m_a = 12$~GeV and $m_a=50$~GeV) and compare to the SM backgrounds predicted by CMS and ATLAS.

\begin{figure}
\centering
\includegraphics[width=1.0\textwidth]{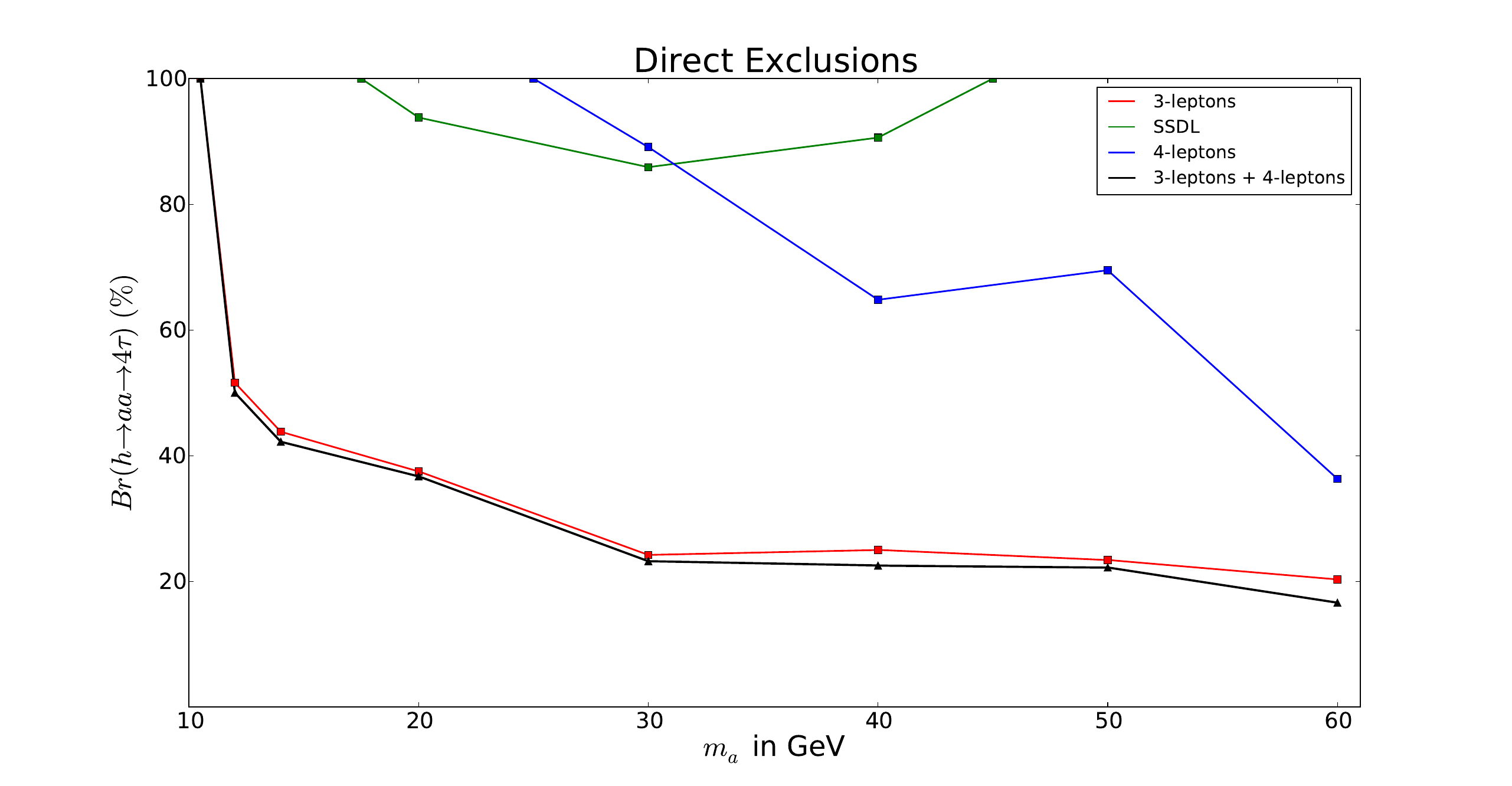}
\caption{\small Estimated exclusion of $\Br(h\to2a\to4\tau)$ from LHC multilepton and same-sign dilepton searches:  (1) CMS 3-lepton from~\cite{CMS026} in red, (2) CMS 4-lepton from~\cite{CMS13010} in blue, (3) ATLAS same-sign dilepton from~\cite{ATLAS:2012mn} in green.  The black line shows a combination of the multilepton searches (1) and (2).  (The combination of all channels, including (3), is less constraining by several percent.)
}
\label{Fig:tauexclusions}
\end{figure}

We estimate 95\% confidence constraints on $\Br(h\to2a\to4\tau)$ using a simple $CL_S$ analysis.  Signal rates in the various experimental analysis bins come from our simulations.  Backgrounds rates, their systematic errors, and observed counts come from the experiments.  We do not apply a systematic error to the signal, as we cannot fully quantify the reliability of our modeling of the detection and reconstruction steps.  (It should be understood that our signal predictions are merely a guide.)  For our test statistic, we use the Poisson likelihood ratio between $S+B$ and $B$ hypotheses, constructed using the central $B$ expectation values. Within each pseudoexperiment, we vary the bin-by-bin expectation values for $B$ according to the reported systematic errors, treating them as independent and gaussian-distributed.\footnote{Negative expectation values are reset to zero when they arise in the pseudoexperiments.}

Fig.~\ref{Fig:tauexclusions} shows the limits that we obtain from the individual analyses, as well as from a combination of the CMS analyses.  It can be seen that $\Br(h\to2a\to4\tau)$ can be excluded at the 20--40\% level provided $m_a \gsim 15$~GeV, and that these limits are dominated by the CMS 3-lepton bins.  Below 15~GeV, standard quarkonium vetoes begin to make all of the searches very inefficient.  Below about 10~GeV, isolation cuts also begin to have a major impact, though less significantly for analysis (2).  We conclude that tight limits can already be placed with existing data, provided that $a$ is massive enough and has small couplings to quarks so that $a\to b\bar b$ does not compete.  However, this leaves fully open the interesting NMSSM-motivated region with $m_a \lsim m_\Upsilon$.

\subsection{Proposals for New Searches at the LHC}

We have focused on multilepton searches because they are relatively clean and because existing limits could be quickly estimated.  These results can be considered an update and extension of some of the strategies proposed in~\cite{Graham:2006tr}.  The other strategies discussed in \S\ref{sec:4tauProposals} can also have a significant role, and we might expect versions of these searches in the near future from the LHC experiments using the 2012 data set.  It will be interesting to see how these extend the limits that we have estimated, especially for lighter $m_a$.  However, looking ahead to possible future searches, we can concretely suggest a novel strategy:  exploit the $2\tau2\mu$ final-state within 3- and 4-lepton events.\footnote{A similar strategy was also discussed for associated production of $a$ with a heavy Higgs (via $q\bar q \to Z^* \to Ha$) in the lepton-specific 2HDM~\cite{Kanemura:2011kx}.  That study was aimed at $m_a,m_H \gsim 100$~GeV.}  This would supplement the more inclusive $2\tau2\mu$ search proposed in~\cite{Lisanti:2009uy} and implemented in~\cite{Abazov:2009yi}, representing an analysis channel with extra-low backgrounds.  Given the shrinking range of viable $\Br$, and the relatively high rate for the $2\tau$ side of the event to produce a lepton, this type of search should offer good long-term prospects.

We have observed in our own simulations that a surprisingly large fraction of 3-lepton and 4-lepton events passing experimental cuts come from the $2\tau2\mu$ channel.  For example, for the point $m_a = 60$~GeV within the bin ``3-lepton, OSSF1, below-$Z$, $H_T < 200$~GeV, $\met < 50$~GeV, $0\tau$, $0b$'' (1e), about 20\% of the events contain $a\to2\mu$.  Since $S/B$ will improve by far more than a factor of 5 by focusing in on a tight resonance peak, this suggests that a powerful search could be constructed by utilizing $m(\mu^+\mu^-)$ spectral information within high-multiplicity lepton events.  The resonance also offers a much safer way to search within the $m_a \lsim 10$~GeV region, where leptonic $a$ decays are expected to dominate for a broader class of models.

To construct an example of such a search, we can follow the reconstructions of the CMS 4-lepton analysis~\cite{CMS13010} (search (2) above), but removing their restriction $m(\ell^+\ell^-) > 12$~GeV and allowing events with three or more leptons instead of exactly four.  Crucially for the low-mass region, this search uses a full particle-flow form of isolation, and does not count leptons towards each others' isolation cones.  We include a $Z$-veto to help reduce $Z$+jets and diboson backgrounds.  We also focus on ``below-$Z$'' events, where the $\ell^+\ell^-$ pair closest to the $Z$ mass is below 75~GeV.  These vetoes have little effect on the signal efficiencies.\footnote{It might also be possible to apply a $\met$ discriminator for this search, though we have not attempted this.  The $\met$ in signal events tends to be below 50~GeV.  An accurate understanding of the efficacy of a $\met$ cut would require a resolution model, as well as a model for the $\met$ distribution of backgrounds.  An approximate reconstruction of the Higgs resonance might also be possible, and usable either for further discrimination or for verification of the source of a possible signal.}

In reconstructing the $\mu^+\mu^-$ resonance, there remains a combinatoric issue when more than one pairing of this type is possible.  This ambiguity afflicts the majority of 3-lepton and 4-lepton events containing at least one $\mu^+\mu^-$ pair, since muons are reconstructed with higher efficiency than electrons.  (E.g.,  $\mu^+\mu^-\mu^\pm$ is found more often than $\mu^+\mu^-e^\pm$.)  In practice, it is possible to pick the smallest-mass pairing for $m_a \ll m_h/2$ and the largest-mass pairing for $m_a \simeq m_h/2$.  However, for $m_a \simeq m_h/4$, neither of these options is ideal.  Instead, we can construct a third option by using the fact that $m_h \simeq 125$~GeV, that the Higgs decays isotropically, and that it is usually produced with little transverse boost: we pick the $\mu^+\mu^-$ pair whose trajectory would make the largest opening angle with the beam in the Higgs rest frame, assuming $p_T(h) = 0$.  For each $m_a$, we use the pairing choice that gives the strongest resonance peak.\footnote{The crossover between smallest-mass and largest-mass choices being the most effective is at $m_a \simeq 40$~GeV, and in this region the largest-opening-angle choice keeps about 15\% more events in the peak.  For very low-mass resonances, this choice underperforms the smallest-mass choice by a comparable amount, and similarly for high-mass resonances (near $m_h/2$) relative to the largest-mass choice.}

Estimating backgrounds to such a search can be difficult, as leptons from heavy flavor decays and from fakes can be significant contributions.  We have simulated the contributions from electroweak 3-lepton and 4-lepton production, including taus and allowing for $Z^*/\gamma^*$ down to $m\sim$~GeV.  Given a signal that lives inside of a resolution-limited mass window of approximately $(1\pm 0.01)m_a$, these backgrounds are usually small, tallying to ${\cal O}$(1 event) for any $m_a$ for 2012.  The dominant $Z^*/\gamma^*+$jets background can be coarsely estimated from the sum of ``below-$Z$'' bins of analysis (1), and would constitute approximately 800~events for $m(\mu^+\mu^-) \gsim 10$~GeV with 20~fb$^{-1}$.  (In this estimate, we conservatively do not attempt to remove the $e^+e^-$ events.)    We are not given a spectral shape for this background, but if we assume that it is not very strongly-featured, then we can estimate ${\cal O}$(10~events) per 1~GeV interval.  We also do not know the spectrum for $m(\mu^+\mu^-) \lsim 10$~GeV, though the shrinking absolute resolution on $m(\mu^+\mu^-)$ (down to less than 100~MeV at CMS) allows the differential background rate to grow by an order of magnitude without affecting $S/B$.  Of course, extra care would need to be taken in the vicinity of known hadronic resonances such as the $\Upsilon$'s.

\begin{figure}
\centering
\includegraphics[width=1.0\textwidth]{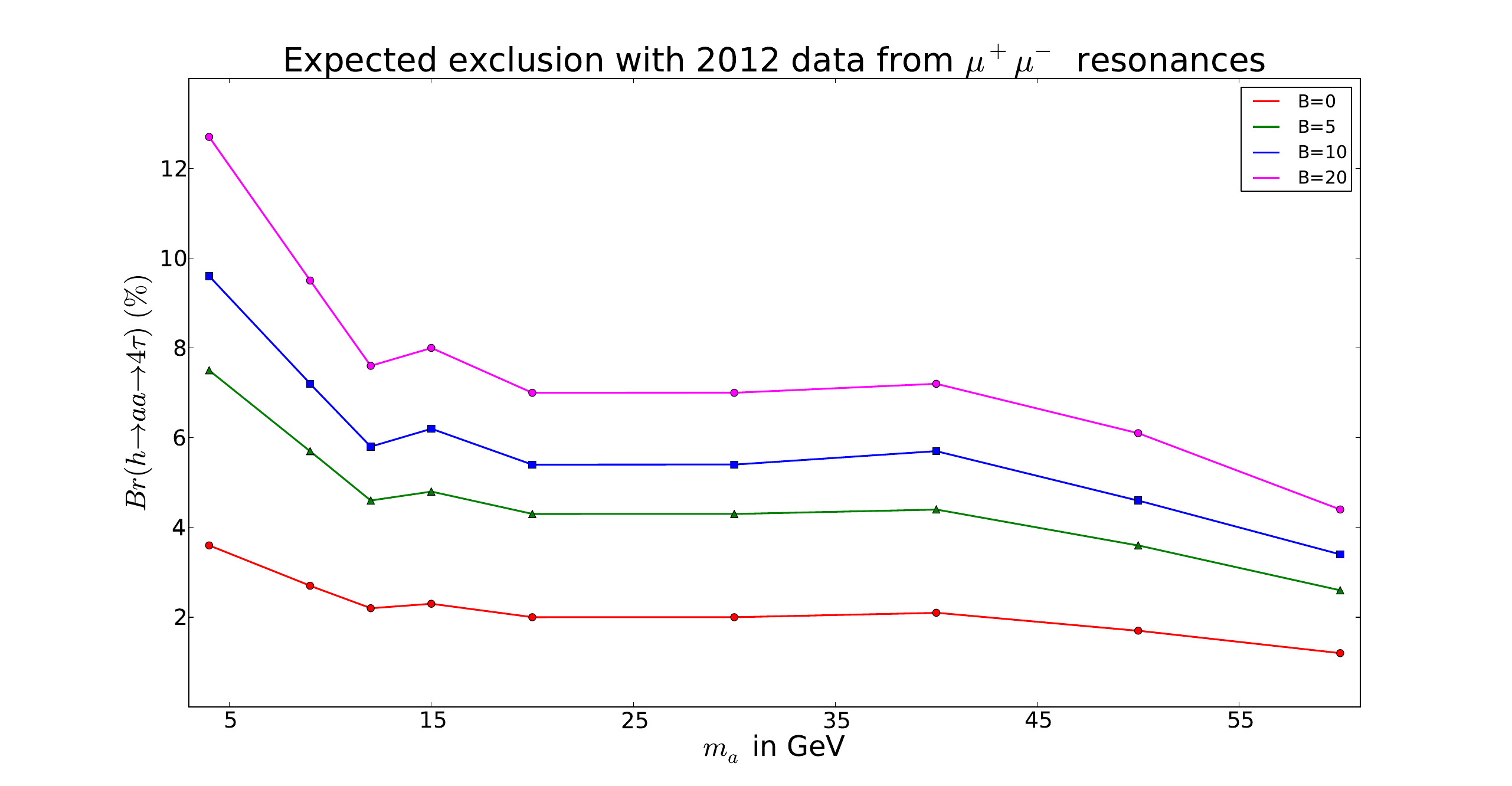}
\caption{\small Median estimates of expected indirect exclusions on $\Br(h\to2a\to4\tau)$ using the subdominant $(a \to 2\tau)(a \to 2\mu)$ channel and exploiting that leptonic branching fractions of $a$ are mass-ordered.  The results are based on a simulated $\mu^+\mu^-$ resonance search in $\ge 3\ell$ events, assuming the 2012 data set.  Since we cannot reliably predict the background under the resonance peak, we show expected exclusions for $B$ = 0, 5, 10 and 20 events respectively.  We neglect systematic uncertainties.  (The lowest displayed mass is 4.0~GeV.)
}
\label{Fig:tauindirect}
\end{figure}

To give a sense of what might be possible with the 2012 data set, we show in Fig.~\ref{Fig:tauindirect} the limits assuming a sequence of possible background levels with $m(\mu^+\mu^-)$ within $\pm1\%$ of $m_a$, and neglecting systematics.  Taking as reference $\Br(h\to2a\to4\tau) = 10\%$, the signal rates inside the peak vary from 8~events for $m_a = 4$~GeV, to 25~events for $m_a = 60$~GeV.  Depending on the background assumption and on $m_a$, the excluded $\Br(h\to2a\to4\tau)$ varies from percent-scale to just above 10\%.  This strong level of exclusion applies even down to $m_a \simeq 2m_\tau$.\footnote{Note that while isolation of a single lepton from the $a\to\tau^+\tau^-$ side of the event becomes progressively more difficult for low-mass points, $\Br(a\to\mu^+\mu^-)$ is also increasing.  At 4~GeV, the rate has doubled.  This effect would be even more pronounced for CP-even scalars.}  We imagine that these results will only improve as data from the next run of the LHC becomes available, provided that the multilepton triggers can be maintained at $p_T$ thresholds comparable to their 2012 values.
	\section{\secsize $\secbold{h \rightarrow 4j}$}
\label{sec:h4j}

Standard Model decays of the Higgs boson can lead to a four-jet final
state via intermediate vector boson decays, $h \rightarrow WW^*/Z Z^*
\rightarrow jjjj$.  Only one of the jet pairs is produced on-resonance
in this process.  In this section, we discuss the distinct possibility
of exotic Higgs decays to $4j$ in a two-step decay process proceeding
through a neutral (pseudo-)scalar field $a$: $h\to a a \to jjjj$.
There are then two jet-pair resonances.  Below, we outline the
theoretical motivations for considering $4j$ decays of the Higgs, and
discuss the LHC phenomenology and future discovery prospects of this
channel.

\subsection{Theoretical Motivation}

The $h \rightarrow jjjj$ channel has been extensively studied in the
context of super Little Higgs models~\cite{Birkedal:2004xi,
Berezhiani:2005pb, Csaki:2005fc} (a brief description of the Little 
Higgs mechanism is given in \S\ref{LHiggs}). The intermediate 
decay product, $a$, is a PNGB and generally very light. In a large 
region of parameter space of these models, $h\to aa \rightarrow jjjj$ 
is the dominant decay mode. 

Given that the Higgs mass of approximately $125 \gev$ requires
fine-tuning of the simplest versions of these models, one may take a
simplified model approach for the cascade decay in the presence of a
light pseudoscalar (or scalar), $a$.  Two possibilities allow for the
decay of $a$ to jets:

(i) The pseudo(scalar) $a$ can mix with another heavier pseudoscalar
if a second Higgs doublet is present, for example in the NMSSM or,
more generally, in the 2HDM + S models, see \S\ref{2HDMS}
and~\S\ref{MSSM}).  This allows for the decay of $a$ to SM fermions,
often (depending on the 2HDM Type) dominated by $a \to b \bar b$ for
$m_a>2m_b$ and $a \to \tau^- \tau^+$ for $2m_\tau<m_a<2m_b$ for a
large or moderate $\tan\beta$.  This leads to $4b$, $2b2\tau$, $2b2\mu$, $4\tau$, and $2\tau2\mu$ signals as 
discussed in \S\ref{sec:hto4b}, \S\ref{sec:h2b2tau}, \S\ref{sec:h2b2mu}, and
\S\ref{sec:hto4tau}.  
However, if $a$ is very light ($3m_\pi<m_a < 2 m_\tau$), it predominantly 
decays to two (merged) light jets as the above channels are not kinematically viable.

If $\tan \beta$ is small ($\tan\beta\lesssim 0.5$), the couplings of
$a$ to the down type quarks and charged leptons can be very
suppressed. In this case, $a$ dominantly decays to light (mostly
charm) jets even if decays to $b$'s or $\tau$'s are kinematically
allowed. Thus, the parameter space of $m_a$ up to $m_h/2$ is available
for the exotic decay mode.  A similar situation also occurs in the
``charming Higgs'' scenario of the Little Higgs
model~\cite{Bellazzini:2009kw}.

(ii) New heavy BSM vector-like fermions can couple to $a$ and,
therefore, allow for its decay into gluons or photons through loop
processes~\cite{Dobrescu:2000jt, Chang:2006bw, Chang:2005ht}. This
scenario can be realized in Little Higgs models and extra dimensional
models. For $m_a$ above a few GeV up to $m_h/2$, $h\to aa\to gggg$
dominates over $h\to aa \to \gamma\gamma gg$ and $h \to aa\to
\gamma\gamma\gamma\gamma$. In general, the signal is hard to find
against combinatorial background. However, large masses of the new
vector-like fermions may lead to visibly displaced vertices of $a\to
gg$, which can enhance the discovery potential of the
channel~\cite{Chang:2006bw}. Studies on related decay modes in this
scenario, $h\to aa\to \gamma \gamma gg$ and $h\to aa \to
\gamma\gamma\gamma\gamma$, can be found in \S\ref{sec:2gamma2jet} and
\S\ref{sec:4gamma}, respectively.

\subsection{Existing Collider Studies}

Before the discovery of the 125~GeV Higgs boson, much of the
phenomenology of the Higgs decaying to four jets was aimed at hiding
the Higgs boson at LEP.  One way to accomplish this was in the
``buried Higgs'' scenario , where the decay $h \rightarrow jjjj$ is
``buried'' in the large QCD background.  Indeed, the LEP bounds for
this scenario are much weaker than the bound on a SM Higgs.  For $m_h
> 90 \gev$~\cite{Abbiendi:2002in}, $m_a$ was studied in a range where
each pair of jets from the pseudoscalar decay would be highly
collimated and appear as a single jet.

There are a few existing collider studies for the 14~TeV LHC run in
the four-jet final state.
In~\cite{Lewis:2012pf} the authors study the $h\to4c$ decay mode
in the context of ``charming Higgs''.  We mention this study
here since it does not use $b$-tagging and hence useful for generic $4j$ decays.
The study uses jet substructure to help identify the pseudoscalar as
a boosted jet while reducing the otherwise overwhelming background.

Other relevant collider studies are \cite{Chen:2010wk}
and \cite{Falkowski:2010hi}, which we briefly summarize below.
(There also exist collider studies that
consider exotic Higgs production modes~\cite{Bellazzini:2010uk},
but we do not consider them here.)

In \cite{Chen:2010wk}, Higgs production in association with a $W$
boson is considered as the production mode for $m_h = 120 \gev$
followed by the Higgs decay, $h \rightarrow a a \rightarrow j j j j$.
The pre-selection cuts in this analysis include isolated leptons with
$p_T > 20 \gev$, at least two jets with $p_T > 40, 30 \gev$, reconstructed leptonic $W$
transverse mass $m_T < m_W$, and a $b$-jet veto to reduce
SM background. Further analysis is divided into categories depending
on the mass of $a$:
\begin{itemize}
\item $m_a = 4 \gev$ : In this case the gluons from $a$ decay appear as a single jet to the HCAL. ECAL variables are imposed to distinguish these merged jets from single-pronged QCD jets.  
  $7\sigma$ significance is possible at the LHC14 with $30 \fb$ data
  assuming $\BR(h\to aa \rightarrow gggg) \sim 100 \%$. However,
  assuming a more realistic branching ratio of $\BR(h \to aa
  \rightarrow gggg) \sim 10 \%$ in the post Higgs discovery era, $2
  \sigma$ exclusion ($3\sigma$ evidence) is possible with $300 \fb$
  ($500 \fb$) of data at LHC14.
\item $m_a = 8 \gev$ : Simple jet substructure techniques can be used
  for discovery. The
  authors find that $\sim 3 \sigma$ statistical significance can be
  reached with $30 \fb$ data assuming $\BR(h\to aa \rightarrow gggg)
  \sim 100 \%$.  With $\BR(h\to aa \rightarrow gggg) \sim 10 \%$,
  however, $2 \sigma$ exclusion ($3\sigma$ evidence) requires $1000
  \fb$ ($3000 \fb$) of data at LHC14.
\end{itemize} 

A separate jet substructure analysis on $h \rightarrow aa \rightarrow
jjjj$ is also presented in \cite{Falkowski:2010hi}, with the inclusion
of the $t \overline{t} h$ production channel besides the $V h$
channel, demonstrating similar discovery potential in both
channels. Here variables sensitive to the soft radiation patterns of
the color singlet $a\to gg$ jet are employed instead of ECAL-based
observables. The authors reach a similar conclusion for discovery
prospects as described above.

The above two analyses \cite{Chen:2010wk, Falkowski:2010hi} have
exploited the fact that very light (pseudo)scalars are boosted,
leading to two fat-jets. A more recent study \cite{Kaplan:2011vf}
explores the $m_a > 15 \gev$ regime. It focuses on the substructure of
fat-jets containing an entire boosted Higgs decay, and that could be
2-, 3-, or 4-pronged. As before, Higgs production in association with
vector bosons is considered. The authors include two cases depending
on the mass of the scalar, $s$: (i) light scalar ($15 < m_s < 30
\gev$) and (ii) heavy scalar ($30 \gev < m_s < m_h/2$).  In the
lighter regime, the $h\to ss \rightarrow jjjj$ signature with 100\%
branching ratio can be observed at a significance of $3\sigma$ with
$100 \fb$ of 14 TeV LHC luminosity, while for the heavy scalar case,
the significance is too small to observe with the same amount data.
For a more realistic $\Br(h\to2a\to4j) = 10\%$, $2 \sigma$ exclusion
for the light scalar case requires $1500 \fb$.  (Note the achievable
limits become much stronger for $h\to 4b$ with b-tags, see
\S\ref{sec:hto4b}.)

\subsection{Existing Experimental Searches and Limits}
There are currently no existing experimental searches looking for a
four-jet resonance in the low invariant mass region, which is
understandable due to the large QCD background. Neither are there any
existing searches that look for fat-jet resonances.

Overall, this is a highly challenging exotic Higgs decay channel. For $m_a \lesssim 5 \gev$, $2 \sigma$ exclusion of $\Br(h\rightarrow 2a \rightarrow 4j) = 10\%$ requires $300 \fb$ of
LHC14 data, while $m_a \gtrsim 5 \gev$ requires more than $1000
\fb$. This search should be undertaken at the 14 TeV LHC (especially
for light $m_a$, where the decay is particularly motivated), but it is
not plausibly part of the LHC7 or 8 physics program. 

\section{\secsize $\secbold{h \rightarrow 2 \lowerGreekBold\gamma 2 j}$}\label{sec:2gamma2jet}

A relatively clean exotic decay mode of the Higgs boson is
$h\to 2\gamma2j$~\cite{Martin:2007dx}.
The SM rate for this signature is negligible:
decays into $2\gamma2q$ are highly Yukawa suppressed while
the $2\gamma2g$ process is loop induced.
However, going beyond the SM, more possibilities arise.  In particular, here we
consider Higgs boson decays to two scalars $ss^{(\prime)}$
which subsequently decay into photons and gluons or quarks.
Below we outline some possible theoretical scenarios leading to such decays and briefly
discuss their collider phenomenology.

\subsection{Theoretical Motivation}
There are several ways in which a SM singlet scalar decays to photons, gluons or quarks.
For example, it can do so via mixing with the Higgs boson, as in the singlet extensions
discussed in \S\ref{SMS} and \S\ref{2HDMS}.
This will generally give a very suppressed rate to
photons compared with that of quarks or gluons, due to the electromagnetic loop factor.

Alternatively, a singlet scalar $s$ may couple to gluons and photons via a dimension-5
operator $sF^{\mu\nu}F_{\mu\nu}$, which arises by introducing new colored
and charged vectorlike states and coupling them to $s$.
Such scenario can easily accommodate larger or even dominant
$s\to2\gamma$ branching ratios, depending on the color vs. electric charge
assignments of the new states.
As a simple example, consider adding new heavy Dirac fermions $\psi_i$ 
along with Yukawa couplings of the form $\lambda_is\bar\psi_i\psi_i$.
The fermions reside in a representation $R_i$ under $SU(3)_C$, have
electric charge $Q_i$ and mass $m_i$.
The scalar $s$ then decays to gluons and photons via heavy fermion loops.
The resulting branching ratios satisfy
\beq
   \rho=\frac{\BR (s\to 2\gamma)}{\BR (s\to 2g)}
   =\frac{1}{8}\lt(\frac{\alpha}{\alpha_s}\rt)^2
    \lt[\frac{\sum\lambda_i\ Q_i^2N(R_i)/m_i}
    {\sum\lambda_i\ C(R_i)/m_i}\rt]^2,
    \label{eq:yyjj_ratio}
\eeq
where $N(R_i)$ and $C(R_i)$ are the dimension and normalization factor of the
representation $R_i$
(the normalization factors of the lowest lying color representations $R=3,6,8$
are $C=1/2, 5/2, 3$).
For example, one heavy down-type quark $b'$ and one heavy charged lepton $\tau'$
(a combination which appears in a single `5' multiplet of $SU(5)$, along with a heavy neutrino),
with masses $m_2$ and $m_3$, and Yukawa couplings $\lambda_2$ and $\lambda_3$,
respectively, would result in
\beq
   \rho=\frac{1}{18}\lt(\frac{\alpha}{\alpha_s}\rt)^2
   \lt(1+3\frac{\lambda_2}{\lambda_3}\frac{m_3}{m_2}\rt)^2
   \simeq0.02\lt(\frac{\lambda_2}{\lambda_3}\rt)^2
   \lt(\frac{m_3}{30\tev}\rt)^2
   \lt(\frac{10\tev}{m_2}\rt)^2.
\eeq
Note that the heavy fermions need not be light in order to induce $2\gamma$ or $2g$ decays,
as long as the singlet $s$ does not mix with the Higgs boson.

In principle, the $4\gamma$ mode (\secref{4gamma}) is much cleaner than $2\gamma2j$,
which is in turn much cleaner than the very difficult $4j$ (\secref{h4j}).
However, since
\beq
   \frac{\BR (h\to 4\gamma)}{\BR (h\to 2\gamma 2g)}\simeq
   \frac{1}{4}\frac{\BR (h\to 2\gamma 2g)}{\BR (h\to 4g)}\simeq
   \frac{1}{2}\frac{\BR (s\to 2\gamma)}{\BR (s\to 2g)}
   =\frac{\rho}{2},
\eeq
for small enough values of $\rho$, as defined in \Eqref{yyjj_ratio}, the $4\gamma$
rate would be too small to be observable for a given integrated luminosity.
In such a situation, which occurs if $b'$ and $\tau'$ are degenerate in mass and couplings,
the $2\gamma2j$ signature may be competitive with $4\gamma$.

Of course, the model described above is just one example of $h\to2\gamma2g$ decays.
Other examples may feature two different states, $s$ and $s'$, allowing for even
more model-building freedom, or decays to quarks instead gluons.
Since the main focus of this section is to explore the $2\gamma2j$ signature and propose
ways to discover it at the LHC, we content ourselves with the model described
above and continue to discuss discovery reach and limits.

\subsection{Existing Collider Studies}

In~\cite{Martin:2007dx}, a search has been proposed for this channel, and the discovery
($5\sigma$) reach at the 14 TeV LHC with $300 \ifb$ was derived as function of the scalar mass $m_s$ and Higgs mass $m_h$. Gluon fusion (ggF) and $W$-associated production ($Wh$) were considered. 
Here we only make use of the latter, both because it provides superior sensitivity in this analysis and because the ggF study, which was conducted before the LHC came online, incorporated di-photon $p_T$ thresholds which are much lower than current triggers. 

The $Wh$ analysis in~\cite{Martin:2007dx} proceeds as follows:
events are required to contain one lepton, two photons
and two jets with $p_T>20$ GeV and $|\eta|<2.5$ for each of these objects.
Moreover, each object pair ($jj,\gamma\gamma,j\gamma,j\ell,\ell\gamma$) is subject
to an angular isolation criterion of $\Delta R>0.4$.
The events are also required to have $\met>20$ GeV.
Additional cuts made were $\Delta\phi_{\gamma\gamma}<1.5$,
$\Delta\phi_{jj}<1.3$, and $\lt|m_{jj}-m_{\gamma\gamma}\rt|\leq15$ GeV.
The Higgs mass resolution was assumed to be $\sim8-10$ GeV. 
The signal efficiency is claimed to be between $3\%$ and $15\%$ within the relevant mass range.

\begin{figure}[t]
\begin{center}
\includegraphics[width=0.47\textwidth]{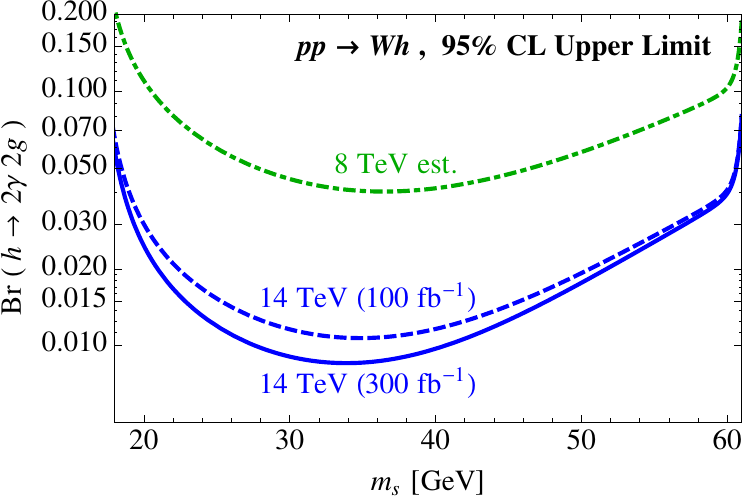}
\caption{\small Projected 95\% CL limits on the branching fraction for $\hsm\to2\gamma2g$
in associated production ($W\hsm$),
as function of $m_s$.
The blue curves refer to $300\ifb$ (solid) and $100\ifb$ (dashed), both at the 14 TeV
LHC.  The dashed-dotted green curve shows a conservative estimate of the sensitivity
for $20\ifb$ at 8 TeV.
All three limits build on the proposed search in~\cite{Martin:2007dx}
($300\ifb$ at 14~TeV LHC), by scaling background with luminosity but not changing its cross section, while signal is rescaled according to both luminosity and cross section. This underestimates the achievable 8 TeV limit. See text for more details.
\label{fig:yygg_estimate}}
\end{center}
\end{figure}

Rescaling the $5 \sigma$ limit at 14 TeV with $300 \ifb$ to 95\% CL yields the sensitivity shown as the solid blue curve in Fig.~\ref{fig:yygg_estimate}. An estimate for the lower luminosity\footnote{Our rescalings include the assumed 10\% systematic errors on the background rate~\cite{Martin:2007dx}.} of $100 \ifb$ is shown as the blue dashed curve. At the 14 TeV LHC, a sensitivity to $\Br(h\to2\gamma 2j)$ below $0.01$ is possible for part of the kinematically allowed $s$ mass range. 
This study can also be used to obtain a \emph{conservative} estimate of the sensitivity
at the 8 TeV LHC.
We scale the production cross section down appropriately without doing so for the
background cross section.
This will underestimate the strength of the limit (assuming the efficiencies do not change by a large amount at 8 vs. 14 TeV). The resulting 95\% CL sensitivity is shown as the green dash-dotted curve in Fig.~\ref{fig:yygg_estimate}. Run I data should be able to set a limit on $\Br(h\to2a\to2\gamma2j)$ as low as $\sim 0.04$ for some scalar masses, and likely better than that, given our pessimistic rescaling.

Two comments are in order:
\ben
\item
Note that the angular isolation cuts reduce the background, but effectively eliminate sensitivity for $m_s \lesssim 20 \gev$. This weakness
of the proposed search might be remedied by means of jet substructure-inspired
techniques~\cite{Ellis:2012sd,Ellis:2012zp} (see also \S\ref{sec:4gamma}).
\item
Since the best limits seem to be given by  associated $Wh$ production, we do not expect too much difficulty
with triggering.  However, since the threshold for the single lepton trigger will be raised for LHC14, it would be helpful to have a trigger that requires a lepton {\it and} a photon.
\een

\subsection{Existing Experimental Searches and Limits}
There are no limits from existing searches.
Potentially relevant searches, such as supersymmetry searches and
isolated photon-pair searches~\cite{Aad:2012tba,Chatrchyan:2011qt}
are generally insensitive to $h\to 2\gamma 2j$, since (a) they employ
relatively hard cuts and (b) without a cut on the total invariant mass,
the QCD background is overwhelming.

The $h\to2\gamma$ search in the VBF mode also cannot be used to place
limits on $2\gamma 2j$, since the VBF dijet tag is targeted at the forward
and high rapidity gap region where a $2\gamma 2j$ signal is faint.

\subsection{Proposals for Future Searches}
Based on the results from \cite{Martin:2007dx}, both the gluon fusion and the $Vh$ associated production mode should be explored for $h\to2j2\gamma$ sensitivity at LHC Run I and II.

An interesting issue arises for very light intermediate resonances, which may result in
unexpected signatures, as follows. As mentioned above, the previous search strategy involved an isolation cut on the
photons. This spoils sensitivity for light $s$ particles, since these would decay
to a collimated pair of photons or gluons.  One would therefore be missing an important portion of
parameter space below $m_s\sim20$~GeV.
Using more sophisticated photon identification inspired by jet substructure techniques will improve the situation.
However, for low enough $m_s\lesssim \gev$ , the two jets cannot be resolved, resulting in a
$j+2\gamma$ signature.

Furthermore, in~\cite{Draper:2012xt} it was shown that for very low $m_s \lesssim100$~MeV
the diphoton system is so collimated that a substantial fraction of the photon pairs would deposit their energy in a single electromagnetic calorimeter cell,\footnote{The study in~\cite{Draper:2012xt}
was geared toward the ATLAS detector, but similar principles may be applied to CMS as well.}
resulting in $h\to4\gamma$ mimicking $2\gamma$ and $3\gamma$ signatures.
While a scalar as light as to induce merged photons is generally not able to decay into gluons
(namely, hadrons), having two different states with different masses may allow for merging both
photons and gluons, resulting in signatures such as $2j+\gamma$ or $j+\gamma$.

It is therefore interesting to consider such topologies, although
they are considered ``impossible'' for Higgs
decays due to the ``wrong'' quantum numbers they seemingly possess.
These subtleties should be taken into account when conducting a future $2\gamma2j$ search.
At the trigger level the two merged photons could pass as one single photon, necessitating the use of a single photon (possibly + jets) trigger.

\section{\secsize $\secbold{h \rightarrow 4\lowerGreekBold\gamma}$}
\label{sec:4gamma}
Here, we consider the decay of a Higgs to four photons.  
In the SM, the branching fraction for this decay is negligible, as it results from a dimension-nine operator and contains an additional factor of $\alpha$ in the amplitude relative to $h\to\gamma\gamma$.  However, it can be important in certain new physics scenarios, as we now discuss.
\subsection{Theoretical Motivation}
The basic decay chain that we consider is $h\to aa^{(\prime)}$, $a^{(\prime)}\to\gamma\gamma$.  Enumerating the possible quantum numbers of the intermediate particles is simple if they decay into two photons and have spin less than two: they must be neutral and spin-0 by the Landau-Yang theorem~\cite{Landau:1948kw,Yang:1950rg}.  The CP phase of the $a^{(\prime)}$ makes no difference phenomenologically as long as the photon polarizations are not measured.

There are a number of theoretically well-motivated candidates for $a$, among them the lighter pseudoscalar of the NMSSM, any pseudoscalar that mixes with the CP-odd Higgses of the (N)MSSM, or a generic SM-singlet boson whose coupling to photons is mediated by a renormalizable coupling to heavy vector-like matter. In the first two cases, the coupling of $a$ to light SM fermions can make the branching for $a\to\gamma\gamma$ subdominant, but the low backgrounds in $4\gamma$ can nonetheless make it an interesting final state. On the other hand, if $a$ couples at the renormalizable level only to the Higgs and to 
heavy vector-like uncolored matter, it may {\it only} be able to decay to $\gamma\gamma$, rendering the $4\gamma$ 
final state extremely important. If, alternatively, the vector-like matter is colored and $a\to gg$ is allowed, 
$h\to gg\gamma\gamma$ can also be important (see \S\ref{sec:2gamma2jet} for details).

It is also worth noting that 
if $m_a < 2 m_\mu$, only the $\gamma\gamma$ and $e^+e^-$ final states may be kinematically allowed. 
The other final states in this case, $4e$ or $2e2\gamma$, are broadly similar phenomenologically to $4\gamma$, since they involve electromagnetically interacting particles.  We do not discuss them further here, leaving a detailed study for 
the future~\cite{MS}. Furthermore, as we show below, for $m_a \lesssim 100~\mev$ with $a$ decaying only to photons, $a$ is typically long-lived on collider scales, potentially leading to displaced vertices or missing energy.  Long-lifetimes are also possible in certain hidden valley models, even for much larger masses \cite{Strassler:2006im,Juknevich:2009gg}.
\subsection{Existing Collider Studies}    
The $h\to aa\to4\gamma$ decay chain was studied in~\cite{Dobrescu:2000jt}, focusing on 
the Tevatron. In this paper, it was pointed out that for $m_a\lesssim0.025\, m_h$ the $a$'s 
are boosted enough that photons coming from their decays are collimated to the extent that they 
will often deposit their energy in a single calorimeter cell, fail isolation cuts and potentially be 
reconstructed as a single photon.
(We discuss some of the experimental issues regarding closely-spaced photons below, focusing on the LHC.)
This light $a$ scenario is motivated if, e.g., $a$ is the lightest pseudoscalar in the R-symmetric limit of the NMSSM (see \S\ref{NMSSMscalar}).  The results of the analysis of~\cite{Dobrescu:2000jt} imply that the full Tevatron dataset is sensitive to branchings of $h\to aa$ at about the 0.5\% level or larger, assuming $\Br\left(a\to\gamma\gamma\right)=1$. 

In~\cite{Chang:2006bw}, a detailed study was performed of the $h\to aa\to4\gamma$ decay at the LHC with $\sqrt s=14$~TeV. The experimental cuts made in this study were that the transverse momenta of the photons were all greater than 20~GeV, the distance between the photons was $\Delta R>0.4$, the photons had rapidity $\left|\eta\right|<2.5$, and there were two separate pairs of photons that reconstructed the same invariant mass (the candidate $a$ mass) to within 5~GeV. Finding backgrounds to be negligible with these cuts, this work indicated that for a Higgs at 125~GeV, 300~fb$^{-1}$ of data at the 14~TeV LHC would allow branchings $\Br\left(h\to aa\right)\simeq5\times10^{-5}$ to be discovered at the $5\sigma$ level for $10~{\rm GeV}\lesssim m_a\lesssim m_h/2$, assuming that the $a$'s decay promptly to photons only. Rescaling this to 100~fb$^{-1}$ would indicate that $\Br\left(h\to aa\right)\simeq9\times10^{-5}$ could be found at $5\sigma$. The isolation cut of $\Delta R>0.4$ is the reason for the lower bound on the $a$ mass that can be accessed. A naive rescaling by the decreased luminosity and Higgs production cross section of the 7 and 8~TeV datasets, assuming that the dominant backgrounds' cross sections do not change appreciably, implies that the current data is sensitive to $\Br\left(h\to aa\right)\sim{\rm few}\times10^{-4}$. As emphasized in~\cite{Chang:2006bw}, the reach is extremely sensitive to the value of the photon $p_T$ cut, especially in the case of a relatively light Higgs with $m_\hsm=125$~GeV.

Closely-spaced pairs of photons in $h\to aa\to4\gamma$ at the LHC when $m_a\ll m_\hsm=125$~GeV were studied recently in~\cite{Draper:2012xt}, 
motivated by early hints at $\sqrt s=7$,~8~TeV that the Higgs rate to diphotons could be larger than in the SM. However, photon pairs that fail
mutual isolation criteria might or might not be detected as a single photon depending on the details of their geometric distribution, as we now explain in detail.    

As mentioned above, it was noted in~\cite{Dobrescu:2000jt} that at the Tevatron, for sufficiently small $m_a$, the pairs of photons from each $a$ decay could be collimated enough 
to appear as a single photon in the detector. If $m_a\lsim 10$~GeV with the $a$'s produced in the decay of a 125~GeV Higgs, the photons that they decay into
will fail the typical isolation cut of $\Delta R=0.4$. However, their energy depositions in the ECAL will normally be broader than that of a true single photon (whose electromagnetic shower has a typical width 
that is material-dependent,
called its Moli\`{e}re radius) and will not be tagged as a single photon. As the mass of the $a$ is pushed down further, the decay photons do eventually become merged enough that their energy depositions are no longer much broader than a single photon's. The value of $m_a$ where this becomes important depends on the spatial resolution of the ECAL in question. The increased granularity of the LHC detectors compared to those at the Tevatron means that $m_a$ must be smaller at the LHC than at the Tevatron for this to be the case. At ATLAS, a single photon's electromagnetic shower deposits its energy in several neighboring cells in the innermost central portion of the ECAL where the cells have a width in the $\eta$ direction of 0.0031 (corresponding to $\sim 0.5$~cm) because the Moli\`{e}re radius of the absorbing material, lead, is ${\cal O}$(cm)~\cite{ATL-PHYS-PUB-2011-007}. In Ref.~\cite{Draper:2012xt}, it was found that requiring $\Delta\eta<0.0015$ (half the smallest cell size at ATLAS) between the two nearby photons from an $a$ decay successfully reproduced the shower shape cuts used to distinguish single photons. For the photons to be this closely separated, $m_a\lesssim100$~MeV.\footnote{This critical value of $m_a$ makes sense since the LHC detectors were designed to be able to tell neutral pions apart from single photons.} In such a case, an apparent increase of $\sim$50\% in the apparent $\hsm\to2\gamma$ rate could be achieved for $\Br\left(h\to aa\right)\simeq10^{-3}$ to $10^{-2}$. Other possible experimental consequences of this scenario mentioned in~\cite{Draper:2012xt} are an increase in the number of events containing a converted photon, a mismatch between the momentum of charged tracks and the energy deposition in the calorimeter in conversions (when one of the two nearby photons converts), or the appearance of apparent $\hsm\to$``$\gamma+j$" events when one pair of photons is very collimated, faking a single photon, while the other is broader, failing isolation requirements for photons and looking like a jet (with large electromagnetic content).

Additionally, the usefulness of jet-substructure-motivated detector variables in distinguishing closely-separated photons (termed photon-jets generically in~\cite{Toro:2012sv}) from single photons and their interplay in $h\to4\gamma$ faking $h\to\gamma\gamma$  at the LHC was studied in detail in~\cite{Ellis:2012sd,Ellis:2012zp}, dealing with both the case where the photons were merged enough to potentially fake a single photon and that in which they are less closely merged but do still fail isolation cuts, potentially looking like a jet. 
Examining $h\to 4\gamma$ with a Higgs mass of 120~GeV, they determined that the use of such variables could decrease the rate of photon-jets faking single photons by a factor of over 10 while preserving at least 80\% of the single photon signal. 

Most of the literature assumes that the photon pairs necessarily reconstruct two equal-mass resonances, however this will not 
be the case when two different particles $a$ and $a'$ are introduced and the 
decay mode $h \to a a'$ is allowed. For an example of such a model which assumes $m_a \approx m_h $ and $m_{a'} \lsim $~GeV, which was originally 
designed to increase an observed $h \to \gamma \gamma $ rate, see~\cite{Batell:2012mj}.  In general, there are no direct 
constraints on $m_a, \ m_{a'}$.       

We pause here to note that if $a$ or $a'$ is light, it is quite natural to get a decay length that is detector-scale.  For example, parametrizing
the coupling of a pseudoscalar to photons as 
\beq\label{eq:affdual}
{\cal L} = \frac{\pi \alpha }{M} a F_{\mu \nu} \tilde F^{\mu \nu } 
\eeq
one gets a decay length, if they are produced in the decay of $\hsm$ at rest, of 
\beq
\gamma c\tau\simeq 0.75~{\rm cm}\left(\frac{M}{5~\rm TeV}\right)^2\left(\frac{1~\rm GeV}{m_a}\right)^{4}\left(\frac{m_h}{125~\rm GeV}\right).
\eeq
It is easy to see that for $m_a\lesssim 100~\mev$ and $M\gtrsim 1~\tev$,\footnote{We would expect such a scale if $a$'s coupling to photons came from integrating out charged matter above the electroweak scale.} $a$'s decay length could be of the order of several meters. Long decay lengths are therefore a generic feature of light pseudoscalars decaying to photons and should be kept in mind when contemplating such signals.\footnote{This conclusion can be modified slightly when other decay channels for $a$ are present or if the operator $a F_{\mu \nu} \tilde F^{\mu \nu }$ is generated below the electroweak scale. See~\cite{Draper:2012xt} for details.}
\subsection{Existing Experimental Searches and Limits}
A search for $h\to aa\to4\gamma$ in the case where $m_a\ll m_\hsm$ leading to very collimated pairs of photons was performed by ATLAS on 4.9~fb$^{-1}$ of 7~TeV data~\cite{ATLAS4gamma}. The search was very similar to the standard one for $h\to\gamma\gamma$ but shower shape variable cuts were relaxed to allow for increased acceptance of the $4\gamma$ signal. This resulted in a very good acceptance for events coming from the $h\to\gamma\gamma$ channel. Results were presented for $m_a=100$, 200, 400~MeV, limiting $\Br\left(h\to aa\right)\Br\left(a\to\gamma\gamma\right)^2\lesssim 0.01$ at $m_h=125$~GeV.\footnote{In the SM $\Br\left(h\to\gamma\gamma\right)\sim 2\times 10^{-3}$. Therefore the impact of the SM diphoton channel on this bound is still rather small.} For larger $a$ masses, there are no limits from collider searches.

Results from low energy experiments (see, e.g. Ref.~\cite{Hewett:2012ns}) are not constraining on this scenario for $m_a\gtrsim10$~MeV so long as the $a$'s decay promptly at the LHC~\cite{Draper:2012xt}.

\subsection{Proposals for New Searches at the LHC}
A search for $h\to4\gamma$ using the full 7 and 8 TeV dataset of both experiments would be highly desirable. Reference~\cite{Chang:2006bw} indicates that 300~fb$^{-1}$ of data at the 14 TeV LHC can access values of $\Br\left(h\to aa\right)\Br\left(a\to\gamma\gamma\right)^2>5\times10^{-5}$ at $5\sigma$ for $m_a\gtrsim 10$~GeV. For $m_a\lesssim 10~\gev$, the $4\gamma$ signal can be hard to disentangle from the large QCD dijet background and for $m_a\lesssim{\rm few}\times 100~\mev$ it can even look very similar to $h\to\gamma\gamma$. In these cases, as shown in~\cite{Ellis:2012sd,Ellis:2012zp}, using detector variables from jet substructure can greatly reduce the QCD dijet backgrounds and help to distinguish these final states, greatly increasing the reach for $h\to4\gamma$. Thus far, most work on this signal has concentrated on either the very light $a$ regime where two photon pairs are very collimated or where $m_a>10~\gev$ and the four photons are well separated. The intermediate mass region is also well motivated and we encourage it to be studied as well.

The assumption that the two intermediate particles have the same mass cuts down on backgrounds but a more general search strategy looking for $\gamma\gamma$ bumps in $h\to4\gamma$ could help to shed light on a scenario where this decay is dominantly mediated by two particles with distinct masses. 

Lastly, macroscopic decay lengths for the particles mediating $h\to aa^{(\prime)}\to4\gamma$ can be naturally realized in simple models, especially when they are light or if they are composites from a hidden valley, which motivates searches for $4\gamma$ events where two pairs of photons each resolve displaced vertices.

	\section{\secsize $\secbold{h\to Z Z_D,  Z a \to 4\lowerGreekBold\ell}$}
\label{sec:htoZa}

Below we discuss decays of the form $h\to Z+X$, where $X$ denotes a
non-SM light boson.  We focus on two possibilities:
\begin{enumerate}
\item $X=Z_D$, a new gauge boson that acquires a mass and mixes with
  the SM gauge bosons, see \S\ref{subsec:SMvector}.
\item $X=a$, a light pseudoscalar as in the 2HDM+S and the NMSSM~\cite{Christensen:2013dra}, see \S\ref{2HDMS}, \S\ref{NMSSMscalar}.
\end{enumerate}
In both cases we are interested in a two-body decay of the Higgs
boson, meaning we require $M_{X}\lesssim34$~GeV.  We outline the
theoretical motivation to consider such decays and discuss the limits
by LEP, Tevatron, and LHC.

\subsection{Theoretical Motivation}

\subsubsection{$h\to ZZ_D$}

As discussed in~\S\ref{subsec:SMvector}, many theories feature a
hidden $U(1)$ sector with small kinetic or mass mixing the the SM
photon and $Z$-boson. This possibility often arises in connection to
dark matter, but similar phenomenology can also arise in more general hidden valley
models, see~\S\ref{sec:hiddenvalley}. The minimal setup
\Eqref{SMVkinmix} to generate $h \rightarrow Z Z_D$ decay involves a
kinetic mixing term between the hypercharge gauge boson and the dark
$U(1)$ gauge boson
\begin{equation}
\label{eq:SMVkinmixshort}
\mathcal{L}_\mathrm{gauge} \supset  \frac{1}{2} \frac{\epsilon}{\cos \theta_W} \hat B_{\mu \nu} \hat Z_D^{\mu \nu},
\end{equation}
where hatted quantities are fields before their kinetic terms are
canonically renormalized by a shift of $B_\mu$. In the canonical
basis, SM matter has a dark milli-charge and there is mass mixing
between the SM $Z$-boson and $Z_D$. The dominantly dark vector mass
eigenstate has photon-like couplings to SM fermions (proportional to
the small mixing $\epsilon$) up to $\mathcal{O}(m_{Z_D}^2/m_Z^2)$
corrections, see \Eqref{SMVcoupling}. If $Z_D$ is the lightest state
in the dark sector it will decay to SM fermions via this
coupling. Prompt decay requires $\epsilon \gtrsim 10^{-5} - 10^{-3}$ 
(depending on $m_{Z_D}$), and the largest
$\Br(h\rightarrow Z Z_D)$ allowed from indirect constraints is $\sim
10^{-3}$, see \Figref{SMVZdbounds}.

It is also possible to have pure mass mixing after EWSB via operators
of the form $h Z^\mu Z^\prime_\mu$, but in this case additional
constraints from parity violating interactions and rare meson decays
apply, see
\cite{Davoudiasl:2012ag,Davoudiasl:2012ig,Davoudiasl:2013aya}. Generically,
new physics similar to that which generates kinetic mixing may also
generate dimension-6 terms of the form $H^\dagger
HB^{\mu\nu}Z_{D\mu\nu}/\Lambda^2$. Once the Higgs acquires a VEV, this
term yields the coupling in~\Eqref{SMVkinmixshort}.

\subsubsection{$h\to Za$}

Next we consider the decay $h\to Z a$. This is motivated by, for
example, the 2HDM+S or the NMSSM, where one of the CP-odd Higgs masses
can be small. The relevant interaction Lagrangian in terms of mass
eigenstates $h$ and $a$ is given by \Eqref{2HDMShzalagrangian} with an
additional Yukawa term: 
\beq 
\mathcal{L}_{int}=g (a\partial^\mu h -
h \partial^\mu a) Z_\mu - g_a \bar f i\gamma_5 f a.
\label{eq:int}
\eeq
with $g = \sqrt{(g^2+g'^2)/2} \ \sin(\alpha - \beta) \sin
\theta_a$. The parameter $\alpha$ is the mixing angle between the
doublet scalars, $\tan \beta = v_u/v_d$, and $\theta_a$ is the mixing
angle between the uneaten doublet pseudoscalar $A$ and the singlet
pseudoscalar. Since the Higgs coupling to $ZZ$ and $W^+ W^-$ is also
proportional to $\sin(\alpha - \beta)$, the SM-like rates in those
channels (as well as the diphoton mode) favor the decoupling limit
$\alpha = \pi/2 - \beta$. $\theta_a$ can be constrained by direct LEP
and Tevatron searches for the CP-odd Higgs, but the SM-like Higgs
could still have large branching fractions to
$Za$~\cite{Christensen:2013dra}. The pseudoscalar coupling to fermions
can be extracted from \tabref{2HDMcoupling},
\begin{equation}
g_a = \sin \theta_a \tan \beta \frac{m_f}{ v} \ , \ \ \mathrm{for} \ \ b , \ \tau , \ \mathrm{and} \ \mu
\end{equation}
and the overall size of $\theta_a$ does not affect its branching ratios. 

For the length of the LHC program it will likely be safe to take
$\Br(h\to Za)=10\%$ as a benchmark point. In the next section, we
discuss the experimental constraints on this mode. Depending on the
mass of this pseudoscalar, the dominant decay mode could be $b\bar
b,~\tau^+\tau^-$, or $\mu^+\mu^-~(s\bar s)$. We consider all of these
cases when proposing search strategies.

\subsection{Existing Collider Studies}

Up to different branching ratios and some angular correlations the
final states for $h \rightarrow Z Z_D$ and $h \rightarrow Z a$ are
identical. As such, collider studies and experimental searches for one
channel generally apply to both. The two relevant parameters to define
a simplified model for this channel are
\begin{equation}
m_X \ \ \ \ \mathrm{and} \ \ \ \  \Br(h \rightarrow Z X \rightarrow Z y \bar y)
\end{equation}
for $X = a, Z_D$ and $y = $ some SM particle, where the different $a,
Z_D$ branching ratios lend different importance to different choices
of $y$.

There have not been many collider studies specifically performed for
the $h \rightarrow Za$ mode.  Ref.~\cite{Christensen:2013dra} pointed
out that this channel may be very large in the context of the
NMSSM. Ref~\cite{Mahlon:2006zc,Chang:2012qu,Almarashi:2011qq}
discussed heavy non-SM-like Higgs decaying into $Za$.

More searches have been inspired by looking for a $Z_D$.  The
phenomenology of a $Z_D$ with mass mixing to the $Z$ has recently been
discussed
in~\cite{Davoudiasl:2012qa,Davoudiasl:2012ag,Davoudiasl:2012ig,Davoudiasl:2013aya}
(see also,
e.g.,~\cite{Hook:2010tw,Gopalakrishna:2008dv,Schabinger:2005ei,Babu:1997st}
for earlier work), including collider phenomenology of $h\to ZZ_D$,
$h\to \gamma Z_D$, and $h\to Z_DZ_D$ decays, as well as low energy
constraints from colliders and fixed-target experiments, $g-2$ of the
muon and electron, rare meson decays, and electroweak precision
observables (see \S\ref{sec:hto4l} for the $h\to Z_D Z_D$ mode).

In~\cite{Davoudiasl:2013aya}, the authors designed a search for
$pp\to h\to ZZ_D\to e^+e^-\mu^+\mu^-$.
The backgrounds considered are $Z(\to\ell^+\ell^-)jj$, $j$ faking
$\ell$ (probability $ \sim 0.1\%$) and leptonic $t\bar t$ (reducible),
as well as $h\to ZZ^*, Z\gamma^*, ZZ \to 4\ell$ (irreducible).  The
authors of~\cite{Davoudiasl:2013aya} assumed only mass mixing of the
form $\varepsilon_Z m_Z^2 Z^\mu Z_{D\mu}$.  For $m_{Z_D} \sim 5 - 10 \gev$, they find that the 14~TeV LHC has $2\sigma$ sensitivity to $\mathrm{Br}(h \to Z Z_D \to Z \ell \ell) \sim \mathcal{O}(1) \times 10^{-4}$ with $30 \ifb$ of luminosity. 

\subsection{Existing Experimental Searches and Limits}

A light pseudoscalar $a$ can be searched for in $\Upsilon$ decays at
Babar~\cite{Domingo:2010am}, top decay at the
Tevatron~\cite{Aaltonen:2011aj}, and direct single production and
decay to dimuons at the
LHC~\cite{ATLAS-CONF-2011-020,Chatrchyan:2012am}.  These dedicated
searches are discussed in other sections of this document, and their
reach depends on many parameters of the theory. There are also many
constraints (most of them not from high energy colliders) on the
existence of a $Z_D$, see \Figref{SMVZdbounds}, but there are large
regions of parameter space relevant for exotic Higgs decays that are
not excluded.

Our focus is the $hZX$ vertex ($X = a, Z_D$). No direct search for $h
\rightarrow Za$ or $ZZ_D$ has been performed to be best of our
knowledge, but there are several channels and other searches at LEP,
Tevatron, and LHC that are sensitive to this interaction term.

\paragraph*{\bf LEP} 

The $hZX$ vertex not only gives rise to the $h\to ZX$ decay, but also
opens the channel $e^+e^-\to Z^* \to h X$ at LEP. Related searches
include $e^+ e^- \to h a, ZZ^\prime \to 4b$~\cite{Schael:2006cr}, $
4\tau$~\cite{Schael:2006cr} and $2b 2\tau$~\cite{Schael:2006cr}.  For
$\Br(h\to Za)=10\%$, these searches are not constraining because the
cross section for $e^+e^-\to Z^* \to h a$ is at the sub-fb level. Even
without considering any branching fraction suppression to the final
states, LEP's integrated luminosity is still too small to be
sensitive.  One can also imagine more spectacular production modes
such as $e^+ e^- \to h a \to aaa \to 6b$ and $e^+ e^- \to h a \to aaa
\to 6\tau$, which can be recast into $e^+e^-\to h a \to Zaa \to 6b$
and $e^+e^-\to h a \to Zaa \to 6\tau$. These channels yield no
constraints even before taking into account kinematic acceptances.

\paragraph*{\bf Tevatron and LHC}

The most relevant existing search sensitive to $h\to ZZ_D$ and $h\to
Za$ is $h\to ZZ^* \rightarrow 4 \ell$ by CMS~\cite{CMS-PAS-HIG-13-002}
and ATLAS~\cite{ATLAS:2013nma}, where $4 \ell$ stands for electrons
and muons.  The clean $4\ell$ decay makes these existing searches very
sensitive to $ZZ_D$ or $Za$ decaying into leptons.

The leptonic $h \rightarrow ZZ^*$ searches divide the four leptons of
each event into two pairs, the ``leading'' pair (likely to have come
from an on-shell $Z$) and the ``subleading'' pair (from the off-shell
$Z^*$, denoted sometimes as ``$Z2$'' or $m_{34}$).  The subleading
dilepton mass distributions from ATLAS and CMS are shown in Fig.~23
of~\cite{ATLAS:2013nma} and Fig.~9 of \cite{CMS-PAS-HIG-13-002},
respectively, using the full $20 + 5$~$\ifb$ data set of LHC7+8.  With
this information it is easy to estimate limits on $h \rightarrow Z X$
decay.\footnote{The $\ell^+\ell^-$ distribution in $h\to Z \ell\ell$
  events can also be used to search for indirect effects of new physics above
 the Higgs mass \cite{Grinstein:2013vsa, Isidori:2013cla}.}  The new
state $X$ will contribute to $h\to Z\ell\ell$ events in two ways,
firstly through resonant $h\to Z X$ production, and secondarily through
interference with the SM amplitude $h\to Z Z^*$.  Here we consider
only resonant production, obtaining a conservative estimate on
Br($h\to Z X)$; a study incorporating the off-shell contributions will
appear in future work.

A $Z_D$ or $a$ decaying through some small mixing to SM particles will
have a much smaller width than $\Gamma_Z \approx 2.6$~GeV or
$\Gamma_{h_\mathrm{SM}} \approx 4.07$~MeV.  Given the $\lesssim 3\%$
dilepton mass resolution of the experiments and the subleading
dilepton mass ($M_{Z2}$) binning of 1.25 (2.5) GeV by CMS (ATLAS) it
is safe to assume that all of the leptonic $h \rightarrow ZX$ events
land in a single bin $M_{Z2} \approx m_X$. Defining the total expected
number of \emph{produced} $h \rightarrow Z Z^*$ events as
\begin{equation}
N_{prod}^{ZZ^*} = \sigma(p p \rightarrow h) \times L \times \Br(h \rightarrow Z Z^* \rightarrow 4 \ell)
\end{equation}
the detector efficiency for dileptons from $Z_D/a$ decay can be estimated as
\begin{equation}
\epsilon_{\ell \ell} \approx \frac{N_{detect}^{ZZ^*}}{N_{prod}^{ZZ^*}}\,,
\end{equation}
where $N_{detect}^{ZZ^*}$ is the total expected number of
\emph{detected} $h \rightarrow Z Z^*$ events as extracted from the
plots of ATLAS and CMS.\footnote{Due to the $m_{Z2} > 12 \ \gev$
  requirement this may slightly underestimate the efficiency. There
  may also be small differences in isolation for leptonic vector vs
  pseudoscalar decay. However, our method suffices for a conservative
  estimate of constraints.  } Therefore, for a given exotic Higgs
decay branching ratio, the expected number of events contributing to
the $m_{Z2}$ distribution is
\begin{eqnarray*}
N^{ZX}_{detect} &=& \epsilon_{\ell \ell}  \times \sigma(p p \rightarrow h)\times L \times \Br(h \rightarrow Z X\rightarrow 4 \ell)\\
&\approx& N_{detect}^{ZZ^*} \times  \frac{\Br(Z \rightarrow \ell \ell)}{\Br(h \rightarrow Z Z^* \rightarrow 4 \ell)}  \times \Big[ \Br(h \rightarrow ZX) \times \Br(X \rightarrow 2 \ell) \Big]\\
&\approx& N_{detect}^{ZZ^*}  \times 450 \times \Big[ \Br(h \rightarrow ZX) \times \Br(X \rightarrow 2 \ell) \Big]
\end{eqnarray*}
By placing the above number of events in each $m_{Z2}$ bin we extract
$95\%$ CL bounds on the quantity in square brackets for different $m_X
> 12 \ \gev$, see \figref{Zalimit}.

\begin{figure}[tp]
\includegraphics[width=280pt]{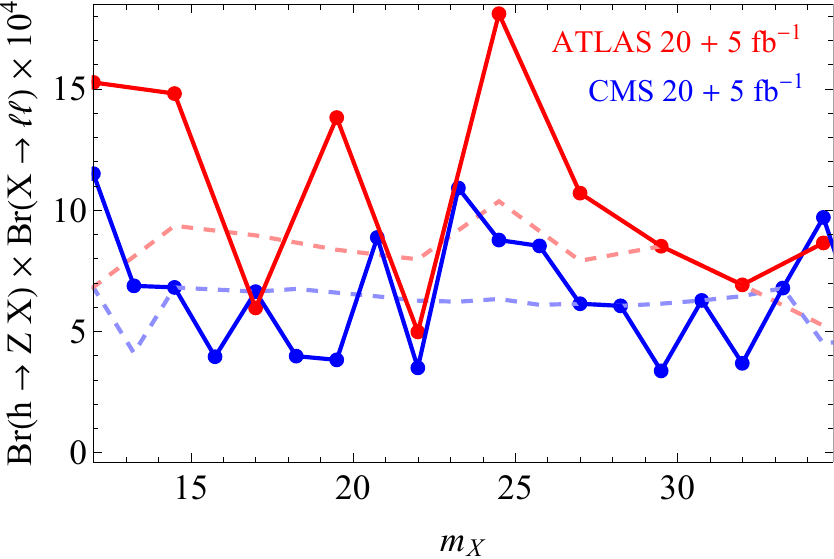}~~~~
\includegraphics[width=180pt]{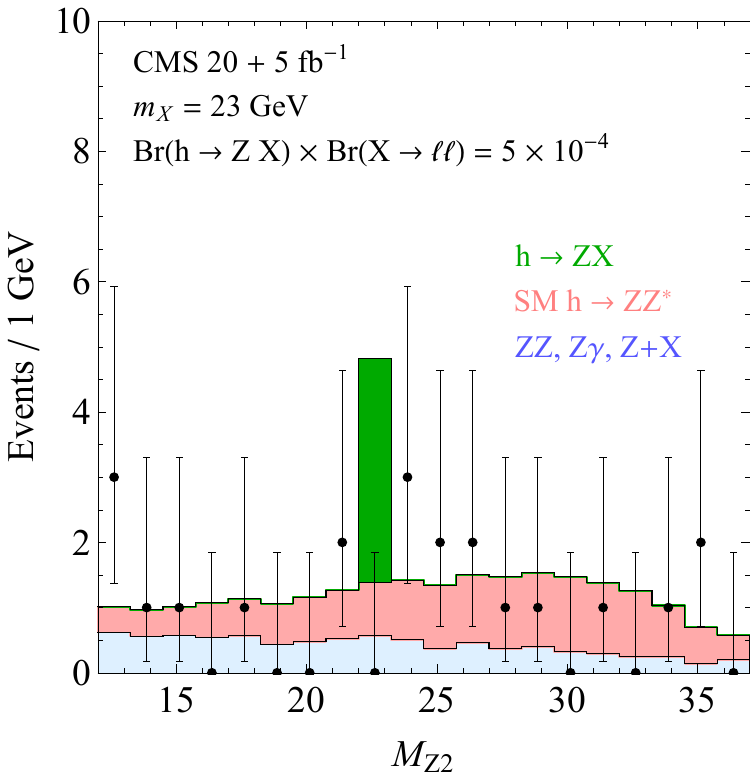}~~~~
\caption[]{{\bf Left}: 95\% C.L. exclusion limit on Br$(h\to Z X)\times$Br$( X\to \ell \ell)$ for $X = Z_D, a$, extracted from the SM $h \rightarrow 4 \ell$ searches ($\ell = e, \mu$) assuming SM Higgs production rate and $\Gamma_X \ll 1 \ \gev$. (The lighter dashed lines indicate the expected limit. The large fluctuations in the observed limit are a consequence of low statistics in each bin.)
{\bf Right}: The CMS distribution of $m_{Z2}$ from~\cite{CMS-PAS-HIG-13-002}, overlaid with a 23 GeV $h\to ZX \to 4\ell$ signal.
}
\label{fig:Zalimit}
\end{figure}

The bound on $\Br(h \rightarrow ZX) \times \Br(X \rightarrow \ell
\ell)$ is $\lesssim 10^{-4} - 10^{-3}$ for $12 \ \gev \lesssim m_X
\lesssim 34 \ \gev$ and $\ell = e, \mu$.  Using \figref{SMVBrZdark} we see
that this already corresponds to $\Br(h \rightarrow Z Z_D) \lesssim 2 \times
10^{-3}$, which represents a new direct constraint on dark photons by
the LHC, see \Figref{SMVZdbounds}. This limit can be optimized with a dedicated analysis, which would make LHC measurements the most sensitive probe of dark vector kinetic mixing in the mass range $10 \gev \lesssim m_{Z_D} \lesssim m_h/2$.

The situation is more ambiguous for pseudoscalars. Their branching
ratios are more model-dependent in general, and their Yukawa couplings
usually imply that $a \rightarrow \tau \tau$ is enormously preferred
over $e, \mu$.  Typical branching ratios to $4\ell$ ($\ell = e, \mu$)
are $10^{-4}-10^{-3}$, depending on the pseudoscalar mass.  Bounds for
$X \rightarrow \tau \tau$ could also be derived from the leptonic $h
\rightarrow Z Z^*$ searches but would be much weaker. Nevertheless
this may be the preferred discovery channel for 2HDM+S and NMSSM type
models, where $\Br(h \rightarrow Z a)$ could easily be $10\%$ and
$\Br(a \rightarrow \tau \tau)$ is generally $\mathcal{O}$(0.05--1),
see \S\ref{2HDMS}.

\subsection{Proposals for New Searches at the LHC}

For $m_{a, Z_D} > 12 \ \gev$ it seems likely that LHC14 searches
inspired by $h\rightarrow ZZ^*$ will constrain $h \rightarrow Za$ in
the $a\rightarrow 2\tau$ modes, while LHC7+8 already gives significant
\emph{direct} bounds to $h \rightarrow Z Z_D \rightarrow 4 \ell$. A
$Z$ $+$ lepton-jet search would be able to set strong limits in
particular for very light $Z_D$. Care must be taken to correctly
account for challenging quarkonium backgrounds. Identifying promising
search strategies will be the subject of future work.

	
\section{\secsize $\secbold{h \to Z_D Z_D \rightarrow 4\lowerGreekBold\ell}$}\label{sec:hto4l}

\subsection{Theoretical Motivation}
Similarly to the discussion in the previous section, two classes of
models can give a Higgs to four-lepton signature, with two pairs of
electrons and/or muons reconstructing the same resonance:

\begin{itemize}
\item As discussed in \S\ref{subsec:SMvector},
models with an
  additional $U(1)_D$ gauge group may lead to the $h\to Z_D Z_D$ decay,
  followed by $Z_D\to\ell^+\ell^-$. In the minimal model, the dark
  $U(1)_D$ is broken by a dark scalar that does not mix with the SM
  Higgs. Then the kinetic mixing operator involving the hypercharge
  gauge field $B_\mu$ and the $Z_D^\mu$ field leads to only a small
  branching ratio of the Higgs to two $Z_D$ gauge bosons, since it is
  suppressed by the fourth power of the kinetic mixing parameter
  $\epsilon$ in Eq.~(\ref{eq:SMVkinmix}). Much larger branching ratios
  can be obtained by introducing a mixing term between the scalar that
  breaks the $U(1)_D$ symmetry and the Higgs of the SM: $\zeta
  |\sing|^2|H|^2$. In these models, even $\zeta\sim 10^{-2}$ can lead
  to branching ratios for $h\to Z_D Z_D$ as large as $\sim 10\%$ in 
  certain regions of parameter space (see left panel of
  Fig.~\ref{fig:SMVdelta}). Furthermore, more extended Higgs sectors
  can also lead to sizable branching ratios. In particular,
  in~\cite{Lee:2013fda} it has been shown that $\Br(h\to Z_D Z_D)\sim
  10\%$ is possible in 2HDM+S models where the SM singlet and one of
  the two Higgs doublets is charged under $U(1)_D$.

\item Many hidden valley models \cite{Strassler:2006im, Strassler:2008bv} (see \S\ref{sec:hiddenvalley}), with either fundamental or composite spin-one bosons, can lead to the same final state.

\item Models predicting a sizable branching ratio for $h\to aa$, where
  $a$ is CP-odd scalar, can also lead to the $4\ell$ signature. As
  presented in \S\ref{2HDMS}, such pseudoscalars can arise in
  2HDM+S models, as for example in the approximately R-symmetric NMSSM
  scenarios (see \S\ref{NMSSMscalar}). However, as shown in the
  figures of \S\ref{2HDMS}, if the pseudoscalar is above the tau
  threshold, it will preferentially decay into two taus, two gluons,
  or two quarks. More specifically, for $m_a>2m_\tau$,
  $\Br(a\to\ell^+\ell^-)/\Br(a\to\tau\tau)\sim m_\ell^2/m_\tau^2\sim
  3\times 10^{-3}$ $(8\times 10^{-8})$ for $\ell = \mu$ $(e)$. For
  this reason, in the discussion of \S.~\ref{sec:ZDZDCurrent} below
  for the collider constraints on the $4\ell$ signature, we will focus
  on models with dark gauge bosons.  Searches that exploit the more
  dominant $4\tau$ and $2\tau 2\mu$ decay modes of the pseudoscalar
  pair are discussed in \S\ref{sec:hto4tau}.

\end{itemize}

\subsection{Existing Collider Studies}    
The authors of~\cite{Gopalakrishna:2008dv} investigate the feasibility
of probing $h\to Z_D Z_D\to 4\ell$ at Tevatron and at the LHC. In
particular, they perform an estimation of the reach at the 14 TeV LHC
for several benchmark scenarios: the most interesting for us are the
scenarios ``A'' and ``B'' with $m_h=120$ GeV and $m_{Z_D}=5\,(50)$
GeV, respectively. They show that there are very good prospects for
detecting this Higgs decay mode, even for small Higgs branching
ratios. In particular, they focus on a Higgs produced in gluon fusion
followed by the decay $h\to Z_D Z_D\to e^+e^-\mu^+\mu^-$. For
$\Br(h\to Z_D Z_D)\sim \mathcal O(1)$, basic cuts on the $p_T$ and
$\eta$ of the leptons, and the requirement that the 4-lepton invariant
mass is close to $m_h$, are sufficient to lead to $S/B\sim
10^4\,(10^3)$ (with $S\sim$ hundreds (tens) of fb in the case of
$m_{Z_D}=5\,(50)$ GeV).  Here $B$ is simply given by the leading
diboson background. Additionally, they comment on the fact that the
reach can be improved further by vetoing events with opposite sign,
same-flavor (OSSF) lepton pairs reconstructing the $Z$
resonance. 

Furthermore, Ref.~\cite{Martin:2011pd} shows that a light Higgs boson
could have been discovered sooner in $h\to Z_D Z_D\to 4\ell$ than in
the traditional decay modes, $\gamma\gamma$, $\tau\tau$, with the
7~TeV LHC data. In particular, the authors claim that, even for
$\Br(h\to Z_D Z_D)\sim \mathcal O(1\%)$, one could have expected 5
events with the first $\ifb$ of 7~TeV LHC data.

\subsection{Existing Experimental Searches and Limits}\label{sec:ZDZDCurrent}

Searches for $\hsm\to aa\to 4\mu$ were performed by the CMS
collaboration with 5~fb$^{-1}$ of data at $\sqrt
s=7$~TeV~\cite{Chatrchyan:2012cg} and 20~fb$^{-1}$ at $\sqrt
s=8$~TeV~\cite{CMS-PAS-HIG-13-010}. For these searches, $a$ refers to
a spin-0 boson with a mass between 250~MeV and $2m_\tau$. Differences
in the acceptance between this signal and $\hsm\to Z_D Z_D\to4\mu$
should be modest for this range of boson masses, and the limits from
these searches at CMS are directly applicable. The 8~TeV
search~\cite{CMS-PAS-HIG-13-010} is more sensitive and results in a
limit $\Br\left(\hsm\to Z_D Z_D\to4\mu\right)<4.7\times10^{-5}$ for
$m_\hsm=125~\gev$ and $250~\mev<m_{Z_D}<2m_\tau$.

For the mass range $5\gev < m_{Z_D} < m_h/2$, limits can be obtained from SM Higgs searches as well as from a plot
reported as part of a $ZZ$ cross section measurement.  To estimate
limits on exotic Higgs decays to four leptons, we use MadGraph to
generate Higgs decays to dark photons, $h\to Z_D Z_D$, followed by
$Z_D\to\ell^+\ell^-$, using FeynRules~\cite{Christensen:2008py} to
construct the dark photon model of \S\ref{subsec:SMvector}.
Gluon fusion signal events are generated in MadGraph 5 and matched up
to one jet, with showering in Pythia.

We begin by considering the SM $h\to Z Z^*$ analyses, which are
conducted with the full 7+8 TeV datasets in both experiments.  The CMS
search \cite{CMS-PAS-HIG-13-002} requires four isolated leptons within
kinematic acceptance, forming two OSSF pairs.  The invariant mass of
the OSSF pair that minimizes $|m_{\ell\ell} - m_Z|$ is denoted $m_1$,
while the remaining OSSF pair invariant mass is denoted $m_2$.  The
pair invariant masses must satisfy
\beq
40\,\gev < m_1 < 120\,\gev,  \phantom {space or} 12\,\gev < m_2 < 120 \,\gev.
\eeq
Events in which any OSSF pair has invariant mass $m_{\ell\ell}<4$ GeV
are rejected, to suppress backgrounds from quarkonia.  To compare to
public data, we study the set of four-lepton events with four-lepton
invariant mass in the range $m_{4\ell}\in (121.5, 130.5)$ GeV.

We estimate signal acceptance using the lepton efficiencies reported
in~\cite{CMS-PAS-HIG-13-002}.  Lepton energies are smeared according
to the resolutions tabulated in the Appendix of that work.  Comparing
our own event yield from SM $h\to Z Z^*\to 4\ell$ events to the
experimental expectations in Table 2 of~\cite{CMS-PAS-HIG-13-002}
determines a final efficiency correction factor for electrons and
muons separately.

The requirement that one OSSF pair of leptons lies within a $Z$ window
means that frequently $h\to Z_D Z_D$ events are not reconstructed as a
pair of resonances: if $m_{Z_D} = 20$~GeV, for instance, a lepton pair
with invariant mass near $m_Z$ can only be obtained by taking one
lepton from each $Z_D$ decay. Since events with two electrons and two
muons cannot be mispaired in this way, for $m_{Z_D} <40$ GeV,
$ee\mu\mu$ events cannot contribute to the reach at all.  In
Fig.~\ref{fig:VdVd_CMS_M1M2} we show the signal $4e$ and $4\mu$ events
as they would appear in the $m_1$-$m_2$ plane, both for $m_{Z_D}=20$
GeV and $m_{Z_D}=40$ GeV.  As $m_{Z_D}$ increases, the fraction of
events which are reconstructed as a pair of resonances increases, so
that when $m_{Z_D}= 60$ GeV, nearly all leptons are correctly paired.

\begin{figure}[t]
\begin{center}
\includegraphics[width=0.42\textwidth]{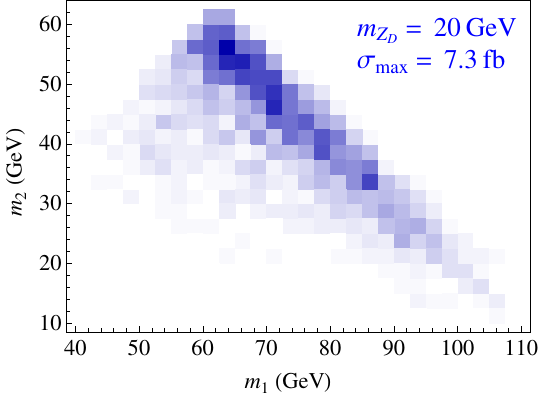}
\includegraphics[width=0.42\textwidth]{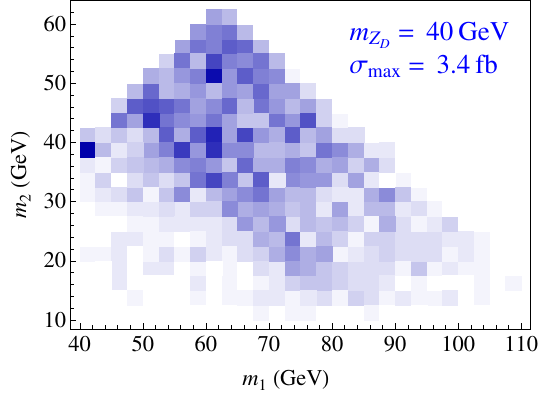}\\
\includegraphics[width=0.40\textwidth]{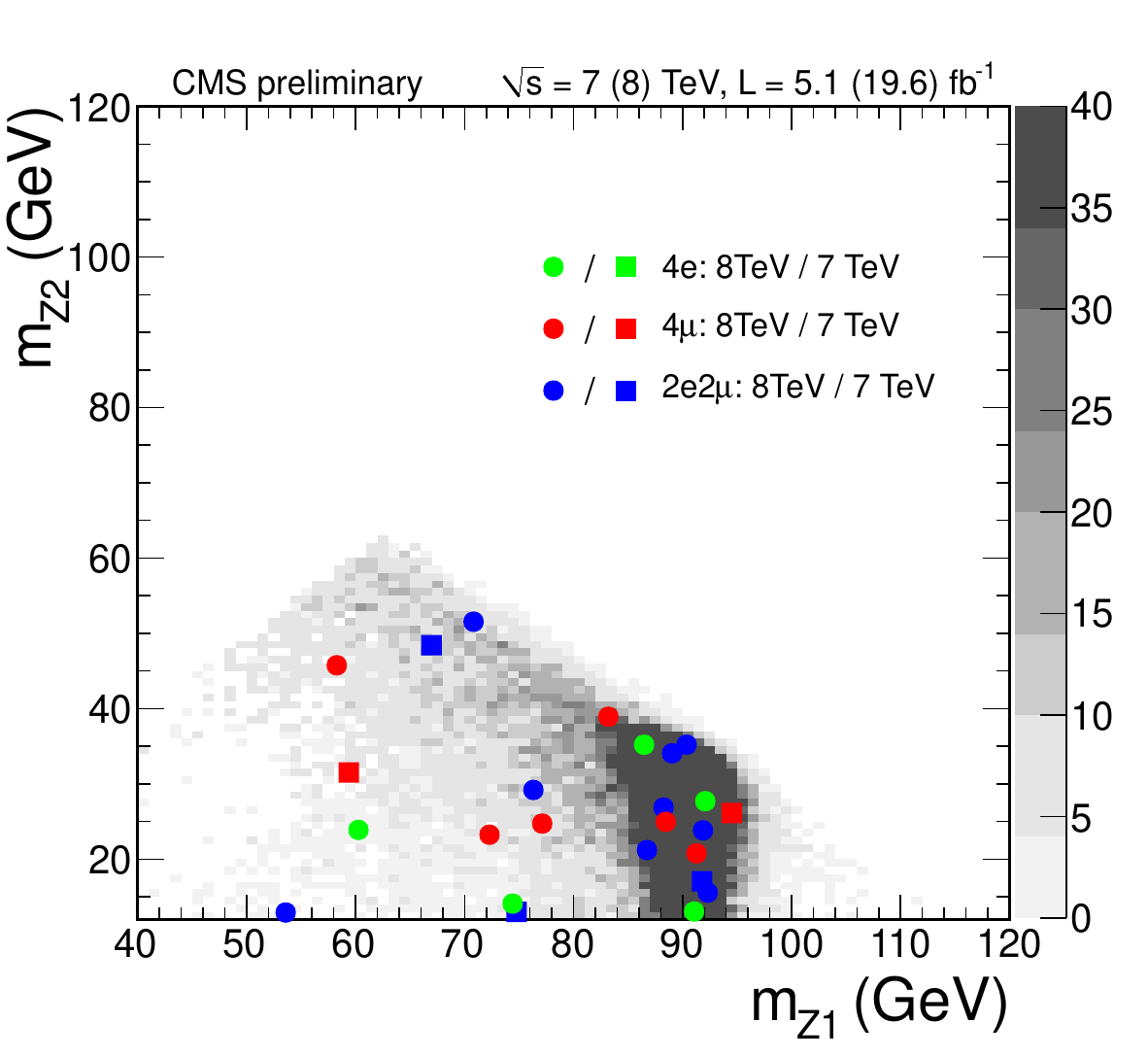}
\end{center}
\caption{{\small {\bf Top left and right:} distribution of lepton pair
    invariant masses in $4e$ and $4\mu$ events according to the event
    selection and reconstruction criteria of
    \cite{CMS-PAS-HIG-13-002}. The maximum cross section (taking
    $\Br(h\to Z_D Z_D) = 1$) in any $2.5\times 2.5$ GeV square is
    indicated in each plot to establish a scale. {\bf Left:} with $m_{Z_D} =
    20$~GeV, only mispaired $4e$ and $4\mu$ events pass the event
    selection criteria.  {\bf Right:} with $m_{Z_D} = 40$~GeV,  both mispaired
    and correctly-paired events are evident, with accumulation of
    events at the mass of the vector boson visible on the far left
    edge of the plot.  (In this case, $2e2\mu$ events, not shown, also
    pass the selection criteria, and accumulate at the mass of the
    vector boson.)  {\bf Bottom:} Expected distribution of lepton pair
    invariant masses for $h\to ZZ^{*} \to 4\ell$ 
    with $m_{4\ell}\in (121.5, 130.5)$, overlaid with
    observed 7 and 8 TeV events, from \cite{CMS-PAS-HIG-13-002}}.
  \label{fig:VdVd_CMS_M1M2}}
\end{figure}

To estimate limits resulting from this search, we perform a simple
counting experiment.  For signals with $m_{Z_D}<40$ GeV, we define a
signal region to be $m_1<80$ GeV, $m_2 > 30$ GeV, and set a 95\% CL
limit by treating all observed events in this region as signal.  In
this signal region, there are one $4\mu$ and one $2e2\mu$ event in the
7 TeV data set, and one $4\mu$ and one $2e2\mu$ event in the 8 TeV
data set.  We consider 6 signal bins, one for each flavor combination
in each CM energy, and define a joint likelihood function as the
normalized product of Poisson likelihood functions $\mathcal{L}(\mu) =
\mbox{Poisson} (N_{obs}| \mu N_{sig})$. When no signal is predicted,
as for the $2e2\mu$ channel for masses $m_{Z_D}<40$ GeV, we do not
include the signal region in the likelihood function.  
The resulting 95\% CL limits are shown in the red line in Fig.~\ref{fig:VdVd_limits}.  For $m_{Z_D}\geq 40$ GeV, we define the
signal region to be $m_{Z_D}-5\,\gev <m_1 <m_{Z_D} +5\,\gev$,
$m_{Z_D}-5\,\gev <m_2 <m_{Z_D} +5\,\gev$.  No observed events fall
inside this signal region for any value of $m_{Z_D}$.  
To translate between limits on $h\to Z_D
Z_D$ and $h\to Z_D Z_D\to 4\ell$ we point out that, as seen in
Fig.~\ref{fig:SMVBrZdark} in \S\ref{subsec:SMvector}, for
$10~\gev\lesssim m_{Z_D}\lesssim 60~\gev$,
$\Br\left(Z_D\to\ell^+\ell^-\right)\simeq0.3$. This implies that
$\Br\left(h\to Z_DZ_D\to4\ell\right)\simeq0.09\times\Br\left(h\to
  Z_DZ_D\right)$.

\begin{figure}[t]
\begin{center}
\includegraphics[width=0.42\textwidth]{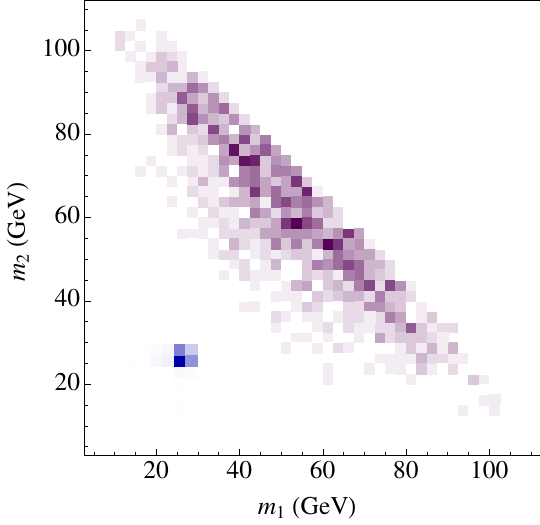} \phantom{mmmm}
\includegraphics[width=0.45\textwidth]{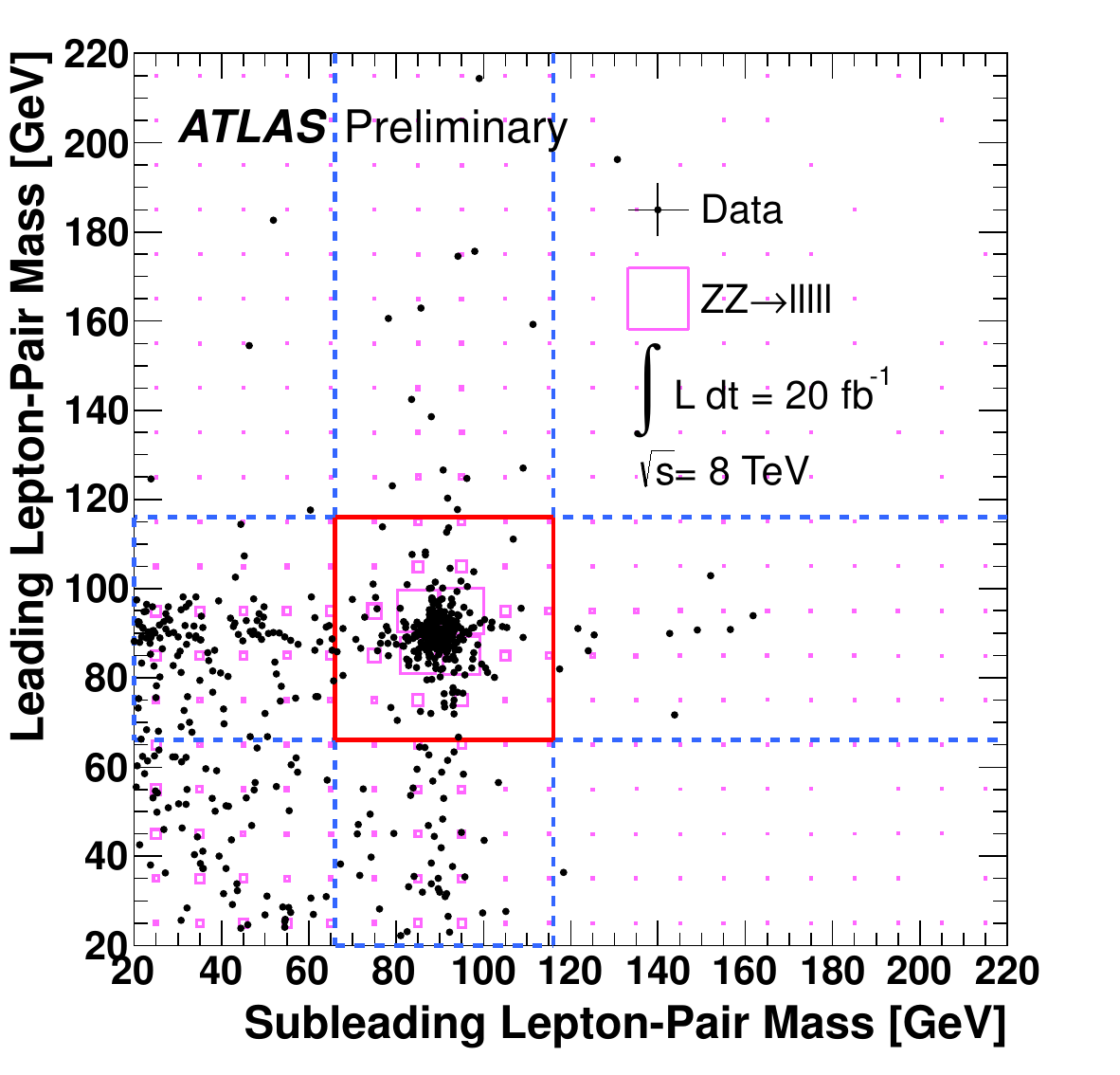}
\end{center}
\caption{{\small {\bf Left:} Distribution of lepton pair invariant
    masses for signal with $m_{Z_D}=25$ GeV for all flavor
    combinations, according to the event selection and reconstruction
    criteria of \cite{ATLAS-CONF-2013-020}.  Correctly paired events
    are shown in blue and make up 55\% of the accepted events, while
    mis-paired events, in purple, make up the remaining 45\%. {\bf
      Right:} Distribution of selected lepton pair invariant masses,
    from \cite{ATLAS-CONF-2013-020}.  Note that the scales of the axes
    differ in the two plots.}
\label{fig:VdVd_ATLAS_ZZ__57}}
\end{figure}

We estimate limits on dark vectors of masses down to 5 GeV.  For
$m_{Z_D}=$ 5 GeV, the daughter leptons are beginning to become
collimated, with a typical $\Delta R_{\ell\ell}\sim 0.2$. Leptons are
not allowed to spoil each other's isolation criteria in
Ref.~\cite{CMS-PAS-HIG-13-002}, and we have therefore applied the same
identification efficiencies and smearings to these semi-collimated
leptons as we use for parameter points with better separated leptons.
If this is a poor approximation, then the exclusion shown for the
range $m_{Z_D}\sim 5$ GeV will prove to be optimistic. Nevertheless,
reductions in electron efficiency of $\mathcal{O}(1)$ still
result in interesting limits, and in the region
$10\,\gev\lesssim m_{Z_D}\lesssim 20$ GeV, the exclusions are robust.

The ATLAS SM $h\to Z Z^*\to 4\ell$ search \cite{ATLAS-CONF-2013-013}
is similar in spirit to the CMS search.  The major difference for our
purposes is that the acceptance is tighter for the OSSF lepton
pair minimizing $|m_{\ell\ell}-m_Z|$,
\beq
50\,\gev < m_1 < 106\,\gev,  \phantom {space or} 12\,\gev < m_2 < 116 \,\gev.
\eeq
This reduces the overall acceptance for the BSM signal, leading to
weaker limits than those from CMS (as both experiments observed 4
total events in the signal region, and as ATLAS does not report flavor
information for these events).

At low masses, the best limits are found from control regions in the
ATLAS $ZZ$ cross section measurement with 20 fb$^{-1}$ of 8 TeV data
\cite{ATLAS-CONF-2013-020}.  Here, events are again required to have
exactly four leptons, which can be paired into two OSSF pairs.  Now
when there is a choice of possible OSSF pairings, the assignment which
minimizes $|m_1 - m_Z | + |m_2-m_Z|$ is chosen.  This still has some
probability of mis-pairing $h\to Z_D Z_D$ events, as can be seen in
Fig.~\ref{fig:VdVd_ATLAS_ZZ__57}.  The invariant mass of the lepton pair
with higher $p_T$ is assigned to be $m_1$.  Note that, unlike the SM
$h\to Z Z^*$ analyses, there is {\em no restriction} on the invariant
mass of the four leptons.

\begin{figure}[t]
\begin{center}
\includegraphics[width=0.6\textwidth]{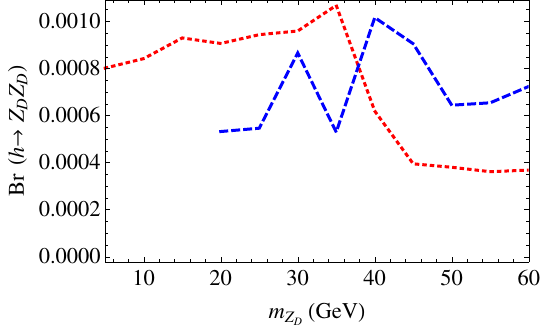}
\end{center}
\caption{{\small 
    Estimated 95\% CL limits on the branching fraction $\Br(h\to Z_D
    Z_D)$ coming from CMS $h\to Z Z^*$ \cite{CMS-PAS-HIG-13-002} (red,
    dotted) and ATLAS $ZZ$ cross section \cite{ATLAS-CONF-2013-020}
    (blue, dashed) measurements. Note that, as seen in
    Fig.~\ref{fig:SMVBrZdark} in \S\ref{subsec:SMvector}, for this
    range of $m_{Z_D}$, $\Br\left(Z_D\to\ell^+\ell^-\right)\simeq 0.3$
    which implies that $\Br\left(h\to Z_DZ_D\to4\ell\right)\simeq
    0.09\times\Br\left(h\to Z_DZ_D\right)$.}
  \label{fig:VdVd_limits}}
\end{figure}

We now set limits by defining a signal region for each mass,
$m_{Z_D}-2\,\gev <m_1 < m_{Z_D} + 2$ GeV, $m_{Z_D}-2\,\gev < m_2 <
m_{Z_D} + 2$ GeV.  Lepton efficiencies are modeled with a
$p_T$-dependent parameterization for electrons \cite{Aad:2009wy,
  Aad:2011mk} and a flat efficiency for muons, and validated against
the fiducial acceptances for $ZZ$ events quoted in
\cite{ATLAS-CONF-2013-020}.  At most one event is observed in each 4
GeV $\times$ 4 GeV signal bin. Treating any observed event in the
signal region as signal, we obtain 95\% CL limits as before.

Fig.~\ref{fig:VdVd_limits} shows the resulting limits (along with
those from CMS's $\hsm\to ZZ^\ast$ search), of order $10^{-3}$, on
Higgs branching fractions to dark vector bosons that further decay to
lepton pairs.  These limits, while impressive, are easy to improve at
low masses by simply looking for OSSF pairs which minimize
$|m_1-m_2|$, instead of a distance from the $Z$ peak.  As backgrounds
are already zero for most bins, improving signal acceptance is the
most likely to improve reach.


\section{\secsize $\secbold{ \hsm \rightarrow \lowerGreekBold\gamma+\met}$}
\label{sec:1gaMET}

We consider here the signature $\hsm\to \gamma+\met$.  This signature
can be usefully represented through the decay of the Higgs into two
neutral fermions, $\hsm\to\chi_1\chi_2$, followed by the decay
$\chi_2\to\gamma \chi_1$.

\subsection{Theoretical Motivations}

While our focus here is on decays to BSM particles, it is worthwhile
to observe that the signature $\hsm \to \gamma +\met$ arises as a rare
decay in the SM, through the loop-induced $\hsm\to\gamma Z$, followed
by $Z\to\nu\bar\nu$.  The SM branching fraction is thus $\Br (h\to
\gamma+\nu\bar\nu)|_{SM}=1.54\times 10^{-3} \times 0.20 = 3.08\times
10^{-4}$ \cite{Heinemeyer:2013tqa}.  Searches for potential
enhancements in $\hsm\to\gamma Z$ are sensitive to the potential
presence of new physics running in the loop, making this rare Higgs
decay signature one of interest for several reasons.  The decay
$\hsm\to\gamma Z$ implies specific kinematics for the photon and
missing energy, however, which do not hold in more general models.

One class of models that gives rise to a $\hsm\to\gamma+\met$
signature are those with very low-scale supersymmetry breaking
\cite{Djouadi:1997gw}. Here the Higgs decays into a gravitino and a
neutralino that is dominantly bino, $\hsm\to\tilde G \tilde B$,
followed by the prompt decay $\tilde B \to \gamma \tilde G$
\cite{Petersson:2012dp}.  As the gravitino is effectively massless,
this model is parameterized by one mass $m_{\tilde B}$. This mass
should lie in the range $m_\hsm/2< m_{\tilde B} < m_\hsm$ to obtain a
large branching ratio to $\hsm\to\gamma+\met$, as for $m_\hsm/2 >
m_{\tilde B} $, the decay $\hsm\to \tilde B\tilde B$ will dominate,
leading to a $\hsm\to 2\gamma+\met$ signature.

This signature can also be realized in the PQ-limit of the NMSSM (see
\S\ref{NMSSMfermion}).  Here the lighter fermion $\chi_1$ is
dominantly singlino, and the heavier fermion is dominantly bino.  The
mass splitting between the two fermions is now much more free.
However, in the PQ-symmetric limit, a light singlino is always
accompanied by a light scalar $s$, and for the loop-induced branching
fraction $\Br(\chi_2\to\chi_1\gamma)$ to be sizable, the tree level
decays $\Br(\chi_2\to s^{(*)}\chi_1\to f\bar f \chi_1)$ must be
phase-space suppressed.  Thus one generically expects mass splittings
between the two neutralino species of no more than 10-20 GeV for the
rate into $\hsm\to\gamma+\met$ to be appreciable.  Outside the
PQ-symmetric limit of the NMSSM, or in other extensions of the MSSM
\cite{Suematsu:1997au}, special parameter cancellations are required
to obtain substantial branching fraction for the radiative decay
$\chi_2\to \gamma \chi_1$.

A more bottom-up approach extends the SM by two Majorana fermions,
$\chi_2$ and $\chi_1$, with a dipole coupling
\beq
\delta\mathcal{L} =
\frac{1}{\mu} \bar{\chi}_2 \sigma_{\mu\nu} B^{\mu\nu}\chi_1 .
\eeq
Note that the presence of the hypercharge field strength $B$ would
predict a $Z+\met$ signal as well, if phase space allowed it; however,
in many UV completions of the dipole operator, the mass-splitting between 
the fermionic states arises due to some symmetry breaking which 
makes it challenging to realize $m_{\chi_2}-m_{\chi_1}\gtrsim m_Z$, 
and the $Z$ mode will typically be highly suppressed.  
The simplified model is then characterized by two parameters 
$m_1$ and $m_2$, where $m_1<m_2$ and $m_1+m_2 < m_h$.

Finally, the $\gamma+\met$ signature also appears as a subleading
decay mode in models of Higgs decay to right-handed neutrinos $N $
\cite{Graesser:2007pc}.  Here the signature arises from $\hsm\to N N$,
followed by the decay of $N\to\gamma\nu$ on one side of the event and
$N\to \nu\bar\nu \nu$ on the other.  In the realization of
\cite{Graesser:2007pc}, both of these $N$ decay modes are highly
subdominant, and the photonic decay may be displaced.

\subsection{Existing Collider Studies}    

An LHC study was carried out at parton level in
\cite{Petersson:2012dp}.  This study targets Higgs bosons produced in
gluon fusion and estimates that 20 fb$^{-1}$ of 8 TeV data would allow
95\% CL sensitivity to branching fractions ranging between Br$(h\to
\gamma+\met) < 0.002$ for $m_{\chi_2} = 120$~GeV, and Br$(h\to
\gamma+\met) < 0.010$ for $m_{\chi_2} = 60$~GeV.  These results are
based on selection criteria that are not obviously compatible with
current LHC triggers, however, as the selection of
Ref.~\cite{Petersson:2012dp} requires
\beq
 45 \mbox{ GeV} < p_{T\gamma} < \frac{m_h}{2} 
\eeq
and no other triggerable objects.  Current monophoton triggers require
$p_{T,\gamma}>80$ GeV, although trigger cuts for CMS parked data are
more relaxed, $p_{T,\gamma}>30$ GeV and $\met > 25$~GeV for central
photons, and therefore could be relevant for this decay channel.

Replacing the cut on photon $p_T$ with one on the transverse mass of
the photon and the missing momentum gives a good separation between
signal and backgrounds.  Trigger thresholds ensure that the dominant
contribution to the reach comes from the high-$p_T$ tail of the Higgs
production spectrum, where the Higgs recoils against one or more hard
ISR jets.  Depending on the mass difference between $\chi_1$ and
$\chi_2$ and the analysis threshold achieved in parked monophoton
$+\met$ triggers, the best signal acceptance may be achieved in
monojet$+\met$-triggered events rather than
monophoton$+\met$-triggered events.

\subsection{Existing Experimental Searches and Limits}
\label{sec:1gamet-limits}

In (N)MSSM realizations of the nonresonant signature, there are {\em
  indirect} limits on the Higgs branching fraction into
neutralino-gravitino from electroweak-ino searches at Tevatron and at
LHC (see also the nonresonant 2$\gamma+\met$ signature in \S\ref{sec:2gaMET}, where similar
considerations apply).  In the case of the neutralino-gravitino
realization, the lightest neutralino $\chi_1$ must have some
Higgsino component in order for the coupling $\hsm\chi_1 \tilde G$ to
be present.  In the neutralino-singlino realization, the heavier
fermion $\chi_2$ is typically dominantly $\tilde B$, with $\chi_1$
dominantly singlino, and the vertex $\hsm\chi_2 \chi_1$ again
proceeds through the Higgsino component of $\chi_2$.  In both
scenarios the non-zero Higgsino component implies the bino-like state
should be produced directly at hadron colliders via Drell-Yan
\cite{Mason:2009qh}, which may or may not lead to constraints
depending on the ensuing decay modes of the bino.  While it is of
interest to work out these indirect limits, the surviving parameter
space is multidimensional, and in more general models, where the
coupling $\hsm n_2 n_1$ arises from a dimension-five Higgs portal
coupling, the new neutral fermions do not need to have tree level
couplings to the $Z$ boson, and no such indirect limit applies.

Very few existing collider searches place any limits on $\Br(\hsm\to
\gamma+\met)$.  Searches for a hard photon plus $\met$, designed to
pick up invisible particles recoiling aginst a hard ISR photon
\cite{Aad:2012fw, ATLAS-CONF-2012-085, Chatrchyan:2012tea}, target
very different kinematic configurations and are not constraining.
Similar conclusions apply to the $Z\gamma$, $Z\to \nu\bar\nu$
cross-section measurements \cite{Aad:2013izg,Chatrchyan:2013nda},
which also target high-$p_T$ photons recoiling against $\met$.

Searches for supersymmetry in final states with $\gamma+\ell +\met+$jets at
the LHC \cite{ATLAS-CONF-2012-144, Chatrchyan:2011ah} and the Tevatron
\cite{Abulencia:2007zi} can be sensitive to $W\hsm$ associated
production when the $W$ decays leptonically.  Acceptance for the Higgs
signal in these supersymmetry searches is small, due to the hardness demanded
of both the $\gamma$ and the $\met$.  No limit is placed by the LHC
$W\hsm $ searches in any part of the $m_1$-$m_2$ simplified model
parameter space.  The Tevatron searches likewise place no limits,
partially due (particularly for large $m_2-m_1$) to a 1$\sigma$ excess
of observed events relative to expectation.  This quick limit check
assumes 100$\%$ photon efficiency; incorporating realistic photon
efficiency would further weaken the search.  The general CDF search
for anomalous $\gamma + \met +$at least one jet also does not
constrain the Higgs branching fraction~\cite{CDF-10355}.

CMS' supersymmetry search in the $\gamma+\met+$jets final state
\cite{CMS-PAS-SUS-12-013} comes closer to being constraining; again,
no limits are placed anywhere in the $m_1$-$m_2$ simplified model
parameter space, but as before this lack of constraint is partially
due to a 1.3$\sigma$ excess of events observed over background
expectation (assuming 100$\%$ photon efficiency).  An updated search
in the same final state \cite{CMS:2012kwa} with 4.04 fb$^{-1}$ of 8
TeV data requires all events to have $H_T>450$~GeV, giving punishingly
small signal effiency. Despite the harshness of this cut, this
analysis is beginning to gain sensitivity to the $\gamma+\met$ decay
mode, as shown in Fig.~\ref{fig:1photon}.  The reported limits from
\cite{CMS:2012kwa} are difficult to recast due to the existence of
signal contamination in a region $\met < 100$ GeV used to model the
dominant QCD background.  The light 125 GeV Higgs contributes
proportionately more to the control region $\met < 100$ GeV than do
the pair produced neutralinos with mass 375 GeV for which the
background predictions are shown.  The limits found by recasting the
analysis for a light Higgs are likely overconservative to an extent
that is difficult to estimate.  In Fig.~\ref{fig:1photon} we show
the result of performing this simple recast.  The signal region is
divided into multiple exclusive bins in $\met$, with background
predictions as reported for the pair-produced neutralinos.  We place
limits by combining the limits from each individual bin using a
Bayesian algorithm with flat priors, and marginalize over background
uncertainty according to a lognormal distribution.  With perfect
photon efficiency, the 95$\%$ CL limits obtained on
$\Br(\hsm\to\gamma+\met)$ is approximately unity in a large range of
parameter space, suggesting that an analysis more tailored to the
signal kinematics could place meaningful limits on the branching
fraction for this channel.

\begin{figure}[t]
\begin{center}
\includegraphics[width=0.6\textwidth]{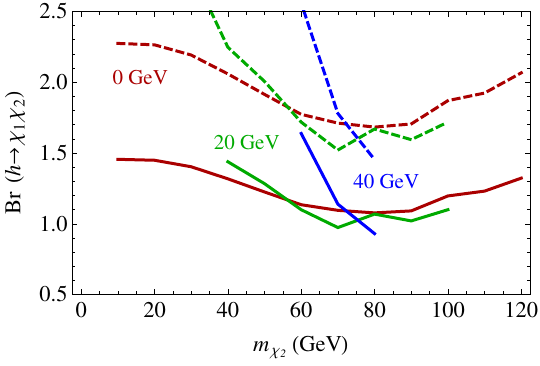}
\end{center}
\caption{{\small Approximate 95\% C.L. upper limit on
    $(\sigma/\sigma_{SM}) \times \Br\left(h\to \chi_1\chi_2\to \gamma
      +\met\right)$ from the results of Ref.~\cite{CMS:2012kwa}, for
    $m_{\chi_1}= (0 \gev, 20\gev,40\gev) < m_{\chi_2}$.  Solid lines
    correspond to 100\% photon efficiency, and dashed lines to a
    (flat) 80\% photon efficiency.}
  \label{fig:1photon}}
\end{figure}

As with all semi-invisible signals, collider reach could be extended
by forming the transverse mass of the visible decay product(s), here
the photon, with the missing transverse momentum vector, and requiring
this to be bounded from above as consistent with production from an
initial resonance.  Much better sensitivity could be achieved if the
prohibitively hard cut on $H_T$ could be relaxed.  This $H_T$ cut is
necessitated by the $\gamma+H_T$ trigger used to select the data in
the current analysis, and is not suited well to the study of the
relatively low-$p_T$ Higgs events.  Somewhat better signal acceptance
is realized for the monophoton$+\met$ triggers in current use for dark
matter searches, though the degree of improvement depends on the
spectrum; again, monojet$+\met$ triggers may provide better
sensitivity. 

	\section{\secsize $\secbold{\hsm \rightarrow 2 \lowerGreekBold\gamma + \met}$}
\label{sec:2gaMET}

In this section we consider the decay $\hsm\rightarrow 2\gamma +\met
$. This signature can be realized in several ways.
\begin{itemize}
\item First, consider the non-resonant signature where the photons come from
opposite sides of the initial two-body decay, $\hsm\to X X$, followed
by $X\to \gamma Y$ on each side of the event with $Y$ a
detector-stable, neutral particle.

\item Second is the case where the photons reconstruct an intermediate
resonance, $\hsm\to X X$, with $X\to \gamma \gamma$ on one side and
$X\to $ invisible on the other.

\item The last decay topology we consider involves the initial decay
$\hsm\to XY$, followed by $X\to Y\phi$, $\phi\to\gamma\gamma$ with $Y$ again
appearing as missing energy in the detector.
\end{itemize}
These different cases may arise in different theoretical models, and
require related but distinct strategies to observe at colliders, as we
discuss below.

\subsection{Theoretical Motivation}

\subsubsection{Non-Resonant}

The non-resonant decay of the Higgs boson to two photons and missing
energy may be realized in several theoretical scenarios.

As a first example, consider gauge-mediated supersymmetry-breaking models.
Here the lightest neutralino is mainly bino, and decays via $\chi_1 ^
0\to \gamma \tilde G$.  Minimal models of gauge mediation make it
difficult to obtain a bino with $m_{\tilde B} < m_h/2$ while keeping
winos sufficiently heavy to satisfy LEP bounds on the charginos as
well as gluinos sufficiently heavy to avoid LHC constraints.  However,
more general models of gauge mediation \cite{Meade:2008wd} can allow
this spectrum to be realized \cite{Mason:2009qh}.

Another realization of the non-resonant $2\gamma+\met$ signature may
be obtained in the PQ limit of the NMSSM (see \S\ref{NMSSMscalar} and \S\ref{NMSSMfermion} for more details), where a light singlino $\tilde s$ replaces the
gravitino.  In this case the photonic signature is realized through a
loop-induced dipole coupling $\tilde B^\dag \sigma^{\mu\nu}B_{\mu\nu}
\tilde s$.  There are typically several other decay modes available to
the $\tilde B$ in these NMSSM models, in particular
\beq
\tilde B \to Z^{(*)}\tilde s, \phantom{space} \tilde B \to a^{(*)}\tilde s, 
      \phantom{space} \tilde B\to s^{(*)}\tilde s,
\eeq
where $a$, $s$ are light, dominantly singlet CP-odd and CP-even
scalars.  The radiative decay $\tilde B \to \gamma\tilde s$ is
typically significantly subdominant to the tree-level decays.  The
$2\gamma+\met$ signature is thus typically 
small compared to other exotic decay modes in the PQ NMSSM.

More generally, this signature may be realized by having two new (Majorana)
fermions $\chi_1$ and $\chi_2$, with a dipole coupling
\beq
\label{eq:dipoleDK}
\delta\mathcal{L} =
\frac{1}{\mu} \chi_2^\dag \sigma_{\mu\nu} B^{\mu\nu} \chi_1
\eeq
and a dimension five Higgs portal coupling $ c_{22} | H | ^ 2( \chi_2
\chi_2 + \chi_2^\dag \chi_2^\dag)$.  In this case, both $m_{\chi_1}$
and $m_{\chi_2}$ are parameters of the model.  It is natural to extend
this simple model to include in addition off-diagonal couplings $
c_{12} | H | ^ 2( \chi_2 \chi_1 + \chi_2^\dag \chi_1^\dag)$ and
couplings of the Higgs directly to the lighter of the two new
fermions, $c_{11}| H | ^ 2( \chi_1 \chi_1 + \chi_1^\dag \chi_1^\dag)$.
This generic model would then also yield $h\to 1 \gamma + \met$ and
$h\to \met$ signatures with relative branching fractions uniquely
determined by the $c_{ij}$.  Previous study of this topology in the
MSSM has been performed in \cite{Mason:2009qh} and, for the heavier
MSSM Higgses, in \cite{DiazCruz:2003bx}; see also \cite{Strassler:2006qa}

\subsubsection{Resonant}

The $2\gamma + \met$ final state can also occur for the decay chain $\hsm\to a a$,
with one intermediate state decaying to photons, $a\to\gamma\gamma$,
and the other decaying invisibly, $a\to{\rm inv}$. This can be simply
realized in a bottom-up fashion by introducing a renormalizable Higgs
portal interaction leading to a coupling of $a$ to $\hsm$,
$\lambda\left|H\right|^2a^2$, and also coupling $a$ to photons and to
a neutral, detector-stable particle $\chi$ via, e.g.,
\begin{equation}
\frac{\alpha}{4\pi M}aF^{\mu\nu}\tilde F_{\mu\nu}
+\frac{\partial_\mu a}{M^\prime}\bar\chi\gamma^\mu\gamma^5\chi.
\end{equation}
$M$ and $M^\prime$ are the scales of the two dimension-five operators,
and we have assumed that $a$ is a real pseudoscalar and that $\chi$ is
a Dirac fermion for definiteness.  For some regions of parameter
space, $a\to\gamma\gamma$ and $a\to\bar\chi\chi$ can have comparable
branching fractions, making $\hsm\rightarrow 2\gamma +\met$ an
important final state. Another possibility arises from the decay chain 
$\hsm\to \chi_1\chi_2\to a \chi_1\chi_1$, where $a$ decays via the first 
dimension-five operator and $\chi_1$ is stable. Note, though these two 
decay topologies can be achieved in the R- and PQ-limits in the NMSSM 
(see \S\ref{NMSSMscalar} and \S\ref{NMSSMfermion}), 
the branching fraction of $a\to\gamma\gamma$ tends to be small.  Alternatively, 
$a$ may be light enough so that $a\to ff$ is kinematically suppressed, in which case 
the lifetime is so long that $a$ would decay outside the detector.  
More general models may give a larger $a\to\gamma\gamma$ coupling than the NMSSM.

Unlike the non-resonant case, the resonant signature has the useful
additional handle that the two photons should reconstruct $m_a$,
improving the search prospects. Additionally, as $m_a$ is decreased and the
intermediate particles become more boosted, a larger fraction of the
photon pairs will fail isolation cuts. For $m_\hsm=125$~GeV, this becomes
important for $m_a\lesssim{\rm few}$~GeV. In this case, the
signal would have some overlap with that from $h\rightarrow \gamma
+\met$ considered in
\S\ref{sec:1gaMET}~\cite{Draper:2012xt}.\footnote{In the $m_a\ll
  m_\hsm$ regime, the relationship between the $\hsm\rightarrow 2\gamma
  +\met$ and $\hsm\rightarrow \gamma +\met$ signals parallels that
  between $\hsm\rightarrow 2\gamma$ and $\hsm\rightarrow 4\gamma$. See
\S\ref{sec:4gamma} for further details.}

This simplified model can be trivially generalized to the case that
the Higgs decays to two distinct states, $a_1$ and $a_2$, with $a_1\to
\gamma\gamma$ and $a_2\to{\rm inv.}$ This can proceed through a
dimension-four Higgs portal interaction, $\lambda_{12} | H | ^ 2a_1
a_2$, if $a_1$ couples to photons while $a_2$ decays invisibly. This
decay mode can dominate over $\hsm\to{\rm inv.}$ or $\hsm\to4\gamma$
if $\lambda_{12}\gg \lambda_{11,22}$ where $\lambda_{11,22}$ are the
coupling constants of the other allowed Higgs portal interactions,
$\lambda_{11} | H | ^ 2a_1^2+\lambda_{22} | H | ^ 2a_2^2$. While, in
this resonant case, we limit our study to the situation $m_{a_1}\simeq
m_{a_2}\equiv m_a$, the two intermediate particles having different
masses is a well-motivated possibility.

\subsubsection{Cascade}

The $\hsm\to2\gamma+\met$ decay can proceed through
$\hsm\to\chi_1\chi_2$, with $\chi_2\to s\chi_1$, $s\to\gamma\gamma$ if
$\chi_1$ is neutral and stable on detector scales. It is easy to write
down a simple model that gives rise to this decay chain. We can couple
(Majorana) fermions $\chi_1$ and $\chi_2$ to the Higgs through a
dimension-five Higgs portal coupling as in the non-resonant case
above, $c_{12}|H|^2(\chi_2\chi_1+\chi_2^\dagger\chi_1^\dagger)$, as
well as to the scalar $s$ through a Yukawa interaction,
$y_{12}s(\chi_2\chi_1+\chi_2^\dagger\chi_1^\dagger)$. Furthermore, $s$
can decay to two photons through the dimension-five operator 
$sF_{\mu\nu}F^{\mu\nu}$.\footnote{The $s F_{\mu\nu}F^{\mu\nu}$ operator could
  arise through mixing between $s$ and $\hsm$, see for example
  \S\ref{SMS}, although that would lead to a very suppressed
  $\hsm\to2\gamma+\met$ branching ratio compared to final states like
  $b\bar b+\met$. For $2\gamma+\met$ to be dominant, the 
  $sF_{\mu\nu}F^{\mu\nu}$
  operator would have to be generated by a direct coupling of $s$ to
  electrically-charged matter, e.g.~(heavy) vector-like leptons.  For
  a similar model, see~\secref{2gamma2jet}.}

\subsection{Existing Experimental Searches and Limits}

In (N)MSSM realizations of the non-resonant signature, there are {\em
  indirect} limits on the Higgs branching fraction into neutralinos
from general electroweak-ino searches at the Tevatron and at the LHC.
These limits arise because the lightest neutralino $\chi^0_1$ must
have some Higgsino component in order for the coupling $\hsm\chi^0_1
\chi^0_1$ to be present. Because of this non-zero Higgsino component,
the lightest neutralino couples to the $Z$ and can be produced
directly at hadron colliders via Drell-Yan.  Model-dependent indirect
limits on Higgs branching fractions arising from Drell-Yan direct
production are nontrivial \cite{Mason:2009qh} and an interesting topic
of study, but in the present work we confine ourselves to considering
(model-independent) direct limits, and make no assumptions about other
production modes for the BSM states.  In general (non-MSSM) models,
where the coupling $\hsm \chi_2 \chi_2$ arises from a dimension five
Higgs portal coupling, the new neutral fermion $\chi_2$ does not need
to have tree-level couplings to the $Z$ boson, and those indirect
limits do not apply.

In GMSB realizations of the non-resonant signal, sufficiently high
SUSY-breaking scales lead to a macroscopic decay length for the
neutralino.  This can also occur in the general Higgs portal
simplified model, for sufficiently large dipole suppression scales
$\mu$ in the decay vertex of Eq.~(\ref{eq:dipoleDK}).  In such cases,
non-pointing photon searches may be motivated or necessary.  Displaced
signatures are beyond the scope of the present work, but are an
interesting and natural avenue for future exploration.
\begin{figure}[t]
\begin{center}
\includegraphics[width=0.49\textwidth]{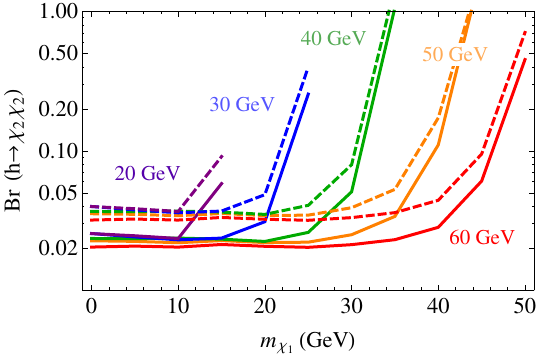}
\caption{{\small Approximate 95\% C.L. upper limit on $(\sigma/\sigma_{SM}) \times \mbox{Br}\left(\hsm\to
    \chi_2\chi_2\to 2\gamma +\met\right)$ from the
  results of Ref.~\cite{CMS:2012kwa}, for multiple values of $m_{\chi_2}$ as indicated
 by the text labeling the different curves.  Solid lines
  correspond to 100\% photon efficiency, and dashed lines to a (flat) 80\%
  photon efficiency.} 
  \label{fig:2gaMETnonresAugLabel__61}}
\end{center}
\end{figure}

GMSB searches at the LHC have good prospects for discovering or
excluding exotic Higgs decays into $2\gamma+\met$, in both the
resonant and non-resonant scenarios.  The ATLAS search for
$2\gamma+\met$ using 7~TeV data~\cite{Aad:2012zza} has some
sensitivity, setting limits of $\lesssim 15\%$ on the exotic Higgs
branching fraction over much of the parameter space.  The more recent
CMS study using 4.04 fb$^{-1}$ of 8 TeV data \cite{CMS:2012kwa} sets
the current best limits.  This search selects events with at least two
photons and at least one central jet, and bins events in 5 exclusive
$\met$ bins beginning from a minimum of $50$~GeV.  We show the reach
of this search in the resonant and non-resonant cases in
Figs.~\ref{fig:2gaMETnonresAugLabel__61} and~\ref{fig:2gaMETres__62} (left), as a
function of $m_{\chi_1}$ in the non-resonant topology and $m_a$ in the
resonant topology. In Fig.~\ref{fig:2gaMETres__62} (right), we show the reach in
the case of the cascade topology as a function of $m_s$, setting
$m_{\chi_1}=0$ and $m_{\chi_2}=60$~GeV. We find that the limit
obtained in this case is not very sensitive to the value of
$m_{\chi_2}=60$~GeV chosen.  In all three topologies the Br$(h\to
2\gamma+\met)$ can be constrained at the level of a few percent over
much of the parameter space.  Higgs signal events are generated in
MadGraph with showering in Pythia, and jet clustering is done with
FastJet.  Gluon fusion is matched out to one jet, and cross-sections
for both gluon fusion and vector boson fusion processes are set to the
values determined by the LHC Higgs Working Group
\cite{LHCHiggsCrossSectionWorkingGroup:2011ti}. VBF production is
responsible for $20$-$25\%$ of the signal.  To obtain limits we combine
individual 95$\%$ CL limits from each of the 5~$\met$ bins according to a
Bayesian algorithm with flat priors, marginalizing over the background
uncertainty according to a log-normal distribution.
\begin{figure}[t]
\begin{center}
\includegraphics[width=0.48\textwidth]{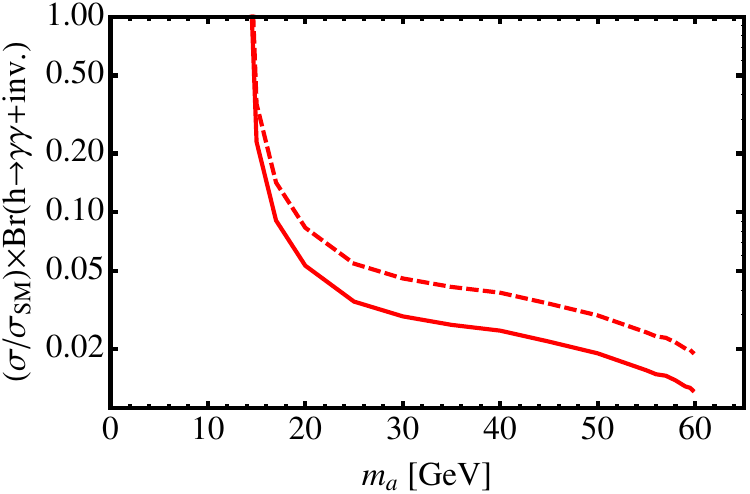}
~~~
\includegraphics[width=0.48\textwidth]{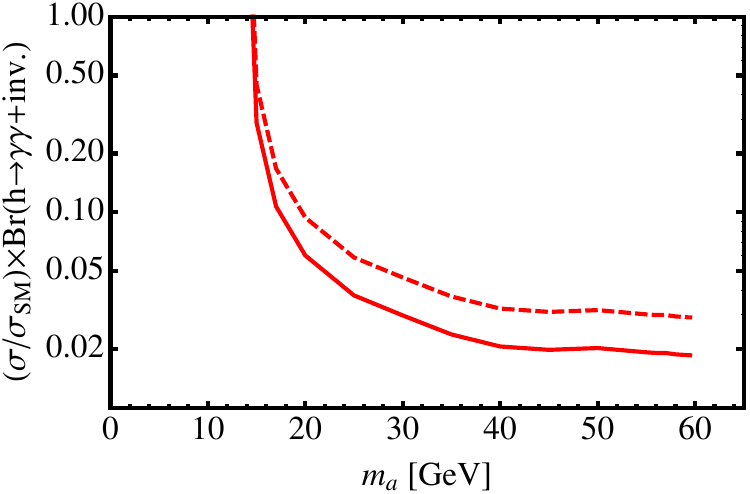}
\caption{{\small Approximate 95\% C.L. upper limit on $(\sigma/\sigma_{SM})\times \mbox{
      Br}\left(\hsm\to 2\gamma +\met\right)$ from the $2\gamma+\met$
    search in~\cite{CMS:2012kwa}. The solid lines correspond to
    100\% photon efficiency, and the dashed lines to a (flat) 80\%
    photon efficiency.  
      {\bf Left:} Resonant case, where $h\to aa$, one $a$ decays to $\gamma\gamma$ and the other
    decays invisibly.  
    {\bf Right:} Cascade case, where  $\hsm\to\chi_1\chi_2$, $\chi_2\to s\chi_1$, $s\to\gamma\gamma$.  
    Here $m_{\chi_1}=0$ and $m_{\chi_2}=60$~GeV (although the limit is
    insensitive to the particular value of $m_{\chi_2}$ as long as it
    is kinematically allowed).}
  \label{fig:2gaMETres__62}}
\end{center}
\end{figure}

Since searches using only 4 fb$^{-1}$ of 8 TeV data and optimized for
other signatures are already able to place limits as stringent as
$\mathcal{O}( 5\%)$ on the Higgs branching fraction into this mode, 
$2\gamma + \met$ is a good candidate for searches in the near future.
The reach could be easily extended by requiring the transverse mass of the
photons and $\met$ to be bounded from above, as consistent with
resonant origin from the 125 GeV Higgs. In the resonant case, looking for a 
peak in the $\gamma\gamma$ spectrum could offer another useful handle.  

\section{\secsize $\secbold{h \rightarrow}$ {\normalsize 4 Isolated Leptons + }$\secbold{\met}$}
\label{sec:IsoLeptonMET}

Exotic Higgs decays into multiple charged leptons together with
missing energy are less frequently motivated by top-down model
building than (e.g.) $h\to a a$ cascade decays, but on the other hand,
they offer excellent discovery potential at the LHC, as we will
demonstrate in this and following sections.

There is some overlap between the theoretical motivations and decay
topologies for different $h\to \geq 2\,{\rm charged \ leptons} + \met
+ X$ signatures. Here we briefly discuss all the cases we consider in
this document before treating the $4 \ell + \met$ case in detail.

Depending on the specific model under consideration, the
characteristic predictions for leptonic final states can be very
different. Exotic Higgs decays $h\to X_1 X_2$ (where $X_{1,2}$ may or
may not be distinct species) can be divided into two main classes of
topologies:

\begin {enumerate}

\item $\ell^+ \ell^-  +\met$, which involves the topologies: 
\begin{itemize}
\item I:  $X_1 \to$ non-leptonic$+ \met$, \ $X_2\to \ell^+\ell^-  +\met$
\item II:  $X_1 \to$ non-leptonic$+ \met$, \ $X_2\to  \ell^+\ell^- $
\end{itemize}
where the the non-leptonic part is typically either nothing (i.e.,
$X_1$ stable and invisible) or hadronic (i.e., $X_1\to$ soft jets$+\met$);\footnote{Charged $X$'s each decaying to $\ell+\met$ are
  highly constrained, and not considered here.}  and
\item $2 \times \ell^+ \ell^-  +\met$, which can be achieved via 
the topology
\begin{itemize}
\item III: $X_1 \to  \ell^+\ell^-   + \met$, \ $X_2\to  \ell^+\ell^-  + \met$ \ .  
\item IV:  $X_1 \to  \ell^+\ell^-$, \ $X_2\to  \ell^+\ell^-  + \met$.
\end{itemize}

\end{enumerate}
Further, the cascade decays of $X_2$ in topologies I and III may
either be three-body, or they may involve an on-shell intermediate
state so that the leptons reconstruct a resonance.  Depending on the
mass of this resonance, and similarly on the mass of the $X_2$
resonance in topologies II and IV, the leptons may be either isolated
or collimated.

This gives us a plethora of experimental signatures, all of which
present interesting targets with the existing LHC dataset.  We discuss
theoretical models and experimental prospects for these leptonic
signatures here and in the following two sections.  In the current
section we discuss final states with four isolated leptons plus
missing energy; in \S\ref{sec:2lMET} we discuss final states
with two isolated leptons plus missing energy; in
\S\ref{Clepton} and \S\ref{sec:collep_nomet} we consider final states that include one or two lepton-jets, respectively;  
decays to leptons without $\met$ are discussed in \S\ref{sec:htoZa} and \S\ref{sec:hto4l}.

\subsection{Theoretical Motivation}

Several classes of models can give rise to Higgs decays to 4 isolated
leptons$+\met$.  First, consider models with weak-scale neutral
states that have non-vanishing couplings to the $Z$ boson, such as
exotic neutrinos or neutralinos.  In this case, leptons can arise from
the three-body decay of one neutral fermion $\chi_2$ to a lighter one
$\chi_1$ through an off-shell $Z$ boson, appearing as an
opposite-sign, same flavor pair.  The $4\ell+\met$ signal then arises
from cascades of the form $\hsm\to\chi_2\chi_2\to\chi_1 Z^* \chi_1
Z^*$ with both $Z^*$ leptonic.  In fourth-generation neutrino models,
$\chi_2$, $\chi_1$ are the two Majorana-split halves of a Dirac
neutrino state; in MSSM-like realizations, $\chi_2$, $\chi_1$ are
neutralinos.  The branching fraction into $4\ell+\met$ is small
compared to the total branching fraction into $\chi_2\chi_2$:
$\Br(h\to 4\ell+\met)/\Br(h\to\chi_2\chi_2)=\Br(Z\to\ell\ell)^2
\approx 0.011$ (including $\tau$s).  Despite the small relative
branching fraction, we will see that
the $4\ell+\met$ final state is typically more constraining than final
states with fewer leptons, due to the low backgrounds for
multi-leptonic final states.

Hidden sectors with a kinetically mixed dark vector boson $Z_D$ can also
realize this decay chain \cite{Strassler:2006im,Batell:2010bp}. For instance, a hidden
sector with meson-like pseudoscalar states $K_v$, $\pi_v$, may have a
spectrum such that the heavier meson may only decay via $K_v\to Z_D^*
\pi_v \to f\bar f \pi_v $, and the lighter meson $\pi_v$ is
collider-stable.  The width for this $K_v$ decay scales like
\beq
\Gamma_{K_v}\approx
\alpha_D\alpha_{EM}\frac{\epsilon^2}{15 \cos\theta_W^2}\, \frac{(m_{K_v}-m_{\pi_v})^5}{m_{Z_D}^4},
\eeq
where $\epsilon$ is the kinetic mixing between hypercharge and the
dark vector boson (see \S\ref{subsec:SMvector}). The $K_v$ meson
decay can be prompt provided the ratio of the dark meson mass
splitting to the dark photon mass, $ (m_{K_v}-m_{\pi_v})/m_{Z_D}$, is
not particularly small.  The branching fractions into leptonic final
states are much larger here than in the case where the three-body
decay is mediated by a virtual $Z$.  For a dark vector with $m_{Z_D} >
2 m_b \gtrsim 10$ GeV, the branching fraction into leptonic final
states (including taus) is $\Br(Z_D\to\mathrm{leptonic}) \approx 45\%$,
as discussed in \S\ref{subsec:SMvector}.

Another realization of this type of decay chain with an off-shell
kinetically mixed dark photon occurs in supersymmetric hidden sectors,
with one or more hidden neutralinos.  In this case the Higgs cascade
decay could begin with a Higgs decay to bino-like neutralinos $\tilde
B$, which in turn decay via $\tilde B\to Z^*_D \chi_1^0$, where
$\chi_1^0$ is a hidden sector neutralino \cite{Strassler:2006qa,Baumgart:2009tn,
  Chan:2011aa}.

If the dark photon is sufficiently light, the decay $K_v\to Z_D
\pi_v\to \ell\ell\pi_v$ can be allowed, and the leptons reconstruct a
resonance at $m_{\ell\ell} = m_{Z_D}$. In the PQ-symmetric limit of the
NMSSM, light (pseudo)scalars in the spectrum similarly enable the
on-shell decay $\chi_2\to s (a) \chi_1\to \ell\ell\chi_1$.  However,
in the NMSSM, the branching fractions to light leptons are suppressed
by small Yukawa couplings, and Br$(h\to 4\mu + \met)$ is cripplingly
small unless the scalar is below the $\tau$ threshhold, $m_{s(a)}< 2
m_\tau$.  When the scalar is this light, it is often produced with
$p_{T,s} \gg m_s$, leading to collimated muons, but this is
spectrum-dependent.  Collimated lepton pairs (lepton-jets) are discussed in
\S\ref{Clepton} and \S\ref{sec:collep_nomet}.

In models with a nontrivial flavor structure, {\it flavor-violating}
decays of the form $h\to\chi\chi\to 4\ell+ 2\nu$ can occur. A familiar
example is Higgs decay into R-parity violating neutralinos $\chi_1$,
where $\chi_1$ decays through the leptonic $L_iL_je_k$ operator.  In
this case the two charged leptons from the decay $\chi_1\to
\ell'\ell\nu$ need no longer necessarily form same-flavor pairs.

Finally, another realization of the same final state occurs when the
Higgs decays into two heavy neutrinos $N$, which then each decay
through $N\to W^*\ell \to \nu\ell'\ell$ \cite{1107.2123}.  Similar phenomena and final states can arise in scotogenic models \cite{Ho:2013hia,Ho:2013spa}.

\subsection{Existing Experimental Searches and Limits}

Several LHC searches give interesting bounds on the exotic decay $h\to
4\ell+\met$. The best bounds when the leptons are non-resonant come
from 8 TeV LHC multi-lepton searches.  In order to highlight the
strong dependence on the exotic spectrum, we will present bounds for
two benchmark models where $h\to\chi_2\chi_2$ and $\chi_2 \to \chi_1 Z^*$:
\begin{itemize}

\item An ``optimistic'' benchmark scenario with relatively large mass
  splitting between $\chi_2$ and $\chi_1$, with $M_2=55$ GeV and
  $M_1=20$ GeV. Generally, models of this type are allowed by the LEP
  precision measurement of the $Z$ width, as long as the coupling of
  the $Z$ boson to $\chi_1\chi_2$ is smaller than $\sim
  0.05$.\footnote{This number has been found under the assumption
    $g_V=g_A$ where $g_V$ and $g_A$ are the vector and axial-vector
    couplings $g_V Z^\mu \chi_2\gamma_\mu\chi_1$ and $g_A Z^\mu
    \chi_2\gamma_\mu\gamma_5\chi_1$, respectively.  Similar limits can
    be found for $g_V\neq g_A$.}  Even for couplings $\mathcal
  O(0.01)$, the decay $\chi_2\to \ell\ell\chi_1$ is prompt.  In
  general the Drell-Yan production of $\chi_2\chi_2$ will yield an
  additional and model-dependent contribution to the leptons$+ \met$
  signature. For simplicity, throughout our analysis, we will always
  assume that the $Z$ coupling to $\chi_2\chi_2$ is sufficiently small
  that the Drell-Yan contribution is much smaller than the
  contribution coming from Higgs decay.
  
\item A ``pessimistic'' benchmark scenario with a smaller mass
  splitting, $M_2 = 55$ GeV and $M_1=35$ GeV.  This particular
  parameter point is consistent with LEP data when $\chi_2$, $\chi_1$
  have the $Z$ couplings of fourth-generation
  neutrinos~\cite{Carpenter:2010sm}.  The relatively small mass
  difference between the exotic final states renders the final state
  leptons softer and makes the benchmark more challenging at the LHC.

\end{itemize}
In both cases we take 
\beq
\Br(\chi_2\to \ell^+\ell^-\chi_1)=\Br(Z^{(*)}\to \ell^+\ell^-) .
\eeq
For Higgs bosons produced in gluon fusion and assuming a reference
10$\%$ branching ratio for $h\to \chi_2 \chi_2$, the initial signal
cross section for
\beq
\label{eq:targetdecay}
pp\to h\to \chi_2 \chi_2\to 4\ell \chi_1 \chi_1
\eeq
is approximately $10$~$\rm{fb}$, giving already $\sim 200$ events
in the present LHC data set.  Below we will indicate the excellent
potential of the LHC to set bounds on the optimistic benchmark by
recasting existent searches in multi-leptons.  To indicate the
sensitivity of these searches to the mass splitting between
$\chi_1,\chi_2$ we also show that the more pessimistic benchmark, with
its much softer daughter leptons, is as yet unconstrained.  Dark
photon models, with larger branching fractions to leptonic final
states, face more stringent limits.

The multilepton analysis strategy pursued by both ATLAS and CMS
divides events into several exclusive bins depending on multiple
variables. The variables most notable for our purposes are: lepton
counts $N_\ell$; OSSF lepton pair invariant masses; and either (1) the
value of $\met$ and $H_T$ (defined as the scalar sum of the transverse
energies of all jets passing the preselection
cuts)~\cite{CMS022,CMS026}, or (2) the value of $S_T$ (the scalar sum
of $\met$, $H_T$, and the $p_T$ of all isolated leptons)~\cite{CMS027,
  ATLAS36}, or (3) the value of $m_T$ in three-lepton
searches~\cite{ATLAS035}. A more inclusive strategy is pursued in
\cite{CMS-PAS-SUS-13-010}, which uses only $N_\ell$ and lepton pair
invariant masses to define the several signal regions, while
\cite{CMS-PAS-SUS-13-006} introduces more specialized kinematic
constraints to target specific models of electroweak production.  All
of these analyses set limits on models beyond the SM by combining
individual limits from all bins, both high-background and
low-background.  As reinterpreting multi-lepton searches is highly
sensitive to the details of modeling lepton acceptance, our aim here
is principally to demonstrate the interesting level of sensitivity
already available to non-resonant multi-leptonic Higgs decay.

In order to estimate signal efficiency, we generate inclusive Higgs
events with at least 2 leptons\footnote{We include taus in the
  generation of the events. Taus are decayed using the Tauola plugin
  within Pythia.} in MadGraph 5, shower them in Pythia, and cluster
them in FastJet.  We generate gluon fusion production matched to one
jet, VBF, and $Wh$ associated production. The signal production
cross-sections are normalized to the values reported by the LHC Higgs
Working Group \cite{Dittmaier:2011ti} (see Table~\ref{tab:expected-exotic-decays}).

For CMS multilepton analyses, we are able to make a fairly precise
approximation of the signal efficiency by passing signal events
through the version of PGS tuned by the Rutgers theory group
\cite{ContrerasCampana:2011aa, Craig:2012pu} to more exactly simulate
the CMS detector.\footnote{Thanks in particular to M.~Park and
  S.~Thomas.} We employ in addition the modified $b$-tagging routines
and the correction factors for electron, muon, and hadronic tau
efficiencies as established in \cite{cpstth}.

For the ATLAS multilepton analyses, we approximate signal acceptance
using the $p_T$-dependent lepton identification efficiencies quoted in
Refs.~\cite{Aad:2011mk,Aad:2009wy}. Since our signal is characterized
by relatively soft leptons, it is important to note that the electron
efficiency drops below $70\%$ for $p_T^e \lesssim {\cal O}(10)$ GeV
while the muon identification efficiency remains high even for very
soft muons ($\sim 90\%$ for $p_T^\mu \gtrsim 7$ GeV).

To set limits we treat each bin as a single Poisson counting
experiment, marginalizing over background uncertainty according to a
log-normal distribution, and combine bins according to a Bayesian
algorithm with flat priors on signal strength.  We quote 95$\%$ CL
upper bounds.

The best limits on the optimistic benchmark come from recasting the
19.6 fb$^{-1}$ search performed by CMS in four-lepton final states
\cite{CMS-PAS-SUS-13-010}.  This search requires exactly four light
leptons in the final state, forming at least one OSSF pair. Denoting
the invariant mass of the OSSF lepton pair with mass $m_{\ell\ell}$
closest to $m_Z$ as $M_{\ell\ell 1}$ and the invariant mass of the
remaining lepton pair as $M_{\ell\ell 2}$, the events are divided into
9 exclusive categories depending on whether $M_{\ell\ell 1}$ and
$M_{\ell\ell 2}$ are below, above, or inside the $Z$ window $90\pm
15$~GeV.  The vast majority of exotic Higgs decays fall in the bin
$M_{\ell\ell 1} < 75$~GeV, $M_{\ell\ell 2}<75$~GeV.  Indeed, this is
the {\it only} bin populated by gluon fusion and VBF; $Wh$ associated
production is the only contributing process in the other bins.  The
combined limit from all populated bins is
\beq
\Br(h\to \chi_2 \chi_2)< 11\% ,
\eeq
which is also the 95$\%$ CL limit set by the single dominant bin.
This translates into the limit $\Br(h\to 4\ell+\met) < 1.2\times
10^{-3}$, with $\ell = (e,\mu,\tau)$\footnote{Note that this limit translates into $\Br(h\to 4\ell+\met) <5.4\times
10^{-4}$ considering simply $\ell=e,\mu$.}
 for dark vectors with $Br(Z_D \to
\ell\ell) = 3 \times 0.15$, $Br(h \to K_vK_v) < 6.1 \times 10^{-3}$.
We show predicted signal events for this bin together with the
expected and observed number of events in
Table~\ref{tab:multileptons}. To show the steep dropoff in signal
acceptance when the mass splitting in the cascade decay becomes
smaller, we also show signal predictions in the same bin for the
pessimistic benchmark, where the acceptance in gluon fusion has almost
entirely disappeared.

\begin{table}
\begin{center}
\begin{tabular}{|l|l|c|c|}
\hline \hline
Model & Mode & CMS bin Prediction \cite{CMS-PAS-SUS-13-010}& ATLAS bin Prediction \cite{ATLAS035} \\
\hline

``Optimistic'' & gluon fusion & 50.4 & 2.4\\ 
 ($M_1 = 20$ GeV, & VBF &  56.2 & 7.6\\ 
  $M_2 = 55$ GeV) & $Wh$ & 2.1 & 14\\ \cline{2-4}
                              & total & 109 & 24\\
  
\hline

``Pessimistic'' & gluon fusion & -- & 0.6\\  

 ($M_1 = 35$ GeV, & VBF &  2.2 &   2.2            \\
  $M_2 = 55$ GeV) &  $Wh$ & 0.2 & 3.6\\ \cline{2-4} 
  & total & 2.4 & 6.4\\
\hline \hline
\end{tabular}
\end{center}
\caption{{\small Benchmark predictions for the number of events in the dominant bin (see text) 
    in the most constraining 
    CMS multi-lepton search \cite{CMS-PAS-SUS-13-010} (third column) and ATLAS 
    three-lepton search \cite{ATLAS035} (fourth column), for the optimistic 
    and pessimistic benchmarks 
    defined in the text, with $\mbox{Br}(h\to \chi_2\chi_2)=1$ and $\mbox{Br}(\chi_2\to\chi_1\ell\ell)=\mbox{Br}(Z\to\ell\ell)$. 
    In the CMS bin, 14 events are observed
    and $10.4\pm 2.0$ are expected. In the ATLAS bin, 41.8 events are 
    excluded at the $2\sigma$ level.
    Signal expectations are reported separately for gluon fusion, VBF, and associated $Wh$ production.}
}
\label{tab:multileptons}
\end{table}%

The CMS three- and four-lepton channel search of Ref.~\cite{CMS027},
done with 9.2 fb$^{-1}$ of 8 TeV data, places a similar limit of
\beq
\mbox{Br}(h\to \chi_2 \chi_2)< 14\% .
\eeq
The signal dominantly populates the lowest bin in $S_T$, namely $0 <
S_T<300$ GeV, for all lepton multiplicity channels; VBF production
also contributes secondarily to the next-highest bin, $300~\gev <
S_T<600$~GeV. The bin with the single greatest signal contribution is
that with three identified leptons and one OSSF pair with mass below the
$Z$ window.  However, the signal-to-background ratio is better in the
bin with the second-largest number of signal events, namely the bin
with four identified leptons and two OSSF pairs below the $Z$ window, no
$b$'s, and no hadronic taus. This bin dominates the limit combination.

For the pessimistic benchmark,  Ref.~\cite{CMS027} limits
\beq
\frac{\sigma (p p\to h)}{\sigma (pp\to h)|_{SM}}\Br(h\to \chi_2 \chi_2)< 1.04  ,
\eeq
or Br$(h\to 4\ell+\met)< 0.011$.  The reach is almost entirely from
VBF production, with several bins contributing significantly to the
limit.

The CMS search of Ref.~\cite{CMS026} uses the same data set as
Ref.~\cite{CMS027} but bins events in $\met$ and $H_T$ instead of in
$S_T$, and sets comparable limits.  Finally, the CMS searches
performed in Ref.~\cite{CMS-PAS-SUS-13-006} use kinematic
discriminants which are tailored to the electroweak production of
heavy states, and are not sensitive to the kinematics of our exotic
Higgs decay signal.

ATLAS multilepton searches~\cite{ATLAS035,ATLAS36} are less sensitive
than the CMS searches we have just discussed, mainly because of the
missing energy requirement (at least 50~GeV in all the signal
regions). In particular, the most sensitive search is the three-lepton
search of~\cite{ATLAS035} performed with 20.7~fb$^{-1}$ of 8 TeV data.
The most constraining bin is the so-called {\it {SRnoZa}} that
requires $\met >50$ GeV and all OSSF lepton pairs to have a invariant
mass below 60 GeV. As shown in Table~\ref{tab:multileptons}, the main
contribution to this bin comes from a Higgs produced in association
with a $W$ boson. Assuming Br$(h\to \chi_2 \chi_2)=1$, the optimistic
benchmark model leads to only $\sim 24$ events, to be compared to the
41.8 events ATLAS can exclude in this bin.

We have checked that $Zh$ associated production does not yield a
sizable contribution to the CMS and ATLAS multilepton analyses. In
particular, these events dominantly populate the CMS $4\ell$ bin with
$75\,{\rm{GeV}}<M_{\ell\ell 1}<105$ GeV and $M_{\ell\ell 2}<75$
GeV~\cite{CMS-PAS-SUS-13-010}, in which the signal would only be $\sim
0.2$ events.

The inclusive multilepton search strategy pursued by CMS does a
reasonable job of constraining multileptonic Higgs decays when the
mass splitting in the cascade decay is sufficiently large that all
four leptons can be identified at a reasonable rate.  However the
rapid degradation of these limits as the mass splitting is squeezed
suggests that further adapting multilepton searches to the kinematics
of exotic Higgs decays would be beneficial in order to recover
sensitivity to cascade decays with smaller mass splittings.

As the mass splitting is decreased, VBF and $Wh$ associated production
become more important relative to gluon fusion.  Although VBF
production yields slightly higher-$p_T$ final states than either gluon
fusion or $Wh$, the Higgs exotic decay is still a lower-$p_T$ signal
than most BSM signals sought in multi-lepton searches.  An analysis
more tailored to the specific kinematics of a 125 GeV Higgs could
improve the reach.  Imposing cuts on the transverse mass of the
leptons and the $\met$ could efficiently separate the Higgs signals
from top and fake backgrounds, so long as VBF is more important than
$Wh$; it may also be beneficial to target VBF production directly, by
requiring the presence of tagging jets.  In the CMS multi-lepton
searches, regardless of the mass splittings in the cascade, $Wh$
production dominantly populates the bin with three identified leptons,
one OSSF pair with invariant mass below the $Z$ window, and zero
$\tau$s and $b$-jets, in the lowest $S_T$ $(H_T)$ bin. This is the
same bin that receives the greatest single contribution from gluon
fusion as well.  The background composition in this bin contains a
larger proportional contribution from fake leptons than in bins with
higher $S_T$ \cite{CMS027}, suggesting tighter lepton ID may be
beneficial in optimizing search strategies for the relatively
low-$p_T$ Higgs signal, as well as more aggressive $b$-jet rejection
to suppress backgrounds from top pair production.  Further, $S_T$
regions designed for SM Higgs production mechanisms could help by
concentrating the VBF signal in a single bin (as gluon fusion and $Wh$
already are).

Finally, we comment on the case where the leptons form resonant pairs.
In particular let us consider the decay chains $h\to K_v K_v\to 2Z_D
2\pi_v \to 4\ell+\met$, so that $\Br(h\to 4\ell+\met)/
\Br(h\to\mbox{BSM}) = \Br(Z_D\to\ell ^ +\ell ^ -) ^2 $.  In general,
the signal acceptance in the above multi-lepton searches does not
change substantially relative to the nonresonant signals.  However,
the presence of the leptonic resonances makes these decays much easier
to constrain.  Once again, limits will be highly sensitive to the BSM
mass spectrum, which controls the lepton $p_T$s.  In spectra giving
rise to decays with little to no $\met$, exclusions on the parent
exotic decay could approach the $\lesssim 10^{-3}$ level obtained for
$h\to 4\ell$ decays with no $\met$ (see \S\ref{sec:hto4l}), with
the sensitivity dropping rapidly as the spectrum is squeezed and the
lepton acceptance drops.

	
\section{\secsize $\secbold{h \rightarrow 2\lowerGreekBold\ell+\met}$}\label{sec:2lMET}

In this section, we study exotic Higgs decays to final states that 
contain two isolated leptons and missing energy, where the leptons do
not reconstruct a resonance (we also comment briefly on the case where
they do). Models which realize these decays often also realize decays
with 4 leptons and missing energy, covered in
\S\ref{sec:IsoLeptonMET}. 

\subsection{Theoretical Motivation}

In \S\ref{sec:IsoLeptonMET}, we outlined many classes of
theories where an initial decay $h\to X X$ is followed by the decay
$X\to \ell\ell\met$.  One example, which produces an OSSF lepton pair,
is the decay of a neutralino $\chi_2$ through an off-shell $Z$ boson
to $\ell\ell\chi_1$.  Similarly, a hidden sector meson $K_v$ could
decay through an off-shell dark vector boson $Z_D$ into OSSF leptons
plus a lighter, detector-stable hidden meson, $\ell\ell\pi_v$.

Decays where $h\to 2\ell+\met+X$ can arise in these theories in two
ways. First, in a decay that begins via $h\to\chi_2\chi_2$, one of the
$\chi_2$'s can decay to $2\ell+\met$ while the other decays to $2 j
+\met$ or $2\nu+\met$.  Second, the Higgs will frequently also have
the off-diagonal decay $h\to\chi_2\chi_1$, giving $h\to 2\ell+\met$.
All of these decay chains result in an OSSF lepton pair together with
missing energy and potentially extra soft jets \cite{0710.4591}.

Another realization of the signature $h\to 2\ell+\met$ is found in
theories with a light sterile neutrino, where the coupling $y_i N H
L_i$ gives rise to the decays $h\to \nu N$, followed by both $N\to
\ell_i W^{(*)}\to \ell_i \ell_j \nu$ and $N\to \nu Z^{(*)} \to \nu
\ell\ell$ \cite{0706.1732,0710.4591}.  Decays through the (virtual)
$W$ could yield opposite-sign dileptons with no flavor correlation,
unlike the OSSF pair of leptons generated through $Z^{(*)}$ and
$Z_D^{(*)}$.  These Higgs decays would also be accompanied by
Drell-Yan production of $N\nu$, which yields a non-resonant
contribution to the same final states.

As discussed in \S\ref{sec:IsoLeptonMET}, if there is a light bosonic
state, the decay $\chi_2\to\chi_1 \ell\ell$ can proceed via an
intermediate on-shell state, $\chi_2\to Z_D\chi_1, a\chi_1, s\chi_1$,
such that the leptons reconstruct a resonance.  For dark vector
bosons, the branching ratio to light leptons is appreciable for any
$m_{Z_D}<m_h /2$.  For (pseudo-)scalars with mass-weighted couplings,
such as can appear in the the PQ-NMSSM \cite{HLWY}, we need
$m_{(a,s)}\lesssim 2m_\tau$ for muonic branching fractions to be
significant.  This does not necessarily imply that the muons will be
collimated, as the $a(s)$ is coming from a cascade decay, and
depending on the particular values of $m_2$, $m_1$, may be produced at
relatively low $p_T$.  Nevertheless the experimental searches for
high-$p_T$ isolated leptons almost invariably require $m_{\ell\ell}>
(10$-$12)$ GeV for all OSSF pairs in order to suppress quarkonia
backgrounds, making such searches insensitive to light bosons
regardless of their $p_T$.  We discuss the case of $h\to
(\ell\ell)+\met$ through a low-mass boson like $a$ or $s$ in
\S\ref{Clepton}.

Finally, we also comment that flavor-violating decays $h\to\chi\chi$
followed by $\chi\to \ell q q'$ yield two leptons plus additional soft
jets, albeit no missing energy.  These decays can arise from Higgs
decay to neutralinos, which decay through R-parity violating operators
such as $L_i Q_j d_k$.  They also occur in models where the Higgs
decays to two heavy right-handed neutrinos, followed by $N\to W^{(*)}
\ell\to q q' \ell$ \cite{1107.2123}.  Similar final states can arise in scotogenic models \cite{Ho:2013hia,Ho:2013spa}. 
When the neutrino or neutralino is Majorana, the leptons may have the same sign, yielding a
distinctive signature.

\subsection{Existing Experimental Searches and Limits}    

The signature of $\geq 2$ leptons together with missing energy occurs
in the SM decays of a 125 GeV Higgs boson: the decays of a
Higgs into $WW^*$, $\tau\tau$ and $ZZ^*$, with subsequent decays of
$W/Z$ bosons and taus into leptons and neutrinos give rise to this
final state.  While the decay $h\to Z^{(*)} Z^*\to\ell\ell+\nu\nu$
suffers from a disadvantageous signal-to-background ratio, both $h\to
W W^*\to 2\ell+\met$ and $h\to \tau \tau\to 2\ell + \met$ are standard
SM Higgs search channels.  These SM leptons + invisible Higgs decays
can, depending on kinematics, present an important background for BSM
Higgs searches in leptons plus missing energy final
states. Conversely, existing SM Higgs searches have sensitivity to
begin to constrain BSM leptons + invisible Higgs decays, though the
tailoring of SM Higgs searches to SM decay kinematics reduces their
reach for BSM multi-lepton + missing energy decays
\cite{Chang:2009zpa}.  Associated $Wh$ production also yields
three-lepton final states, but at rates too small to be constrained by
both ATLAS and CMS multilepton searches~\cite{ATLAS035,ATLAS36,CMS026,
  CMS027}.

We will estimate the limits on a benchmark decay chain that begins
with the off-diagonal decay $h\to \chi_1\chi_2$, followed by
$\chi_2\to \chi_1+2\ell$ through an off-shell $Z$,
\beq
\label{eq:targetDK2l}
h\to\chi_1\chi_2\to 2\ell+2\chi_1.
\eeq
We will show results for the optimistic reference working point
presented in the previous section, where $m_{\chi_1}=20$ GeV,
$m_{\chi_2} = 55$ GeV.  Limits for $h\to\chi_2\chi_2\to 2\ell+\met+X$
cascade decays will be less constraining than those for the
off-diagonal decay due to the reduced $\met$.

For the decay $h\to\chi_2\chi_1$, depending on the masses $m_2,m_1$,
the kinematics of the daughter leptons and $\met$ are often broadly
similar to the SM $h\to W W^*$ decay.  Recalling that $\mbox{Br}(h\to
WW^*\to 2\ell 2\nu)\approx 0.26\times 0.103$ and that
$\mbox{Br}(Z\to\ell\ell)\approx 0.102$ (we include $\tau$s), a Higgs
with 10\% branching fraction to $\chi_1 \chi_2$ contributes roughly
40\% the rate of the SM $WW^*$ dileptonic decay mode before acceptance
is taken into account.

Performing a careful recast of SM $h\to WW^{*}$ searches is
challenging as the sensitivity to exotic signals is not
straightforward to extract from the published experimental analyses.
CMS' SM searches use multivariate discriminants to separate signal
from background, rendering a careful recast challenging except in the
earliest analyses (such as \cite{CMS-PAS-HIG-13-022}), which are not
constraining.  Meanwhile, ATLAS's full 7+8 TeV results \cite{ATLAShWW}
extract the SM signal using a multichannel likelihood, and a recast
would require use of the full likelihood function.  Here our main aim
is to estimate the BSM branching fraction into dileptonic modes, which
is allowed by SM Higgs searches. To this end we approximate the BSM
acceptance to be equal to the SM acceptance in the multivariate
discriminants.  This is a conservative choice, but likely to be the
correct order of magnitude for the particular benchmark model we
consider.  For more general choices of $m_1$, $m_2$, the acceptance
will often be significantly reduced relative to this benchmark, as the
daughter leptons may be much softer.

\begin{figure}[t]
\begin{center}
\includegraphics[width=0.5\textwidth]{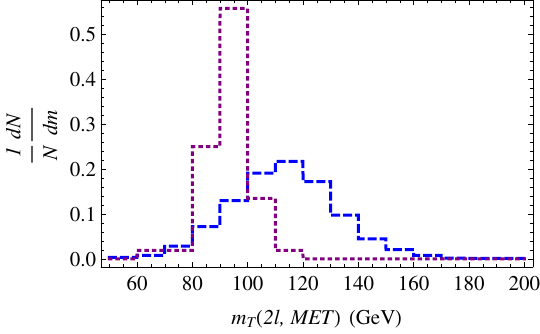}~~~~
\caption{\small Unit-normalized distributions of $m_T (2\ell, \met)$.
  The blue dashed line shows the ATLAS prediction for SM $h\to W W^*$
  events passing all selection criteria in both 7 and 8 TeV data sets~\cite{ATLAShWW}. 
  The purple dotted line shows the distribution for
  the BSM $h\to 2\ell+\met$ events arising from $h\to\chi_2\chi_1$ at
  the 8 TeV LHC in the benchmark model described in the text. }
\label{fig:2lMETmTdist}
\end{center}
\end{figure}

As in the previous section, to obtain these limits we use MadGraph 5
and Pythia 6 to generate gluon fusion Higgs signal events, matched out
to one jet.  For CMS searches, we employ a version of PGS tuned to
CMS' operating parameters.  For ATLAS searches, we use parameterized
lepton efficiencies as reported in the searches under consideration,
with jet clustering performed in FastJet. We neglect VBF production,
as well as the VBF-like event categories in the ATLAS and CMS
searches.

The ``cut-based'' analysis of the full 7+8 TeV CMS $0j$ and $1j$ $h\to
WW^*$ analysis \cite{CMS-PAS-HIG-13-003} employs a multivariate
discriminant in states with same-flavor leptons to separate $h\to
WW^*$ signal from Drell-Yan pair production.  Approximating the
efficiency of this multivariate discriminant at the SM Higgs-like
value $\epsilon\approx 0.5$ on the BSM decay mode $h\to \chi_2
\chi_1\to 2\ell +\met$, and combining the effect of this multivariate
cut with the rest of the analysis selection, we can estimate the ratio
of the BSM signal to the SM signal.  Using CMS' best fit for the SM
signal strength $\mu$ in the $h\to WW^*$ mode in the 0 and 1 jet
categories,
\beq
\left.\mu\right|_{fit} = 0.79\pm 0.38,
\eeq
we estimate 
\beq
\label{eq:approxbound2lmet}
\frac{\sigma(pp\to h)}{\left.\sigma(pp\to h)\right|_{SM}}\mbox{Br}(h\to \chi_1 \chi_2) \lesssim 1.0 
\eeq
for the reference benchmark point. Again, this limit includes an
assumed factor of Br$(Z\to \ell^+\ell^-)\sim 0.102$; decay chains with
off-shell dark photons, which have leptonic branching fractions
roughly 4 times larger, are subject to the tighter constraint Br$(h\to
\chi_2\chi_1)\lesssim 0.24$.

Meanwhile, in the ATLAS analysis \cite{ATLAShWW}, the final step in
the analysis is fitting SM signal and background distributions in the
transverse mass variable $m_T(2\ell, \met)$.  ATLAS'
background-subtracted predictions for the SM signal strength are shown
in Fig.~\ref{fig:2lMETmTdist}, together with the prediction from the
BSM benchmark, to indicate the degree of similarity between the two
signals in the final discriminating variable.  The cuts employed in
the ATLAS analysis give comparatively less sensitivity to the BSM
signal than do the CMS cuts.  As a consequence, under the simplifying
assumption that the SM and BSM signals are extracted with similar
efficiency in the final fit, no limit is placed on the branching
fraction into the BSM final state.

Since the signal investigated in this section contributes almost
entirely to same-flavor final states, better sensitivity could be
obtained by considering different-flavor and same-flavor final states
separately.  As our recasting is highly approximate due to the lack of
information about the multivariate discriminants employed in the
same-flavor final states, we will simply mention this as one obvious
avenue for improving on the approximate bound shown in
Eq.~(\ref{eq:approxbound2lmet}).  In cases where the two leptons
reconstruct a resonance, significantly better limits may be possible.
Meanwhile the heavy neutrino decay through a (virtual) $W$, which does
contribute to different-flavor final states, would show interesting
departures from flavor universality depending on the flavor mixings in
the neutrino sector; this heavy neutrino model should be looked for
simultaneously in Drell-Yan and Higgs decays as the ratio of the two
signals is fixed.

	\section{\secsize $\secbold{h \rightarrow}$ {\normalsize One Lepton-jet +} $\secbold{X}$}

\label{Clepton}

In this and the following section, we study exotic Higgs
decays to final states that contain one or two {\em low-mass}
resonant lepton pairs.  Higgs decays to collimated pairs of leptons
(here $\ell=e,\mu$ but not $\tau$) have been a focus of much
experimental and theoretical work.  Searches for collimated pairs of
leptons are typically carried out inclusively, that is, no
attempt to reconstruct the Higgs mass is made.  Thus the same searches
constrain decays both with and without the presence of $\met$,
although events with $\met$ (or other Higgs daughter products, such as
soft jets) will typically have reduced acceptance.  In this section, we
consider Higgs decays to one lepton-jet$+X$, and in the following
section we consider Higgs decays to two lepton-jets$+X$.  For
simplicity we focus on {\em simple} lepton-jets, consisting of a
collimated pair of either electrons or muons; {\em complex}
lepton-jets, which have a larger and more variable particle content
that can involve hadrons and detector-stable states as well as
leptons, are important and interesting signals, but less transparent
to survey.

In the current section we study Higgs decays to one (simple)
lepton-jet$+X$.  Because experimental backgrounds for a single
lepton-jet are higher than those for two, traditionally the focus has
been on signals with two lepton-jets.  In this section we emphasize,
firstly, that there are well-motivated signals that produce a {\em
  single} lepton-jet only or dominantly, and secondly, that exclusive
analyses targeting these states can yield meaningful sensitivity to these
decays.

The opening angle of two partons coming from a parent particle $X$ can
be roughly estimated as $\Delta R\simeq 2 m_X/p_{T,X}$.  We can
estimate $p_{T,X} \sim 50$ GeV, for a particle $X$ coming from the
decay of a 125 GeV Higgs produced at rest. Partons from the $X$ decay
are then typically separated by $\Delta R < 0.2$ when $m_X \lesssim 5 $ GeV.
Therefore, we expect to have a Higgs decaying into collimated leptons 
that fail typical isolation cuts requiring $\Delta R > 0.4$
if the parent particle $X$ has a mass of the order of 10 GeV or
less. Meanwhile if the parent particle $X$ is produced in a cascade
decay instead of directly, it will be less boosted.  Clearly the
transition between having isolated leptons and collimated leptons
happens smoothly as a function of the parent particle mass $m_X$.
The reader may also be interested in 
\S\ref{sec:hto4l}, which considers isolated leptons with
$m_{\ell\ell}>4$ GeV.

\subsection{Theoretical Motivation}

One theory that realizes the decay $h \to (\mu\mu) +\met$ is is
the PQ-symmetric limit of the NMSSM~\cite{draper2011dark,HLWY}.  In
this limit, the degrees of freedom $(s, a, \chi_1)$ (scalar,
pseudo-scalar, and fermion, respectively) comprising the singlet
multiplet are all light.  Decays of the Higgs to $h\to\chi_1\chi_2$ or
$h\to\chi_2\chi_2$, with subsequent decays $\chi_2 \to \chi_1 s,
\chi_1 a$, give Higgs decay signatures with missing energy in the
final state.  In an appreciable portion of parameter space, these
decays can dominate the exotic Higgs branching fraction, as detailed
in \S\ref{NMSSMfermion} and Refs.~\cite{draper2011dark,HLWY}.
If $s$ (or $a$) is very light, with mass order $m\lesssim \mathcal{
  O}(1)$ GeV or below, phase space forbids decays to heavier fermions
and the branching fraction into light leptons becomes appreciable
($\mathcal O(10\%)$; see, e.g., Fig.~\ref{fig:fBRs2HDMatype2}).  The
resulting signatures are dileptons + $\met$ for $h\to \chi_1\chi_2$
and four leptons + $\met$ for $h\to \chi_2\chi_2$, which correspond to
the type-I and type-III decay topologies presented in
\S\ref{sec:IsoLeptonMET}.  The $s(a)$ is produced with a $p_T$ that
is dependent on the masses of $\chi_2$ and $\chi_1$, but in the regime
where decays to muons dominate, typically we will have $p_{T(a,s)}\gg
m_{a,s}$, and the daughter muons will be collimated: $\Delta
R_{\ell\ell} \lesssim 0.1$.

Dark vector boson models can also realize the collimated leptons$+\met$
Higgs decay signature.  In a supersymmetric context, $\chi_2$ would
now be mainly bino and $\chi_1$ a dark photino, but in this case the
off-diagonal $h\to \chi_2\chi_1$ decay can only be important if the
decay $h\to \chi_2\chi_2$ is kinematically forbidden.  In a more
general hidden sector, the role of the neutralinos $\chi_i$ may be
played instead by hidden sector mesons $K_v$, $\pi_v$ or similar
states, see \S\ref{sec:hiddenvalley}.  Dark photon models can also yield Higgs
decays of type II topology (see \S\ref{sec:IsoLeptonMET}).  In this case, the
Higgs decays directly to dark vectors, $h\to Z_D Z_D$, followed by
$Z_D\to$ lepton-jet on one side and $Z_D\to$ invisible on the
other. Here the invisible states are detector-stable hidden sector
states, perhaps dark photinos \cite{Strassler:2006qa,Baumgart:2009tn, Falkowski:2010cm,
  Chan:2011aa}; the relative branching fractions to leptons, $\met$,
and other SM partons are model-dependent.  Similar signatures can be
obtained in the R-symmetric NMSSM if the light pseudo-scalar is
coupled to a hidden sector.  Another possible realization of the type
II topology is provided by the decay $h\to Z Z_D$, followed by $Z\to
\nu\nu$.
 
Also a possibility are decays $h\to (\mu\mu) + (jj)$, i.e., where the
lepton-jet recoils against hadronic activity.  This kind of decay
arises in, e.g., the R-symmetric limit of the NMSSM, where $h\to a a$
is followed by $a\to \mu\mu$ on one side of the event, and $a\to$
hadrons on the other. As $\Br(a\to \mu\mu)\lesssim 0.1$ even below the
$\tau$ threshold, $\Br(h\to (\mu\mu)(jj))> \Br (h\to 2(\mu\mu))$;
however the $2(\mu\mu)$ final state has notably lower background, as
well as sharper resolution.  Similarly, $h\to Z_D Z_D \to (\ell\ell)(j
j)$ leads to a lepton-jet balanced against a ``weird'' hadronic jet.

Unlike the NMSSM (pseudo)scalars, dark photons have appreciable
branching fractions to light leptons even for large masses $m_{Z_D}$.
However, possible connections with cosmic ray anomalies
\cite{ArkaniHamed:2008qn,Pospelov:2008jd} and the discrepancy between the 
measured and calculated muon anomalous magnetic moment~\cite{Pospelov:2008zw} 
have stimulated interest in
dark vectors with a mass at or below the GeV scale, thus involving
collimated leptons in the final state.  For discussion of dark vectors
outside the collimated regime, see \S\ref{sec:htoZa}
and \S\ref{sec:hto4l}.

\subsection{Existing Collider Studies}

A dedicated analysis for $h\to \chi_1\chi_2 \to \ell^+\ell^- +\met$ 
is presented in~\cite{HLWY}, which indicates that the 8~TeV LHC could
have good sensitivity to this final state when a targeted search is
performed that exploits the $\met$ in the final state from the Higgs decay. As an
illustration, the analysis focuses on a benchmark inspired by the
PQ-symmetric limit of the NMSSM, with a light scalar(pseudoscalar) resonance $s(a)$ set
to have a mass of 1 GeV (see Table~\ref{table:benchmark1}).

\begin{table} [t]
\begin{center}
\begin{tabular}{c|c|c|c}\hline\hline
 $m_{s}$ &  $m_{h}$  &  $m_{\chi_1}$ & $m_{\chi_2}$  \\ \hline
 1 GeV & 125 GeV & 10 GeV & 80 GeV
  \\ \hline
\end{tabular}
\end{center}
\caption{Mass parameters of the $h \rightarrow $ collimated leptons + $\met$ benchmark model.}
\label{table:benchmark1}
\end{table} 

The analysis focuses on the $W^\pm h$ production mode where the $W$
decays leptonically. The resulting signature contains one hard lepton
($e$, $\mu$) from the $W$ decay, two collimated muons, and $\met$.
Since there are no jets in the hard scattering process, the $W + $jets, $Z + $jets, and $t \bar{t}$ backgrounds can be efficiently
eliminated with a jet veto. The diboson $WZ$ and $ZZ$ backgrounds are
be removed by a dimuon mass window cut.  A muon isolation cut is
applied to remove the low-mass dimuon background from meson decays,
which requires the transverse momentum sum of hadronic jets (excluding
the contribution from any nearby muons) in a cone of $R = 0.4$ around
each muon candidate to be less than 5~GeV. Then the light resonance
can be reconstructed via the two nearby muons, and the main background
is $W \gamma^*/Z$, with $\gamma^*/Z$ decaying into $\mu^+\mu^-$.  A
trilepton trigger is assumed in the analysis, though alternatively,
one can trigger on the single lepton from the $W$ decay.  The analysis
indicates that, with 20~fb$^{-1}$ data, a sensitivity $S/\sqrt{B} > 6
\sigma$ can be achieved at the 8~TeV LHC, with
\begin{eqnarray}
c_{\rm eff} =  \frac{\sigma(h)}{\sigma(h_{\rm SM})}
\times {\rm Br}(h \to \chi_1 \chi_2)  \times {\rm Br}(\chi_2 \to s \chi_1)
\times {\rm Br}(s \to \mu^+\mu^-) = 0.1
\end{eqnarray} 
assumed. Details of the analysis can be found in~\cite{HLWY}.

This analysis for searching for a dimuon resonance with $\met$ can be
easily generalized to other related possibilities.  If the light
resonance is a vector, then a wider range of masses should be
considered, which would result in a larger average separation between
the two daughter leptons. Another possibility arises from the decay chain 
$h\to \chi_2\chi_2 \to (\mu\mu)(\tau\tau)+\met {\rm~or~} (\mu\mu)(bb) +\met$
(for details, see \S\ref{NMSSMfermion}). Obviously such decay chains 
can be picked up also by the proposed collider search. Further, although 
in this analysis only $Wh$ events are considered, it is straightforward to generalize the
analysis to $Zh$ events that trigger on the leptons from the $Z$
decay.  It is also of interest to consider gluon fusion and VBF
production, where lepton-jet or even dilepton triggers may yield a
reasonable acceptance for this decay mode. We leave this question for
future work.

\subsection{Existing Experimental Searches and Limits}
\label{ExpBounds}

Leptons arising from very light parents will typically fail standard
isolation requirements, and isolated leptons + $\met$ searches at LHC
are not very sensitive to such scenarios.  Even in searches where
lepton isolation criteria are relaxed, typically a cut is placed on
the invariant mass of any opposite-sign, same-flavor lepton pair in
the event, usually $m_{\ell\ell} > 10-12$~GeV (in some cases
$m_{\ell\ell}>$4 GeV), in order to suppress backgrounds from
quarkonia.  Thus even if a light boson were produced with moderate to
low $p_T$, it would be missed by most searches in leptonic final
states.  The potentially significant bounds come from dedicated
searches for lepton-jets, where modified lepton isolation criteria are
applied, and low mass ranges are considered.  Searches for lepton-jets
have been pursued by both CMS~\cite{Chatrchyan:2011hr} and
ATLAS~\cite{Aad:2012kw,Aad:2012qua,Aad:2013yqp}.

In the ATLAS analyses, either a displaced vertex for the
lepton-jets~\cite{Aad:2012kw}, or at least four muons within a single
lepton-jet~\cite{Aad:2012qua}, or at least two lepton-jets are
required~\cite{Aad:2012qua,Aad:2013yqp}. All of these three features
are absent in the scenario
\beq
\label{eq:PQasymmdecay}
h\to \chi_1\chi_2 \to \ell^+\ell^- +\met \, .
\eeq
The most relevant search is from the CMS search for light resonances
decaying into pairs of muons~\cite{Chatrchyan:2011hr}, which sets an
upper bound on the cross-section for $p p \to \mathrm{(\phi\to
  \mu^+\mu^-)}+X$ for new bosons $\phi$ with masses below 5 GeV, using
35 pb$^{-1}$ of data collected at the 7 TeV LHC.  Selection cuts of
$|\eta_{\mu\mu}| < 0.9$ and $p_{T,\mu\mu} > 80$ GeV are applied for
the muon pair. As indicated by the study in~\cite{HLWY}, most events
arising from the decay mode of Eq.~(\ref{eq:PQasymmdecay}) cannot pass
the CMS selection cuts because the $s$-originating dimuon pairs are
too soft, with an average $p_T \sim 40$ GeV. The signal efficiency of
the CMS selection cuts is $ \epsilon \lesssim 0.7\%$ for the benchmark
introduced below, and roughly of the same order for a lighter $s$.
Then the signal cross section is given by $\sigma_{h_{\rm SM}}\times
c_{\rm eff} \times \,\epsilon\, \sim (0.1\;{\rm pb}) \times c_{\rm
  eff}$, with $c_{\text{eff}} = \frac{\sigma(h)}{\sigma(h_{\rm SM})}
\times {\rm Br}(h \to \chi_1 \chi_2) \times {\rm Br}(\chi_2 \to s
\chi_1) \times {\rm Br}(s \to \mu^+\mu^-)$, which well satisfies the
$0.15-0.7$ pb limit for masses $\lesssim 1$ GeV at 95\% C.L. (at the
mass point $m_{\ell\ell} \sim 1$ GeV, the limit is $\sim 0.4$ pb)
obtained in~\cite{Chatrchyan:2011hr}.  This CMS analysis is
not updated yet to use the full LHC Run 1 data set.  The experimental
bounds obtained by the LHC searches therefore do not place any limits
on the branching fraction Br$(h\to \mu^+\mu^- +\met)$ in the
collimated/low-mass regime.

\subsection{Proposals for New Searches at the LHC}

A search for $h\to$ one lepton-jet (or one light resonance)+$\met$ is highly motivated  
on both theoretical and experimental sides. Theoretically, such a decay topology can 
arises in a couple of well-motivated scenarios. Experimentally, $\met$ and the light  
resonance reconstruction bring new inputs for exploring new physics. Using the full 7 and 8 TeV 
dataset of both experiments, strong constraints or discovery-level sensitivity might be achieved. 
As is illustrated in~\cite{HLWY}, for $h\to$ one lepton-jet($\mu\mu$) +$\met$ and 
$c_{\rm eff} = 0.1$,  a sensitivity $S/\sqrt{B} > 6 \sigma$ can be achieved, using 
20 fb$^{-1}$ of data at the 8 TeV LHC. Though the light resonance is assumed to 
be $\sim 1$~GeV, a good sensitivity for probing a wider range of masses should be expected.

\section{\secsize $\secbold{h \rightarrow}$ Two Lepton-jets + $\secbold{X}$}
\label{sec:collep_nomet}

Here we consider Higgs decays to 2 lepton-jets$+X$; see also the
previous section for related signatures.  Again, for simplicity we
concentrate on {\em simple} lepton-jets, consisting of a single
collimated electron or muon pair.

\subsection{Theoretical Motivation}

As mentioned in the previous section, one well-studied model for a
Higgs decaying to pairs of collimated muons is the NMSSM.  Here the
Higgs decays via $h\to a a$, with $a$ subsequently decaying to pairs
of SM partons according to the Yukawa couplings of a Type II 2HDM 
model plus a singlet.  The branching ratios of $a$ to SM
partons are shown in Fig.~\ref{fig:fBRs2HDMatype2}.  Notably, in the
NMSSM, branching fractions of $a$ into a muon pair only reach the
$\mathcal{O}(\mbox{few~}\%)$ level even below the mass threshold $m_a <
2 m_\tau$.  This necessarily places the pseudoscalar $a$ in the mass
range to produce collimated daughter muons.  Another way to realize
$h\to 2(\mu\mu)+X$ arises in the PQ-symmetric limit of the NMSSM
(\S\ref{NMSSMfermion}), where the initial Higgs decay is into
neutralinos, producing light (pseudo)scalars in subsequent cascade
decays, $h\to \chi_2\chi_2$, $\chi_2\to (a)s \chi_1$ (see also
\S\ref{Clepton}). In this case the light scalar will typically be less
boosted, but in the mass range where decays to muons are relevant, the
muons will generally still be collimated.

In any singlet-augmented 2HDM model, once $m_a>2m_\tau$, the branching
fraction for $a\to \mu\mu$ will always be suppressed by the small
ratio $m_\mu^2/m_\tau^2 \sim 3.5\times 10^{-3}$.  As discussed in
\S\ref{sec:hto4tau}, the tiny branching fraction into $h\to 4\mu$ is
not competitive with $4\tau$, $2\mu 2\tau$.  Thus if $a$ couples proportional to 
mass, only the range $2 m_\mu < m_a < 2 m_\tau$ is of interest for the
decay $h\to 4\mu$.  Decays to electron pairs are always negligible
(unless $m_a \lesssim 2m_\mu$, which we do not consider
comprehensively here).  

Higgs decays to collimated lepton pairs may also arise in models with
light vector bosons $Z_D$ that mix with the SM hypercharge gauge boson
(see \S\ref{subsec:SMvector}). The motivation to consider $m_{Z_D} \ll
m_h$ has been driven by dark matter models that require $m_{Z_D}\sim$
GeV or below \cite{ArkaniHamed:2008qn,Pospelov:2008jd}.  In these
models, the branching fractions of $Z_D$ depend on the SM fermion
gauge couplings, rather than on Yukawas, and therefore electron and
muon pairs are produced with comparable branching fraction unless
$Z_D$ is extremely light, $m_{Z_D}\lesssim 2 m_\mu$.  Importantly  \cite{Strassler:2006im}, the
branching fraction for $h\to 2 (\ell\ell)$ remains large even when
$m_{Z_D}>2m_b$, motivating searches for both electrons and muons in
this mass range.

Dark photon models can give $h\to 2$ lepton-jets directly, via an
initial decay $h\to Z_D Z_D$, as well as $h\to 2 \,
\mathrm{lepton-jets} +\met$.  There are two distinct possibilities for
obtaining $\met$.  One possibility is that non-trivial showering of
the dark vector boson occurs, resulting in the production of
detector-stable states in the dark sector together with leptons
\cite{Strassler:2006im,Baumgart:2009tn, Falkowski:2010cm}, yielding complex lepton-jets
containing $\met$.  Another possibility is that the Higgs decays first
to (e.g.) bino-like neutralinos $\chi_2$, which then (similar to \cite{Strassler:2006qa}) decay to a dark
vector and a dark photino, $\chi_2\to \chi_1 Z_D$.  Since bino-dark
photino mixing is proportional to the kinetic mixing parameter
$\epsilon\ll 1$, off-diagonal decays $h\to \chi_2\chi_1\to Z_D 2
\chi_1$ are negligible in comparison to the unsuppressed $h\to
\chi_2\chi_2\to 2 Z_D 2\chi_1$ as long as both decays are
kinematically available.  In a non-supersymmetric case, the role of
the neutralinos $\chi_i$ may be played instead by hidden sector mesons
$K_v$, $\pi_v$ or similar states \cite{Strassler:2006im}, and the off-diagonal decays may not be suppressed; see \S\ref{sec:hiddenvalley}.

\subsection{Existing Collider Studies}

A collider search for $h\to 2a\to 4\mu$ was first proposed in
\cite{Zhu:2006zv}, which took $m_a \approx 215$ MeV, as motivated by an
excess in HyperCP measurements of $\Sigma^+\to p\mu^+\mu^-$
decay~\cite{Park:2005eka}.  This study pointed out that modifications
of the (then-)standard muon isolation algorithms would be required to
preserve the signal, and concluded that as long as reasonable
efficiencies for muon identification could be maintained, the signal
had excellent prospects for detection.  However the dominant QCD
background to this signal was not identified.  A more careful
treatment of the dominant QCD backgrounds was carried out in
\cite{Belyaev:2010ka}, which concluded that the signal would still be
nearly background-free, with excellent prospects for discovery in
early 14 TeV LHC running (considering exotic branching fractions of
tens of percent).

Ref.~\cite{Falkowski:2010gv} performed a collider study of the
Higgs decaying to multiple electron-jets plus $\met$ through a
$100\mev$ $Z_D$. Production in association with a leptonic $W$ or $Z$
was identified as the most promising channel, in which the dominant
background is $W$ or $Z$ plus QCD jets. Ref.~\cite{Falkowski:2010gv}
found that an analysis distinguishing electron-jets from QCD jets
using the electromagnetic fraction and charge ratio of the jet
candidates could discover the Higgs with $1~{\rm fb}^{-1}$ of 7~TeV
LHC data at 95\% CL with $\Br\left(\hsm\to {\rm
    electron~jets}+\met\right)=1$ for $m_\hsm<135\gev$.

\subsection{Existing Experimental Searches and Limits}

The $h\to 2(\mu\mu)$ signature has become established in experimental
programs, beginning with the D0 search \cite{Abazov:2009yi}.  The most
stringent constraints on $h\to 2(\mu\mu) +X$ are set by the LHC, where
several searches have been carried out, looking for Higgs decays to
both prompt \cite{Chatrchyan:2011hr, Chatrchyan:2012cg,
  CMS-PAS-HIG-13-010} and displaced \cite{Aad:2012kw} dimuon jets.  As
this final state is extremely clean, these searches are carried out
inclusively, and in particular do not require $m_{4\mu}=m_h$.  Thus
these searches are sensitive to both the NMSSM-like $h\to aa \to
2(\mu\mu)$ decay topology and the SUSY-dark vector-like topology $h\to
\chi_2\chi_2\to 2(\mu\mu) 2\chi_1$, where the dimuon jets are
accompanied by missing energy.

The best existing limits on prompt $h\to 2 (\mu\mu)+X$ come from the
recent CMS analysis \cite{CMS-PAS-HIG-13-010}, which was performed
with the full 8 TeV data set. This search, like the previous CMS
search \cite{Chatrchyan:2012cg}, only covers the range $2 m_\mu < m_a
< 2 m_\tau$.\footnote{It also requires the two lepton-jet masses to be
  within 0.1 GeV of each other, meaning it is insensitive to decays $h
  \rightarrow a_1 a_2$ with $a_1 \neq a_2$.}  This search limits
\beq
\label{eq:dilepjetbound}
\sigma (p p \to 2a+X)\Br (a\to\mu\mu)^2 \alpha_{\rm gen} < 0.24\; \mathrm{fb}
\eeq
at 95\% CL over almost all of the mass range in consideration, where
$\alpha_{\rm gen}$ is a (model-dependent) fiducial acceptance. This
translates to a limit 
\beq
\Br\left(\hsm\to aa\right)\Br (a\to \mu\mu)^2 < 1.2\times 10^{-5}
\eeq
for $m_\hsm=125~\gev$.  Outside this mass range, the 35 pb$^{-1}$
search of Ref.~\cite{Chatrchyan:2011hr} extends to 5 GeV, placing
limits of $\sigma (p p \to 2a+X)\Br (a\to \mu\mu)^2 \epsilon < 125$
fb, where $\epsilon$ is again an acceptance.

The analysis of Ref.~\cite{CMS-PAS-HIG-13-010} has been presented in a
way that is particularly easy to recast.  Limits are shown as a
function of the parameter $\alpha_{\rm gen}$, which represents the
{\em generator level} efficiency for a given signal to have at least
four muons satisfying $p_T > 8$ GeV, $|\eta|<2.4$ and at least one
muon to have $p_T>17$ GeV, $|\eta|<0.9$.
Ref.~\cite{CMS-PAS-HIG-13-010} estimates a systematic uncertainty on
the relation of $\alpha_{\rm gen}$ to the full efficiency of
approximately $7.9\%$.  We show some reinterpretations of the bound of
Eq.~(\ref{eq:dilepjetbound}) for the cascade decay $h\to \chi_2\chi_2$,
$\chi_2\to a (Z_D)\chi_1$, $a (Z_D)\to \mu\mu$ in Fig.~\ref{fig:2lj}.
Gluon fusion Higgs events are generated in MadGraph 5 and showered in
Pythia 6, matched out to one jet.  Our signal model contains no spin
correlations; a proper treatment of spin would yield small corrections
to the muon acceptance.  We show results for masses $m_a (m_{Z_D}) =$
0.4 GeV (blue), 1 GeV (green), and 3 GeV (red).  Dark vector branching
fractions to muons are taken according to the tree-level computation
of \S\ref{subsec:SMvector}, while a reference branching fraction
$\Br(a\to \mu\mu)=0.1$ is assumed.  Caution should be used in
interpreting the recast limits for the smallest values of $m_2-m_1$,
which is furthest from the spectra considered in
Ref.~\cite{CMS-PAS-HIG-13-010}, as in this region the linear relation
between $\alpha_{gen}$ and the full experimental efficiency may no
longer hold.

\begin{figure}
\begin{center}
\includegraphics[width=0.49\textwidth]{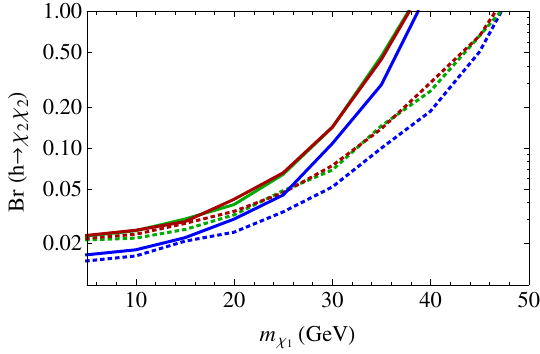}
\includegraphics[width=0.49\textwidth]{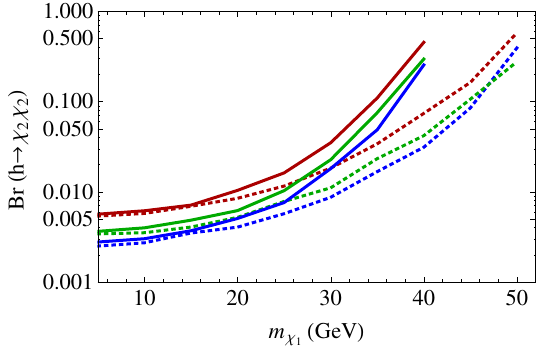}
\end{center}
\caption{ {\small Approximate bounds on the branching fraction for
    $h\to \chi_2\chi_2$, assuming {\bf (left)} $\Br(\chi_2\to a \chi_1)
    = 1$, and {\bf (right)} $\Br(\chi_2\to Z_D \chi_1) = 1$, as a
    function of $m_{\chi_1}$, from \cite{CMS-PAS-HIG-13-010}.  Here
    solid lines indicate $m_{\chi_2}= 50$ GeV and dotted lines
    $m_{\chi_2}=60 $ GeV, while red, green, and blue correspond to
    $m_{a, Z_D} = $ 3 GeV, 1 GeV, and 0.4 GeV respectively.  We use
    tree-level results for $\Br (Z_D\to \mu\mu)$ (see Fig.~\ref{fig:SMVBrZdark}) and a reference
    $\Br(a\to \mu\mu) = 0.1$ (which can occur in Type IV 2HDM+S models, see Fig.~\ref{fig:fBRs2HDMatype4}). } }
\label{fig:2lj}
\end{figure}

Searches in electron-jets are more challenging, as backgrounds from
QCD jets with a large electromagnetic fraction are significant, and as
identifying collimated electrons from BSM physics is complicated by
photon conversions.  Nonetheless, searches for $h\to 2$ electron-jets
have been carried out, targeting $Wh$ associated production first at
CDF with 5.1 fb$^{-1}$ data~\cite{Aaltonen:2012pu} and later at ATLAS
with 2.04 fb$^{-1}$ of 7 TeV data \cite{Aad:2013yqp} and inclusively
for pairs of electron-jets with 5 fb$^{-1}$ of 7 TeV data
\cite{Aad:2012qua}.  It is challenging to reinterpret either of these
searches as a limit on Higgs decays to {\em simple} electron-jets, as
both require $> 2$ tracks per electron jet, to better reject photon
conversions. 
	\section{\secsize $\secbold{h\to b\bar b+\met}$}
\label{bbmet}

Decays of the form $h\to b\bar b+\met$ can be classified into two main
types, assuming a primary two-body decay stage $h\to X_1 X_2$:
\begin{enumerate}[I.]
\item $X_1 \to \met$, \ $X_2\to  b\bar b+\met$, 
\item $X_1 \to \met$, \ $X_2\to  b\bar b$.   
\end{enumerate}
Here, $X_{1,2}$ are intermediate on-shell particles (possibly the same
particle undergoing different decays), and $X_1$ is either stable and
invisible, or decays invisibly.\footnote{A logical third option that
  leads to this final state would be a decay into a pair of
  bottom-partners, that each subsequently decay to $b+\met$.  However,
  this option is now almost entirely ruled out \cite{Batell:2013psa}.}
The $b\bar b$ pair may either be resonant or nonresonant in general
for first class of decays, though we will mainly assume that it is
resonant.  The second class is resonant by definition.  Below,
theoretical motivations and experimental search strategies will be
discussed.  As we will see, decays with a $b\bar b$ resonance might
lead to an observable signal at the 14~TeV LHC.

\subsection{Theoretical Motivation}
\label{subsec:2bmettheorymotivation}

\begin{itemize}

\item NMSSM in PQ-symmetry limit: $h\to \chi_1 \chi_2$ (topology I,
  resonant); see also Hidden Valleys (\S\ref{sec:hiddenvalley});

  Here, $X_1$ and $X_2$ represent the lightest and the
  next-to-lightest neutralinos $\chi_1$ and $\chi_2$, respectively,
  with $\chi_2$ decaying to $\chi_1$ plus a scalar or pseudoscalar of
  the extended Higgs sector.  For details on the decay $h\to
  \chi_1\chi_2$ (and $h\to \chi_2\chi_2$) and example parameter
  points, see \S\ref{NMSSMfermion} or~\cite{draper2011dark,HLWY}.  If
  the scalar is heavier than $2 m_b$, its decays are typically
  dominated by $b\bar b$.  The signatures at colliders will then be
  one or two $b$-jets + $\met$, depending on how collimated the two
  $b$ quarks are.

  If $m_{\chi_2} - m_{\chi_1}>m_Z$, the decay $\chi_2 \to \chi_1 Z$ is
  open and the $Z$-boson can further decay into a $b \bar b$ pair.
  However, this decay tends to be kinematically disfavored.

\item $\nu$SM: $h\to \nu N$ (topology I, resonant or non-resonant)

  In the $\nu$SM, the Higgs can decay into an active neutrino and a
  sterile neutrino via the neutrino portal Yukawa interaction,
  Eq.~(\ref{eq:YukawaNeutrino}).  In this case, we identify $X_1 =
  \nu$ and $X_2 = N$, and the topology is the same as in the
  PQ-symmetric NMSSM.  The mass mixing between RH sterile neutrinos
  and LH active neutrinos allow the RH neutrinos to decay via $N \to
  \nu Z^{(*)}\to \nu b \bar b$. For more details, refer to
  \S\ref{subsub:SMF}.

\item Other models: $h\to a a, Z_D Z_D, \phi_1 \phi_2$ (topology II)

  In the PQ limit of the NMSSM (\S\ref{NMSSMscalar}) it is possible
  for $a$ to decay competitively into singlinos as well as bottom
  quarks. In that case, the decay $h\rightarrow 2a \rightarrow 2b +
  \met$ may be realized. Dark vector extensions
  (\S\ref{subsec:SMvector}) will usually have an invisible decay mode
  $Z_D \rightarrow \bar \nu \nu$, so the $2b + \met$ final state can
  occur (even if it may not be the first discovery channel for such a
  model). Finally, it is of course possible to imagine a more
  complicated hidden sector (see e.g. \S\ref{sec:hiddenvalley}) where
  $h\rightarrow \eta_1 \eta_2$ and $\eta_1 \rightarrow \bar b b$ but
  $\eta_2$ is invisible or decays invisibly.

\end{itemize}

\subsection{Existing Collider Studies}

As the kinematics of $h\to b\bar b+\met$ can be significantly
different from the standard $h\to b\bar b$ decay, dedicated analyses
are required to search for it.  Inspired by the PQ-limit of the NMSSM,
a dedicated study of this process has recently been
performed~\cite{HLWY}. The signals from gluon fusion and vector boson
fusion production would be overwhelmed by QCD backgrounds (similar to
SM $h\to b\bar b$), even if they could be triggered on, so the
analysis focuses on vector boson associated production, triggering on
leptonic boson decays.  As an illustration, $Zh$ with $Z\to
e^+e^-/\mu^+\mu^-$ is considered. In addition to two neutralinos
$\chi_1,\chi_2$, the final state includes a spin-0 state $s$ (either
scalar or pseudoscalar) that decays to $b\bar b$. The study is based
on a benchmark model in the PQ-limit of the NMSSM, with its
parameters presented in Table~\ref{table:bbMETbenchmark2w}.  The main
backgrounds include $Z b \bar{b}$, $Z c \bar c$, $Z c + Z \bar c$ and
$t \bar{t}$+jets.
\begin{table} [h!]
\begin{center}
\begin{tabular}{c|c|c|c} \hline 
   $m_h$ &  $m_{\chi_2}$ & $m_{\chi_1}$ & $m_s$   \\ \hline
 125 GeV &     80 GeV   &   10 GeV    & 45 GeV
  \\ \hline
\end{tabular}
\end{center}
\caption{Benchmark  masses used for the $h \to b\bar b + \met$ collider analysis of~\cite{HLWY}.}
\label{table:bbMETbenchmark2w}
\end{table} 
The analysis includes basic detector effects but no pile-up
simulation. Jet substructure tools~\cite{Butterworth:2008iy} are also
applied to investigate $b$-tagged fat-jets. The analysis indicates
that $\sim 2 \sigma$ sensitivity to $\Br(h\rightarrow X_1 X_2
\rightarrow 2b + \met) = 0.2$ may be possible at the 14 TeV LHC with
$300~\ifb$, though it is very challenging, and more realistic studies
are needed.

\subsection{Existing Experimental Searches and Limits}

Although the signature $h \rightarrow b\bar b + \met$ is
well-motivated, dedicated experimental searches have not yet been
performed. There are similarities to the SM Higgs decay $h\rightarrow
\bar b b$, but the generally softer bottom quarks and lower rate make
this a more challenging signal to detect. The $h\rightarrow b \bar b$
searches from $(W\to \ell \nu)h$, $(Z\to \ell\ell)h$ and $(Z\to \nu
\nu)h$ production by both the CMS and the ATLAS
collaborations~\cite{CMS-PAS-HIG-13-012,ATLAS-CONF-2013-079} have only
recently achieved SM sensitivity, yielding no constraints on the rarer
$2b + \met$ final state. The $(Z\to \nu \nu)h$ search could in
principle be sensitive to this exotic Higgs decay from ggF and VBF
production channels, with the orders-of-magnitude larger production
rate offsetting the subdominant exotic $\Br$.  However, the jet $p_T$
and $\met$ cuts in the standard $Zh$ analysis are quite high and would
likely eliminate almost all of the signal. This underlines the need
for dedicated searches. 
	\section{\secsize $\secbold{h \rightarrow  \lowerGreekBold\tau^+ \lowerGreekBold\tau^- + \met}$}
\label{tatamet}

$h \rightarrow X_1 X_2 \rightarrow \tau^+ \tau^- + \met$ is another
new class of exotic Higgs decays.  As for the $2b + \met$ final state
of \S\ref{bbmet}, the two most important non-excluded topologies
are \begin{enumerate}[I.]
\item $X_1 \to \met$, \ $X_2\to \tau^+ \tau^- +\met$
\item $X_1 \to \met$, \ $X_2\to  \tau^+ \tau^-$.
\end{enumerate}
Here $X_{1,2}$ are intermediate particles, which can be either the
same or different, and the $\tau^+ \tau^-$ pair can be either resonant
or non-resonant (though this resonance would be difficult to
reconstruct with taus).

\subsection{Theoretical Motivation}

\begin{itemize}

\item The PQ-limit of the NMSSM: $h\to \chi_1 \chi_2$ (topology I, resonant)

  As discussed in detail in \S\ref{NMSSMfermion} (see
  also~\cite{Strassler:2006qa,draper2011dark,HLWY}), $X_{1,2}$ represent the lightest
  and next-to-lightest neutralinos in this limit, and we can get decay
  chains similar to those that lead to $h \rightarrow b \bar b + \met$
  (see \S\ref{subsec:2bmettheorymotivation}).  The second neutralino
  $\chi_2$, which will often be mostly bino, decays into
  $\chi_1s$ and/or $\chi_1a$.  If $s$ or $a$ have a mass $ 2 m_\tau<
  m_{s/a} < 2 m_b$, they dominantly decay into $ \tau^+ \tau^-$ via
  mixing with the MSSM Higgs doublets. In this case, the $\tau^+
  \tau^-$ pair is resonant.

\item $\nu$SM: $h\to \nu N$ (topology I, non-resonant)

  Neutrino models can also give rise to this signature. For example,
  in the $\nu$SM, the Higgs can decay into an active neutrino and a
  sterile neutrino via Yukawa interaction~\cite{0706.1732}. The mass
  mixing between RH sterile neutrinos and LH active neutrinos then
  make the RH neutrinos decay via $N \to \tau^+ W^{-(*)}\to \tau^+
  \tau^- \bar v_\tau$ and its conjugate (given Majorana $N$), or/and
  $N \to \nu Z^{(*)}\to \nu \tau^+ \tau^-$.  Here the $\tau^+ \tau^-$
  are generally non-resonant, though in some cases they could sit on
  the $Z$ resonance.  For more details, see \S\ref{subsub:SMF}.

\item Other models: $h\to a a, Z_D Z_D, \phi_1 \phi_2$ (topology II)

  As explained in \S\ref{subsec:2bmettheorymotivation}, it is possible
  to realize topology II as a possibly subdominant mode in dark vector
  models (\S\ref{subsec:SMvector}), in the PQ-NMSSM
  (\S\ref{NMSSMscalar}) via $a$ decaying to singlinos and taus if it
  satisfies $2 m_\tau < m_a < 2 m_b$, or in a more complicated hidden
  sector (\S\ref{sec:hiddenvalley}).

\end{itemize}

\subsection{Existing Collider Studies}

A preliminary analysis for the type-I topology is in progress, based
on a benchmark model inspired by the PQ-limit of the NMSSM, which is
presented in Table~\ref{table:benchmark3}~\cite{HLWY2}.
\begin{table} [h!]
\begin{center}
\begin{tabular}{c|c|c|c} \hline
   $m_h$ &  $m_{\chi_2}$ & $m_{\chi_1}$ & $m_s$   \\ \hline
 125 GeV &     80 GeV   &   10 GeV    & 8 GeV
  \\ \hline
\end{tabular}
\end{center}
\caption{Mass parameters used for the $h \rightarrow  \tau^+ \tau^- + \met$ collider analysis.}
\label{table:benchmark3}
\end{table}
Given the large mass hierarchy between $\chi_2$ and its decay products
$\chi_1$ and $s$ (here a scalar or pseudoscalar), as well as the fact
that $m_s/2m_\tau$ is only ${\cal O}(1)$, the $\tau^+\tau^-$ pair
produced in this decay tends to be highly collimated, forming a
``ditau-jet'' (much like some of the cases discussed in
\S\ref{sec:hto4tau} and references therein). The study is focused on
Higgs events from associated production with a leptonic $Z$ boson
($Z\to e^{+}e^{-}$, $\mu^{+}\mu^{-}$, and $\tau^{+}\tau^{-}$), due to
the very large expected QCD backgrounds for other production
modes. The distinguishing features of this signal are therefore two
leptons with their invariant mass falling in the $Z$ mass window, one
ditau-jet, and a moderate amount of $\met$. The dominant backgrounds
in this analysis are $Z$+jets, $t\bar t$+jets, and diboson+jets. They
can be greatly reduced by cutting on the number of tracks in the
ditau-jet candidate (QCD jets have more tracks than ditaus) and
requiring the reconstructed $h$ to be back-to-back with the $Z$. This
preliminary analysis suggests that extracting the $h \to X_1 X_2
\rightarrow 2\tau + \met$ signal is extremely challenging at the 14
TeV LHC, although more study is in progress~\cite{HLWY2}.

\subsection{Existing Experimental Searches and Limits}

Though these decays are motivated in several theoretical contexts,
there are no dedicated experimental searches yet, and the allowed
parameter space is still mostly open. The main constraints could come
from $h\to \tau^+ \tau^-$ searches~\cite{CMShditau,Aad:2012mea}, $h\to
WW^*$ searches~\cite{CMShWW,ATLAShWW}, and the $\tilde \tau \tilde
\tau$ search by ATLAS~\cite{ATLASstau}. However, in all of these
analyses the selection cuts are too aggressive to pick up the exotic
Higgs decay efficiently.  Some LHC searches might partly pick up some
special corners, though we will not attempt to delineate these regions
here. Dedicated searches are clearly needed, although, as mentioned
above, very challenging.


	~\def\Unknwn{?}

\def\highlumi{*}
\def\superscriptlegend{An asterisk indicates that 300 $\ifb$ was assumed; otherwise all estimates for 14 TeV assume 100 $\ifb$.}

\section{Conclusions \& Outlook}
\label{sec:conclusions}

We now summarize our results from various perspectives.  Our main
goals are to help experimentalists choose which analyses to undertake,
and to guide both theorists and experimentalists in understanding
which feasibility studies would be well-motivated but have not been
done.

In \S\ref{subsec:theory}, we considered various theories in which
non-SM Higgs decays arise.  Some of these are simplified models,
others more fully established theoretical structures, such as the
NMSSM.  Within any one of these models, certain classes of decays tend
to occur with definite relative probabilities.  If we are in such a
model, we may ask: which of the various decay modes offers the best
sensitivity to the presence of the exotic decays?  More precisely,
given the limits on $\frac{\sigma}{\sigma_\mathrm{SM}} \cdot \Br(h\to
{\cal F}_i)$ that can be obtained for the various exotic final states
${\cal F}_i$, which search gives us the strongest limit on
$\frac{\sigma}{\sigma_\mathrm{SM}} \cdot \Br(h\to {\text{non-SM \
    decays}}) = \frac{\sigma}{\sigma_\mathrm{SM}} \cdot \sum_i
\Br(h\to {\cal F}_i)$?

For instance, as we will see in a moment, the case of $h\to Z_D Z_D$,
where $Z_D$ is a spin-one particle decaying to fermion pairs, leads to
many final states, ranging from $jjjj$ to $b\bar b\ell^+\ell^-$ to
$\ell^+\ell^-\ell^+\ell^-$.  Not surprisingly, although
$\ell^+\ell^-\ell^+\ell^-$ only appears in about 10\% of $h\to Z_DZ_D$
decays, searches for it are so sensitive that it provides the best
limit in $\frac{\sigma}{\sigma_\mathrm{SM}}~\cdot~\Br(h~\to~Z_DZ_D)$.
As another example, if $h\to aa$, $a$ a pseudoscalar decaying to
$\tau^+\tau^-,\mu^+\mu^-$, the decay $\tau^+\tau^-\mu^+\mu^-$ provides
the greatest sensitivity; a decay to four muons is too rare.

We now proceed to organize our results along these lines.  Initially
we will limit ourselves to cases without very low-mass particles that
result in highly collimated pairs (or more) of jets, leptons, or
photons.  The collimated cases will be addressed separately.

At the end of this section we provide a final summary of our findings.

\subsection{How to interpret the tables}
\label{subsec:interprettables}

Below we organize our results into tables to allow for certain
comparisons to be made easily.  These tables are presented to guide
the reader, but necessarily suppress many essential details, all of
which are to be found in the main sections of our text.  {\bf It is
  important not to over-interpret the numbers presented in the
  tables}; the interested reader who is considering what searches or
studies should be undertaken must rely on the longer descriptions in
the main text in order to obtain the full picture.

We consider a number of different ``simplified model'' scenarios
below. For each one, we consider different final states ${\cal F}_i$
to which the Higgs may decay.  In the main text, we have obtained
information from several different types of sources: from existing
theoretical studies of a search for $h \to {\cal F}_i$ in the
literature; from our own studies of this decay mode; from existing
experimental searches for $h\to {\cal F}_i$; and from existing
searches for other processes that we reinterpret as limits on $h\to
{\cal F}_i$.  Whichever of these gives the best {\it current or
  potential} limit is listed in the tables; we indicate with a
superscript whether the limit is current or potential and whether it
arises from a theory study or from published LHC data.  If no limit is
known to us, we indicate it is ``unknown'' with the symbol
``\Unknwn''.

Importantly, {\bf the numbers presented in the tables are merely
  representative.}  The limits that can be obtained from any search
depend on the masses of new particles to which the Higgs is decaying,
and so in general they cover a range, sometimes a very wide
one. Because our goal is to point out where searches may be worth
performing, the tables present values at or near the {\it optimistic}
end of the range.  For example, if we show potential sensitivity at
the 1\% level, this means that there is a significant range of masses
in which such a branching fraction would be experimentally accessible,
though in other ranges sensitivity might be much less. Conversely our
numbers are in many cases {\it conservative}, because they are often
from theoretical studies that may not use optimal methods, or from
reinterpreting experimental searches that were optimized for something
other than Higgs decays.  The reader is urged to look at the relevant
sections in the main text to properly appreciate these subtleties.

\subsection{Final States Without $\met$}

\subsubsection{$h\to aa\to {\rm fermions}$}
\label{subsec:conclusions:haa}

In the simplified model of the SM coupled to a real or complex
SM-singlet scalar (\S\ref{SMS}), in certain regimes of the two Higgs
doublet model with an extra singlet (\S\ref{2HDMS}), and in regimes of
the NMSSM (\S\ref{NMSSMscalar}), Little Higgs models (\S\ref{LHiggs}),
and Hidden Valley models (\S\ref{sec:hiddenvalley}), one often finds
the phenomenon of a Higgs decaying to two particles that in turn decay
to SM fermions {\it with couplings weighted by mass} (though sometimes
separately for up-type quarks, down-type quarks, and leptons). We
write this as $h\to aa$ for short.

\nc{\mtrb}[1]{\multirow{2}{*}{#1}}
\nc{\mtrJ}[9]{\mtrb{#1} & #2 & #4 & #5 & #7 &  #8 \\
                        & #3 &       &   & #6 &     &        & #9 \\ }

\nc{\rowA}[8]{\mtrb{#1} & #2 & \mtrb{#3} & \mtrb{#4} & #5 & \mtrb{#6} & #7 &  #8 \\ }
\nc{\rowB}[4]{          & [ #1 ] &           &           & [ #2 ] &           & [ #3 ] &  #4 \\ \hline}
\nc{\rowC}[6]{\mtrb{#1} & #2             & \mtrb{#3} & \mtrb{#4} & #5       &  #6 \\ }
\nc{\rowD}[3]{          & [ #1 ]       &           &           & [ #2 ] &   #3 \\ \hline}
\nc{\rowDx}[4]{{\scriptsize \ \  #1}&        [ #2 ]       &           &           & [ #3 ] &   #4 \\ \hline}

\nc{\athstudy}[1]{Our study, \S\ref{#1}}
\nc{\alitsearch}[2]{Theory study \cite{#1}, \S\ref{#2}}
\nc{\alitsearchnosec}[1]{Theory study \cite{#1}}
\nc{\arecast}[2]{Recast of expt.~result \cite{#1}, \S\ref{#2}}
\nc{\aexplimit}[2]{Expt.~limit \cite{#1}, \S\ref{#2}}
\nc{\aexplimitnosec}[1]{Expt.~limit \cite{#1}}

\begin{sidewaystable}
\begin{center}
\begin{tabular}{|c|c|c||c|c||c|c||c|}\hline
& Projected/Current& & \multicolumn{2}{c||}{quarks allowed} & \multicolumn{2}{c||}{quarks suppressed}  & Comments\\
Decay & 2$\sigma$ Limit  &Produc- & & Limit on & &  Limit on &  \\
Mode & on $\Br({\cal F}_i)$ &tion& $\frac{\Br({\cal F}_i)}{\Br({\text{non-SM}})}$ & $\frac{\sigma}{\sigma_\mathrm{SM}} \cdot \Br$(non-SM) & $\frac{\Br({\cal F}_i)}{\Br({\text{non-SM}})}$ & $\frac{\sigma}{\sigma_\mathrm{SM}} \cdot \Br$(non-SM)  &  \\
${\cal F}_i$  & 7+8 [{14}] TeV &Mode& & 7+8 [{14}] TeV & & 7+8 [{14}] TeV &
\\ \hline  \hline
\rowA{$b\bar b b \bar b$}{$0.7 $}{$W$}{0.8}{0.9}{0}{--}{\arecast{Chatrchyan:2013zna}{sec:hto4b}}
\rowB{$0.2$}{0.2}{--}{\alitsearch{Dermisek:2006wr,Kaplan:2011vf}{sec:hto4b}} 
\rowA{$b\bar b \tau\tau$}{$>1$}{$V$}{0.1}{$> 1$}{0}{--}{} 
\rowB{$0.15$}{1}{--}{\alitsearch{Adam:2008aa}{sec:h2b2tau}}
\rowA{$b\bar b \mu\mu$}{$(2-7)\cdot 10^{-4} $}{$G$}{$3\cdot 10^{-4}$}{$0.5-1$}{0}{--}{\athstudy{sec:h2b2mu}}
\rowB{$(0.6-2)\cdot 10^{-4} $}{$0.2-0.8$}{--}{\athstudy{sec:h2b2mu}}
\rowA{$\tau\tau \tau\tau$}{$0.2-0.4 $}{$G$}{0.005}{$40-80$}{1}{$0.2-0.4$}{\arecast{CMS026,CMS13010}{sec:hto4tau}}
\rowB{\Unknwn}{\Unknwn}{\Unknwn}{}
\rowA{$\tau\tau \mu\mu$}{$(3-7)\cdot 10^{-4}$}{$G$}{$3\cdot 10^{-5}$}{$10-20$}{0.007}{$0.04-0.1$}{\athstudy{sec:hto4tau}}
\rowB{\Unknwn}{\Unknwn}{\Unknwn}{}
\rowA{$\mu\mu \mu\mu$}{$1\cdot 10^{-4}$}{$G$}{$1\cdot 10^{-7}$}{$1000$}{$1\cdot 10^{-5}$}{$10$}{\arecast{CMS-PAS-HIG-13-002,ATLAS-CONF-2013-020}{sec:hto4l}}
\rowB{\Unknwn}{\Unknwn}{\Unknwn}{}
\end{tabular}
\end{center}
\caption{\small Estimates for current or projected limits on various processes in $h\to aa$, if $a$ couplings are proportional to masses, and either $a\to$ quarks is allowed as in an NMSSM-type model (center columns) or $a\to$ quarks is suppressed relative to $a\to$ leptons (right columns).  If no relevant estimate is known, we indicate this with a ``\Unknwn''. The source of each estimate is listed in the ``Comments'' column.  
Production modes: $G$ for $gg\to h$, $V$ for vector boson fusion, $W,Z$ for $Wh$ and $Zh$.  For 14~TeV, estimates require 100~$\ifb$. 
See \S\ref{subsec:interprettables} for additional information and cautionary remarks.
}
\label{table:haabb}
\end{sidewaystable}

We consider this situation in Table~\ref{table:haabb}. For each decay
mode ${\cal F}_i$ that arises in this context, we list (second column)
the best potential sensitivity to the particular mode, obtained either
from existing papers in the literature, or from our own studies, or
from a reinterpretation of an existing ATLAS or CMS search for some
other phenomenon. In later tables, we will also see {\it current
  limits} from ATLAS and CMS searches for the mode $h\to {\cal F}_i$,
where those exist.  Where possible, we give estimates both for the
existing Run I data (LHC7+8) and for a certain amount of Run II data
(LHC14), taken to be 100~$\ifb$ except where indicated by an asterisk.
In the third column, we indicate by $G,V,W,Z$ whether the best known
limit is obtained through $gg\to h$, Vector Boson Fusion (VBF or
$qq\to qqh$), $Wh$ or $Zh$ production.

We then try to put these results in a model-dependent but broad
perspective.  The relative branching fractions, {\it i.e.} the rates
of particular final states relative to the total rate for all {\it
  non-SM} modes, are shown for two fiducial classes of models: one
(fourth column) where $a$ decays to both quarks and leptons with
relative branching fractions representative of NMSSM-type models, and
a second (sixth column) where quark decays are suppressed either by
couplings (vanishing $aq\bar q$ couplings) or by kinematics
$(m_a<2m_b)$. (In the latter case, our numbers are approximate because
we ignore $a\to c\bar c$, etc.)  Then, by dividing these relative
branching fractions by the potential (or current) limit (second
column), we obtain the sensitivity that this search provides for
$\Br(h\to aa)$, for the two fiducial models (fifth and seventh
columns.)  {\bf We emphasize that some searches could be more
  constraining for other models} (e.g.~for other Types of 2HDM+S), as
we describe below.

With $a\to q\bar q$ allowed and $a\to b\bar b$ dominant, it is notable
that $h\to 4b$ and $h\to b\bar b\mu\mu$ are both potentially promising
in Run II. Furthermore, for this scenario $b\bar b \mu \mu$ is the
only channel that may set marginally relevant limits with Run I data.
The $b\bar b\tau\tau$ mode suffers by comparison from the absence of a
resonance and large $t\bar t$ backgrounds, and analysis improvements
will be necessary if it is to be useful.

In the absence of $a\to$ quarks, or for $m_a< 10$ GeV, the search for
$h\to \tau^+\tau^-\mu^+\mu^-$ is more sensitive than that for $h\to
\tau^+\tau^-\tau^+\tau^-$, but sufficiently close that both should be
investigated further.  It is worth considering both modes in searches
within Run I data.

In some models the ratio of $a\to bb$ to $a\to \tau\tau$ can change
continuously as a function of parameters.  Since the achievable limits
on $\Br(h\to2a\to2b2\mu)$ and $\Br(h\to2a\to2\tau2\mu)$ are very
similar, the former will set a better limit on overall exotic
branching fraction if $\Br(a\to2b) \gtrsim \Br(a\to2\tau)$, and vice
versa. At least one of these two channels should approach a
sensitivity of $\Br(h\to\ aa) \sim 0.1$. Investigating both is
therefore vital to achieve `full coverage' of this scenario.

Our suggestion is that the searches for $b\bar b\mu^+\mu^-$ and
$\tau^+\tau^-\mu^+\mu^-$, assuming a $\mu^+\mu^-$ resonance at the $a$
mass, should be undertaken, even with Run I data.  We note that both
triggering and analysis are far easier for $b \bar b \mu^+\mu^-$ and
$\tau^+\tau^-\mu^+\mu^-$ than for other modes, due to the higher-$p_T$
muons and the narrow peak in the di-muon mass. We also emphasize that
these searches should be carried out {\it with minimal prejudice as to
  the range of $m_a$.}  For $\tau^+\tau^-\mu^+\mu^-$, the common
assumption $m_a<2m_b$ is unnecessary; as we have noted in
\S\ref{2HDMS}, there are many models in which $a\to b\bar b$ is
suppressed not by kinematics but by coupling constants.  Meanwhile,
for $b\bar b\mu^+\mu^-$, the assumption that both fermion-antifermion
pairs come from the same type of particle implies that $m_{\mu\mu} =
m_a > 2 m_b$, but the decay $h\to a a'$ can occur in some non-minimal
models, in which case $m_{a'}= m_{\mu\mu}<2m_b<m_a$ may occur,
possibly with an increased rate.

\subsubsection{$h\to aa\to {\rm SM \ gauge \ bosons}$}

Next we turn to a case where the $a$ does not couple strongly to fermions, and instead decays mainly to gluon pairs and photon pairs through loops of heavy particles.  Such couplings are commonly proportional to gauge couplings squared (i.e. to $\alpha_i$), in which case $\Br(a\to \gamma\gamma)\sim 0.004 \times \Br(a\to gg)$ for a degenerate $SU(5)$ multiplet of fermions coupling equally to $a$ (see \S\ref{sec:2gamma2jet}). But if the masses $M$ of the heavy colored particles in the loops are larger than the masses $m$ of the colorless ones, the rate for photon production may be enhanced by at least a factor of $(M/m)^2$.

\begin{sidewaystable}
\begin{center}
\begin{tabular}{|c|c|c||c|c||c|c||c|}\hline
 & Projected/Current& & \multicolumn{2}{c||}{$\Br(a\to \gamma\gamma) \approx 0.004$} & \multicolumn{2}{c||}{$\Br(a\to \gamma \gamma)\approx 0.04$} & Comments \\
Decay & 2$\sigma$ Limit  &Produc- & & Limit on & &  Limit on &   \\
Mode & on $\Br({\cal F}_i)$ &tion& $\frac{\Br({\cal F}_i)}{\Br({\text{non-SM}})}$ & $\frac{\sigma}{\sigma_\mathrm{SM}} \cdot \Br$(non-SM)  & $\frac{\Br({\cal F}_i)}{\Br({\text{non-SM}})}$ & $\frac{\sigma}{\sigma_\mathrm{SM}} \cdot \Br$(non-SM)   & \\
${\cal F}_i$  & 7+8 [14] TeV &Mode& & 7+8 [14] TeV & & 7+8 [14] TeV & \\ \hline  \hline
\rowA{$jjjj$}{$>1$}{$W$}{0.99}{$> 1$}{0.92}{$> 1$}{} 
\rowB{$0.1^{\highlumi}$}{$0.1^\highlumi  $}{$0.1^\highlumi  $}{\alitsearch{Chen:2010wk,Falkowski:2010hi}{sec:h4j}} 
\rowA{$\gamma\gamma jj$}{$0.04$}{$W$}{0.008}{$5$}{0.08}{$0.5$}{} 
\rowB{$0.01^{\highlumi}$}{$1^{\highlumi}$}{$0.1^\highlumi$}{\alitsearch{Martin:2007dx}{sec:2gamma2jet}} 
\rowA{$\gamma\gamma\gamma\gamma$}{$2\cdot 10^{-4}$}{$G$}{$1\cdot 10^{-5}$}{$20$}{0.001}{$0.2$}{\athstudy{sec:4gamma}} 
\rowB{$3\cdot 10^{-5  \highlumi}$}{$1^\highlumi  $}{$0.03^\highlumi$}{\alitsearch{Chang:2006bw}{sec:4gamma}} 
\end{tabular}
\end{center}
\caption{\small As in Table \ref{table:haabb}, estimates for various processes in $h\to aa$ if $a$
  decays only to SM gauge bosons through loops.  The central columns
  show the case where the couplings are generated by initially
  degenerate $SU(5)$ multiplets; the right columns show the case where
  the $a\to\gamma\gamma$ rate is enhanced by a factor of 10. {\it An asterisk denotes     that all 14~TeV estimates shown require 300~$\ifb$ of data.} 
}
\label{table:haagg}
\end{sidewaystable}

Estimated limits for this case are shown in
Table~\ref{table:haagg}. If the heavy particles are degenerate and in
complete $SU(5)$ multiplets, then the center columns show that only
the four-jet search has any reach, with phenomenologically relevant
sensitivity possible for $m_a \lesssim 5~\gev$ with $300~\ifb$ of
data. If the branching fraction $a\to \gamma\gamma$ is enhanced by a
factor of 10, as would happen if the colored particles appearing in
the loop graph were about 3 times heavier than the colorless
particles, then the situation is given in the right columns.  In this
case, the four-photon search is clearly superior.

We should of course note that it is possible to have a particle that
dominantly decays to $\gamma\gamma$.  This could occur for a
pseudoscalar $a$ if it couples to the visible sector only through
loops of heavy colorless charged particles.  In this case there would
be only $4\gamma$ decays and no $4j$ or $2j2\gamma$ decays.

With these considerations in mind, it would seem four-jet,
four-photon, and mixed searches are all well-motivated in Run II.
However, for Run I data, a four-jet search is hopeless, while a
four-photon search is already sensitive to models where $a$ has
enhanced decays to photons. We therefore suggest a search for $h\to
4\gamma$ even in the existing Run I data.  We also suggest that
triggers for multiple photons be set so as to retain this signal in
Run II.

\subsubsection{$h\to Z_DZ_D$, $Z Z_D$, $Z a$}

Now we consider the possibility that the Higgs decays either to two
dark vector bosons $Z_D$ or to one $Z_D$ and one SM $Z$. This can
occur in dark vector scenarios (\S\ref{subsec:SMvector}) and more
general hidden valleys (\S\ref{sec:hiddenvalley}).  The main
difference compared to $h\rightarrow a a$ is that $Z_D$ branching
ratios are ordered by SM gauge charge instead of mass, which leads to
large leptonic branching fractions.

The $h\rightarrow Z Z_D$ search can also set limits on the $h\rightarrow Z a$ scenario, where $a$ is a pseudoscalar which decays to fermions in proportion to their masses. If decays to $\bar b b$ are suppressed or forbidden the limits can already be appreciable.

A useful fiducial model is to take $Z_D$ to couple to SM fermions proportional to their electric charge. This is the case if decays occur via kinetic $\gamma-Z_D$ mixing, and if $m_{Z_D}\ll m_Z$ so that photon-$Z$ mixing is unimportant (see Fig.~\ref{fig:SMVBrZdark} in
\S\ref{subsec:SMvector}), but also gives the qualitatively correct picture for more general dark vector scenarios.

\begin{table}[t!]
\begin{center}
\begin{tabular}{||c|c|c||c|c||c|c||c|}\hline
 & Projected/Current& & &  &  \\
Decay & 2$\sigma$ Limit  &Produc- & & Limit  on & Comments \\
Mode & on $\Br({\cal F}_i)$ &tion& $\frac{\Br({\cal F}_i)}{\Br({\text{non-SM}})}$ & $\frac{\sigma}{\sigma_\mathrm{SM}} \cdot \Br$(non-SM) &  \\
${\cal F}_i$  & 7+8 [14] TeV &Mode& & 7+8 [14] TeV & \\ \hline  \hline
\rowC{$jjjj$}{$>1$}{$W$}{0.25}{$> 1$}{} 
\rowD{$0.1^{\highlumi}$}{$0.4^\highlumi  $}{\alitsearch{Chen:2010wk,Falkowski:2010hi}{sec:h4j}} 
\rowC{$\ell\ell\ell\ell$}{$4\cdot 10^{-5}$}{$G$}{0.09}{$4\cdot 10^{-4}$}{\arecast{CMS-PAS-HIG-13-002,ATLAS-CONF-2013-020}{sec:hto4l}}
\rowD{\Unknwn}{\Unknwn}{}
\rowC{$jj\mu\mu$}{$0.002-0.008$}{$G$}{0.15}{$0.01-0.06$}{\athstudy{sec:h2b2mu}}
\rowD{$(5-20)\cdot 10^{-4}$}{$0.003-0.01$}{\athstudy{sec:h2b2mu}}
\rowC{$b\bar b \mu\mu$}{$(2-7)\cdot 10^{-4}$}{$G$}{0.015}{$0.01-0.05$}{\athstudy{sec:h2b2mu}}
\rowD{$(0.6-2)\cdot 10^{-4}$} {$0.003-0.01$}{\athstudy{sec:h2b2mu}}
\end{tabular}
\end{center}
\caption{\small As in Table \ref{table:haabb}, estimates for various processes in $h\to Z_D Z_D$ if $m_{Z_D} > 2m_b$ and couplings are proportional to electric charges. $\ell=e,\mu$ and all numbers represent the {\it sum} of processes involving $e$ and $\mu$; $j$ represents all jets except $b$ quarks.  
\superscriptlegend
}
\label{table:hZDZD}
\end{table}

We first treat the $h\rightarrow Z_D Z_D$ decay, see Table \ref{table:hZDZD}. Not surprisingly, the search for $h\to (\ell^+\ell^-)(\ell^+\ell^-)$, which allows full reconstruction at high resolution, is the most powerful. The published data on four-lepton events used in the Higgs search and in $Z^{(*)} Z^{(*)}$ studies puts tremendous constraints on this decay, already, according to our reinterpretation of the published data, reaching $\Br(h\to Z_DZ_D)< 4\times 10^{-4}$.
It is important to improve on the constraints we found on this well-motivated model; specifically, our reinterpretation did not allow for an optimal constraint, since it does not make full use of the three available mass resonances.

Limits on $\Br(h\to Z_DZ_D)$ from dilepton plus jets searches are
probably in the few times $10^{-2}$ range, see \S\ref{sec:h2b2mu}.  As
the table indicates, our studies suggest that $jj\mu^+\mu^-$ and
$b\bar b \mu^+\mu^-$ would have comparable sensitivity, and this might
also be true for electron final states, though triggering and
reconstruction efficiencies will be lower than for muons in many
cases.  But even combining all of these together, it appears that
dilepton plus jets final states would only be competitive in models
where the branching fractions for leptons is significantly reduced
compared to the case we consider in Table~\ref{table:hZDZD}.

The constraints on $h\to Z Z_D$ and $Za$ are shown in
Table~\ref{table:hZZDZorZa}. The $h\rightarrow Z^*Z$ search sets
powerful constraints.  In the case of $Z Z_D$, they are still one
order of magnitude weaker than indirect constraints from electroweak
precision measurements for $m_{Z_D}\gtrsim 10$~GeV (see
Fig.~\ref{fig:SMVZdbounds}).  (For $m_{Z_D} \lesssim 10$~GeV, the
constraints are even stronger.)  A more optimized search with
sufficient luminosity at the 14 TeV LHC will yield competitive or even
eventually superior limits for $m_{Z_D}\gtrsim 10$~GeV.  The bounds on
$h\to Za$ from four lepton final state are rather weak due to Yukawa
suppression. The decay $h\to Za$ is an example of an asymmetric $h\to
2\to 4$ decay, and other search channels such as $h\to Za\to
(\ell^+\ell^-)(b\bar b)$ may provide better sensitivity in the long
run.

\begin{sidewaystable}
\begin{center}
\begin{tabular}{||l|c|c||c|c||c|c||c|}\hline
 & Projected/Current & & & & \\
Decay & 2$\sigma$ Limit  &Produc- & & Limit  on & Comments\\
Mode & on $\Br({\cal F}_i)$ &tion& $\frac{\Br({\cal F}_i)}{\Br({\text{non-SM}})}$ & $\frac{\sigma}{\sigma_\mathrm{SM}} \cdot \Br$(non-SM)  & \\
${\cal F}_i$  & 7+8 [14] TeV &Mode& & 7+8 [14] TeV &   \\ \hline \hline
\rowC{$ZZ_D\to\ell\ell\ell\ell$}{$4\cdot 10^{-5}$}{$G$}{0.02}{0.002}{\arecast{CMS-PAS-HIG-13-002, ATLAS:2013nma}{sec:htoZa}}
\rowD{\Unknwn}{\Unknwn}{}
\rowC{$Za\to\ell\ell\mu\mu$}{$4\cdot 10^{-5}$}{$G$}{$2\cdot 10^{-5}$}{2}{\arecast{CMS-PAS-HIG-13-002, ATLAS:2013nma}{sec:htoZa}}
\rowDx{$\Br(a\to b\bar b)\sim0.9$}{\Unknwn}{\Unknwn}{}
\rowC{$Za\to\ell\ell\mu\mu$}{$4\cdot 10^{-5}$}{$G$}{$2\cdot 10^{-4}$}{0.2}{\arecast{CMS-PAS-HIG-13-002, ATLAS:2013nma}{sec:htoZa}}
\rowDx{$\Br(a\to b\bar b)=0$}{\Unknwn}{\Unknwn}{}
\end{tabular}
\end{center}
\caption{\small As in Table \ref{table:haabb}, estimates for all-leptonic  processes in $h\to Z Z_D$ and $h\to Z a \to \ell\ell\ell\ell$; other processes were not studied. For $Z_D$ we assume couplings are proportional to electric charges; for $a$ we assume all couplings are proportional to masses, and either that $a\to b\bar b$ is dominant or highly suppressed (as in certain Type III 2HDM+S models described in \S\ref{2HDMS}). Here  $\ell=e,\mu$ and all numbers represent the {\it sum} of processes involving $e$ and $\mu$.  
\superscriptlegend
}
\label{table:hZZDZorZa}
\end{sidewaystable}

We therefore find that searches for four-lepton final states in $h\to
(\ell^+\ell^-)(\ell^+\ell^-)$ via non-SM channels are extremely
well-motivated in Run I.  As we have noted earlier, the available data
as published in the search for the SM $h\to Z Z^*$ mode are not ideal
for the $Z_DZ_D$ or $Z Z_D$ searches, since neither the selection cuts
nor the analysis approach are appropriate to the signal, with some
events unnecessarily discarded and with leptons often systematically
mis-assigned.  The analysis for $Z Z_D$ in particular (but also $Z_D
Z_D$ in general) should preferably also extend to very low $Z_D$ mass
ranges, where isolation cuts and quarkonium backgrounds are an issue.

Triggering is not a problem for these final states because the leptons have relatively high $p_T$.  Multi-lepton triggers where two or three leptons are soft may contribute to sensitivity, a point that deserves further exploration.

\subsection{Final States with $\met$}

In the $h\to 2\to 4$ final states we discussed above, only one unknown
particle need appear, and its decays are often controlled by a single
type of coupling.  By contrast, final states with $\met$ can arise
from multiple decay topologies (see
Fig.~\ref{f.higgsdecaytopologies}), and the type of search required
may depend on whether the energy carried by invisible particles is
large (in the Higgs rest frame) relative to $m_h$.

\subsubsection{Larger $\met$, without resonances}

First we consider cases where the $\met$ in the $h$ rest frame is a
significant fraction of its mass, and the invariant mass of the
visible objects in the Higgs decay lies well below 125~GeV and may be
highly variable.  In general, there may be no resonances among the
visible particles in the non-SM Higgs decay modes.
Fermion-antifermion pairs may be produced in 3-body decays such as
$\psi\to f\bar f \psi'$; in this case there will be kinematic
endpoints, but statistics may be too small to use them.  Branching
fractions are very model-dependent, but tend to be similar either to
the heavy-flavor-weighted or the flavor-democratic cases associated
with (pseudo)scalars $a$ or vectors $Z_D$ discussed above.
Tables~\ref{table:h2bbttMET} and \ref{table:h2ZDMET} show that cases
with leptons are promising, but with $b\bar b$ or $\tau\tau$ the
situation is difficult even if, as in the studies we refer to in the
main text, the $2b$ and $2\tau$ are assumed to be on-resonance.  More
study of the difficult cases is warranted.

\begin{sidewaystable}
\begin{center}
\begin{tabular}{|c|c|c||c|c||c|c||c|}\hline
& Projected/Current& & \multicolumn{2}{c||}{quarks allowed} & \multicolumn{2}{c||}{quarks suppressed} &  \\
Decay & 2$\sigma$ Limit  &Produc- & & Limit  on & &  Limit  on & Comments \\
Mode & on $\Br({\cal F}_i)$ &tion& $\frac{\Br({\cal F}_i)}{\Br({\text{non-SM}})}$ & $\frac{\sigma}{\sigma_\mathrm{SM}} \cdot \Br$(non-SM)  & $\frac{\Br({\cal F}_i)}{\Br({\text{non-SM}})}$ & $\frac{\sigma}{\sigma_\mathrm{SM}} \cdot \Br$(non-SM) &  \\
${\cal F}_i$  & 7+8 [14] TeV &Mode& & 7+8 [14] TeV & & 7+8 [14] TeV & \\ \hline  \hline
\rowA{$(b\bar b)$  $\met$}{$>1 $}{$Z$}{0.9}{$>1$}{0}{--}{\athstudy{bbmet}}
\rowB{$0.2^{\highlumi}$} {$0.2^\highlumi  $}{--}{\alitsearch{draper2011dark,HLWY}{NMSSMfermion}}
\rowA{$(\tau\tau)$  $\met$}{$>1$}{$Z$}{0.1}{$> 1$}{1}{$>1$}{\athstudy{tatamet}}
\rowB{$>1^{\highlumi}$}{$>1^\highlumi  $}{$>1^\highlumi  $}{\alitsearch{draper2011dark,HLWY}{NMSSMfermion}}
\rowA{$\mu\mu$  $\met$}{$0.07$}{$G$}{0.003}{$40$}{0.03}{$4$}{\arecast{CMS-PAS-HIG-13-003,ATLAShWW}{sec:2lMET}}
\rowB{\Unknwn}{\Unknwn}{\Unknwn}{}
\end{tabular}
\end{center}
\caption{As in Table \ref{table:haabb}, estimates for various processes $h\rightarrow \psi \psi'$, where $\psi'$ is invisible and $\psi \rightarrow \psi' + \bar f f$ via an intermediate $a$ coupling to fermions proportional to their masses. In the two cases shown, either $\Br(a\to b\bar b)$ dominates (center columns) or quarks are suppressed relative to leptons (right columns).
The limits for $\bar b b$ and $\tau \tau$ assume an intermediate resonance (indicated with parentheses), while the $\mu \mu$ limits do {\it not} assume a resonance and are artificially weak.  
{\it An asterisk indicates that all 14~TeV estimates shown require 300~$\ifb$ of data.}
}
\label{table:h2bbttMET}
\end{sidewaystable}

In particular, for $b\bar b\met$, $\tau\tau\met$, and even $\mu\mu\met$ where the muons are too soft to pass di-muon trigger thresholds, it may become important to consider VBF production.  Triggering in this case might require combining a VBF dijet requirement, a $\met$ requirement, and a requirement of $b$, $\tau$, or $\mu$ candidates.  This requires further investigation.

Photons, by contrast, may be produced singly, as in $\psi\to
\gamma\psi'$, and thus non-resonant $\gamma$ + $\met$ and
$\gamma\gamma$ + $\met$ final states are possible.  We show results in
Table~\ref{table:h2gammasMET}.  There is no preferred pattern of
branching fractions here; the decay $\psi\to \gamma\psi'$ may have a
branching fraction of 100\%, or may be diluted by other final states,
such as $\psi\to Z^*\psi'$ or $\psi\to Z_D\psi'$.  Existing searches
involving $\gamma$ + $\met$ have a high $H_T$ cut and are very
inefficient for a Higgs signal of this type; see \S\ref{sec:1gaMET}.
Because we do not know the size of fake $\met$ backgrounds in $\gamma$
+ jet events at low photon-$p_T$ and especially low $\met$, we cannot
determine whether a single photon search is well-justified;
experimental studies would be required on this point.  We note that
data from a parked data trigger for $\gamma$ + $\met$, available at
least for CMS \cite{parked-CMS}, may allow for an interesting search.

In any case, a $\gamma\gamma\met$ search in Run I data is certainly
justified.  It is quite reasonable theoretically to have $\Br(h\to
\chi_2\chi_2)\sim 0.1$ and $\Br(\chi_2\to \gamma\chi_1)=1$, and (see
Fig.~\ref{fig:2gaMETnonresAugLabel__61}) a non-optimized GMSB search
already reaches the level of $\Br(h\to\gamma\gamma + \met) \sim 0.05$.
A search more optimized for a Higgs signal should do considerably
better.

\begin{sidewaystable}
\begin{center}
\begin{tabular}{|c|c|c||c|c||c|c||c|}\hline
 & Projected/Current& & \multicolumn{2}{c||}{$h\to \psi \psi' \to \bar f f+ \met$} & \multicolumn{2}{c||}{$h\to\psi \psi \to \bar f_1 f_1 + \bar f_2 f_2 + \met$} & \\
Decay & 2$\sigma$ Limit  &Produc- & & Limit on & &  Limit  on & Comments\\
Mode & on $\Br({\cal F}_i)$ &tion& $\frac{\Br({\cal F}_i)}{\Br({\text{non-SM}})}$ & $\frac{\sigma}{\sigma_\mathrm{SM}} \cdot \Br$(non-SM)  & $\frac{\Br({\cal F}_i)}{\Br({\text{non-SM}})}$ & $\frac{\sigma}{\sigma_\mathrm{SM}} \cdot \Br$(non-SM)  &  \\
${\cal F}_i$  & 7+8 [14] TeV &Mode& & 7+8 [14] TeV & & 7+8 [14] TeV & \\ \hline  \hline
\rowA{$(b\bar b)$  $\met$}{$>1$}{$Z$}{0.05}{$>1$}{0.1}{$>1$} {\athstudy{bbmet}}
\rowB{$0.2^{\highlumi}$}{$4^\highlumi  $}{$2^\highlumi  $} {\alitsearch{Strassler:2006qa}{sec:hiddenvalley}}
\rowA{$(\tau\tau)$  $\met$}{$>1$}{$Z$}{0.15}{$> 1$}{0.28}{$>1$} {\athstudy{tatamet}}
\rowB{$>1^{\highlumi}$}{$>1^\highlumi  $}{$>1^\highlumi  $]} {\alitsearch{Strassler:2006qa}{sec:hiddenvalley}}
\rowA{$\ell\ell$  $\met$}{$0.07 $}{$G$}{0.30}{$0.2$}{0.51}{0.1}{\arecast{CMS-PAS-HIG-13-003,ATLAShWW}{sec:2lMET}} 
\rowB{\Unknwn}{\Unknwn}{\Unknwn}{}
\rowA{$\ell\ell\ell\ell$  $\met$}{$5\cdot 10^{-4} $}{$G,V$}{--}{--}{0.09}{0.005} {\arecast{CMS-PAS-SUS-13-010}{sec:IsoLeptonMET}}
\rowB{\Unknwn}{--}{\Unknwn}{}
\end{tabular}
\end{center}
\caption{As in Table \ref{table:haabb}, estimates for various processes $h\rightarrow \psi \psi'$ (middle column) and $h\rightarrow \psi \psi$ (right column), where $\psi'$ is invisible and $\psi \rightarrow \psi' + \bar f f$ via an intermediate (possibly off-shell) vector boson, which couples to fermions proportionally to electric charges. The limits for $\bar b b$ and $\tau \tau$ assume an intermediate resonance (indicated with parentheses), while the $2\ell, 4\ell$ limits do not, making the limits artificially weak. For $h\to \psi\psi$ (right-most columns), there are four fermions in the final state; we assume here that the limits obtained for $f\bar f+\met$ are not much changed by the presence of the two additional fermions. {\it An asterisk denotes that all 14~TeV estimates shown require 300~$\ifb$ of data.} 
}
\label{table:h2ZDMET}
\end{sidewaystable}

\begin{table}[t!]
\begin{center}
\begin{tabular}{|c|c|c|c|}\hline
Decay & Projected/Current 2$\sigma$ Limit  &Production & Comments
 \\
Mode  & Limit on $\Br({\cal F}_i)$ & Mode & \\
${\cal F}_i$  & 7+8 [14] TeV & &  \\ \hline  \hline
\mtrb{$\gamma$  $\met$}  &  $>1 $  & \mtrb{$G$} &\arecast{CMS:2012kwa}{sec:1gaMET} \\
& \Unknwn & & \\ \hline
\mtrb{$\gamma\gamma$  $\met$}  &  $0.04$ & \mtrb{$G$} &\arecast{CMS:2012kwa}{sec:2gaMET}
\\ 
& \Unknwn & & \\ \hline
\end{tabular}
\end{center}
\caption{\small As in Table \ref{table:haabb}, estimates for $h\rightarrow \psi \psi$ or $\psi \psi'$, where $\psi \rightarrow \psi' + \gamma$ and $\psi'$ is invisible. Note the limits we have obtained do not require a $\gamma\gamma$ resonance.
}
\label{table:h2gammasMET}
\end{table}

\begin{sidewaystable}
\begin{center}
\begin{tabular}{|c|c|c||c|c||c|c||c|}\hline
 & Projected/Current& & \multicolumn{2}{c||}{$\Br(a\to \gamma\gamma)\approx 0.004$} & \multicolumn{2}{c||}{$\Br(a\to \gamma\gamma)\approx 0.04$} & \\
Decay & 2$\sigma$ Limit  &Produc- & & Limit on & &  Limit on & Comments   \\
Mode & on $\Br({\cal F}_i)$ &tion& $\frac{\Br({\cal F}_i)}{\Br({\text{non-SM}})}$ & $\frac{\sigma}{\sigma_\mathrm{SM}} \cdot \Br$(non-SM)  & $\frac{\Br({\cal F}_i)}{\Br({\text{non-SM}})}$ & $\frac{\sigma}{\sigma_\mathrm{SM}} \cdot \Br$(non-SM)  &  \\
${\cal F}_i$  & 7+8 [14] TeV &Mode& & 7+8 [14] TeV & & 7+8 [14] TeV & \\ \hline  \hline
\rowA{$\gamma\gamma$  $\met$}{$0.04 $}{$G$}{0.004}{$10$}{0.04}{1}{\arecast{CMS:2012kwa}{sec:2gaMET}}
\rowB{\Unknwn}{\Unknwn}{\Unknwn}{}
\end{tabular}
\end{center}
\caption{\small As in Table \ref{table:haabb}, estimates for $h \rightarrow a + \met$ with $a\to\gamma\gamma$ if $a$ couplings to gluons and photons are proportional to gauge couplings  (center columns), or with $\Br(a\to \gamma\gamma)$ enchanced by a factor of 10 (right columns). Note the limits we have obtained do not require a $\gamma\gamma$ resonance. 
}
\label{table:h2aMET}
\end{sidewaystable}

\subsubsection{Larger $\met$, with resonances}

If the objects in the final states are produced in resonances, and the resonances in question are from scalar or vector particles, then as in the previous section there are preferred scenarios for their branching fractions.  In these cases, the limits will obviously be stronger than in the non-resonant cases, especially for photons and leptons.  On the other hand, the numbers we have presented in this document are obtained by reinterpreting ATLAS and CMS searches which do {\it not} seek resonances, and are therefore unnecessarily pessimistic.

For instance, in a decay $h\to \psi\psi'$ where $\psi \to a \psi'$, $\psi'$ is invisible, and $a$ decays to gluon and photon pairs, we will potentially have $h\to jj$ + $\met$ and $h\to \gamma\gamma$ + $\met$, see Table~\ref{table:h2aMET}.  The dijet signal has not been studied, and given the difficulty of the search for $h\to bb$ + $\met$ it is not likely to be useful.  We therefore show only the $2\gamma$ + $\met$ case.  Note these numbers are obtained from searches that do not require a di-photon resonance, so the true sensitivity may be significantly higher in a resonance search.  Even so there may only be sensitivity in this channel at present with some enhancement of $\Br(a\to\gamma\gamma)$, but since this is a reasonable possibility, we view a dedicated search in Run I data as well-motivated.  Even though we have not done so in this document, one could also investigate $h\rightarrow \psi \psi \rightarrow 4\gamma + \met$ via two intermediate pseudoscalars.  Aside from a direct search for the final state,  perhaps a $\geq3$-photon search, where one looks for a resonance in nearby photon pairs, is warranted.

Meanwhile, in a decay $h\to \psi\psi'$ where $\psi \to a \psi'$, $\psi'$ is invisible, and $a$ decays to fermions with couplings proportional to masses, we will potentially have $h\to b\bar b$ + $\met$, $jj$ + $\met$, $\tau^+\tau^-$ + $\met$, $\mu^+\mu^-$ + $\met$ final states.  We already showed results for this case in Table \ref{table:h2bbttMET}.  Only the $\mu^+\mu^-$ search will be sensitive in the next few years, and the rate for this final state may be quite low if $m_a \gg 2 m_\tau$, but importantly the search may be quite a bit more sensitive than shown when one requires a resonance.  Admittedly we are quoting numbers for optimistic scenarios; as the $\met$ increases and the $p_T$ of the visible objects decreases, efficiencies and sensitivities may drop rapidly.
  Also shown are the numbers if the decay of the $a$ to $b\bar b$ is suppressed by kinematics or by coupling constants.  Even in this case the decay to $\mu^+\mu^-$ appears too small, but it important to note that the numbers for $\mu^+\mu^-$ + $\met$ are obtained assuming no resonances (see \S\ref{sec:IsoLeptonMET}).  Therefore, in this case a search in the 7+8~TeV data is probably merited.

Note that if $m_h-2m_\psi$ is small, the leptons will have low $p_T$.  Then the search strategy we refer to in Table \ref{table:h2bbttMET}, which relies on $gg\to h$, will not work, because the leptons will lie below di-lepton trigger thresholds.  This is unfortunate, because despite their low $p_T$ the leptons may form a resonance that makes off-line backgrounds small.  So as not to lose the possibility of discovery, it may be essential to trigger on such events produced via VBF. where the trigger combines the jets from VBF with $\met$ and the soft leptons.

Note that if instead of $h\to a$ + $\met$ the decay is to $h\to aa$ + $\met$, via $h\to \psi\psi$ and $\psi\to a\psi'$, the situation is quite similar.  Aside from $\mu^+\mu^-$ + $\met$ inclusive, which has twice as large a branching fraction as in Table \ref{table:h2bbttMET}, no other searches may be sensitive in the near term.  However, some advantage can be obtained from $\tau^+\tau^-\mu^+\mu^-\met$ events, via multi-lepton searches.

Next we turn to the case where $a$ is replaced by $Z_D$. We already showed both the cases where $h\to Z_D$ + $\met$ and $h\to Z_DZ_D$ + $\met$ in Table \ref{table:h2ZDMET}.  Again we emphasize that no resonances are assumed in the leptonic searches, so true sensitivities should be better than shown.  Clearly searches in the dilepton and four-lepton mode are well-motivated by these models.

\subsubsection{Small $\met$}

If the amount of $\met$ is always small, modes with $\met$ may be probed in searches that assume no $\met$, as long as kinematic requirements are loosened appropriately.  These include both searches for SM decay modes and for non-SM $h\to 2\to 4$ modes discussed above.

Small $\met$ arises naturally when the invisible particles are emitted in two-body decays that constrain their $p_T$ to be small, for instance in $h\to a a'$ where $a'$ is invisible and $m_h\sim m_a \gg m_h-m_a\sim m_{a'}$.  It is common for the other particle in the two-body decay to then produce an observable resonance; in the previous example we might have $a\to b\bar b$ resonantly. In addition to this $h\to 2\to 3$ decay, similar features may arise in a $h\to 2\to 4 \to 6$ decay, such as $h\to \psi\psi$ and $\psi\to Y\psi'$, where $m_{\psi'} < m_\psi-m_Y\ll m_h$ and $Y$ decays visibly; in this case there are two $Y$ resonances, plus small $\met$.  Another possibility is for a $h\to 2\to 3\to 4$ decay, for instance if $h\to \psi\psi'$, $\psi\to Y\psi'$, if $m_Y\sim m_\psi \sim m_h$.

For an $h\to 2\to 3$ (or $h\to 2\to 3\to 4$) decay with one (or two) low-$p_T$ invisible particles, SM searches are often sensitive, as long as cuts do not exclude resonances below 125~GeV.  For example, the decay $h\to \psi' \psi \to \psi'\psi' Y\to \psi'\psi'(f\bar f)$, where $f$ is a SM fermion, closely resembles the decay $h\to f\bar f$, except that the mass of the $f\bar f$ lies at $m_Y$, {\it slightly below} $m_h$.  The same applies for a decay to photons.

For a $h\to 2\to 4\to 6$ decay, with two low-$p_T$ invisible particles, the final state of the Higgs resembles an $h\to 2\to 4$ decay, such as we have already discussed extensively in the preceding subsection.  The only new requirement is to allow for the total invariant mass of the two $Y$ resonances to lie between $2m_Y$ and $m_h$.

There are other cases to consider, such as $h\to Z Z_D$, $Z_D\to a_1 s_1$ where $a_1$ decays outside the detector and $s_1\to \mu^+\mu^-$.  The general lesson is the same, however: if the $\met$ is small and the final states are resonant, as is commonly the case, the only necessary change between standard and exotic searches is to relax the requirement, as appropriate, that the invariant mass of the visible objects is 125 GeV.  This loosening of cuts is only relevant in channels where the invariant mass reconstruction has excellent resolution, {\it i.e.} final states containing electrons, muons, or photons.

We therefore find that:
\begin{itemize}
\item
the four-lepton and four-photon searches mentioned in the previous section, aimed at $h\to 2\to 4$ decays, should also be performed so that limits can be obtained on scenarios where the invariant mass of the observed objects lies somewhat below $m_h$, whether or not the leptons or photons form resonances in pairs.
\item
it is useful to study the data from the SM diphoton search for resonances below 125~GeV and for continua that extend from a lower mass limit up to $m_h$.
\end{itemize}
We emphasize that in these cases, a premature invariant-mass requirement in event pre-selection could eliminate a signal. (This same concern applies to these searches for another reason: the possibility of a second Higgs with a different mass, a low cross-section, and unknown branching fractions to SM-like and non-SM-like decays.)

\subsubsection{Summary}

Summarizing the situation for final states with invisible non-SM particles, we suggest searches already in Run I for $\gamma\gamma$ + $\met$, $\ell^+\ell^-$ + $\met$, and $\ell^+\ell^-\ell^+\ell^-$ + $\met$, both with and without requiring pair-wise resonances.  Multi-photon searches may also be warranted, now or in Run II. If possible, various possible simplified models generating these final states should be considered for each search, including ones that have very different kinematics for the observed particles and for the $\met$.  
Experimental studies of the $\gamma$ + $\met$ final state are warranted.  
Study of the final states with pair-wise resonances is lacking and may be useful.

\subsection{Collimated objects in pairs}

Kinematics may force pairs or groups of visible particles to be produced with large $p_T$ compared to their invariant mass, such that they emerge collimated.  In such situations, special search strategies are necessary, since the collimated particles must often be treated as a single special object in order that they be distinguished from a single QCD jet, or be viewed as a pair of objects with special isolation criteria, such that each does not ruin the isolation of the other.  We have briefly discussed a few cases, and summarize them below and in Table \ref{table:collimated}. In contrast to other tables, we do not attempt to interpret the results in terms of models, because for particles of mass $\ll$ 5 GeV, branching fractions to specific final states often vary rapidly as a function of mass.

In this document, collimated leptons are considered in \S\ref{Clepton} (one lepton jet) and \S\ref{sec:collep_nomet} (two lepton jets). We concentrate on simple lepton-jets, consisting of a single lepton-antilepton pair that are collimated, yet isolated from other particles.  Complex lepton-jets, which may contain multiple lepton-antilepton pairs and possibly hadron pairs as well, are not studied here.  Simple lepton-jets may involve both muons and electrons (for a vector $Z_D$), muons almost always (for a scalar or pseudoscalar $a$ with $m_a>2m_\mu$), or electrons only (for $m_a<2 m_\mu$) though we have not considered the latter case.

There have been no searches using more than $35\ipb$ of LHC data for final states with a single lepton jet. However, the study conducted by \cite{HLWY} (see \S\ref{Clepton}) indicates that exotic branching fractions $\sim 10^{-2}$ can be probed if there is additional $\met$ from the Higgs decay.

\begin{sidewaystable}
\begin{center}
\begin{tabular}{|c|c|c|c|}\hline
& Projected/Current& &  \\
Decay & 2$\sigma$ Limit  &Produc- & Comments  \\
Mode & on $\Br({\cal F}_i)$ &tion& \\
${\cal F}_i$  & 7+8 [14] TeV &Mode& \\ \hline  \hline
$\{\mu\mu\}\{\mu\mu\}$ & $1\cdot 10^{-5}$ ($5\cdot 10^{-3}$) [\Unknwn] &$G$& CMS \cite{CMS-PAS-HIG-13-010}, $2 m_\mu < m_a
< 2 m_\tau$ (CMS \cite{Chatrchyan:2011hr} $m_a<5$ GeV) \\ \hline
$\{ee\}\{ee\}$ & limit unclear [\Unknwn]  & $W$, $G$ & reinterpretation of \cite{Aad:2012qua,Aad:2013yqp} needed  \\ \hline
$\{\mu\mu\}X$ &   ~1 [\Unknwn] & $G$ & CMS, \cite{Chatrchyan:2011hr}, $2 m_\mu < m_a < 5 \gev$   \\ \hline
$\{\mu\mu\}$ $\met$ &   $0.03$  [\Unknwn]  & $W$ & Theory study \cite{draper2011dark,HLWY}, \S\ref{NMSSMfermion} and Appendix~\ref{sec:nmssmplots}; our study, \S\ref{Clepton}   \\ \hline
$\{\mu\mu\}\{\mu\mu\}$  $\met$ & $1\cdot 10^{-5}$ ($5\cdot 10^{-3}$) [\Unknwn] &$G$& CMS \cite{CMS-PAS-HIG-13-010}, $2 m_\mu < m_a < 2 m_\tau$ (CMS \cite{Chatrchyan:2011hr} $m_a<5$ GeV) \\
& & & however, see \S\ref{sec:collep_nomet} for important details \\ \hline
$\{ee\}\{ee\}$  $\met$ &  limit unclear [\Unknwn]  & $W$, $G$& reinterpretation of \cite{Aad:2012qua,Aad:2013yqp} needed  \\ \hline
$\{\tau\tau\}\{\mu\mu\}$  &   $(3-7)\cdot 10^{-4}$ [\Unknwn]
& $G$ &  This work, see \S\ref{sec:4tauProposals}  \\ \hline
$\{\gamma\gamma\}\{\gamma\gamma\}$  &  $0.01$ [\Unknwn] & $G$ & ATLAS \cite{ATLAS4gamma}, $m_a<400$ MeV \\ \hline
$\{\gamma\gamma\}$  $\met$  &  \Unknwn [\Unknwn]&  & no studies \\ \hline
$\{gg\}\{gg\}$ &   $>1$ [$0.7$] & $W$ & boosted $Wh$ \cite{Kaplan:2011vf}, $m_a<30$ GeV \\ \hline
$\{b\bar b\}\{b \bar b\}$ &   $0.7 $ [$0.2$]& $W$ & boosted $Wh$ \cite{Kaplan:2011vf}, $m_a\sim 15$ GeV \\ \hline
\end{tabular}
\end{center}
\caption{\small Estimates for sensitivity of certain searches for collimated pairs of objects; collimation is denoted by curly brackets.
See Table \ref{table:haabb} for notation and text for more details. \superscriptlegend}%
\label{table:collimated}
\end{sidewaystable}

Searches for two dilepton jets have been carried out at both the Tevatron and the LHC, as shown in Table \ref{table:collimated}, but there has not been systematic coverage, and existing LHC searches have in some cases been done  with only a small fraction of the existing data set.  There are specifically searches for Higgs decays to two dimuon jets $\{\mu^+\mu^-\}\{\mu^+\mu^-\}$ (here curly brackets denote collimation) without reconstructing the $h$ resonance, so we can use these searches to constrain the cases with and without $\met$ .  There have been searches for lepton jets with $>2$ muons but we do not consider them in our table.  Meanwhile, although there are searches for two electronic lepton-jets, the one search \cite{Aad:2013yqp}
for $h\to$ electron jets looks for two $\{e^+e^-e^+e^-\}$ jets, while the only search for two di-electron jets $\{e^+e^-\}\{e^+e^-\}$ \cite{Aad:2012qua} assumes a large supersymmetric production cross-section.  We have not attempted to reinterpret either search as a limit on $h\to$ two $\{e^+e^-\}$-jets, and so leave these cases blank in our table.  To our knowledge there are no searches for two lepton-jets of different types.

\S\ref{sec:hto4tau} considered collimated $\tau$ pairs in $h\to
\{\tau^+\tau^-\}\{\tau^+\tau^-\}$ decays, as well as
$\{\tau^+\tau^-\}\{\mu^+\mu^-\}$.  We found that a search for the
latter is more powerful, since the collimated muons have higher $p_T$
than any daughters of $\tau$ decays and have a fixed invariant mass.
Our study suggested limits even at Run 1 in the $(3-7)\times 10^{-4}$
range might be possible.  This is much stronger than the previous
measurement from D0 \cite{Abazov:2009yi}, and would put limits on
$\Br( h\to aa)$, assuming $a\to \tau\tau, \mu\mu$ with couplings
weighted by mass, in the range of 5-10\%.
 States such as $b\bar b\tau\tau$ and $b\bar b\mu\mu$ will not have collimated leptons or taus if the $b$ pair and lepton pair come from two particles of the same mass $>2 m_b$; only in more complex models will this arise (though for $m_a\lesssim 25$~GeV, the $2b$'s can merge into a single jet, see below).  For this reason, along with the fact that there are no strong experimental limits on these cases, we have not listed them in our table.

A more complete search program is highly warranted in Run I data looking for simple lepton-jets, both within Higgs searches and beyond.  For reasons that we have outlined, no mass restrictions should be placed on these searches, except those absolutely required by kinematics.  For instance, even if a model has $a\to \mu^+\mu^-$ as motivation, it should not be restricted to $m_a<2m_\tau$ or $m_a<2m_b$, both because such a search has sensitivity to a vector $Z_D$, with substantial leptonic branching fractions at all masses, and because if $a$ couples weakly to $b$ quarks then the $b\bar b$ threshold will have almost no effect on its branching fractions.  Similarly, models with a $Z_D$ vector boson may have electron-positron lepton-jets with arbitrary invariant mass, so such a search should not be limited to extremely low masses.
The range between the obviously collimated region ($m_{\ell\ell}<5$ GeV) and the obviously uncollimated one ($m_{\ell\ell}>20$ GeV) remains almost completely unexplored, and efforts to close this gap would be well-motivated.  Once the simple lepton-jets are fully covered, a program to study more complex lepton-jets will also be a high priority.

Collimated photons can arise if a scalar or pseudoscalar with a substantial coupling to photons has a low mass.  A search for $h\to \{\gamma\gamma\}\{\gamma\gamma\}$ where the photon pairs are very highly collimated, loosely reconstructed as a single photon, has already been done.  A search for $h\to \{\gamma\gamma\}\{\gamma\gamma\}$ with less collimated photon pairs, recognizable as two separate but closely spaced photons that are isolated from other particles, is also well-motivated.  The ability to identify just one such object with low backgrounds is critical for any search for $h\to \{\gamma\gamma\}$ + $\met$; see \S\ref{sec:2gaMET}.  Whether these searches are well-motivated in Run I data needs further study. 

For completeness, we include $4b$ and $4j$ final states in our table. These cases, which are important if $h\to aa$ and then $a\to b\bar b$ or $a\to gg$ are dominant, are effectively collimated if $m_{jj}, m_{bb}<20$ GeV or so, since the jets will typically merge.  Moreover, searches for these modes almost certainly require a boosted $h$, so in the end there will potentially be further merging.

\subsection{For further study}
\label{subsec:furtherstudy}

We note a number of important possible decays that we have not considered in this work, and that merit study.  
First, we did not study two-body decays such as $h\to \tau\mu$ or $h\to Z\gamma$, but these have been studied extensively in the literature.  
More exotic decays that have received varying degrees of attention include
\begin{itemize}
\item $h\to 2\to 6$ e.g. decays of the Higgs to neutralinos that decay via R-parity violation to three jets, etc.
\item $h$ to $>4$ leptons, $\tau$s, $b$s; decays such as $h\to 6\tau$ or $8b$ have been suggested in the literature \cite{Chang:2005ht}, see also \S\ref{subsec:SMvector}, \S\ref{sec:hiddenvalley}, but both theoretical and experimental study has been limited, though CDF has looked for decays of the Higgs to many soft leptons \cite{Aaltonen:2012pu}.  
\item $h$ to complex lepton jets ({\it i.e.} with $>2$ tracks), including both purely electronic, purely muonic, purely leptonic with a mix of muons and electrons, and mixed leptonic/hadronic jets (see for example~\cite{Falkowski:2010gv}).
\item Decays to one or more photonic jets (consisting of $\geq 2$ collimated photons) need more experimental study; theory studies include \cite{Toro:2012sv,Ellis:2012zp,Ellis:2012sd}.
\item $h$ decaying to long-lived particles with decays in flight~\cite{Strassler:2006im,Strassler:2006ri,Carpenter:2007zz}.  There have been a number of searches for specific final states at particular decay lifetimes, but not a coherent program that covers all cases.
\end{itemize}
This is certainly not the complete list; for example one should not forget $h\to 3\to n$, with a 3-body decay $h\to Z_D Z_D^*$ or $h\to a a^*$ (for $m_{Z_D},m_a\geq m_h/2$), though, with the exception of all-leptonic modes, sensitivity to such decay modes needs further study.  Also,
\begin{itemize}
\item Further studies in more difficult channels, such as $b\bar b \tau\tau$, $b \bar b \met$, $\tau\tau\met$, $jj\gamma\gamma$, are needed particularly in the context of VBF production.  If such studies reveal VBF can yield significant improvements in sensitivity, then developing triggers for 2015 aimed at these final states may offer a significant advantage.
\item  Also well-motivated are studies of exotic decays in the $tth$ associated production channel, which can be competitive with $Wh,Zh$ for non-SM Higgs decays.  The
combinatoric backgrounds that make this channel difficult for a SM Higgs may be significantly reduced for certain non-SM decay modes
 \cite{Falkowski:2010hi}, and the hard leptons and $b$ jets from the $t$ decays offer another inclusive trigger pathway.
\end{itemize}

\subsection{Summary of Suggestions}

Based on our results so far, we find that the following searches are highly motivated within the 7 and 8~TeV data set as well as within future data sets.
In some cases, especially in regimes where the objects are collimated, searches have already been done by ATLAS and/or CMS, though not always with the full data set.
\begin{itemize}
\item Search for $h\to Z_D Z_D \to (\ell^+\ell^-)(\ell^+\ell^-)$ across the full range of kinematically allowed $Z_D$ masses, including regimes where the leptons are collimated (forming simple ``lepton-jets'').  This could also be interpreted as a search for $h\to Z_D Z'_D$ if the dilepton pairs have different masses, or as $h\to Z_D Z_D +$ $\met$, for small $\met$, if the condition $m_{4\ell}=m_h$ is relaxed.
\item Search for $h\to Z Z_D \to (\ell^+\ell^-)(\ell^+\ell^-)$ across the full range of kinematically allowed $Z_D$ masses, including regimes where the leptons are collimated (forming a simple ``lepton-jet'').  This search should also be interpreted as a search for $h\to Z a\to (\ell^+\ell^-)(\mu^+\mu^-)$.
\item Search for $h\to\ell^+\ell^- +$ $\met$, including regimes where the leptons are collimated, and including the cases where there is a resonance in $m_{\ell\ell}$.  
Benchmark models include $h\to XY\to Z_D  YY$ or $a Y Y$, $h\to XX\to a a^{(\prime)} YY$ for $m_a < 2m_\tau$, $h\to XX\to Z^* Z^* YY$, where $Y$ is invisible and $Z^*$ is an off-shell $Z$ boson.
\item Search for $h\to\ell^+\ell^- \ell^+\ell^- +$ $\met$, including regimes where the leptons are collimated, and including the cases where there is a resonance in $m_{\ell\ell}$.  Benchmark models include $h\to XX\to Z_D Z_D YY$, $h\to XX\to aa^{(\prime)} YY$ for $m_a<2 m_\tau$, $h\to XX\to Z^* Z^* YY$, where $Y$ is invisible and $Z^*$ is an off-shell $Z$.
\item Search for $h\to aa\to (b\bar b)(\mu^+\mu^-)$ across the full range of kinematically allowed $a$ masses, including regimes where the $b\bar b$ pair tend to merge.  If possible, searches for $h\to a a'$, where $m_a>2m_b>m_{a'}$, could be considered, in which case the leptons may be collimated.
\item Search for $h\to aa\to (\tau^+\tau^-)(\mu^+\mu^-)$ across the full range of kinematically allowed $a$ masses, including regimes where the leptons are collimated.  A search for $h\to aa \to (\tau^+\tau^-)(\tau^+\tau^-)$ may not be as powerful, but deserves to be investigated further.
\item Search for $h\to aa\to (\gamma\gamma)(\gamma\gamma)$, including regimes where the photons are collimated.  This could also be interpreted as a search for $h\to a a'$ if the diphoton pairs have different masses, or as $h\to aa +\met$, for small $\met$, if the condition $m_{4\gamma}=m_h$ is relaxed.
\item Search for $h\to\gamma\gamma +\met$, including the cases where there is a resonance in $m_{\gamma\gamma}$.  Benchmark models include $h\to XY\to a YY$, $h \to XX\to a a^{(\prime)} YY$, $h\to XX\to (\gamma Y)(\gamma Y)$, where $Y$ is invisible.
\end{itemize}

Additional theoretical and experimental studies relevant for 14~TeV and up to 100~$\ifb$ appear warranted for
\begin{itemize}
\item $h\to XY\to \gamma YY$ where $Y$ is invisible, giving $\gamma + \met$. 
\item $h\to aa\to (b\bar b)(b\bar b)$.
\item $h\to aa\to (b\bar b)(\tau^+\tau^-)$, perhaps in VBF production.
\end{itemize}
Note also the other suggestions in \S\ref{subsec:furtherstudy}.

It is important to reemphasize that searches should look for a reconstructucted ``Higgs'' resonance at mass {\it not} equal to 125 GeV.  This is because new Higgs bosons, produced with lower rates and unknown branching fractions, may lie hidden in the data, either at higher or lower masses than the known Higgs.  Also, $h$ decays involving low $\met$ may show up in searches for SM or non-SM decay modes as bumps or broad features below $125$ GeV.

We conclude by noting the implications of our study for triggering in Run II.  \begin{itemize}
\item For several searches, boosted $h$ recoiling against a leptonically-decaying $W$ or $Z$ is expected to be necessary.  Presumably even the higher lepton $p_T$ thresholds required at Run II will not much affect these searches.
\item However, many searches that we have not studied directly (high multiplicity of soft particles, long-lived particles, etc.) will require as many events as possible be retained under triggers on the lepton in $Wh$ (and $t\bar th$) and on the jets in VBF.   Keeping the one-lepton trigger thresholds low, or combining one lepton or VBF dijet triggers  with signatures of unusual Higgs decay final states, is critically important for achieving high sensitivity.
\item Many of our searches involve triggering on two or more leptons, possibly soft and possibly collimated; these issues have been well-explored already in Run I and should remain a priority.
\item For $h\to \ell^+\ell^-\met$, if the leptons are soft and the $\met$ is substantial, then a VBF-based search may be essential, in which triggering off a combination of the VBF jets, the $\met$, and the soft leptons may be needed.
\item The same issues apply to photons; triggering on multiple photons, possibly collimated, and on softer photons in combination with VBF jets and $\met$ may be important.  
\item We have not studied them here, but final states with leptons and at least one photon are possible; this may have trigger implications for any combined lepton and photon trigger pathway.
\item Triggering in the VBF context is also potentially important for other difficult modes, such as $b\bar b\tau\tau$, $b\bar b\met$, etc., but more theory studies are needed.
\item Although the search at CMS for $\gamma + \met$ is expected to benefit from a data parking trigger in the 2012 data, the trigger challenge for this final state in Run II is very severe, and a thorough study is needed to determine if it is both feasible and worth the bandwidth.  The VBF channel may be helpful here. 
\end{itemize}

To conclude, exotic decays of the Higgs represent a unique opportunity to discover new physics.Ê
A large number of experimental searches and additional theoretical and experimental studies are 
highly motivated Êin order to realize the full and exciting physics potential of the LHC.


\vspace{5mm}

\noindent \textbf{Acknowledgements}
\vspace{2mm}

We thank Neil Christensen, Hooman Davoudiasl, Sally Dawson, Albert de Roeck, Adam Falkowski, Yuri Gershtein, Andy Haas, Tao Han,  John Hobbs, Jinrui Huang, Philip Ilten, Greg Landsberg, Hye-Sung Lee, Ian Lewis, Patrick Meade, Maurizio Pierini, George Redlinger, Pedro Schwaller, Robert Shrock, George Sterman, Shufang Su, Scott Thomas, Dmitri Tsybychev, Tomer Volansky, Lian-Tao Wang, and Felix Yu for useful conversations.

D.~C.~and Z.~S.~ are supported in part by the National Science Foundation (NSF) under grant PHY-0969739.
R.~E.~is supported in part by the Department of Energy (DoE) Early Career research program DESC0008061 and by a Sloan Foundation Research Fellowship.
S.~G.~is supported in part by Perimeter Institute for Theoretical
Physics. Research at Perimeter Institute is supported
by the Government of Canada through Industry Canada
and by the Province of Ontario through the Ministry of
Research and Innovation
P.~J.~is supported in part by the DoE under grant DE-FG02-97ER41022.
A.~K.~is supported in part by the NSF under Grant PHY-0855591.
T.~L~is supported in part by his start-up fund at the Hong Kong University of Science and Technology.
Z.~L.~ is supported in part by the DoE under grant DE-FG02-95ER40896, by the NSF under grant PHY-0969510 (LHC Theory Initiative), the Andrew Mellon Predoctoral Fellowship from Dietrich School of Art and Science, University of Pittsburgh, and by the PITT PACC.
D.~M.~is supported in part by NSERC, Canada and the US Dept. of Energy under Grant No. DE-FG02-96ER40956.
J.~S.~is supported in part by NSF grant PHY-1067976 with additional support for this work provided by the LHC Theory Initiative under grant NSF-PHY-0969510.
M.~S.~is supported in part by the DOE under grants DE-FG02-96ER40959 and DE-SC00391
and by the NSF under grant PHY-0904069.
B.~T.~is supported in part by the DoE under grants DE-FG-02-91ER40676 and DE-FG-02-95ER40896, by the NSF under grant PHY-0969510 (LHC Theory Initiative), and by the PITT PACC.
Y.~Z.~is supported in part by the DoE under grant DESC0008061.
We thank variously the SEARCH Workshop; the Aspen Center for Physics under Grant NSF 1066293; the KITP and the National Science Foundation under Grant No. PHY05-25915; 
the Galileo Galilei Institute for Theoretical Physics and the INFN; and Kavli Institute for Theoretical Physics China for hospitality during the completion of this work. 

\appendix

\renewcommand\thesection{\Alph{section}}
\renewcommand\thesubsection{\Alph{section}.\arabic{subsection}}
\renewcommand\thesubsubsection{\Alph{section}.\arabic{subsection}.\arabic{subsubsection}}
\makeatletter
\def\p@subsection{}
\def\p@subsubsection{}
\makeatother

\section{Decay Rate Computation for 2HDM+S Light Scalar and Pseudoscalar}
\label{sec:2HDM+Sapp}

We will now outline how the branching ratios in \S\ref{SMS} (SM+S) and \S\ref{2HDMS} (2HDM+S) are calculated. We mostly 
follow~\cite{Djouadi:2005p17157, Djouadi:2005p17109}, neglecting hadronization effects. This is sufficient for our purposes of demonstrating the range of possible exotic Higgs decay phenomenologies in 2HDM+S.

The relevant part of the Lagrangian is 
\beq
{\mathcal L}\supset -\sum_{f}\frac{m_f}{v} \left[\bar f f\left(H^0_1 g_{H^0_1 f \bar f}+H^0_2 g_{H^0_2 f \bar f}\right)-i\bar f\gamma_5 f A^0 g_{A^0 f \bar f}\right]\,,
\eeq
where $f$ stands for SM charged fermions.
Higgs-vector boson interactions are obtained from the kinematic terms of the vector bosons. The relevant terms are
\beq
{\mathcal L}\supset-\sum_{V} \frac{2 m^2_V}{v}\left[V_\mu V^\mu\left(H^0_1 g_{H^0_1 VV}+H^0_2 g_{H^0_2 VV}\right)\right]
+\sum_{i=1,2}i\frac{m_Z}{v}g_{Z H^0_i A^0}\partial_{\mu} Z^\mu H^0_i A^0\,.
\label{eq:hv}
\eeq
Given the the $A^0, H^0_{1,2}$ content of the singlet-like scalar $s$ and pseudoscalar $a$ in Eqs.~(\ref{eq:2HDMSpseudoscalar}) and 
(\ref{eq:2HDMSscalar}), and the couplings in \tref{tab:2HDMcoupling}, the couplings $g_{s f \bar f}$, $g_{a f \bar f}$, and $g_{s VV}$ can be derived.

The approach for calculating branching ratios is different for light Higgs mass above or below $\sim \gev$. The theoretical uncertainties in the hadronic region of the latter case are very large, and an effective theory computation must be used.

\subsection{Light Singlet Mass Above 1 GeV}

According to the discussion in \S\ref{2HDMS}, the relevant decay channel for the lightest Higgs scalar/pseudoscalar are $a / s \to f \bar{f}$, $a / s \to \gamma\gamma$, and $a / s \to g g$. Ref.~\cite{Djouadi:2005p17109} contains the decay widths for the MSSM Higgs at tree-level and higher orders. We include the relevant formulas here, which are valid for the 2HDM+S and SM+S case after rescaling the Yukawa and gauge couplings by the small singlet mixing angle. 
\begin{enumerate}[(i)]
\item{\bf{Decays to light SM fermion pairs} $a/s\to f \bar{f}$}.

The tree level decay width of $\phi=a, s$ into fermion pairs is given by
\beq
\Gamma(\phi\to f \bar f)= \frac{ N_c G_F }{4\sqrt 2 \pi} {g}^2_{\phi f \bar f}\, m_\phi m_f^2\, \beta_f^p\,,
\eeq
where the phase volume, $\beta$, is 
\beq
\beta_f=\sqrt{1-\frac{4 m_f^2}{m_\phi^2}}
\eeq
with $p=1 (3)$ for $\phi=$ pseudoscalar $a$ (scalar $s$). For quarks, additional QCD radiative corrections are taken into consideration. For light quarks with mass $m_q \ll m_\phi/2$ ($q=u,d,s$ for $m_\phi$ we considered), the $\mathcal{O}(\alpha_s)$ correction is given by
\beq
\Gamma(\phi\to q \bar q)= \frac{3 G_F }{4\sqrt 2 \pi} {g}^2_{\phi q \bar q} m_\phi \bar m_q^2  \left(1+\frac{17}{3} \frac{\bar{\alpha}_s}{\pi} \right).
\label{eq:lightquark}
\eeq
Here $\bar m_q$ stands for the running of the quark mass in the $\overline{\mathrm{MS}}$ scheme with the renormalization scale $\mu=m_\phi$. This redefinition absorbs logarithms of masses of quarks from NLO QCD. $\bar \alpha_s$ stands for the running of strong coupling. Again we choose the renormalization scale $\mu=m_\phi$. Above $\sim$ GeV, $\alpha_s$ is small enough that perturbative QCD can give accurate results.
%

The masses of heavy quarks ($b$ and $c$) can be close to $m_\phi/2$, where Eq.~\ref{eq:lightquark} is no longer applicable. Instead we use the threshold formula for the  QCD correction at $\mathcal{O}(\alpha_s)$~\cite{Drees:1989du, Surguladze:1994gc, Surguladze:1994em}: 
\beq
\Gamma(\phi\to Q \bar Q) = \frac{G_F N_c }{4\sqrt{2} \pi} g_{\phi Q \bar{Q}}^2 m_\phi m_Q^2 \left(1+\frac{\alpha_s}{\pi}\delta_\phi \right) \beta_Q^p.
\label{eq:heavyquark}
\eeq
where $m_Q$ is the quark pole mass. For the pseudo-scalar and the scalar scenarios, $\delta_\phi$ are respectively given by
\begin{align}
\delta_a =\frac{4}{3} \left(\frac{a}{\beta_Q} +\frac{19+ 2 \beta_Q^2 + 3 \beta_Q^4}{16 \beta_Q}\ln \gamma +\frac{21-3 \beta_Q^2}{8}\right),\\
\delta_s =\frac{4}{3} \left(\frac{a}{\beta_Q} +\frac{3+ 34 \beta_Q^2 -13 \beta_Q^4}{16 \beta_Q^3}\ln \gamma +\frac{21\beta_Q^2 -3}{8 \beta_Q^2}\right)
\end{align}
with 
\begin{align}
\gamma ={}& \frac{1+\beta_Q}{1-\beta_Q},\\
a={}& (1+\beta_Q^2)\left[2 \text{Li}_2 (-\gamma^{-1})+4 \text{Li}_2 (\gamma^{-1})-\ln \gamma \ln \frac{8 \beta_Q^2}{(1+\beta_Q)^3}\right]-\beta_Q \ln\left[\frac{64 \beta_Q^4}{(1-\beta_Q^2)^3}\right].
\end{align}
The relations between Eq.~\ref{eq:lightquark} and Eq.~\ref{eq:heavyquark} are shown in~\cite{Surguladze:1994gc, Surguladze:1994em}. 

\item{\bf Loop induced decays to photon pairs $a/s\to \gamma\gamma$}.

The couplings between Higgs scalars and $\gamma \gamma$ are induced by charged particle loops. The decay widths can be written as 
\begin{align}
\Gamma(a\to\gamma\gamma)={}&\frac{G_F \alpha^2 m_{a}^3}{128 \sqrt 2 \pi^3}\Big|\sum_f N_c Q^2_f g_{a f \bar f}\,A^{a}_{1/2}\left(\frac{m_{a}^2}{4 m_f^2}\right)\Big|^2\\
\Gamma(s\to\gamma\gamma)={}&\frac{G_F \alpha^2 m^3_{s}}{128\sqrt 2 \pi^3}\Big|\sum_f N_c Q_f^2 g_{s f \bar f}\, A^{s}_{1/2}\left(\frac{m^2_{s}}{4 m^2_f}\right)+g_{s VV}\,A^{s}_1\left(\frac{m^2_{s}}{4 m^2_W}\right)\Big|^2\,,
\label{eq:loopyy2}
\end{align}
where $Q_f$'s are electric charges in units of $e$. The form factors for spin half and one particles, $A_{1/2}$ and $A_{1}$, are given by
\begin{align}
A^{a}_{1/2}(x)={}&2 x^{-1} f(x)\\
A^{s}_{1/2}(x)={}&2[x+(x-1) f(x)] x^{-2}\\
A^{s}_{1}(x)={}&-[2x^2+3x+3(2x-1)f(x)]x^{-2}
\end{align}
with

\begin{equation}
f(x)=
\left\{
\begin{array}{ll}
\arcsin^2 \sqrt x & \quad x \leq 1\\
-{\frac{1}{4}}\left[\log\frac{1+\sqrt{1-1/x}}{1-\sqrt{1-1/x}}-i\pi\right]^2& \quad x>1\\
\end{array}
\right. \,.
\end{equation}

In the limit $x\to 0$
\begin{align}
A^{a}_{1/2}\to{}&2\\
A^{s}_{1/2}\to{}&4/3\\
A^{s}_{1}\to{}&-7
\end{align}
We neglect the contributions of possible heavy BSM charged particles, which are generically highly suppressed. 

\Eref{eq:loopyy2} shows that the dominant contribution to $s \to \gamma\gamma$ for SM-like fermion couplings comes from $W$- and $t$-loops. The top loop also dominates $a \rightarrow \gamma \gamma$ but there is no $W$ contribution. However, $\alpha'$ and $\beta$-dependent factors in the couplings can also make the $b$ loop important. This occurs in Type II and Type IV models when $\tan \beta\times \tan\alpha'$  or $\tan \alpha$ is large for $s$ or $a$, respectively. The QCD corrections can be found in ~\cite{Djouadi:2005p17109}.


\item{\bf Loop induced decays to gluon pairs $a, s \to g g$.}

Gluons are massless particles that couple to the Higgs dominantly via heavy quark loops, $Q=t, b, c$. The decay widths are given by 
\begin{align}
\Gamma(\phi \to gg)={}&\frac{G_F \bar \alpha^2_s m_{\phi}^3}{36\sqrt 2 \pi^3}\Big|\frac{3}{4}\sum_{Q=t,b,c} g_{\phi Q\bar Q} A^{\phi}_{1/2}\left(\frac{m_{\phi}^2}{4 m_Q^2}\right)\Big|^2\,.
\end{align}
Other potential heavy particle contributions are neglected. The QCD corrections are shown in ~\cite{Djouadi:2005p17109}.

\item{\bf Other Decay Channels of the lightest Higgs.}

Decays to $\gamma+$quarkonium final states are enhanced for pseudoscalar masses near the $2c, 2b$ thresholds. These are challenging to calculate~\cite{Baumgart:2012pj, and the other}, and we neglect them along with hadronization effects, which likely invalidates our quantitative results near the $B$/$D$-meson and quarkonia thresholds. 

\end{enumerate}

\subsection{Light Singlet Mass Below 1 GeV}

For a sub-GeV (pseudo)scalar Higgs, hadronization effects dominate and the perturbative analysis is not valid above the pion threshold. The calculation of decay widths in this region is extremely difficult due to the QCD uncertainties in the hadronic final states. Light (pseudo)scalars that decay to two (three) pions would look similar to hadronic taus in an experimental analysis, and care would have to be taken not to reject them based on track quality requirements. 

We now outline our methods for estimating the branching ratios in this low-mass regime. 

\begin{enumerate}[(i)]

\item{\bf Singlet-like scalar $s$}

For $m_{s}<2m_e\simeq 1.02\, \mev$, $\gamma\gamma$ decay is the only available channel. In the region $2m_e\leq m_{s}< 2m_\mu\simeq 211\, \mev$, $e^+ e^-$  rises and competes with $\gamma\gamma$. $\BR$'s of $\gamma \gamma$ may be enhanced in  Type II, III, and IV by appropriate choice of $\tan \beta$ and $\alpha'$. In the region $2m_\mu\leq m_{s}<2m_{\pi^0}\simeq 270\, \mev$, $\mu^+\mu^-$ decay appears and replaces $e^+e^-$ to compete with $\gamma\gamma$. 

Branching ratios are most difficult to estimate accurately in the mass window from the $\pi\pi$ threshold to about 1 GeV. $\mu^+\mu^-$ competes with $\gamma\gamma$, $\pi \pi $,  $K\overline{K}$, and $\eta\eta$. Several methods are available for the estimation in this region, such as soft pion theory and the chiral Lagrangian method. All suffer from significant final-state uncertainties. According to Ref.~\cite{gunion2000higgs}, the perturbative spectator approximation  gives a reasonable and relatively simple approximation of decay widths. They are given by\footnote{Here ``s"  stands for the strange quark in order to differentiate with the singlet-like scalar, $s$.} 
\begin{align}
\Gamma(s \to\gamma\gamma)={}&\frac{G_F \alpha^2 m^3_{s}}{128\sqrt 2 \pi^3}\Big|\sum_f N_c Q_f^2 g_{s f \bar f}\, A^{s}_{1/2}\left(\frac{m^2_{s}}{4 m^2_f}\right)-7 g_{s VV}\Big|^2\\
\Gamma(s\to\mu \bar \mu, e \bar e)={}&\frac{G_F}{4\sqrt 2 \pi} m_{s} g^2_{s \mu \bar \mu, e \bar e} m_{\mu, e}^2\beta_\mu^3\\
\Gamma(s \to u \bar u, d \bar d)={}&\frac{ 3 G_F}{4\sqrt 2 \pi} m_{s} g_{s u\bar u,  d \bar d}^2m^2_{u,d}\beta_\pi^3\\
\Gamma(s\to \text{s} \bar {\text{s}})={}&\frac{ 3 G_F}{4\sqrt 2 \pi} m_{s} g^2_{s \text{s} \bar {\text{s}}}m_{\text s}^2 \beta_K^3\\
\Gamma(s \to g g)={}& \frac{G_F \alpha_s^2 m^3_{s}}{36\sqrt 2 \pi^3}\left(\sum_q  g_{s q \bar q}-(g_{s u \bar u}+g_{s d \bar d})\beta^3_\pi-g_{s \text{s} \bar {\text{s}}}\beta^3_K\right)^2
\end{align}
and we define the non-charm hadron decay width as
\beq
\Gamma(s \to had.)=\Gamma(s \to u\bar u)+\Gamma(s \to d \bar d)+\Gamma(s \to \text{s} \bar {\text{s}})+\Gamma(s \to g g).
\eeq

Another source of uncertainty in the $\BR$ estimation lies in the definition of the light quark mass. Different definitions render different $\BR$'s, especially to $\gamma\gamma$. For our computation, we use $m_u=m_d=40\,\mev$, $m_{\text s}=450\,\mev$, and $\alpha_s/\pi=0.15$ as~\cite{gunion2000higgs}. The values are chosen such that results from the spectator approximation method match results from the chiral Lagrangian method, but we emphasize that the uncertainties remain very large above the pion threshold.

\item{\bf Singlet-like pseudoscalar $a$}

Below the $3\pi$ threshold ($m_{a}<3m_{\pi^0}\simeq 405\,\mev$), $\BR$'s of $a$ are similar to $\BR$'s of $h$ and dictated mostly by thresholds (and possibly a competitive decay to $\gamma\gamma$). Above the $3\pi$ threshold,  decays of $a$ to $3 \pi,  \rho^0 \gamma, \omega \gamma, \theta \pi\pi$  arise as $m_{a}$ increases and competes with $\mu^+\mu^-$ and $\gamma\gamma$ decays. 
We apply a similar spectator approximation as for the scalar case, with a threshold of twice the Kaon mass, $2 m_K$, for strange quark final states~\cite{Dermisek:2010p17405}, 
\begin{align}
\Gamma(a\to\gamma\gamma)={}&\frac{G_F \alpha^2 m_{a}^3}{128 \sqrt 2 \pi^3}\Big|\sum_f N_c Q^2_f g_{a f \bar f}\,A^{a}_{1/2}\left(\frac{m_{a}^2}{4 m_f^2}\right)\Big|^2\\
\Gamma(a\to\mu \bar \mu, e \bar e)={}&\frac{G_F}{4\sqrt 2 \pi} m_{a} g^2_{a \mu \bar \mu, e \bar e } m_{\mu, e}^2\beta_\mu\\
\Gamma(a \to u \bar u, d \bar d)={}&\frac{ 3 G_F}{4\sqrt 2 \pi} m_{a} g_{a u\bar u,  d \bar d}^2m^2_{u,d}\beta_\pi\\
\Gamma(a\to  \text{s} \bar {\text{s}})={}&\frac{ 3 G_F}{4\sqrt 2 \pi} m_{a} g^2_{a \text{s} \bar {\text{s}}}m_{\text s}^2 \beta_K\\
\Gamma(a \to g g)={}& \frac{G_F \alpha_s^2 m^3_{a}}{16\sqrt 2 \pi^3}\left(\sum_q  g_{a q \bar q}-(g_{a u \bar u}+g_{a d \bar d})\beta_\pi-g_{a \text{s} \bar {\text{s}}}\beta_K\right)^2\\
\Gamma(a\to had.)\equiv{}&\Gamma(a\to u\bar u)+\Gamma(a\to d \bar d)+\Gamma(a \to \text{s} \bar {\text{s}})+\Gamma(a\to g g)\,.
\end{align}

\end{enumerate}
\section{Surveying Higgs phenomenology in the PQ-NMSSM}
\label{sec:nmssmplots}

As the exotic Higgs decay phenomenology of the PQ-limit of the NMSSM
may not be as well-known as the $h\to a a$ decays familiar from the
NMSSM in the R-symmetric limit, we provide in this Appendix some
quantitative illustrations of the phenomenology discussed in
\S\ref{NMSSMfermion} (also see~\cite{draper2011dark,HLWY}).

\begin{figure}[htp!]
\begin{center}
\begin{tabular}{c}
\includegraphics[width=0.32\textwidth]{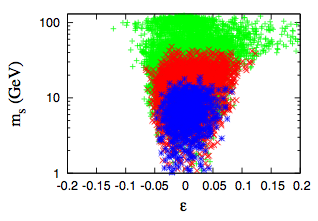} 
\includegraphics[width=0.32\textwidth]{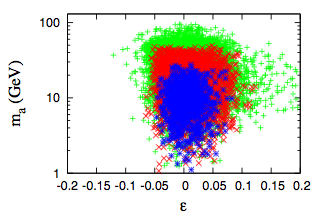} 
\includegraphics[width=0.32\textwidth]{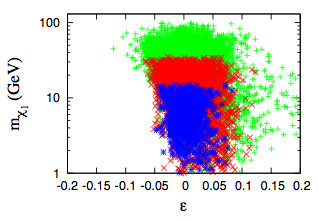} \\
\includegraphics[width=0.32\textwidth]{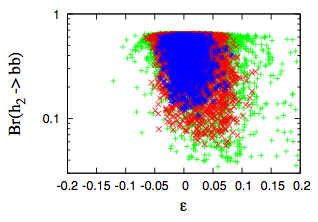} 
\includegraphics[width=0.32\textwidth]{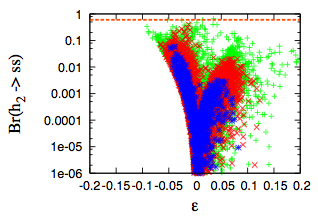} 
\includegraphics[width=0.32\textwidth]{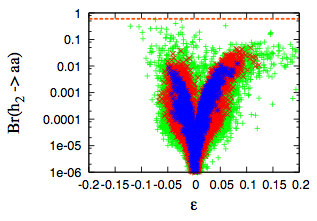} \\
\includegraphics[width=0.32\textwidth]{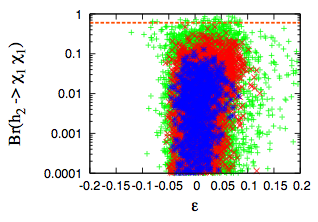}
\includegraphics[width=0.32\textwidth]{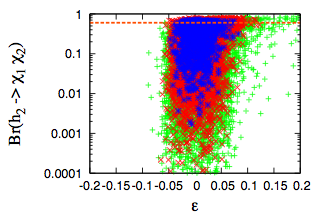}
\includegraphics[width=0.32\textwidth]{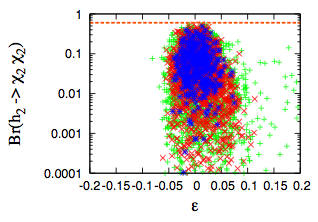}\\
\includegraphics[width=0.32\textwidth]{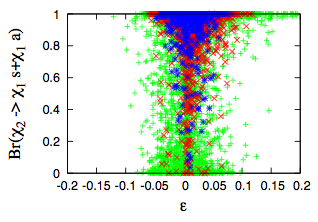}
\includegraphics[width=0.32\textwidth]{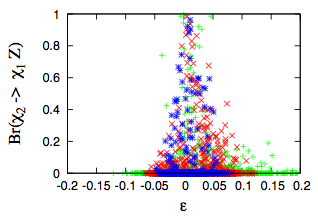}\\
\end{tabular}
\end{center}
\caption{{\small Higgs phenomenology in the PQ-symmetry limit of the NMSSM, as
  discussed in \S\ref{NMSSMfermion} \cite{HLWY}.  {\bf Top row:} Masses of
  $s$, $a$, and $\chi_1$, respectively. {\bf Second and Third rows:}
  Branching ratios of the SM-like Higgs $h$ (denoted here as $h_2$) to
  $b\bar{b}$, $s\bar{s}$, $aa$, $\chi_1\chi_1$, $\chi_1\chi_2$, and
  $\chi_2\chi_2$, respectively.  {\bf Bottom row:} Branching ratios of
  the next-to-lightest neutralino $\chi_2$ to on-shell $\chi_1 s
  +\chi_1 a$ and $\chi_1 Z$, respectively.  All points are required to
  have a mass $124-126$ GeV for their SM-like Higgs boson. Green
  (light gray) points are sampled in the ranges $3 \leq \tan \beta
  \leq 30$, $0.015 \leq \lambda \leq 0.5$, $0.0005 \leq \kappa \leq
  0.05$, $-0.8 \leq \varepsilon' \leq 0.8$, $-50 \text{ GeV} \leq
  A_\kappa \leq 0$, and $0.1 \text{ TeV} \leq \lambda v_s \leq 1
  \text{ TeV}$, Green (light gray) points cover the whole scan range,
  red (medium gray) points correspond to the subset satisfying
  $\lambda < 0.30$, $\kappa / \lambda < 0.05$ and $\lambda v_s < 350$
  GeV, while blue (dark gray) points satisfy $\lambda < 0.15$, $\kappa
  / \lambda < 0.03$ and $\lambda v_s < 250$ GeV.}}
\label{fig:NMSSMPQpheno}
\end{figure}

\begin{table} 
\begin{center}
\begin{tabular}{|c|c|c|c|}\hline
   {\rm Examples}  &  $h \to \chi_1 \chi_2$ &  $h \to \chi_1 \chi_2$ &  $h \to \chi_2 \chi_2$ \\ \hline \hline
 $\lambda$  & 0.18 & 0.064  & 0.02\\ 
  $\kappa$  &  $3.4 \times 10^{-3}$ &  $9.0 \times 10^{-3}$& $1.2 \times 10^{-3}$  \\
   $\tan \beta$  & 9.0 &12.5 & 10 \\
 $\lambda s ({\rm GeV})$   & 326& 138 & 160 \\
 $A_\lambda ({\rm GeV})$  &  2960 &1700& 1800 \\
$A_\kappa ({\rm GeV})$  &   -43.5 & -17 & -7 \\
 $M_1 ({\rm GeV})$  & 85 & 80 &  55 
 \\ \hline \hline
 $m_{s} ({\rm GeV})$ & 23.0 &  34.6 & 17.4 \\ \hline
 $m_{h} ({\rm GeV})$ &124.7 &  125.3 & 124.9 \\ \hline
 $m_{a} ({\rm GeV})$ &28.7 &  31.6 & 14.2 \\ \hline
 $m_{\chi_1} ({\rm GeV})$ &12.7 &  39.1 & 19.7 \\ \hline
  $m_{\chi_2} ({\rm GeV})$ &80.8 &  66.4 & 47.3 \\ \hline 
 BR$(h\to a a)$   & $<0.01$ & $< 0.01$ & $< 0.01$\\ \hline
  BR$(h\to \chi_1 \chi_1)$ &  $< 0.01$& 0.04 &  $< 0.01$\\ \hline
   BR$(h\to \chi_1 \chi_2)$ & 0.28 & 0.27 & 0.05\\ \hline
   BR$(h\to \chi_2 \chi_2)$  & $< 0.01$ &  $< 0.01$ & 0.31\\ \hline
   BR$(\chi_2 \to \chi_1 (a, s)$& 0.92+0.08 & $< 0.01$ & 0.09 + 0.60 \\ \hline  
   BR$(\chi_2 \to \chi_1 (a, s)^* )$&  $< 0.01$ & 0.96 & 0.30 \\ \hline 
   BR$(\chi_2 \to \chi_1 \gamma )$ & $< 0.01$ & $0.04$ & 0.01 \\ \hline  
\end{tabular}
\caption{Example models illustrating the main exotic decay modes of
  the SM-like Higgs boson in the PQ-symmetry limit of the
  NMSSM~\cite{HLWY}. Here soft squark masses of 2 TeV, slepton masses
  of 200 GeV, $A_{u,d,e} = -3.5$ TeV, and wino and gluino soft masses
  250 and 2000 GeV are universally assumed.} 
\label{table:NMSSMbenchmark}
\end{center}
\end{table} 

Fig.~\ref{fig:NMSSMPQpheno} shows the results of parameter scans run
with the package NMSSMTools \cite{NMSSMTools, Ellwanger:2006rn,
  Muhlleitner:2003vg, Das:2011dg}.  All points in this scan are
required to to have a SM-like Higgs in the mass window $m_h\in
(124,126)$ GeV.  We assumed soft squark masses of 2 TeV, slepton
masses of 200 GeV, $A_{u,d,e} = -3.5$ TeV, and bino, wino and gluino
masses of 30-120, 150-500 and 2000 GeV, respectively.  Scans are done
over the parameter $\varepsilon \equiv \lambda \mu_{\rm eff}/m_Z
\times \varepsilon'$, with $\varepsilon'$ given by
Eq.~\ref{eq:varepsdef} and $\mu_{\rm eff}\equiv \lambda v_s$.

The simultaneous smallness of the $s$, $a$, and $\chi_1$ masses and
the generic suppression of Br$(h \to a a, s s)$ are shown in the first
and the second rows of Fig.~\ref{fig:NMSSMPQpheno}. The branching
ratios of $h$ into $\chi_1\chi_1$, $\chi_1\chi_2$, and $\chi_2 \chi_2$
as well as the branching ratios of $\chi_2$ into $\chi_1 s + \chi_1 a$
(on-shell) and $\chi_1Z$ (on-shell) are presented in the third row.
These plots clearly indicate that, although $h \to \chi_1\chi_1$ has a
larger available phase space, that branching fraction tends to be
suppressed compared to $h \to \chi_2 \chi_2$ and especially $h \to
\chi_1 \chi_2$.  Almost all points in the blue region have $m_{\chi_2}
- m_{\chi_1} > {\rm min} \{m_{s}, m_{a}\}$.  Thus $\chi_2$
overwhelmingly decays into on-shell $s$ or $a$ and $\chi_1$, while
both $\chi_2 \to \chi_1 Z$ and three-body decays are suppressed.  In
the red and green regions, the ${\rm min} \{m_{s}, m_{a}\}$ values
increase. Some points (mainly green ones) have $m_{\chi_2} -
m_{\chi_1} < {\rm min} \{m_{s}, m_{a} \}$, so that $\chi_2 \to \chi_1
\gamma$ may become significant. On-shell $\chi_2 \to \chi_1 Z$ 
can occur in a small sliver of the $m_1, m_2$ plane.

We present three example model points in Table~\ref{table:NMSSMbenchmark}, which
represent the main exotic Higgs decay modes in this limit: (1) $h\to
\chi_1\chi_2$, with $\chi_2 \to \chi_1 a, \chi_1 s$; (2) $h\to
\chi_1\chi_2$, with $\chi_2$ mainly decaying to $\chi_1 a^*$ or
$\chi_1 s^*$ with $a^*\to {\rm SM}$ and $s^*\to {\rm SM}$; (3) $h\to
\chi_2\chi_2$, with $\chi_2$ decaying to $\chi_1 a, \chi_1 s$.

\bibliography{bibliography}
\bibliographystyle{JHEP}

\end{document}